\documentclass[12pt]{article}
\usepackage{epsfig}
\def\Journal#1#2#3#4{{#1} {#2} (#4) #3 }

\def\PHYS{{\em Physica}}

\def\NPB{{\em Nucl. Phys.} B}

\def\PLB{{\em Phys. Lett.} B}

\def\PRL{\em Phys. Rev. Lett.}
\def\PREV{\em Phys. Rev.}
\def\PREP{\em Phys. Rep.}

\def\PRD{{\em Phys. Rev.} D}
\def\PRC{{\em Phys. Rev.} C}

\def\ZPC{{\em Z. Phys.} C}

\def\ANNP{\em Ann. Phys. (N.Y.)}
\def\RMP{{\em Rev. Mod. Phys.}}

\def\JHEP{{\em JHEP}}
\def\PPNP{{\em Prog. Part. Nucl. Phys. }}
\def\NPHYS{{\em Nucl. Phys.}}
\def\RMF{{\em Rev. Mex. Fis.}}
\def\MPLA{{\em Mod. Phys. Lett.} A}
\def\NCIM{{\em Nuovo Cimento}}
\def\EPJC{{\em Eur. Phys. J.} C}
\def\CMP{{\em Comm. Math. Phys.}}
\def\r{\vec r}

\newcommand{\be}{\begin{equation}}
\newcommand{\ee}{\end{equation}}
\newcommand{\ba}{\begin{array}}
\newcommand{\ea}{\end{array}}
\newcommand{\bea}{\begin{eqnarray}}
\newcommand{\eea}{\end{eqnarray}}
\newcommand{\no}{\nonumber}
\newcommand{\nn}{\nonumber}

\newcommand{\bi}{\begin{itemize}}
\newcommand{\ei}{\end{itemize}}
\newcommand{\bal}{\begin{aligned}}
\newcommand{\eal}{\end{aligned}}
\newcommand{\wplus}{W^{+}}
\newcommand{\wminus}{W^{-}}
\newcommand{\edote}[2]{ (\epsilon_{#1} \cdot \epsilon_{#2} ) }
\newcommand{\edotqmp}[3]{ (\epsilon_{#1} \cdot (q_{#2} -p_{#3})) }
\newcommand{\pdote}[2]{ (p_{#1} \cdot \epsilon_{#2} ) }

\newcommand{\eeee}[4]{ (\epsilon_{#1} \cdot \epsilon_{#2} ) (\epsilon_{#3} \cdot \epsilon_{#4}) }
\newcommand{\eepepe}[6]{ (\epsilon_{#1} \cdot \epsilon_{#2} ) (p_{#3} \cdot \epsilon_{#4} )(p_{#5} \cdot \epsilon_{#6}) }
\newcommand{\fracp}[2]{ \left( \frac{#1}{#2} \right)}
\newcommand{\lrp}[1]{ \left( #1 \right) }

\usepackage{a4wide}
\usepackage{epsfig,epsf,graphics}
\usepackage{times}
\usepackage{authblk}
\usepackage{color}
\usepackage{amsfonts,amssymb,latexsym,amsmath} %stmaryrd,
\usepackage{bm}
\usepackage{enumerate}
\usepackage{enumitem}
\usepackage{hhline}
\usepackage{longtable}
\usepackage{slashed}
\usepackage{array} % for extrarowheigh
\usepackage[%dvipdfm, % not for pdflatex
colorlinks=true,
linkcolor=black,
breaklinks=true,
urlcolor=green,
citecolor=blue]{hyperref}
\usepackage{cite}
\usepackage{epstopdf}
\def\significance{\sigma^{\rm stat}}

\newcommand{\Tr}{\textrm{Tr}}

\newcommand{\ew}{\end{widetext}}

\newcommand{\bse}{\begin{subequation}}
\newcommand{\ese}{\end{subequation}}
\newcommand{\eei}{\end{eqnarray}\indent\indent}
\newcommand{\ber}{\begin{eqnarray}}
\newcommand{\eer}{\end{eqnarray}}

\def\case#1/#2{\textstyle\frac{#1}{#2} }

\newcommand{\bver}{\begin{verbatim}}
\newcommand{\ever}{\end{verbatim}}

\newcommand{\Imag}{\mathop{\mathrm{Im}}}

\newcommand{\Od}{{\cal O}}

\newcommand{\tr}{\mbox{tr}}

\newcommand\tabref[1]{Table~\ref{#1}}

\newcommand{\mD}{\mathcal{D}}

\newcommand{\mL}{\mathcal{L}}
\newcommand{\mM}{\mathcal{M}}
\newcommand{\mN}{\mathcal{N}}
\newcommand{\mO}{\mathcal{O}}
\newcommand{\mQ}{\mathcal{Q}}

\newcommand{\mU}{\mathcal{U}}

\begin{document}
\title{ \vspace{1cm} Strongly coupled theories beyond the Standard Model}
\author{Antonio Dobado$^1$ and Dom\`enec Espriu$^2$
\\
$^1$Departamento de F\'\i sica Te\'orica and Instituto IPARCOS,\\
Universidad Complutense de Madrid,
 Plaza de las Ciencias 1, 28040 Madrid, Spain\\
$^2$Departament de F\'\i sica Qu\`antica i Astrof\'\i sica\\ 
and Institut de Ci\`encies del Cosmos (ICCUB), \\
Universitat de Barcelona, Mart\'i Franqu\`es 1, 08028 Barcelona, Spain}
\maketitle
\begin{abstract} 
This article presents a number of technical tools and results that may be instrumental to discern the nature 
of the Higgs particle. In scenarios where an additional strongly interacting sector is present in the 
electroweak theory resulting in a composite Higgs and longitudinal components of the massive gauge bosons, 
unitarity, analyticity and related techniques will be crucial to understand the properties of such a sector.
The situation today may be reminiscent of the pre-QCD days: a strongly interacting theory governs the short-distances,
but we have only access to long-distance experiment involving Nambu-Goldstone or pseudo Nambu-Goldstone bosons. Like
in those days we  can only rely on symmetry and general properties of field theory. Luckily, unlike in the 
pre-QCD days, we have now a much clearer idea of what we may be after.

After presenting a classification of the various types of effective theories, we establish the criteria as to under which
conditions different representations are possible and their equivalence. We discuss in detail the implications of 
analyticity, causality and unitarity; describe various unitarization methods and establish the properties of dynamically
generated resonances and form factors. The relation to effective Lagrangians with explicit resonances is explained in the 
context of beyond the Standard Model (BSM) physics and the Higgs effective field theory (HEFT). 
We discuss how various BSM models may be reduced to
the HEFT as well as implications from holography and lattice studies in establishing BSM phenomenology. The methods presented
are then applied to various processes relevant to disentangle the existence and nature of an extended electroweak symmetry 
breaking sector visible in $VV$ fusion: two-Higgs production, vector resonances, $\gamma\gamma$ physics and top-antitop production.

\end{abstract}
%\eject
%\tableofcontents
\section{Introduction}
Assessing the nature of the Higgs-like boson discovered at the LHC~\cite{atlas,cms} in 2012
is probably the most urgent task that particle theorists face in our time. Is the
Higgs particle truly elementary?  Or maybe, is there a new scale of compositeness associated to it?
Is the symmetry breaking sector (SBS) of the Standard Model the ultimate origin of the masses
of the elementary particles we know today or, on the contrary, there is yet another layer of structure.  
If the latter possibility is realized in nature, then there should exist a 
new strongly interacting sector, an extension of the Minimal Standard Model (MSM). Conventionally 
the term extended electroweak symmetry breaking sector (EWSBS) has been coined to describe
this possibility. The study of the EWSBS using some basic principles of Quantum Field Theory (QFT)
and the analysis of  potential experimental consequences are the purpose of this work. 

The lack of direct evidence  for new states at the energies explored so far suggests that
the scale associated to the EWSBS may be substantially larger than the electroweak scale, the 
Fermi scale given by $v=246$ GeV. But naturalness also suggests that the new EWSBS scale should 
not be too large. Otherwise the mass of its lightest scalar resonance would become very difficult to 
sustain~\cite{scaleofcompositeness}. This makes studying the possible experimental relevance of 
an EWSBS compelling at this time. 

First of all let us review some reasons that make us believe that the existence of new physics
beyond the MSM in the form of an EWSBS is quite plausible. 

Nothing would prevent the Higgs from being truly elementary and  the MSM being realized in nature, 
but then some fundamental questions of elementary particle physics would remain unanswered: there would be no natural
dark matter candidate ---not even an axion, no hope of understanding the flavor puzzle and perhaps even the
vacuum of the theory could be unstable and jeopardize our whole picture of the universe (see \cite{degrassietal}
for updated results). A frequently quoted argument is the hierarchy problem(see e.g. \cite{hierarchyprob}), i.e. the fact that the 
quadratically divergent Higgs mass is not protected by any symmetry and it would naturally be of the order of the higher
scale present in the theory when radiative corrections act via quantum loops are taken into account. Actually the problem poses itself in 
its crudest version when the MSM is embedded in a larger theory where higher mass states would be present. 
Compositeness has the capability of solving this puzzle in principle, and this means new interactions and new
strongly interacting degrees of freedom.

It is therefore reasonable to entertain the idea that a new EWSBS may exist in nature and that
this sector could contain new strong interactions, presumably implemented by some non-abelian gauge group,
not necessarily QCD-like. As a result of these strong interactions some of the particles that we see today 
could be composite. The Higgs, for instance, or the longitudinal components of the $W$ and $Z$ weak gauge bosons. 
This possibility has some obvious aesthetical appeal as it would allow us to dispense with elementary 
spin zero fields, for instance.
The Higgs mechanism \cite{Hi64} would then be an effective description, not a fundamental microscopic one, similarly
to what we known to occur in superconductivity or superfluidity\cite{superfluidity}.

Strong interactions surely means the presence of a mass gap. Namely, there should be composite states 
that are not continuously connected
to the MSM degrees of freedom. This is certainly the case in QCD where nucleons and mesons are substantially heavier than their
constituents. The exception are of course Nambu-Goldstone bosons (NGB). In the presence of global symmetries that are spontaneously
broken, scalar massless states appear. In the MSM the longitudinal components of the $W$ and $Z$ bosons are (in a sense
that we will qualify later) just that, NGBs of a $SU(2)_L \times SU(2)_R$ global symmetry, spontaneously
broken to its diagonal $SU(2)_{L+R}$ subgroup. The Higgs could be another NGB as well; then its low mass, when compared
to the compositeness scale, could be understood.

However, the Higgs particle is not massless. It can not be an exact NGB then. In exactly the same way as
pions are pseudo Nambu-Goldstone bosons (pNGB) because quarks are not exactly massless and therefore the global $SU(2)_L \times SU(2)_R$ symmetry is not an exact one, the underlying global symmetry group in the extended EWSBS need not be an exact invariance
of the theory; some soft breaking terms might occur and those would eventually provide a mass for the Higgs. Once
the possibility of soft symmetry breaking terms is admitted, it is clear that the compositeness scale cannot be arbitrarily
high. If this were the case, the mass of the pNGB ---the Higgs in this case--- would be 
phenomenologically inviable.    

It is clear that there are very many possibilities for the microscopic theory 
underlying an extended EWSBS. For instance, the Higgs could indeed be elementary and compositeness affects  only
the longitudinal degrees of the massive weak vector bosons, like in the old technicolor 
scenario \cite{Weinberg:1975gm,Susskind:1978ms}, where one could
dispose of a Higgs particle altogether. This possibility is actually ruled out by electroweak precision parameter
($S, T, U,...$) measurements \cite{TCexcl}. On the other hand, there could be more than one Higgs doublet, even in composite scenarios,
thus leading to many possibilities and a rich phenomenology. The lightness of the Higgs could perhaps be understood 
as being due to the breakdown of conformal symmetry too. In addition, the microscopic theories 
leading to these phenomena at the weak scale are multiple. It is very challenging to try to learn about a possible EWSBS
from the existing data. While certainly no smoking gun exists, and it may be quite some time before it  is possibly found,
it is very important to be very careful about what the signals could be and whether they might be already hidden in 
present data and become visible when a larger luminosity is accumulated.  

In this review we shall try to introduce the reader to a variety of methods that may be useful to understand to what extend
an EWSBS may be viable when the present experimental bounds are taken into account. The techniques introduced here
are to a very large extent borrowed from the pre-QCD days and this is no surprise. After all, we may be in
a situation where all we would see from the EWSBS are the pNGB bosons (the Higgs, the longitudinal $W$'s), 
while the production of vector resonances or other heavier states would not yet be manifest. These techniques based on the use of
effective Lagrangians ---the modern version of current algebra--- combined with the requirements of causality, analyticity 
and unitarity may pave the way to unveil the fundamental microscopic degrees of freedom, to set limits
on such theories, or at least, to find hints that the Higgs particle might be composite.

%%%%%%%%%%%%%%%%%%%%%%%%%%%%%%%%%%%%%%%%%%%%%%%%%%%%%%%%%%%%%%%%%%%%%%%%%%%%%%%%%%%%%%%%%%%%%%%%%%%%%%%%%%%%%%%%%
%%%%%%%%%%%%%%%%%%%%%%%%%%%%%%%%%%%%%%%%%%%%%%%%%%%%%%%%%%%%%%%%%%%%%%%%%%%%%%%%%%%%%%%%%%%%%%%%%%%%%%%%%%%%%%%%%

\section{Effective theories, gauge interactions and global symmetries}
We have just seen that there are many possibilities for a microscopic theory underlying an extended EWSBS. 
Therefore trying to make definite phenomenological predictions out of this myriad of possibilities would look
absolutely hopeless. Luckily, this is not the case. The convenient tool to try to test the predictions of 
possible new physics (NP), without actually having to run through all possible models are effective theories.

Effective theories (see e.g. \cite{Pich:2018ltt, Manohar:2018aog}) consist in proposing a Lagrangian density that contains all the 
degrees of freedom that are present (i.e. that can be produced) at a given energy. This Lagrangian density 
is organized as an expansion in powers of energy and momentum, normalized by some characteristic 
high scale. This scale may represent
the scale of new physics, or some other technically convenient scale. In any case, a perturbative treatment
of this Lagrangian requires that the energies involved are below this scale.

The information about the microscopic degrees of freedom or in fact about any state above the characteristic 
high energy scale is encoded in a number of low energy constants, i.e. coefficients of the local operators
in the effective Lagrangian. At a given energy, typically only a certain number of operators need to be included
and accordingly, only a hopefully manageable number of low-energy constants play a relevant role. All the information
experimentally accessible could then be predicted by this finite number of constants. Reciprocally, all we can
learn from experiments in a given energy range will translate into measurements for the set of relevant low-energy constants. 

It is then a task for the theorist to try to relate the measure values of those coefficient to various families of
underlying theories. For instance, microscopic models based on the exchange of scalars lead to a pattern of low-energy 
constants in QCD that is very different\cite{EGPR} from those where interactions are dominated by the exchange of vector mesons.
The latter turns out to be the case in QCD. 

Of course one would have never guessed QCD from the low energy interactions of pions. But the QCD chiral Lagrangian
historically played a pivotal role in systematizing in a simple Lagrangian a plethora of results known as current 
algebra at the time and were extremely useful to establish the global symmetry group of strong interactions, 
semi-phenomenological models such as vector-meson dominance (VMD)\cite{VMD} or hidden symmetry models\cite{hidden} that 
represented decisive steps in unveiling the properties of strong interactions and also good calculational tools at 
low energies, uncalculable in perturbation theory in QCD. Effective QCD models have also allowed for very 
many determinations of basic QCD parameters such as the number of colors, current quark masses, QCD condensates and so on.
There is thus much to be learned from effective theories if one knows where to look for the relevant information.

In fact, at a given order in the energy expansion not all predictions of effective theories depend on the low energy
constants that collect information from the ultraviolet. There are universal predictions that actually test the group
structure and symmetries of the theory. They are very robust, but on the other hand they provide no information on
the microscopic realization on these symmetries. Examples can be found e.g. in \cite{EspMatPL}.     

Symmetries, both gauge and global and their realization, are fundamental to establish an appropriate effective theory 
together with obvious requirements such as locality, Lorentz invariance and discrete symmetries.
On the other hand, the concrete realization of the effective theory is not that important
in principle. By definition, any
local Lagrangian containing the relevant degrees of freedom and the correct symmetries is equivalent to any other, 
being related by field redefinitions that leave $S$-matrix elements invariant \cite{Kamefuchi:1961sb}.
In practice of course some parametrizations are more useful than others. We will return to this point later.

We shall assume from now on that the only gauge symmetry that is manifest in the effective theory is the Standard
Model (SM) gauge group $SU(3)_c\times SU(2)_L\times U(1)_Y$. An eventual confining gauge group responsible for the
strongly interacting EWSBS will not be explicitly considered as it is expected that only singlet bound states, if any, appear 
at low energies, exactly as the chiral Lagrangian of QCD contains only the (color-singlet) pions. 

Next one has to discuss the global symmetry of the theory as well as its vacuum breaking pattern
that, we just have emphasized, plays a 
fundamental role in setting up its low energy effective theory. We shall denote by $G$ the global
symmetry group and by $H$ its unbroken subgroup. As befits a relativistic theory\cite{LeutLect}, 
the number of NGB will be $n_{NGB}= \dim  G - \dim  H$.

There are some obvious restrictions on the global group. First of all it is obvious that it should include the MSM global 
group $SU(2)_L\times SU(2)_R$. Secondly, we need at least three NGB to reproduce the MSM field content. 
A further requirement that we will impose upon the effective Lagrangians describing an extended EWSBS is the existence
of an unbroken $SU(2)$ subgroup. This is the `custodial symmetry', for which there is ample experimental evidence.
Within the larger global symmetry $SU(2)_L\times SU(2)_R$, custodial symmetry is just $H_C= SU(2)_{L+R}=SU(2)_V$.    
We of course know that custodial symmetry is not exact. It explicitly is broken by the electromagnetic interactions, 
that involve just one generator of $SU(2)_V$ and also by the Yukawa couplings as $M_b \neq M_t$, etc. Still the
fact that the $T$ parameter is quite close to zero\cite{custodial_evidence} allows us to make  
this simplifying hypothesis that makes sense at the present level of precision.

A rather unnatural situation would be to have three NGB, providing the longitudinal polarization of the 
weak gauge bosons, generated through some dynamical mechanism (technicolor like) and yet a light Higgs as an 
elementary singlet. From this point of view, given that the Higgs is light, it is presumed to be a pNGB  
itself, and thus four is the minimal number of broken 
generators required. The Higgs itself would not be a fundamental degree of freedom and the family of models where
this situation takes place are termed Composite Higgs Models (CHM).

%%%%%%%%%%%%%%%%%%%%%%%%%%%%%%%%%%%%%%%%%%%%%%%%%%%%%%%%%%%%%%%%%%%%%%%%%%%%%%%%%%%%%%%%%%%%%%%%%%%%%%%%%%%%%%%%
%%%%%%%%%%%%%%%%%%%%%%%%%%%%%%%%%%%%%%%%%%%%%%%%%%%%%%%%%%%%%%%%%%%%%%%%%%%%%%%%%%%%%%%%%%%%%%%%%%%%%%%%%%%%%%%%

\section{The structure of composite Higgs models}
Composite model extensions of the SM consist in a dynamical system which induces the 
electroweak spontaneous symmetry breaking (SSB) on the rest of the SM fields (gauge bosons, quarks and leptons).  
This SSB is assumed to 
have a group of global symmetries $G$ which, by means of some  unspecified mechanism, is spontaneously  broken to some 
subgroup $H \subset G$. According to the Nambu-Goldstone  (NG) theorem,  the space of degenerate vacua is isomorphic to 
the coset space $M= G/H$. The massless $m=\dim M$ NGB fields can be understood as properly normalized 
coordinates on this coset, and they dominate the low-energy dynamics of the SBS. In addition, the (local) reparametrization 
invariance of QFT guarantees that this low-energy dynamics  can depend only on the geometrical 
properties of the coset (scalar) space $M$ which, as we will see, are completely determined by $G$ and $H$. 

On the other hand, in order to make contact with the SM, we need to gauge the electroweak group $G_{EW}= SU(2)_L \times U(1)_Y$ 
which is required to be a subgroup of $G$, i.e. $G_{EW} \subset G$. Then it is possible to gauge $G_{EW}$ in the 
standard way. However, by doing so, the global $G$ symmetry is explicitly broken, since gauge currents and not gauged 
currents play a different role in the SBS dynamics. If we define $S$ as the maximal subgroup of $G$ whose generators 
commute with all the $G_{EW}$ generators it turns to be the case that in gauging $G_{EW}$ the global symmetry $G$ is explicitly 
broken to  $G_{EW} \times S$. Therefore the $G_{EW}$ gauged group $G$ is spontaneously broken to $H$ and 
explicitly to $G_{EW} \times S$\cite{Pokorski}.  

\begin{figure}[tb]
  \begin{center}
    \includegraphics[width=1\textwidth]{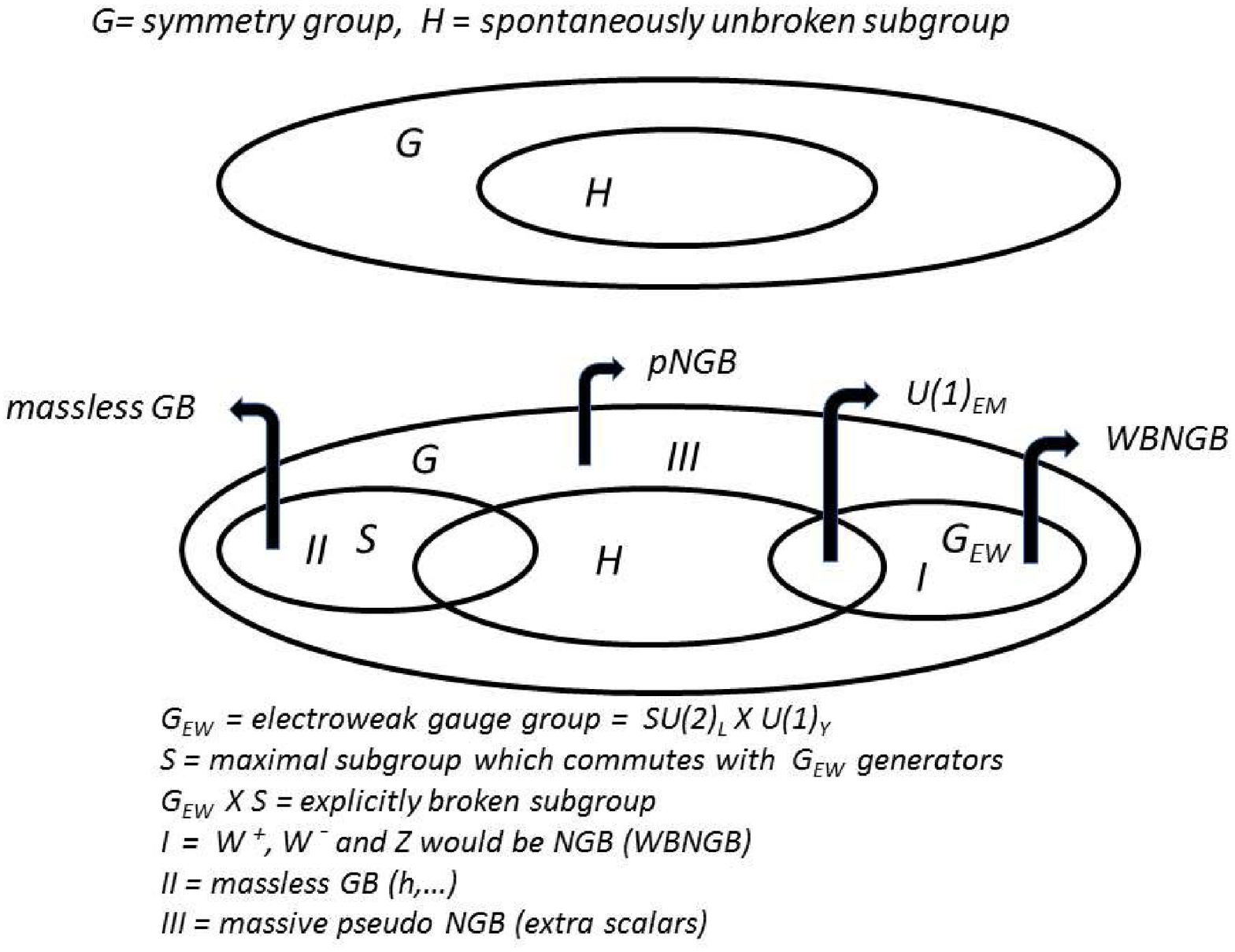}
    \caption{\label{fig:SBSdiagram} Symbolic representation of the SBS patterns in composite Higgs models. 
Upper and lower diagrams correspond to the  ungauged and gauged cases respectively.}
  \end{center}
\end{figure}

The most general situation is symbolically represented  in Fig.~\ref{fig:SBSdiagram}. Initially we have the exact 
global group (with algebra $\cal G$) $G$ which is spontaneously broken to the subgroup $H$ (with subalgebra $\cal H$), giving rise to the 
corresponding NGB parametrizing the coset $G/H$ of degenerated vacua. When the electroweak group $G_{EW}$ 
is gauged, the group $G$ is also explicitly broken to $G_{EW} \times S$. In the general case we may have a non empty  
intersection between the $H$ and $G_{EW}$ generators and also between the $H$ and $S$ generators, but, by construction, not 
between those of $S$ and $G_{EW}$. The original $m=n_{NGB}$ NGB now fall in three categories denoted by $I$, $II$ and $III$ 
in the figure. The NGB in region $I$ are eaten by the electroweak gauge bosons and become the would-be NGB (WBNGB) of the 
Higgs mechanism. The generator in $H \cap G_{EW}$ is associated to the unbroken gauge group, which in this case is just 
the electromagnetic $U(1)_{EM}$. The NGB in region $II$ are not effected by the gauging and continue being genuine NGB. 
In composite models the Higgs field $h$ belong to this region (its mass is eventually generated by a Weinberg-Coleman 
potential induced by one-loop interactions with the gauge and fermionic sector or by other mechanism). 
Finally the NGB of region $III$ are affected by the explicit symmetry breaking and acquire masses of the order of the 
gauge coupling times a characteristic scale.  In any case, the explicit symmetry breaking produced by gauging 
$G_{EW}$ breaks the degeneracy of the vacua and singles out an unique vacuum. This vacuum correspond to the $H$ 
and $G_{EW}$ orientation which minimize the energy of the system (vacuum alignment). 

\begin{figure}[tb]
  \begin{center}
    \includegraphics[width=1\textwidth]{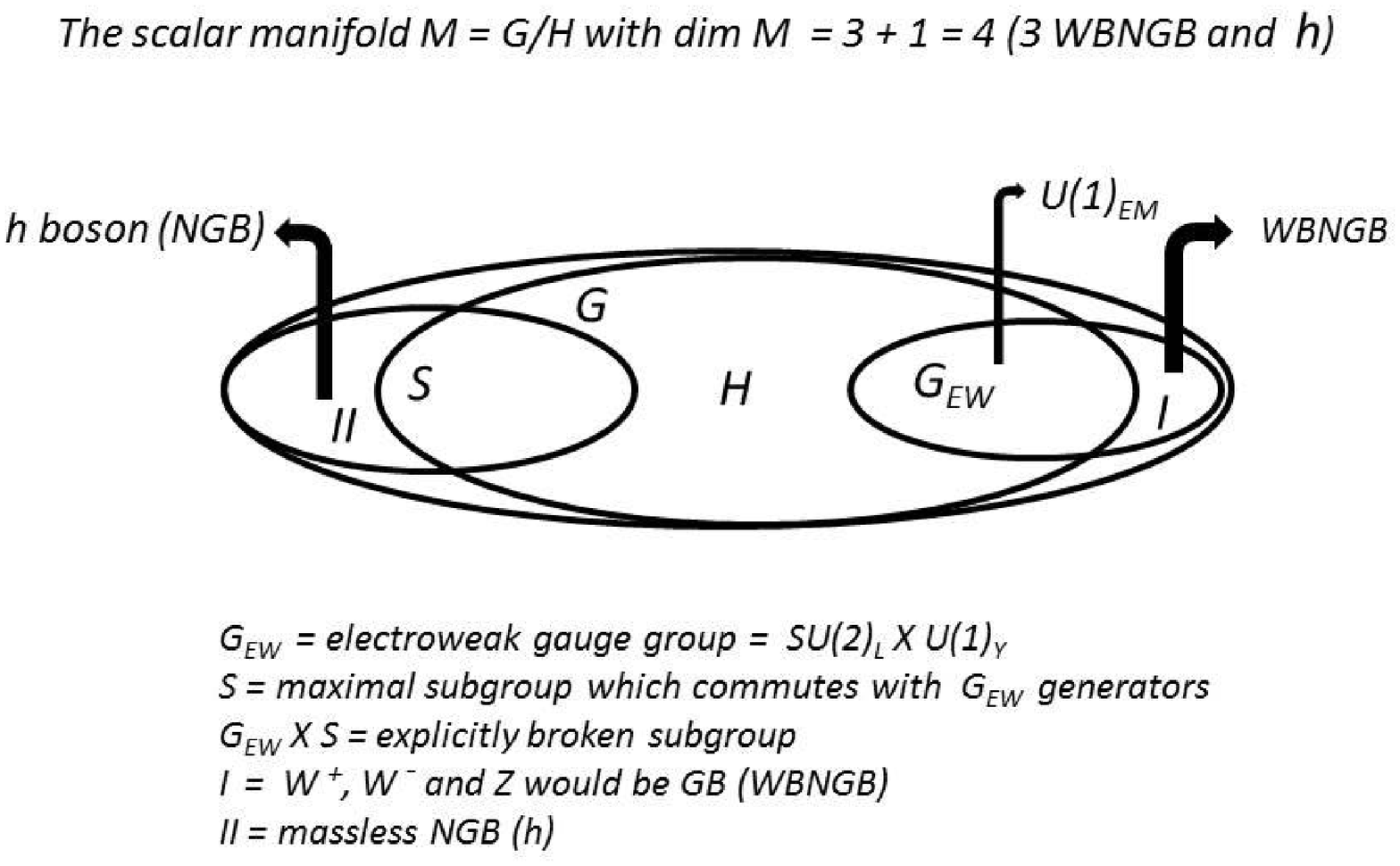}
    \caption{\label{fig:SBSdiagram2} Symbolic representation of the SBS pattern in minimal composite models.}
  \end{center}
\end{figure}

When dealing with composite Higgs models (CHM) the minimal case corresponds to having only one NGB in region $II$, namely the Higgs 
field, and then the total number of original NGB is just 4 (Fig.~\ref{fig:SBSdiagram2}). In  \tabref{tableBMP} 
it is possible to find some examples of 3 (no-Higgs electroweak model) and 4-dimensional composite Higgs  coset 
spaces which could give rise to minimal composite Higgs models (MCHM). As discussed above, one interesting property that 
the unbroken $H$ may possess is having the custodial (isospin) symmetry group $H_C= SU(2)_{L+R}$ as a subgroup, 
i .e. $H_C \subset H$. This guarantees $T \simeq 0$ as emphasized above. 

%%%%%%%%%%%%%%%%%%%%%%%%%%%%%%%%%%%%%%%%%%%%%%%%%%%%%%%%%%%%%%%%%%%%%%%%%%%%%%%%%%%%%%%%%%%%%%%%%%%%%%%%%%%%%%%%%%
%%%%%%%%%%%%%%%%%%%%%%%%%%%%%%%%%%%%%%%%%%%%%%%%%%%%%%%%%%%%%%%%%%%%%%%%%%%%%%%%%%%%%%%%%%%%%%%%%%%%%%%%%%%%%%%%%%

\section{The geometry and dynamics of composite Higgs models}\label{sec:dchm}
Once the groups $G$ and $H$ are given, the low-energy dynamics of the NGB is almost completely determined in terms of the 
$M=G/H$ geometry \cite{CoWeZu69}. Let $H^i$ be the $H$
generators  ($i=1,2,...h$), $X^a$ ($a=1,2,...,m$) the broken generators associated to the NGB 
and $T=(H,X)$ the complete set of $G$ generators denoted by $T^A$ ($A=1,2,..g$). Assuming $G$ and $H$ to be compact and
simple groups the generator commutation relations can be written as:
\begin{eqnarray}
[H^i,H^j]  &=&iC_{ijk}H^k  \\ \nonumber 
[H^i,X^a]&=&iC_{iab}X^b  \\ \nonumber 
[X^a,X^b]&=&iC_{abi}H^i+iC_{abc}X^c
\end{eqnarray}

In most of the physics applications the $M$ space is symmetric. This means that there is a $G$ automorphism 
 $\Sigma$ which is involutive $(\Sigma^2=1)$ with $H$ being  
$\Sigma$-invariant (maximal if $G$ is compact). Then it is possible to arrange  things 
 so that the $C_{abc}$ structure constants vanish, i.e. the commutator of any two $X$
generators is a linear combination of  $H$  but not of  $X$ generators. Obviously $\Sigma$ has two 
eigenvalues ($1$ and $-1$) corresponding to the $H$ and $X$ generators.
In the case of chiral perturbation theory (ChPT) or the massless electroweak Lagrangian the physical interpretation 
of $\Sigma$ is spatial parity. In the general case custodial symmetry implies $H_C= SU(2)_{L+R} \subset  H$ with $R$ 
and $L$ having the standard parity related interpretation.

If the low-energy dynamics of the CHM model is determined entirely by $G$ and $H$ it should be possible to describe 
it in terms of geometrical quantities only, and in particular, in a way independent of the coordinates 
chosen to parametrize the scalar manifold $M=G/H$. This is consistent with the reparametrization invariance of the 
S-matrix elements in QFT. In the following we are going to find a Lagrangian (action)  depending 
only on the NGB fields and having the minimum possible number of derivatives since we are interested in the 
low-energy regime.  In order to build such a Lagrangian first of all  we choose some coordinates 
 $\pi^{\alpha} (\alpha=1,2,...,m= \dim M)$ on $M$, such that $\pi^{\alpha}=0$ 
corresponds to the chosen vacuum state in $M$ (after the $G$ symmetry is explicitly broken). A $M$  point 
with coordinates  $\pi^{\alpha}$ will correspond to some  $G$ element $l(\pi)$ which is the canonical representative  of 
 element (a class) of $M$. Notice that we have many ways to assign a
representative to a class in $M$, and each choice defines a map $l(\pi):G/H \rightarrow G$. 
For example a typical choice can be $l(\pi)=\exp( i \pi^{\alpha}\delta^{\alpha a}X^a)$
but there are infinite many other possibilities, all them  related by some analytical change of
coordinates on the $M$ scalar manifold.

Each element $g\in G$ admits a left decomposition of the form $g= l h$, where $l$ is a representative of a class in $G/H$ and $h\in H$ (a similar 
right factorization is also possible). The left action of an element $g\in G$  on $l(\pi)$ can be written as
\begin{equation}
gl(\pi)=l(\pi')h(\pi,g)
\label{E6}
\end{equation}
with $h(\pi,g)\in H$. Now if  $p\in M$ is the point labeled with coordinates
$\pi$ and  $p'$ (labeled by $\pi'$) is the result of the transformation
$g\simeq 1+i\theta^AT^A \in G$ on $p$. Then there must be some
non-linear transformation
\begin{equation}
\pi'^{\alpha}=\pi^{\alpha}+\xi^{\alpha}_A(\pi)\theta^A
\end{equation} 
where $\xi^{\alpha}$ are the Killing vectors associated with the infinitesimal symmetry action of $G$ on  
the scalar manifold $M$.  In particular this  means  that they fulfill the $\cal G$ algebra
 \begin{equation}
\{\xi_A,\xi_B\}=C_{ABC}\xi_C
\end{equation} 
where
$\{X,Y\}$ is the Lie bracket
\begin{equation} \{  X, Y  \}^\alpha=X^{\gamma}\frac{\partial
Y^{\alpha}}{\partial\pi^{\gamma}}  -Y^{\gamma}  \frac{\partial
X^{\alpha}}{\partial\pi^{\gamma}}.
\end{equation} 
introducing the infinitesimal transformation $h(\pi,g)\simeq 1+i\theta^A\Omega_A=1+i\theta^A\Omega_A^iH^i$
it is not difficult to obtain
\begin{equation}
\xi^{\alpha}_A(\pi)\frac{\partial l(\pi)}{\partial \pi^{\alpha}}=iT^Al(\pi)-il(\pi)\Omega_A.
\end{equation}
Following the Callan, Coleman, Wess and Zumino (CCWZ) method\cite{CoWeZu69}  we introduce the $\cal G$ algebra element $l^{-1}dl$
which can be written as
\begin{equation}
l^{-1}dl
=(\omega_{\alpha}^i(\pi)H^i+e^a_{\alpha}(\pi)X^a)d\pi^{\alpha}=\omega^iH^i+e^aX^a=\omega+e.
\end{equation} 
Notice that  $e^a$ are $m= n_{NGB}$ independent forms that can be understood as  a
vielbein defined on the $M$ manifold. $\omega^i$ is usually known as the canonical $H$ connection and will 
be useful later to couple fermions to the NGB. From Eq.\ref{E6} it is quite easy to obtain the transformation 
equations for $\omega$ and $e$
\begin{eqnarray}
\omega'(\pi)&=&h(\pi,g)\omega (\pi)h^{-1}(\pi,g)+h(\pi,g)dh^{-1}(\pi,g) \\ \nonumber
e'(\pi)&=&h(\pi,g)e(\pi)h^{-1}(\pi,g)
\label{E12}
\end{eqnarray}
i.e. the canonical connection transforms like a  gauge field under the $H$ group.

From the vielbein $e^ a= e^a_\alpha d \pi^\alpha$ we can define a metric on $M$ as usual  
\begin{equation}
 g_{\alpha\beta}= f^2  e_{\alpha}^ae_{\beta}^a 
\end{equation}
which, due to the transformation equations above, is  $G$ invariant. This means that this metric has $G$ as  isometry
group and thus the $\xi_A$ are the corresponding Killing vectors. Thus for this metric the  $G$
transformations are:
\begin{equation}
g'_{\alpha\beta}(\pi)=g_{\alpha\beta}(\pi)
\end{equation} 
Under some technical conditions to be given below and given the value of the dimensional constant $f$ 
(setting the size of the manifold in energy  units), the metric is essentially unique.
Now it is very easy to find the $G$ invariant and covariant  (in the $M$ and Minkowski space sense) 
Lagrangian describing the low-energy NGB dynamics  
\begin{equation}
{\cal L}_0=\frac{1}{2}g_{\alpha\beta}\partial_{\mu}\pi^{\alpha}\partial^{\mu}\pi^{\beta}
\end{equation} 
where the metric has been normalized  so that   $g_{\alpha\beta}=\delta_{\alpha\beta}+\Od(\pi^2)$ and then the first 
term is the properly normalized NGB kinetic term. A system described by this Lagrangian is called non-linear 
$\sigma$ model (NLSM). The corresponding action is also $G$ invariant meaning
\begin{equation}
S_0[\pi]=\int dx \frac{1}{2}g_{\alpha\beta}(\pi)\partial_{\mu}\pi^{\alpha}
\partial^{\mu}\pi^{\beta} =S_0[\pi'] 
\end{equation}
It is important to stress that, in spite of its simple form, this action gives rise to  a very complex non-trivial 
dynamics since  it  contains interactions for any even number of NGB. All these
couplings are proportional to $p^2/f^2$ where $f$ is some dimensional parameter  appearing in 
the metric $g_{\alpha\beta}(\pi)$ which provides the energy scale.

Therefore, the very low energy NGB dynamics  of two theories with the same spontaneous symmetry breaking  
pattern $G \rightarrow H$ is 
essentially the same. At higher energies, when other $p^2/f^2$ powers become relevant, we have to consider new terms 
in the effective Lagrangian \cite{We79}. For example, at order $(p^2/f^2)^2$ one can consider
terms like  \begin{equation}
 {\cal L}={\cal L}_0+
c_1(g_{\alpha\beta}\partial_{\mu}\pi^{\alpha}
\partial^{\mu}\pi^{\beta})^2+
c_2(g_{\alpha\beta}\partial_{\mu}\pi^{\alpha}
\partial^{\nu}\pi^{\beta})^2+...
\end{equation}

When we move out from the low-energy limit, the NGB dynamics is no longer fully determined by the $M$ geometry  since it  
depends as well  on the couplings $c_1, c_2 ,...$ which encode the
 information about the microscopic underlying physics.   In principle, these couplings could be obtained 
by matching  a more fundamental UV completion theory at some scale $\Lambda$, if it is available and computable. 
Alternatively,  they could also be fitted from the experiment.

From the commutation relations for the generators of two compact groups $G$ and $H$, it is easy to realize that 
the broken generators $X^a$ span a linear representation of the subgroup $H$ and as a consequence  the NGB 
transform linearly under the $H$,  but in general non-linearly
under the whole $G$.  However, this linear $H$   could be  reducible. If this is the case, the $X^a$ generators 
decompose in $r$ irreducible sectors and the same happens with the associated vielbein which also transform 
independently under $H$. In this case the most general $G$ invariant $M$ metric is:
\begin{equation}
 g_{\alpha\beta}=\sum_{i=1}^r f_i^2  e_{\alpha}^{a_i}e_{\beta}^{a_i} .
\end{equation}
Then the metric is determined uniquely by $G$, $H$ and the set of dimensional constants $f_1,f_2,...,f_r$.

One can also consider the possibility of having non-compact $G$ or $H$. When $G$ is not compact,  its generator 
metric is not positive defined in general, i. e.  $tr T^AT^B=\epsilon_A\delta_{AB}/2$ with $\epsilon_A=\pm$. In this case, 
provided $H$ is compact, it is possible to
adjust the signs of the $f_i$ parameters so that the NGB kinetic term is well defined (this is the situation in the 
last example in  \tabref{tablaBMP} where $ M=H^4 = SO(1,4)/SO(4)$). However, for non-compact $H$, this is not 
possible in general.

The geometric formalism introduced above makes it possible to couple fermionic matter fields to NGB  in a $G$ invariant 
way as follows. Let $\psi_i$ be some fermionic fields belonging to some linear representation of the subgroup $H$ 
with generators ${\cal H}^i$. Then the Lagrangian:
\begin{equation}
{\cal L}_m=\bar \psi i\gamma^{\mu}(D_{\mu}+\omega_{\mu}) \psi, 
\end{equation} 
with $\omega_{\mu}=\omega_{\alpha}^i\partial_{\mu}\pi^{\alpha}{\cal H}^i$,  is
$G$ invariant.  However, in the case where the fermions fields are chiral,  some subtleties related to 
reparametrization anomalies must be taken  into account  for $n_{NGB} \ge 4$  in the quantum version of the theory 
(see last reference in  \cite{CoWeZu69} for a complete exposition of this issue).

\begin{table}[t!]
\begin{center}
\vspace{.2cm}
\begin{tabular}{ |c|c|c|c|c|c|c| }
\hline
\rule{0pt}{1ex}
{\footnotesize {\bf $G$}} & {\footnotesize {\bf $H$}}  & {\footnotesize  {\bf $M=G/H$}}  & {\footnotesize  Cus. Sym.}  & {\footnotesize NGB}  
\\[5pt] \hline
\rule{0pt}{1ex}
$SU(2)_L \times SU(2)_R$  & $SU(2)_{L+R}  $ & $S^3 $   & Yes  &  3
\\[5pt] \hline
\rule{0pt}{1ex}
$SU(3)$  & $SU(2)_L \times U(1)_Y $ & $SU(3)/SU(2)_L \times U(1)_Y $   & No  &  4
\\[5pt] \hline
\rule{0pt}{1ex}
$SU(2)^2\times U(1)$  & $ SU(2)_{L+R}$ & $SU(2)^2\times U(1)/  SU(2)_{L+R} $ &  Yes  & 4
\\[5pt] \hline
\rule{0pt}{1ex}
$SU(3)\times SU(2)_R$  & $SU(2)^2 \times U(1)$ & $SU(3) \times SU(2)_R/SU(2)^2\times U(1) $   & Yes  &  4
\\[5pt] \hline
\rule{0pt}{1ex}
$SO(5)$ & $SO(4)$ & $S^4$ & Yes &   4
\\[5pt] \hline
\rule{0pt}{1ex}
$SO(1,4)$ & $SO(4)$ & $\mathbb{H}^4$ & Yes &   4
\\[5pt] \hline
\end{tabular}
\caption{\small Possible coset spaces with 3 and 4 NGB's ($SU(2)^2= SU(2)_L \times SU(2)_R$).   }
\label{tableBMP}
\end{center}
\end{table}

Concerning the quantum version of the NLSM some comments are in order. First the generating functional $W[J]$ of 
the connected $\pi$ Green functions is
\begin{equation}
e^{iW[J]}=\int[d\pi\sqrt{g}]e^{i(S[\pi]+<J\pi>)} 
\label{E22}
\end{equation}
where $S[\pi]$ is the NLSM action (including higher order terms), 
$g$ is the  $M$  metric determinant   \cite{Cha70} and  $<J\pi>= \int dx J_\alpha \pi^\alpha$ .  
Obviously, with these definitions both the action and the path integral measure are $M$ covariant and $G$ invariant. On 
the other hand, it should be noted that the term $\langle J\pi\rangle$ is not $M$ covariant. However, it can be 
replaced by $\int dx \Gamma_{\alpha}J^{\alpha}$ where $\Gamma_{\alpha}$ is defined as  $\Gamma_{\alpha}=\partial
S/\partial\pi^{\alpha}$ where $S$ is the geodesic distance from the $M$ origin (vacuum) to the point with 
coordinates $\pi^\alpha$. Then both $\Gamma$ and the external source $J$ transforms like $M$ vectors and $W[J]$ are 
completely defined as an $M$ covariant and $G$ invariant object. In the general case 
$\pi^{\alpha}=\Gamma^{\alpha}+\Od(\pi^2)$ so that one can always  chose $\pi^{\alpha}=\Gamma^{\alpha}$ which amounts to 
a change of $M$ coordinates with no effect on the on-shell S matrix elements.

By exponentiating the $\sqrt{g}$ factor included in the measure one gets an extra piece in the NLSM Lagrangian given by
\begin{equation} 
\Delta{\cal L}=-\frac{i}{2}\delta^D(0)\tr\; \log g 
\end{equation} 
Where the  $\delta^D(0)$ is to be understood  as $\int d^Dk/(2\pi)^D$ with $D$ being the space-time dimension 
$\int dx = \int d^Dx$. However, by using dimensional regularization  ($D=4-\epsilon$) the above $d^Dk$ integral vanishes and  
 we can simply forget about the measure factor in  the path integral \cite{Ta75}. In addition dimensional 
regularization preserves the gauge (BRST) invariance to be introduced in the next section thus simplifying 
enormously the renormalization program.  

From the above generating functional it is possible to derive the Feynman rules for perturbation theory. 
Using dimensional regularization, and working up to some given number of derivatives, the number of counter 
terms needed for the renormalization program is finite and they are $G$ invariant as in ordinary massless chiral perturbation
theory.   Also $G$ invariance appears as the level of Green functions in the form of a set of Ward identities 
satisfied by these functions. 

%%%%%%%%%%%%%%%%%%%%%%%%%%%%%%%%%%%%%%%%%%%%%%%%%%%%%%%%%%%%%%%%%%%%%%%%%%%%%%%%%%%%%%%%%%%%%%%%%%%%%%%%%%%%%%%%
%%%%%%%%%%%%%%%%%%%%%%%%%%%%%%%%%%%%%%%%%%%%%%%%%%%%%%%%%%%%%%%%%%%%%%%%%%%%%%%%%%%%%%%%%%%%%%%%%%%%%%%%%%%%%%%%

\section{Including electroweak interactions}
In order to establish contact with the SM it is necessary to gauge the electroweak group $G_{EW}=SU(2)_L \times U(1)_Y$ 
in the NLSM described by the general Lagrangian
\begin{equation}
{\cal L}=\frac{1}{2}g_{\alpha\beta}(\pi)\partial_{\mu}\pi^{\alpha}\partial^{\mu}\pi^{\beta}
+\hbox{higher  derivative  terms }
\end{equation} 
which is globally $G$ invariant and  covariant in the space-time and in the coset  space $M$ 
sense. Gauging $G_{EW}$  can be achieved in the standard way by introducing the
appropriate gauge boson fields and covariant derivatives. As discussed above this will produce an explicit 
breaking of the $G$ symmetry to the $G_{EW}\times S$ group, with $S$ being the maximal $G$ subgroup commuting 
with the $G_{EW}$ generators. The resulting spectrum will be a consequence of the relative orientation 
between $H$ and $G_{EW}$ after the  alignment produced by this explicit symmetry breaking and other possible 
physical effects (one-loop coupling to gauge bosons, fermions, resonances,...). A  successful CHM  must produce 
massive $W^\pm$ and $Z$ and a massless photon out of this alignment by means of the Higgs mechanism \cite{Hi64}. 

Let us start from a SBS system with spontaneous symmetry breaking $G \rightarrow  H$ (we assume for
simplicity $G$ and $H$ to be compact) so that the dynamics of the corresponding NGB is described
by the globally $G$ invariant Lagrangian above. Now we consider the electroweak gauge 
subgroup $G_{EW}=SU(2)_L \times U(1)_Y \subset G$
with generators $T_L^{\bar a}$ ($\bar a =1,2,3$) and $Y$ respectively with commutation relations
\begin{equation}
[T_L^{\bar a},T_L^{\bar b} ]=i\epsilon_{\bar a\bar b\bar c}T_L^{\bar c}.    
\end{equation}
and 
\begin{equation}
[T_L^{\bar a},Y ]=0.    
\end{equation}
Following the standard procedure, we introduce the $W^{\mu}=-igT_L^{\bar a}W_{\mu}^{\bar a}$ and the $Y_{\mu}=-ig' B_\mu Y$ 
connections and then  $W_\mu^{\bar a}$ and $B_\mu$ are the $SU(2)_L$ and $U(1)_Y$ gauge bosons.
Then $G_{EW}$ gauge transformations of the NGB and the electroweak gauge boson
fields are
 \begin{eqnarray}
\pi'^{\alpha}(x) & = &\pi^{\alpha}(x)+l^{\alpha}_{\bar a}(\pi)\epsilon_L^{\bar a}(x)+y^{\alpha}(\pi)\epsilon_Y(x) \\ 
\nonumber  W_{\mu}^{\prime\bar a}(x)& = &W_{\mu}^{\bar a}(x)+\frac{1}{g}\partial_{\mu} \epsilon_L^{\bar
a}(x)-\epsilon_{\bar a\bar b\bar c}\epsilon _L^{\bar b}(x)W^{\bar c}_{\mu}(x) \\ 
\nonumber 
 B_{\mu}^{\prime}(x)& = &B_{\mu}(x)+\frac{1}{g'}\partial_{\mu} \epsilon_Y(x)
\label{E37}
\end{eqnarray}
where $l^{\alpha}_{\bar a}(\pi)$ and $y^{\alpha}$ are the Killing vectors associated to the  electroweak group $G_{EW}$ 
and thus their Lie brackets (the closure relations) are given by
\begin{equation}
\{l_{\bar a},l_{\bar b}\}=\epsilon_{\bar a\bar b\bar c}l_{\bar c}
\end{equation} 
and
\begin{equation}
\{l_{\bar a},y\}=0.
\end{equation} 

Now the non-linear $\sigma$ model Lagrangian above can be made $G_{EW}$ gauge invariant by replacing the 
NGB field derivatives by covariant derivatives defined as:
\begin{equation}
\partial_{\mu}\pi^{\alpha}\rightarrow D_{\mu}\pi^{\alpha}=\partial_{\mu}\pi^{\alpha}-
gl^{\alpha}_{\bar a}(\pi)W_{\mu}^{\bar a}-g'
y^{\alpha}(\pi)B_{\mu}.
\end{equation}

The $G_{EW}$ gauged non-linear $\sigma$ model can be described by the Lagrangian
\begin{equation}
{\cal L}_{\bar G}={\cal
L}_{YM}+\frac{1}{2}g_{\alpha\beta}(\pi)D_{\mu}\pi^{\alpha}D^{\mu}\pi^{\beta} +\hbox{ higher
covariant derivative terms}
\end{equation} 
where the pure Yang-Mills term ${\cal
L}_{YM}$ for the gauge bosons has been added. 
By using the Killing vector commutators above  it is easy to show that the above Lagrangian is $G_{EW}$
gauge invariant. If the CHM  is well  defined (having an appropriate $G_{EW}$ and $H$ missalignement)  one 
can recover the well known mass matrix eigenstates $W^\pm_\mu$, $Z_\mu$ and $A_\mu$ from the gauged Lagrangian, 
in terms of $g$, $g'$ and $v$.

%%%%%%%%%%%%%%%%%%%%%%%%%%%%%%%%%%%%%%%%%%%%%%%%%%%%%%%%%%%%%%%%%%%%%%%%%%%%%%%%%%%%%%%%%%%%%%%%%%%%%%
%%%%%%%%%%%%%%%%%%%%%%%%%%%%%%%%%%%%%%%%%%%%%%%%%%%%%%%%%%%%%%%%%%%%%%%%%%%%%%%%%%%%%%%%%%%%%%%%%%%%%

\section{The Higgs Effective Field Theory}\label{sec:heftheory}

Let us start from the MSM Lagrangian turning off the gauge fields and the Yukawa couplings
\be\label{eq:heft01}
{\cal L}=\frac{1}{2}\partial_\mu  H^\dagger \partial^\mu   H- \lambda (  H^\dagger  H -\frac{v^2}{2})^2
\ee
where $\lambda$ is the Higgs self-coupling, $v  \simeq 246 $ GeV, is the vacuum expectation value (VEV) and $H$ is the Higgs doublet 
that can be parametrized in terms of four real scalar fields $\phi_i$ ($i=1,2,3,4$) as
\be
 H=\frac{1}{\sqrt{2}}\begin{pmatrix}
 \phi_2 + i \phi_1\\
 \phi_4-i \phi_3\\
\end{pmatrix}.
\ee
Introducing the multiplet $\phi^T=(\phi_1,\phi_2,\phi_3,\phi_4)$ the Lagrangian above can be written as
\be
{\cal L}=\frac{1}{2}\partial_\mu \phi^T \partial^\mu \phi- \frac{\lambda}{4} (\phi^T \phi -v^2)^2.
\ee
This Lagrangian is $SO(4)$ invariant with $\phi$ belonging to the fundamental representation and has an energy  
minimum on the constant field configurations $\phi^T \phi = v^2$ which defines the manifold $M=S^3$ associated 
to the SSB pattern $SO(4)/SO(3)= SU(2)_L \times SU(2)_R/ SU(2)_{L+R}$ where $ SU(2)_L \times SU(2)_R$ is the chiral group 
containing the electroweak gauge group $SU(2)_L \times U(1)_Y$ and $SU(2)_{L+R}$ is the custodial group. 
By choosing the vacuum state as $\phi^T = (0,0,0,v)$ and introducing the Higgs field $\sigma$  as $\phi_4= v+\sigma$ we have
\be\label{eq:heft02}
{\cal L}=\frac{1}{2}\partial_\mu \vec \phi \cdot \partial^\mu \vec \phi  + \frac{1}{2}\partial_\mu \sigma \partial^\mu \sigma
-\frac{\lambda}{4}(\sigma^2+2 v \sigma + \vec \phi \cdot \vec \phi)^2, 
\ee
where $\vec \phi = (\phi_1,\phi_2,\phi_3)$ and we get a Higgs mass $M_\sigma^2 = 2\lambda v^2$.

Another very interesting field parametrization of the SBS of the SM is the following. Let
\be
\phi = (1+ \frac{h}{v})\Pi
\ee
with
\be
\Pi=\begin{pmatrix}
 \vec \pi\\
 \sqrt{v^2- \pi^2}\\
\end{pmatrix}.
\ee
Clearly $\Pi^T \Pi= v^2$ so that $\vec \pi$  parametrizes a $S^3$ sphere of radius $v$ and consequently the three 
$\vec \pi $ fields can be understood as NGB associated to the SSB $SO(4) \rightarrow SO(3)$. Notice that 
they transform linearly under the unbroken $SO(3)$. Also $\phi^T \phi = (v+h)^2$ so that $h$ 
is a  radial excitation of the $S^3$ sphere which is just a reparametrization of the Higgs $H$ field (polar Higgs). 
In terms of these new fields the Lagrangian reads
\be
{\cal L}=\frac{1}{2}(1+ \frac{h}{v})^2g_{ab}\partial_\mu  \pi^a\partial^\mu  \pi^b  + \frac{1}{2}\partial_\mu h \partial^\mu h
-\frac{\lambda}{4}(h^2+2 v h )^2,
\ee
where $a=1,2,3$ and $g_{ab}$ is the $S^3$ metric
\be
g_{ab}(\pi)= \delta_{ab}+\frac{\pi_a\pi_b}{v^2-\pi^2}.
\ee
From this point of view the Higgs Effective Field Theories (HEFT) can be considered as a generalization of the MSM   
\cite{Alonso:2016oah}. The (ungauged and not coupled to fermions) HEFT Lagrangian has the general form
\be
\label{HEFT}
{\cal L}=\frac{1}{2}F(h)g_{ab}(\pi)\partial_\mu  \pi^a\partial^\mu  \pi^b  + \frac{1}{2}\partial_\mu h \partial^\mu h-V(h)
\ee
where $F(h)$ is an analytical function of $h$ with $F(0)=1$
\be
F(h) = 1 + 2 a \frac{h}{v}+ b (\frac{h}{v})^2+...\ ,
\ee
and obviously
\be
F_{SM}(h)= (1+ \frac{h}{v})^2.
\ee
As in the MSM case $g_{ab}$ is typically the $S^3$ metric (we will see one exception later) and $V(h) $ is an 
analytical arbitrary Higgs potential
\be
V(h) = v^4 \sum _{n= 3}^\infty V_n \left (\frac{h}{v}\right)^n.
\ee
The HEFT Lagrangian can be written in a more compact and geometrical suggestive form as follows. Let us define 
$\pi_4=h$ and $\alpha, \beta,...=1,2,3,4)$. Then
\be
{\cal L}=\frac{1}{2}g_{\alpha\beta}(\pi)\partial_\mu  \pi^\alpha\partial^\mu  \pi^\beta  -V(h)
\ee
with the new metric
\be
g_{\alpha\beta}(\pi)=\begin{pmatrix}
 F(h)g_{ab}(\pi)  &  0    \\
 0   &  1  \\
\end{pmatrix}.
\ee
This is the metric of the scalar space $M$ which has a coordinate $\pi_4 = h $ with fiber $S^3$.  The Lagrangian 
above is just the Lagrangian of a NLSM considered in previous sections. The MSM corresponds to the flat 
case $M= \mathbb{R}^4$ but more generally $M$ can be an homogeneous space $M=G/H$ corresponding to some CHM. 
For example if one considers the minimal Higgs composite model (MHCM) based on the groups $G=SO(5)$ and $H=SO(4)$,  then $M=G/H=S^4$.

In many situations it is possible to consider an apparently different generalization of the SM model called 
Standard Model effective field theory (SMEFT). It consists of an operator expansion written in terms of the $SU(2)$ 
Higgs doublet $ H$ (or the $\phi$ $SO(4)$ multiplet) including all the higher dimension operators $O^{(d)}_i$ of 
arbitrary dimension $d$ according to some UV scale $\Lambda$. Keeping only the terms with two field derivatives one has
\begin{eqnarray}
{\cal L}=\frac{1}{2}\partial_\mu  H^\dagger \partial ^\mu  H +\frac{1}{\Lambda^{d-4}}\sum_i  c_i^{(d)}O^{(d)}_i
\end{eqnarray}
which in terms of the $\phi$ fields can be written as:
\be
{\cal L}=\frac{1}{2}\left[ J(\frac{\phi^T \phi}{ \Lambda^2} )\partial_\mu \phi^T \partial ^\mu \phi + \frac{1}{\Lambda^2}
   H(\frac{\phi^T \phi}{ \Lambda^2} )( \phi^T \partial_\mu \phi) (\phi^T \partial ^\mu \phi)      \right].
\ee
By introducing the metric tensor
\be
g_{\alpha\beta}(\phi)= J(\frac{\phi^T \phi}{ \Lambda^2} )\delta_{\alpha\beta}+  
H(\frac{\phi^T \phi}{ \Lambda^2} )\frac {\phi_\alpha\phi_\beta}{\Lambda^2}
\ee
the two-derivative part of the SMEFT can be written as
\be
{\cal L}=\frac{1}{2}g_{\alpha\beta}(\phi)\partial_\mu  \phi^\alpha\partial^\mu  \phi^\beta 
\ee
with $J(0)=1$ and $H(0)=0$, which is  the NLSM or the two-derivative part of  some HEFT. Therefore the SMEFT can be 
considered as a particular case of HEFT (this is true also taking into account higher derivative terms). Then the 
point is: when can a particular HEFT  be written in the form of a SMEFT? Or in other words: when a particular HEFT can   
be written in terms of the  $SU(2)$ doublet $ H$ or the $SO(4)$ multiplet $\phi$?  The answer is the 
following \cite{Alonso:2016oah}: given some four-dimensional HEFT scalar manifold with metric $g_{\alpha\beta}(\pi)$ 
(with $h=\pi^4$), it is possible to find a field reparametrization so that the Lagrangian can be written in terms 
of the doublet $H$ or multiplet $\phi$ whenever there exists a $SO(4)$ invariant point on $M$. Equivalently 
this happens if and only if there is a $h_*$ root of $F(h)$. This is not always the case, but it happens in 
many important situations. For example, in the case of the MSM $F_{SM}(h_*)=0$ has the solution $h_*=-v$ and clearly 
it can be written in terms of $H$ transforming linearly under $SU(2)_{L+R}$.

Finally it is interesting to mention that the HEFT Lagrangian can be extended to include additional scalar fields which are singlets under the $SU(2)_{L+R}$ custodial symmetry by passing from $h$ to $h^I, (I=1,2,3..,L)$ \cite{Alonso:2016oah}. The generalization of Eq. \ref{HEFT} is in this case:
\be
\label{HEFTG}
{\cal L}=\frac{1}{2}F(h)g_{ab}(\pi)\partial_\mu  \pi^a\partial^\mu  \pi^b  + \frac{1}{2}g_{IJ}(h)\partial_\mu h^I \partial^\mu h^J-V(h)
\ee
where $h$ now refers to the set of custodial symmetry singlets $h=(h^1, h^2,...,h^L)$, $g_{IJ}(h)$ is a new metric in the $h$ space with $g_{IJ}(0,0,...,0)=\delta_{IJ}$, $ F(h)$ an analytical function of $h^I/v$ with $F(0,0,...,0)=1$ and $V(h)$ is some appropriate function on $h^I$. In a similar way we can introduce the coordinates $\pi^\alpha$ for $\alpha=I+3$  so that 
$\pi^\alpha=(\pi^1, \pi^2, \pi^3,h^1,h^2,...,h^L)$. Then the extended HEFT Lagrangian can be written  \cite{Alonso:2016oah} as follows:
\be
{\cal L}=\frac{1}{2}g_{\alpha\beta}(\pi)\partial_\mu  \pi^\alpha\partial^\mu  \pi^\beta  -V(h)
\ee
where $\alpha,\beta=1,2,...,3+L$ and the metric:
\be
g_{\alpha\beta}(\pi)=\begin{pmatrix}
 F(h)g_{ab}(\pi)  &  0    \\
 0   &  g_{IJ}(h)  \\
\end{pmatrix}.
\ee

%%%%%%%%%%%%%%%%%%%%%%%%%%%%%%%%%%%%%%%%%%%%%%%%%%%%%%%%%%%%%%%%%%%%%%%%%%%%%%%%%%%%%%%%%%%%%%%%%%%%%%%%%%%%%
%%%%%%%%%%%%%%%%%%%%%%%%%%%%%%%%%%%%%%%%%%%%%%%%%%%%%%%%%%%%%%%%%%%%%%%%%%%%%%%%%%%%%%%%%%%%%%%%%%%%%%%%%%%%%

\section{The MCHM}\label{sec:mchm}
One paradigmatic example of the above ideas is provided by the MCHM \cite{SO(5)} with scalar manifold $M=G/H=SO(5)/SO(4)=S^4$. 
In this model the gauge group is contained in the subgroup $H'= SU(2)_L\times SU(2)_R = SO(4)'$
$(G_{EW} \subset H')$, which is different from the spontaneously unbroken subgroup $H= SO(4)$ (missalignment). 
The $H'=SO(4)'$ group is defined as the invariant group of the $\mathbb{R}^5$ vector

\be
\Phi_0'=f\begin{pmatrix}
  0   \\
    0    \\
      0   \\
        0 \\
        1 \\
\end{pmatrix}.
\ee
with $f$ being the scale of the $G$ to $H$ symmetry breaking. On the other hand $H=SO(4)$ is defined as the 
invariant group of a different vector $\Phi_0$ pointing to another direction
\be
\Phi_0= f\begin{pmatrix}
  0   \\
    0    \\
      0   \\
        s   \\
        c \\
\end{pmatrix}.
\ee
where $s= \sin \theta$ and $c=\cos \theta$ and
\be
\sin \theta= \frac{v}{f} 
\ee
with $v < f$ and  $\theta$ being the missalignment angle. 

The scalar Lagrangian can be written in terms of the 
real multiplet $\Phi \in \mathbb{R}^5$ as:
\be
{\cal L}=\frac{1}{2}\partial_\mu  \Phi^T\partial^\mu  \Phi
\ee
with the constraint $\Phi \in S^4$ ($\Phi^T \Phi = f^2$). Introducing the coordinates 
$\omega^\alpha$ ($\alpha=1,2,3,4$), $\Phi \in S^4$ can be written as
\be\label{eq51}
\Phi= f\begin{pmatrix}
  \omega^1  \\
     \omega^2  \\
        \omega^3  \\
  c   \omega^4 + s \chi   \\
      -s \omega^4 + c \chi \\
\end{pmatrix}.
\ee
where
\be
\chi = \sqrt{f^2- \sum_\alpha (\omega^\alpha) ^2}
\ee
and then the Lagrangian is: 
\be
{\cal L}=\frac{1}{2}g_{\alpha\beta}(\omega)\partial_\mu  \omega^\alpha\partial^\mu  \omega^\beta 
\ee
with
\be
g_{\alpha\beta}(\omega)= \delta_{\alpha\beta}+ \frac{\omega_\alpha\omega_\beta}{f^2-\sum_\alpha (\omega^\alpha) ^2}.
\ee
Finally, by introducing the fields $\pi^a$ ($a=1,2,3$) and $h$ as
\be
\omega^a= \pi^a \frac{f}{v} \sin(\theta + \frac{h}{f})
\ee
and
\be
\omega^4=f \left(  c \chi \sin(\theta + \frac{h}{f})  - s \cos(\theta + \frac{h}{f}) \right)
\ee
we have
\be
{\cal L}=\frac{1}{2}F(h)g_{ab}\partial_\mu  \pi^a\partial^\mu  \pi^b  + \frac{1}{2}\partial_\mu h \partial^\mu h
\ee
where $a=1,2,3$  $g_{ab}$ is the $S^3$ metric
\be
g_{ab}(\pi)= \delta_{ab}+\frac{\pi_a\pi_b}{v^2-\pi^2}
\ee
provided
\be
F(h)= \frac{f^2}{v^2} \sin ^2( \theta + \frac{h}{f}).
\ee
Therefore the low-energy dynamics of the MCHM is a particular instance of HEFT (the particular form of 
the potential $V(h)$ depends on the details of the model). On the other hand, it is possible to solve the 
equation $F(h_*)=0$ to find $h_*= - \theta f$. Therefore, the MCHM is also a SMEFT, i.e. its scalar sector 
can be written completely in terms of the $SU(2)$ doublet $ H$.

%%%%%%%%%%%%%%%%%%%%%%%%%%%%%%%%%%%%%%%%%%%%%%%%%%%%%%%%%%%%%%%%%%%%%%%%%%%%%%%%%%%%%%%%%%%%%%%%%%%%%%%%%%%%%%%
%%%%%%%%%%%%%%%%%%%%%%%%%%%%%%%%%%%%%%%%%%%%%%%%%%%%%%%%%%%%%%%%%%%%%%%%%%%%%%%%%%%%%%%%%%%%%%%%%%%%%%%%%%%%%%%

\section{Geometry and low energy dynamics: some examples}\label{sec:geometry}

From the Lagrangian of the (ungauged) HEFT in Eq. \ref{HEFT} it is possible to obtain very important information 
about the low-energy  NGB  behavior.  As commented above, the NGB dynamics in that regime can only depend on the geometry 
of the scalar space $M=G/H$, in particular it depends on the metric $g_{\alpha\beta}$ which in turn depends only 
on the $g_{ab}(\pi)$ the and the function $F(h)$ appearing in the HEFT Lagrangian. For example, it is possible 
to compute the scalar curvature $R$ of the four dimensional manifold $M$ which is given by \cite{Alonso:2016oah}: 
\be
R = \frac{6}{F(h)}\left[  \frac{1}{v^2} - \frac{1}{2}\frac{d^2 F(h)}{dh^2}  \right].
\ee
On the other hand it is also possible to compute the tree level (low-energy) amplitude for elastic NGB elastic scattering
\be
T(\pi_a\pi_b \rightarrow \pi_c\pi_d)=\frac{s}{v^2}(1-a^2) \delta_{ab}\delta_{cd}+\frac{t}{v^2}(1-a^2) 
\delta_{ac}\delta_{bd}+\frac{u}{v^2}(1-a^2) \delta_{ad}\delta_{bc}
\ee
or the inelastic process
\be
T(\pi_a\pi_b \rightarrow hh)=\frac{s}{v^2}(a^2-b)\delta_{ab}.
\ee
These amplitudes can be considered as the ChPT Weinberg Low-energy theorems translated to the SBS of the SM in the HEFT 
framework. Obviously the $M$ geometry and the low-energy dynamics are strongly correlated \cite{Alonso:2016oah}.
In order to clarify further this issue in the following we will  consider some particular examples of HEFT.
\begin{itemize}

\item
The Higgsless Model.

It corresponds to the degenerate case where $G=SU(2)_L \times SU(2)$ and $H= SU(2)_{L+R}$ and therefore the scalar 
manifold $M= S^3$ is three dimensional (no $h$ field). This is the case of  the old electroweak chiral 
Lagrangians \cite{Appelquist} which had a structure very close to two-flavors ChPT. Now $F=0$ which means $a=b=0$ and then
\be
T(\pi_a\pi_b \rightarrow \pi_c\pi_d)=\frac{s}{v^2} \delta_{ab}\delta_{cd}+\frac{t}{v^2} 
\delta_{ac}\delta_{bd}+\frac{u}{v^2} \delta_{ad}\delta_{bc}
\ee
and obviously
\be
T(\pi_a\pi_b \rightarrow hh)=0.
\ee

\item
The MSM.

Next example is the MSM itself. In this case the scalar manifold is flat ($M= \mathbb{R}^4$) and the scalar 
curvature vanishes ($R=0$). Also we have $a=b=1$, the SBS dynamics is weakly interacting and the amplitudes of the 
elastic and inelastic processes vanish at lowest order in the chiral expansion
\be
T(\pi_a\pi_b \rightarrow \pi_c\pi_d)=T(\pi_a\pi_b \rightarrow hh)=0.
\ee

\item
Dilaton models.

This is a very general class of models where the Higgs field $h$ is assumed to be the NGB (dilaton) associated 
to the spontaneous breaking of the scale symmetry \cite{Grinstein}. In many of these models one has
\be
F(h)= (1+ \frac{h}{f})^2
\ee 
which means 
\be
a^2= b = \frac{v^2}{f^2}<<1
\ee
and a non-constant scalar curvature
\be
R(h)= \frac{6}{v^2(1+\frac{h}{f})^2}(1- \frac{v^2}{f^2}).
\ee
In particular
\be
R(0)= \frac{6}{v^2}(1- \frac{v^2}{f^2}) >0
\ee
for $v<f$. The tree-level amplitudes are
\be
T(\pi_a\pi_b \rightarrow \pi_c\pi_d)=\frac{s}{v^2}(1-\frac{v^2}{f^2}) \delta_{ab}\delta_{cd}+\frac{t}{v^2}(1-\frac{v^2}{f^2}) \delta_{ac}\delta_{bd}+\frac{u}{v^2}(1-\frac{v^2}{f^2}) \delta_{ad}\delta_{bc}
\ee
and
\be
T(\pi_a\pi_b \rightarrow hh)=0.
\ee

\item
The MCHM.

In this class of models $M= SO(5)/SO(4)=S^4$. This is a maximally symmetric space having $g= \dim SO(5) =10$ 
isometries which is the maximum number for a four-dimensional space. The constant scalar curvature is:
\be
R= \frac{12}{f^2} >0
\ee
and from the $F(h)$ one has
\begin{eqnarray}
a^2 & = & 1- \frac{v^2}{f^2}     \nonumber \\
b & = & 1- 2 \frac{v^2}{f^2}  .
\end{eqnarray}
and therefore the tree-level amplitudes
 \be
T(\pi_a\pi_b \rightarrow \pi_c\pi_d)=\frac{s}{f^2} \delta_{ab}\delta_{cd}+\frac{t}{f^2} \delta_{ac}\delta_{bd}+\frac{u}{f^2} \delta_{ad}\delta_{bd}
\ee
and
\be
T(\pi_a\pi_b \rightarrow hh)=\frac{s}{f^2} \delta_{ab}.
\ee
Notice that both amplitudes have the same form due to the scalar space $M=S^4$ symmetries.

\item
The $M= \mathbb{H}^4= SO(1,4)/SO(4)$ model:

This is a model based on the MCHM but with some relevant differences. Trading the $G$ group $SO(5)$ by $SO(1,4)$ 
we have two important effects. First $SO(1,4)$ is not compact and second the $M= \mathbb{H}^4= SO(1,4)/SO(4)$ scalar 
space is also unbounded and has negative curvature \cite{Alonso:2016btr}. The Lagrangian has the same form than 
the general HEFT, but now the metric is
\be
g_{ab}(\pi)= \delta_{ab}-\frac{\pi_a\pi_b}{v^2+\pi^2}
\ee
where $a,b=1,2,3$. Notice that the NGB kinetic term is well defined. On he other hand, in this model we have
\be
F(h)= \frac{f^2}{v^2} \sinh ^2( \sinh^{-1} \frac{v}{f} + \frac{h}{f})
\ee
and then
\begin{eqnarray}
a^2 & = & 1+\frac{v^2}{f^2}     \nonumber \\
b & = & 1+2 \frac{v^2}{f^2}  .
\end{eqnarray}
This space is also maximally symmetric with constant negative scalar curvature
\be
R= -\frac{12}{f^2}<0.
\ee
The low-energy amplitudes are
 \be
T(\pi_a\pi_b \rightarrow \pi_c\pi_d)=-\frac{s}{f^2} \delta_{ab}\delta_{cd}-\frac{t}{f^2} \delta_{ac}\delta_{bd}-\frac{u}{f^2} \delta_{ad}\delta_{bc}
\ee
and
\be
T(\pi_a\pi_b \rightarrow hh)=-\frac{s}{f^2} \delta_{ab}.
\ee
Therefore, negative curvature produces a completely  different behavior of the low-energy amplitudes 
so that dynamics is closely tied to scalar-space geometry. Formally, moving form $M=S^4$ to   $M= \mathbb{H}^4$ amounts 
to the substitution $f^2$ by $-f^2$
\end{itemize}

%%%%%%%%%%%%%%%%%%%%%%%%%%%%%%%%%%%%%%%%%%%%%%%%%%%%%%%%%%%%%%%%%%%%%%%%%%%%%%%%%%%%%%%%%%%%%%%%%%%%%%%%%%%%%%%%%%%%%%%
%%%%%%%%%%%%%%%%%%%%%%%%%%%%%%%%%%%%%%%%%%%%%%%%%%%%%%%%%%%%%%%%%%%%%%%%%%%%%%%%%%%%%%%%%%%%%%%%%%%%%%%%%%%%%%%%%%%%%%%

\section{Gauging the Electroweak group in HEFT}
 Now it is possible to introduce the electroweak $G_{EW}=SU(2)_L \times U(1)_Y$  gauge bosons as usual. 
Using the real multiplet $\phi$ in the MSM we have
\ber
\mathcal{L}_g=\mathcal{L}_0 +\mathcal{L}_{YM}
\eer
with
\ber \label{eq:gaugedlsmlagrangian}
\mathcal{L}_0= \frac{1}{2} (D_{\mu} \phi)^T  D^{\mu} \phi -\frac{\lambda}{4}(\phi^T \phi -v^2)^2.
\eer
The covariant derivative is defined by
\ber
D_{\mu} \phi = (\partial_{\mu}   -\frac{1}{2} g M^k_L W^k_{\mu}  - \frac{1}{2} g' M_Y B_{\mu})\phi=(\partial_\mu + V_\mu)\phi
\eer
where $W_{\mu}^k$ and $B_\mu$ are the $SU(2)_L$ and $U(1)_Y$ gauge fields respectively and $g$ and $g'$ are 
the corresponding gauge couplings. The generators can be represented by the  matrices
\begin{equation}\label{ML1}
  M_L^1 = 
   \begin{pmatrix}
     0   &  0     & 0  &-     \\
     0   & 0     & -  &0        \\
      0        & +       & 0  & 0  \\
        +   &     0   &  0 & 0 
  
   \end{pmatrix},\quad
  M_L^2 = 
   \begin{pmatrix}
     0   &  0     & +  &0     \\
     0   & 0     & 0  &-        \\
      -        & 0      & 0  &0   \\
        0     &   +     &0   &0  
   \end{pmatrix}, \quad
  M_L^3 = 
   \begin{pmatrix}
     0   &  -     & 0  &0    \\
     +   & 0     & 0  &0        \\
      0      & 0      & 0 &-  \\
      0     &  0    & +  & 0  
   \end{pmatrix}
\end{equation}
and
\begin{equation}\label{MY}
  M_Y = %
   \begin{pmatrix}
     0   &  -     & 0  &0    \\
     +   & 0     & 0  &0      \\
      0      & 0      & 0  &+  \\
        0    & 0   & - &0 
   \end{pmatrix}.
\end{equation}

Then it is easy to check $[T^i_L, T^j_L]= i \epsilon_{ijk}T_L^k$  and  $[T_L^k,T_Y]=0$. The Yang-Mills Lagrangian is 
defined as usual as
\ber
\mathcal{L}_{YM} = -\frac{1}{4}W^i_{\mu\nu}W^{i\mu\nu} -\frac{1}{4}B_{\mu\nu}B^{\mu\nu}
\eer
with
\ber
W^i_{\mu\nu}=\partial_\mu W_\nu^i-\partial_\nu W_\mu^i+g \epsilon_{ijk}W_\mu^j W_\nu^k
\eer
and
\ber
B_{\mu\nu}=\partial_\mu B_\nu-\partial_\nu B_\mu.
\eer
Using again the spherical-polar coordinates $\phi'^\alpha =(\pi^a,h)$ the covariant derivative becomes
\be
D_\mu \phi =D_\mu \left(    (1+ \frac{h}{v}) \Pi  \right)   =   \frac{\Pi}{v}\partial_\mu h+ (1+ \frac{h}{v})\partial_\mu \Pi +   
(1+ \frac{h}{v})V_\mu \Pi
\ee
so that it is obvious that, unlike the scalar doublet $\tilde H$, the polar Higgs field $h$ is a $SU(2)_L \times U(1)_Y$ singlet. 
This covariant derivative transforms like a vector under space-time and $M$ space transformations. In particular we have
\be
D_\mu \phi^\alpha = \frac{\partial\phi^\alpha}{\partial \phi'^\beta }D_\mu \phi'^\beta.
\ee 
with $D_\mu \phi'^\alpha =  (D_\mu \pi^a,\partial_\mu h)$. In terms of these coordinates, the general Lagrangian 
for the gauged HEFT is
\be
{\cal L}=\frac{1}{2}F(h)g_{ab}(\pi)D_\mu  \pi^a  D^\mu  \pi^b  + \frac{1}{2}\partial_\mu h \partial^\mu h-V(h)
\ee 
or
\be
{\cal L}=\frac{1}{2}F(h) \left[       D_\mu  \vec \pi \cdot  D^\mu  \vec   \pi + 
\frac{(\vec \pi \cdot D_\mu \vec \pi)(\vec \pi \cdot D^\mu \vec \pi)}{v^2 - \pi^2}  \right]  + 
\frac{1}{2}   \partial_\mu h \partial^\mu h-V(h)
\ee 
This Lagrangian is both space-time and $M$ scalar, and it is $G_{EW}$ gauge invariant too. Now using the diffeomorphism $S^3 \simeq SU(2)$ 
it can be written as
\be
\label{GaugedHEFT}
\mathcal{L}  =   \frac{v^{2}}{4} F(h) {\rm Tr}\, D_{\mu}U^{\dagger}D^{\mu}U+ \frac{1}{2} \partial_{\mu} h \partial^{\mu} h  -V(h)
\ee
where
\be
U(x) = \sqrt{1-\frac{\pi^2}{v^2}}+ i \frac{\pi^a\tau^a}{v}.
\ee
and
\be
 D_{\mu} U =  \partial_{\mu}U +
\frac{1}{2} i g W_{\mu}^{i} \tau^{i} U - \frac{1}{2} i g' B_{\mu}^{i} U \tau^{3},
\ee
where $\tau^a$ are the familiar Pauli matrices.

Parametrizing the $U(x)$ $SU(2)$-fields in a different way, as for example
\be
\label{eq:2}
U = \exp \left( i~\frac{w^a \tau^a}{v} \right)
\ee
boils down to a change of coordinates on the scalar manifold $S^3$. This form of the Lagrangian is particularly 
useful in the unitary gauge where the NGB fields $\pi^a$ (or $w^a$) are absorbed by a gauge transformation 
so that $U(x)= 1$. In this gauge it is very simple to read  the SM tree-level 
mass matrix from the gauged HEFT Lagrangian. The well known eigenstates are $W^{\pm}_\mu=(W^1_\mu \mp iW^2_\mu)/\sqrt{2}$ with mass 
$M_W= gv / 2$,  $Z_\mu = \cos \theta_W W_\mu^3- \sin \theta_a B_\mu $ with mass $M_Z= M_W/ \cos \theta_W$, 
and the massless photon $A_\mu= \sin \theta_W W^3_\mu+ \cos \theta_W B_\mu$ where $\tan \theta_W= g'/g$. 

Including the gauge pieces and expanding the function $F(h)$ as well as the leading terms in the Higgs
potential one gets
\bea
\label{eq:1}
\mathcal{L} & = &  - \frac{1}{4} W_{\mu\nu}^i W^{i\ \mu\nu} - \frac{1}{4} B_{\mu\nu} B^{\mu\nu}
+ \frac{1}{2} \partial_{\mu} h \partial^{\mu} h  -
\frac{M_H^2}{2} h^{2} - d_{3} (\lambda v)  h^{3} -  d_{4} \dfrac{\lambda}{4} h^{4} \\ \nn
& & + \frac{v^{2}}{4} \lrp{1+2 a\fracp{h}{v}+ b \fracp{h}{v}^{2}+...} {\rm Tr}\, D_{\mu}U^{\dagger}D^{\mu}U
+ \sum_{i=0}^{13} a_{i}(h) \mathcal{O}_i\,.
\eea 
Recall that in this parametrization the Higgs field $h$ is a gauge and $SU(2)_L \times SU(2)_R$ singlet. 
We have also included higher
dimensional operators constructed using the same symmetry principles.

These operators $\mathcal{O}_{i}$ include the complete set of   ${\cal O}(p^4)$ operators defined
in~\cite{ECHL,Espriu:2012ih}. For later use we note that only two $O_4$
and $O_5$ will contribute to $W_LW_L$ scattering in the custodial limit:
\be\label{4scatteringops}
\mathcal{O}_{4} = {\rm Tr}\left[ V_{\mu}V_{\nu} \right]{\rm Tr}\left[ V^{\mu}V^{\nu} \right]  \qquad
\mathcal{O}_{5} = {\rm Tr}\left[ V_{\mu}V^{\mu} \right]{\rm Tr}\left[ V_{\nu}V^{\nu} \right],
\ee
where $V_{\mu} = \left( D_{\mu} U \right) U^{\dagger}$. When writing Eq.~(\ref{eq:1}) we have assumed
the well-established chiral counting rules to limit the number of operators to the ${\cal O}(p^4)$ ones.
Note that because the Higgs field in this representations is an $SU(2)$ singlet, the coefficients
$a_i$ can actually be arbitrary functions of $h$ as indicated in Eq.\ref{eq:1}.

%%%%%%%%%%%%%%%%%%%%%%%%%%%%%%%%%%%%%%%%%%%%%%%%%%%%%%%%%%%%%%%%%%%%%%%%%%%%%%%%%%%%%%%%%%%%%%
%%%%%%%%%%%%%%%%%%%%%%%%%%%%%%%%%%%%%%%%%%%%%%%%%%%%%%%%%%%%%%%%%%%%%%%%%%%%%%%%%%%%%%%%%%%%%%

\section{Equivalence of representations}

Effective Lagrangians for describing the Higgs and gauge bosons were first introduced in 
its present form in the LEP days \cite{Appelquist}, where they were mostly used to parametrize 
the so-called oblique corrections.

The effective Lagrangian approach has the advantage of being  model-dependent in a very controlled 
way, as the model dependence enters only via a finite number of low energy constants. A drawback is
that number of constants beyond the leading order operators is usually large and the choice
of a convenient basis is often subject of intense debate \cite{debate}.

In practice in the literature one encounters basically two types of effective theories: those based in a linear
realization (SMEFT) and those where the Golstone bosons transform non-linearly, as is generically
the case in HEFT. We have already seen when the SMEFT is possible, as it is the case in the MSM. In practice then
in the MSM case or simple modifications of it, linear and non-linear parametrizations 
of the effective theory coexist.
As befits a field redefinition, for on-shell S-matrix elements all parametrizations give
formally identical results \cite{Kamefuchi:1961sb}. What is the point then of the debate?

This coexistence of linear and non-linear sigma models predates QCD and goes back actually to the 60s where the 
linear and non-linear sigma models were introduced \cite{gellmannlevy,weinbergNL}
to describe low-energy strong interactions between pions and nucleons.
Ignoring here the fermionic fields altogether a simplified version of the Lagrangian, with a $SU(2)_L\times SU(2)_R$ 
invariance, is
\be\label{linearsigma}
\mathcal{ L}=i\bar \psi_L \not\! \partial \psi_L
+i\bar \psi_R \not\! \partial \psi_R -g\bar \psi_L\Sigma\psi_R- 
g\bar\psi_R\Sigma^\dagger \psi_L - V({\rm Tr\Sigma^\dagger \Sigma})
\ee
where 
\be\label{sigmadecompo}
\Sigma(x)=\sigma(x)I + i\pi^a(x) \tau^a
\ee
This Lagrangian is invariant under the linear transformations
\be 
\Sigma(x)\to \Omega_L\Sigma(x) \Omega^\dagger_R, \qquad \psi(x)_L\to \Omega_L\psi(x)_L,\qquad \psi(x)_R\to \Omega_R\psi(x)_R.
\ee 
The ground state of the potential 
\be
V(\rho)= \frac{\lambda}{4}(\rho^2 -v^2)^2,\qquad \rho\equiv \frac12 {\rm Tr\ \Sigma^\dagger \Sigma}
\ee
is achieved when $ \rho=v$, implying from (\ref{sigmadecompo}) $\sqrt{\sigma^2 + \vec\pi^2}= v$. 
Alternatively, we could have
parametrized the $\Sigma$ matrix as $\Sigma= \rho U(w)$, where $U$ is a unitary matrix depending 
on three unconstrained parameters as
in Eq.\ref{eq:2}. At the minimum of the potential $\Sigma= v\ U(w)$. Of course
there are infinitely many other parametrizations. 

Of all possible $SU(2)$ related vacua one conventionally chooses the  
one with $\sigma(x)=v$, $\pi^a(x) =0$ in one case, 
or $w^a(x)=0$, i.e. $U(w)=I$, in the other one.

When fluctuations above the vacuum state are taken into account, $\rho$ becomes a dynamical
variable and one can replace  $v \to v+\rho$  (non-linear representation), or $v \to v +\sigma$
(linear representation), implying that $(v+\sigma)^2 + \vec\pi^2$ 
is now unconstrained and the redefined $\sigma$ and $\pi^a$ are all dynamical. 
Plugging the previous expressions back in Eq.\ref{linearsigma} we arrive at two different Lagrangians, 
depending on the chosen representation 
\be\label{laglinear}
\mathcal{ L}=i\bar \psi \not\! \partial \psi -gv\bar\psi\psi- g\sigma \bar\psi\psi + ig \vec\pi
\bar\psi\vec\tau\gamma_5 \psi - V(\sigma,\pi),
\ee
or
\be\label{lagnonlinear}
\mathcal{ L}=i\bar \psi \not\! \partial \psi -g(v+\rho)\bar\psi_L U(w)\psi_R- g(v+\rho)\bar\psi_R U^\dagger(w)\psi_L ,
- V(\rho),
\ee
respectively. Note that $m_\sigma= m_\rho =\sqrt{2\lambda v^2}$. In the first case the symmetry is linearly realized, but 
requires the additional field $\sigma$, even if it happens to be very massive when $\lambda$ is large. 
In the second case
the symmetry is non-linearly realized (on the fields $w^a$) and $\rho$ is inessential in the large $\lambda$ limit
and could be removed from the Lagrangian.

In the limit where $\lambda$ becomes very large, the $\sigma$ particle is not in
the low-energy spectrum and the non-linear sigma model, where the chiral symmetry is non-linearly
realized, emerges naturally as a natural descripction. 
Yet, the spectrum of Eq.\ref{laglinear} and Eq.\ref{lagnonlinear} are identical as
are all observables, no matter the value of $\Lambda$. This is, as it should be,  the relation 
between the two sets of variables which  is just a field redefinition.
However it is clear from the previous discussion that Eq.\ref{lagnonlinear} provides a more suitable description when the
$\sigma$ particle is heavy because decoupling occurs naturally in those variables. Note that in the original 
description, the spectrum consists of one complex $SU(2)$ doublet, while in Eq.\ref{laglinear} we have one triplet and one
singlet (the $\sigma$). The situation becomes reversed if the $\sigma$ particle is light; a linear realization
becomes then natural. 

For any value of the $\sigma$ mass, being related by a field redefinition, the two models are strictly equivalent. The
information on the mass and couplings of the $\sigma$ particle is encoded in the respective coefficients of 
higher dimensional operators. When all terms are considered, both descriptions, the linear and the non-linear one, are absolutely 
equivalent and logically provide identical $S$-matrix elements. 

However, any truncation at a given dimensionality necessarily leads to different results. This may or may not
be relevant depending on the observable. For instance in the SMEFT effective 
Lagrangian, an anomalous (i.e. resulting from a possible EWSBS) contribution to the triple gauge boson vertex
appears at dimensionality 6, while in order to find a contribution to the quadruple vertex one has to
go to dimension 8. This may (incorrectly) make us believe that the latter is expected to be smaller
than the former. Of course this need not to be so. In fact in the non-linear realization both contributions
appear in operators of chiral (and engineering) dimension four. The 
quartic vertex is described by the operators $\mathcal{O}_{4}$ and $\mathcal{O}_{5}$ of Eq.\ref{4scatteringops}, 
while the triple gauge boson vertex gets a contribution from the operators
\be
{\mathcal O}_2= i\frac{g^\prime}{2} B_{\mu\nu} {\rm Tr}\ [T [V^\mu, V^\nu]],  
\qquad {\mathcal O}_3= -i \frac{g}{2} {\rm Tr}\ [W_{\mu\nu} [V^\mu,V^\nu]],
\ee
with $T= U \tau^3 U^\dagger$ and $W_{\mu\nu}= \vec W_{\mu\nu} \frac{\vec \tau}{2}$. These are all operators 
of dimension four, as indicated. This suggests that the corresponding coefficients $a_2$, $a_3$, $a_4$ and $a_5$ 
have a similar origin and size, and this is indeed the case in many situations (technicolor-like models, heavy Higgs,...
\cite{EHRM}). In models where the underlying physics is left-right symmetric, $a_2= -a_3$.

Another interesting operator of dimension four in the non-linear realization is 
\be
{\mathcal O_1} = \frac{i}{2} g g^\prime B_{\mu\nu} {\rm Tr}\ [T W^{\mu\nu}].
\ee
This operator, that contributes to the $S$ parameter of the oblique corrections\cite{DEH,Peskin:1990zt}, has a
counterpart of dimension 6 in the linear realization, a fact that might hint that
its contribution to the $S$ parameter might be small \cite{DeRuj}. In fact this is not the case, as
this operator corresponds to a `blind direction' in the SMEFT (but it is perfectly fine and of 
natural size in the non-linear HEFT).
The interested reader is invited to see the quoted references for more details on the effective
electroweak chiral Lagrangian, as the previous construction is known.

Thus in the same way that in QCD the low energy constants are to a large extent saturated by vector
and axial-vector exchange rather than by scalar exchange, and  the non-linear representation seems better
suited to describe pion physics than the linear sigma model, one should expect that
the nature of NP to some extent may privilege some effective descriptions in front of others.

A relevant discussion on the previous points can be found in \cite{yellowrep}.

%%%%%%%%%%%%%%%%%%%%%%%%%%%%%%%%%%%%%%%%%%%%%%%%%%%%%%%%%%%%%%%%%%%%%%%%%%%%%%%%%%%%%%%%%%%%%%%%%%%%%%%%%%
%%%%%%%%%%%%%%%%%%%%%%%%%%%%%%%%%%%%%%%%%%%%%%%%%%%%%%%%%%%%%%%%%%%%%%%%%%%%%%%%%%%%%%%%%%%%%%%%%%%%%%%%%

\section{HEFT for other breaking patterns}
Now we would like to address the question as to how general is the construction presented in previous sections
to describe an extended EWSBS. We have seen in the previous discussion that larger symmetry groups
  could be adopted and consequently additional Goldstone bosons may exist, while in the previous developments, 
both in the linear and non-linear realizations, we have just
written the fields that remain light at low energies and the only symmetry that is manifestly incorporated
into the effective Lagrangian is (apart from the electroweak gauge symmetry) the global $SU(2)_L\times SU(2)_R$
symmetry of the MSM. Indeed, should more Goldstone bosons exist, they could be included in a larger
unitary matrix. However, all these would-be
Goldstone bosons eventually should acquire masses, drop from an extended unitary matrix and could
be parameterized by a polynomial expansion.

In order to set the discussion, let us review in general terms the construction of effective
Lagrangians in other cases.

We assume that the underlying dynamics is governed by some confining gauge group $ G_H$ (from hypercolor group not to be
confused with the global group $G$). This gauge group acts on some microscopic degrees of freedom (some
elementary scalars, or, more naturally, fermions). These degrees of freedom belong to some representation
$\cal R$ of $G_H$. If $\cal R$ is real or pseudo-real, the fermions are chiral and the global
flavor symmetry group is $SU(N)$. A condensate of the form $\langle \psi^i \psi^j\rangle$
can get a vacuum expectation value.
This leads to the possible breaking patterns (we assume that condensation takes place for all ``flavors'', 
simultaneously; other partial breaking patterns would certainly be possible):\\

\begin{itemize}
\item
If $\langle \psi^i \psi^j\rangle$ is symmetric:  $SU(N) \to SO(N)$  \\

\item
If $\langle \psi^i \psi^j\rangle$ is antisymmetric:  $SU(N) \to Sp (N)$   \\ 

\item
Otherwise, if  $\cal R$ is complex,  fermions are not chiral, the global symmetry group $G$ is $SU(N)_L\times
SU(N)_R$  and the unbroken group is $SU(N)_{L+R}\equiv SU(N)_V$.
\end{itemize}

Let us start with the $SU(N)_L\times SU(N)_R\to SU(N)_V$ pattern. This case is analogous to the SSB pattern 
that we have described in detail in previous sections when the Higgs is absent. If $N=2$ or
$N=3$ they are rather familiar breaking patterns in QCD. More NGB than those required for the electroweak
theory are generated if $N>2$.  We will choose as building block of the corresponding effective
theory the matrix-valued field $H\equiv \xi \Sigma \xi$, where $\xi$ is an element of the corresponding
coset $SU(N)\times SU(N)/ SU(N)$. The matrix $\Sigma$ describes the vacuum; in this case $\Sigma= f I_N$ is
the conventional choice (all are related by $SU(N)$ transformations). However, one may include scalar fields
in the unbroken $SU(N)_V$ sector too
\be
\Sigma = fI_N + i T^a \sigma^a ,
\ee
also including singlet perturbations $f \to f + \sigma$. 
Note that the subindices $L$ and $R$ do not necessarily have 
the familiar meaning, at least unless fermions are introduced.

The most general lagrangian that can be constructed with the matrix field $H$ and that contains terms
up to dimension four is of the type:
\begin{align}
\nonumber \mathcal L=&\frac 14 \text{Tr}\left (D_\mu HD^\mu H^\dagger\right )+
\frac b2 \text{Tr}\left [M(H+H^\dagger)\right ]+
\frac{M^2}2 \text{Tr}\left (H H^\dagger\right ) -
\frac{\lambda_1}2 \text{Tr}\left [(HH^\dagger)^2\right ]-\frac{\lambda_2}4 \left [\text{Tr}\left (H H^\dagger\right )\right ]^2\\
&+\frac c2 (\text{det}H + \text{det}H^\dagger) +
\frac{d_1}2 \text{Tr}\left [M(H H^\dagger H+H^\dagger H H^\dagger)\right ]+
\frac{d_2}2 \text{Tr}\left [M(H+H^\dagger)\right ] \text{Tr}\left (H H^\dagger\right )\label{lageff}.
\end{align}
The terms proportional to $b, d_1$ and $d_2$ are soft breaking terms (proportional to a mass matrix), while the one
proportional to $c$ is of dimension four only for $SU(2)$. Of course higher dimensionality terms can be constructed. 
The fields $\sigma, \sigma^a$ are heavy. Integrating them out eventually leads to an effective theory of the non-linear
sigma model type, Eq.\ref{eq:1}. 

We note that, excluding the mentioned terms,  this Lagrangian is not very different from the one presented in 
Eqs. (\ref{eq:heft01}) and (\ref{eq:heft02}), but written in a matrix language that allows for the inclusion
of more degrees of freedom.

Let us now consider the case $SU(N)\to SO(N)$ or $SU(N)\to Sp(N)$. In this case we define $H\equiv \xi \Sigma \xi^{-1}$
with the following transformations:
\be
\xi \to \xi^\prime = g \xi h^{-1}, \quad \Sigma \to \Sigma^\prime= h\Sigma h^{-1}, \quad H \to H^\prime = g H g^{-1},
\ee
where $h=h(\pi, g)$ is an element of the unbroken subgroup (see Sect. \ref{sec:dchm}) and $g$ an element of
the global group $G$. 

The minimal case corresponding to this breaking pattern is $SU(4)\to Sp(4)$. $Sp(4)$ is in fact isomorphic to $SO(5)$ and 
 contains an $SU(2)\times  SU(2)$ subgroup. The breaking (see Table 1) produces five Goldstone bosons. One of them
(singlet under the $SU(2)\times SU(2)$ subgroup) needs to be made massive (it might be considered as
a suitable dark matter candidate) while the other four would give rise
to the three $W_L$ plus the (initially massless) Higgs \cite{Cacciapaglia:2014}.  In the minimal case 
\be
\Sigma=  f\  \begin{pmatrix} 0 & I_2 \\
                  -I_2 & 0 \end{pmatrix}  \equiv f J.
\ee
The $SU(2)\times SU(2)$ is made manifest in this choice of the vacuum by selecting among the 10 generators of $Sp(4)$ 
that have to fulfill the condition $J^{-1} X^\intercal J= -X$ the following six (three antisymmetric and three symmetric)
\be
J_1 = \frac{i}{2}\begin{pmatrix} 0 & \sigma_3 \\
                          -\sigma_3 & 0 \end{pmatrix},
\qquad
J_2= \frac{i}{2} \begin{pmatrix} 0 &  \sigma_1 \\
                          -\sigma_1 &  0 \end{pmatrix},
\qquad
J_3= \frac12  \begin{pmatrix}\sigma_2  &  0 \\
                        0 &   \sigma_2 \end{pmatrix} ,
\ee
\be
K_1 = \frac{-1}{2} \begin{pmatrix} 0  & \sigma_1 \\
                             \sigma_1 & 0 \end{pmatrix},
\qquad
K_2 = \frac{i}{2}  \begin{pmatrix} 0  &  \sigma_3 \\
                             \sigma_3  &  0 \end{pmatrix},
\qquad
K_3= \frac{i}{2} \begin{pmatrix}  I_2  &  0 \\
                            0  &  I_2 \end{pmatrix}.
\ee
The combinations $L_i= J_i \pm i K_i$ form two independent representations of the $SU(2)$ algebra. The matrix-valued
field $\Sigma$ also contains ``modulus'' ($f \to f + \sigma$) as well as other excitations in the unbroken
directions, similarly to the case previously discussed above.

Like in the MCHM, the presence of a misalignement in the gauge group with respect to $\Sigma$
will produce the breaking of the gauge symmetries (see next section).

Note that the minimal $SO(5) \to SO(4)$ does not correspond to any of the above, as it cannot be realized
with fermions as microscopic degrees of freedom. This will be discussed in the context of holography in the next
section.

We now refer the reader to Fig 1. In the so-called non-minimal models (such as the ones we have just described) region 
II in that plot may contain one or more additional 
massless Goldstone boson (that is, additional to the Higgs), and region III in the same figure may contain 
pseudo-Goldstone bosons. Having states in region II or III
is contingent upon the corresponding commutation relations among the various generators. The situation 
is described in Sect. 3.

It is well known that the higgs in the MSM is a singlet under the custodial group $SU(2)_{L+R}$; therefore lies in region II. 
This is also the case in the $SO(5)\to SO(4)$ model where the custodial group is the $SO(3)$ subgroup, whose generators 
commute with the one associated to $h$ as can be
seen e.g. in Eq.\ref{eq51}. 

Phenomenologically, having states in region II other than the Higgs could be worrisome from a phenomenological 
point of view because it would be very difficult or impossible to
find mechanisms that would make them so massive to be able to escape detection. If these states were to exist, they 
could be described by a HEFT of the type described at the end of 
Sect.\ref{sec:heftheory}. Of course, like in the Higgs case in CHM, there is the possibility that these states may 
acquire a mass through explicitit breaking terms that render
the initial global symmetry $G$ an approximate one. 

On the contrary, possible PNGB in region III would be naturally massive, with masses proportional to the electroweak 
gauge couplings, as the corresponding generators do not 
commute with $G_{EW}$, times a large scale and they would not be present in the low-energy HEFT..

%%%%%%%%%%%%%%%%%%%%%%%%%%%%%%%%%%%%%%%%%%%%%%%%%%%%%%%%%%%%%%%%%%%%%%%%%%%%%%%%%%%%%%%%%%%%%%%%%%%%%%%%%
%%%%%%%%%%%%%%%%%%%%%%%%%%%%%%%%%%%%%%%%%%%%%%%%%%%%%%%%%%%%%%%%%%%%%%%%%%%%%%%%%%%%%%%%%%%%%%%%%%%%%%%%%

\section{The input from Holography}\label{sec:holog}
It is a fact that our knowledge about the dynamics and spectrum of 
strongly interacting theories such as the ones that may describe 
an extended EWSBS at the microscopic level is rather fragmentary. An exception would
scaled-up QCD models, even though these are not particularly favoured phenomenologically.
This lack of knowledge is particularly acute for models where the global symmetry  $G$ cannot be
realized with Dirac fermions at the microscopic level, such as e.g. 
$G=SO(5)$ and $H=SO(4)\simeq SU(2) \times SU(2)$, i.e. the MCHM
that provides the most economical way to preserve the custodial symmetry. Yet it is often implicitly assumed 
that the spectrum in such models can be inferred from what we have learned in QCD.

One can get a relatively accurate description of QCD using the so-called bottom-up holographic 
models, where space-time is extended with an additional dimension $z$, and assumed to be described by an 
anti-de Sitter (AdS) metric. The value $z=0$ corresponds to the UV brane, where the theory is assumed 
to be described by a conformal field theory (CFT), 
as befits a critical point of QCD at short distances. In the IR the holographic model
should reproduce the fact that QCD breaks conformality, becoming a confining theory.
Strictly speaking the obtained results correspond to the large $N_c$ (or $N_{tc}$) limit, but they are assumed 
to remain valid at finite values of $N_c$.

Originally inspired by formal developments in supergravity \cite{Maldacenaetal}, 
the bottom-up holographic models remain to this day conjectural. Nevertheless, they 
can provide a relatively accurate description of several facets of QCD. It  
seems reasonable to try these techniques to get some information on
theories whose dynamics is unknown.

In order to implement the holographic treatment one should introduce an infrared brane, i.e. 
to restrict the metric of a model to be a slice of the AdS metric; this is
the hard wall (HW) proposal \cite{HW_2005}. The second way is to make the AdS metric smoothly cut-off
at large $z$ \cite{SW_2006}. This is referred to as the soft wall (SW) model. 

The holographic MCHM with breaking pattern $SO(5)\to SO(4)$ was first proposed in \cite{ACP_2005} using a HW approach. 
In these models the gauge symmetry of the SM is extended to the bulk and the symmetry breaking pattern relies on 
the two branes being introduced.
The Higgs is associated with the fifth component of gauge fields in the direction 
of the broken gauge symmetry. The Higgs  potential  is absent at the tree-level and is  determined  by  quantum
corrections (dominantly gauge bosons and top quarks at one-loop level).  In \cite{ACP_2005,Bellazzini2014,Panico2016}
a complete calculation of the Higgs potential and analysis of several electroweak
observables ($S,\ T,\ Z\rightarrow b\overline{b}$) was done, with an emphasis on the way 
SM matter fields are embedded into the $5D$ model.

Here we shall report on a SW model introduced in \cite{katanaeva}, laying emphasis on an alternative way
to realize the global symmetry breaking pattern and to introduce spin zero fields.
The $SO(5) \rightarrow SO(4)$ breaking happens in the bulk Lagrangian of the scalar fields,
reminiscent of the one of the generalized sigma model used for QCD~\cite{gasiorowicz}.
The Goldstone bosons are introduced explicitly and there is no gauge-Higgs unification
characteristic to the former studies in the SW framework\cite{Falkowski2008}. 
Quite differently from the methods of \cite{ACP_2005, Falkowski2008},
the dynamics responsible for the $SO(5)\to SO(4)$ breaking is entirely `decoupled' 
from the SM gauge fields. The latter are treated as external sources that do not 
participate in the strong dynamics (except eventually through mixing of fields
with identical quantum numbers). In fact, promotion of the electroweak bosons into the bulk
may result in tension with the holographic treatment being supposedly valid only 
in the regime of a strong coupling. Although specifics to the particular model are being
discussed, the methods presented here can be generalized to other breaking patterns.

We will not treat the SM matter fields at all and bypass the relevant issue of a possible generation
of a perturbative Higgs potential that turns the originally massless Higgs into a pseudo-Goldstone boson and generates
a value for the misalignment angle $\theta$ introduced in Section \ref{sec:mchm}. We will not consider either  
the naturalness problem \cite{Strumia} or the origin of the hierarchy of the various scales involved because
our purpose is describing the strong 
dynamics behind the composite sector, the resulting spectrum and verifying the fulfillment  the expected current 
algebra properties, together with the existing constraints from electroweak precision measurements. 
The treatment presented here is inspired by various bottom-up holographic approaches to 
QCD \cite{HW_2005, SW_2006, Hirn2005}, but the spectrum and several properties are different. 
Breaking patterns of the form $SO(N) \to SO(N-1)$ can be treated in exactly the same way ---obviously they generate more
than 4 NGB and are therefore not minimal.  
$SO(4) \simeq SU(2)\times SU(2)$ is the  minimal structure to preserve the custodial symmetry and have 
exactly one Higgs doublet in the coset. 

The MCHM is characterized by a missalignement angle between the global unbroken $H$ and the gauged
subgroups. This coupling is relevant for the $hWW$ coupling (see Sect. \ref{sec:mchm}).
The experimental bound  on the misalignment angle in the conventional MCHM is $\sin\theta \leq 0.34$ (see below), 
assuming the coupling of the Higgs to gauge bosons to be $\kappa_v = \cos\theta$, even though this identification will 
need to be revised in the context of this type of models.

As emphasized, we treat the SM gauge fields perturbatively on the UV brane and consider them as sources of the
vector currents of $SO(5)$ with the same quantum numbers, thus
\be\label{lagr1}
\mathcal L_{4D} = \mathcal {\widetilde L}_{str. int.} + \mathcal L_{SM} + \widetilde J^{a\ \mu}  W_\mu^{a}
+ \widetilde J^{Y\ \mu}  B_\mu,
\ee
where the tilde on the Lagrangian of the strongly interacting sector and its currents ($J^{a}_\mu$ and $J^{Y}_\mu$)
signifies the realization of the misalignment through the rotation of the $SO(5)$ generators,
\be
T^A(\theta)=r(\theta)T^A(0)r^{-1}(\theta), \ \text{with} \ r(\theta)=\begin{pmatrix}
          1_{3\times3} & 0 & 0\\
          0 &\cos(\theta)&\sin(\theta)\\
         0 & -\sin(\theta)&\cos(\theta)\\
         \end{pmatrix}, \quad A=1,...,10.
\ee

In order to have less free parameters it is essential to make an assumption on the microscopic
structure of the strongly interacting sector. This can be achieved by constructing the two-point 
correlators of the corresponding operators and matching their short-distance expansion to the holographic result.
At the microscopic level, one cannot reproduce a $SO(5)$ global symmetry 
 with Dirac (techni)fermions \cite{Cacciapaglia:2014}. Therefore in the present case one chooses to define a 
fundamental theory made out of massless scalar fields in order to determine the scalar operators 
and the conserved currents and to match the normalizations of spin zero and spin one sectors. Other possibilities 
for the fundamental degrees of freedom are conceivable, but the one considered here is the simplest one. 

We choose a scalar field of rank 2, $s^{\alpha\beta}$, then the Lagrangian invariant under the global 
$SO(5)$ transformation
is:
\be 
\mathcal L =\frac12 \partial_\mu s_{\alpha\beta}\partial^\mu s^\top_{\beta\alpha}
-\frac12 m^2 s_{\alpha\beta}s^\top_{\beta\alpha}.
\ee
 We can construct a scalar invariant 
$s^{\alpha\gamma}s^{\gamma\alpha}$, giving a scalar operator 
$\mathcal O_{sc}^{\alpha\beta}(x)=s^{\alpha\gamma}s^{\gamma\beta}$ with dimension $\Delta=2$, spin $p=0$; and a Noether current 
$i[T^A,s]_{\alpha\beta}\partial^\mu s^\top_{\beta\alpha}$ giving a vector operator $\mathcal O_{vec}^{A\ \mu}(x)$, with $\Delta=3$, $p=1$.
The $5D$ masses for the fields dual to these operators follow the general formula from the AdS/CFT dictionary 
$M^2R^2=(\Delta-p)(\Delta+p-4)$ \cite{Maldacenaetal}.
In the end holography provides all the necessary $n$-point functions of the composite operators to 
calculate self-energies for the SM gauge bosons and analyze possible effective interactions and mixings
between EW and composite degrees of freedom.

The $5D$ AdS metric is given by 
\be
g_{MN}dx^M dx^N=\frac{R^2}{z^2}(\eta_{\mu\nu}dx^\mu dx^\nu-d^2z),
\ee
where $R$ is the AdS radius and the convention for the Minkowski space is $\eta_{\mu\nu}=\text{diag}(1,-1,-1,-1)$.
The $SO(5)$ invariant action takes the following form 
\begin{align}\label{5Daction}
S_{5D}=&-\frac{1}{4g_5^2}\int d^4xdz\sqrt{-g}e^{-\Phi(z)}\Tr F_{MN}F_{MN} \\ \nn
&+\frac1{k_s}\int d^4xdz\sqrt{-g}e^{-\Phi(z)}\Tr\bigg[D_M H^\top D^M H-M^2 HH^\top- M^2 (HD^\top+H^\top D) \bigg]. \label{5Daction}
\end{align}
The normalization constants have the dimensionality $[g_5^2]=[k_s]=L$; and following the SW holographic approach 
we have introduced a dilaton exponent with $\Phi(z)$. The $5D$ mass of the scalar field $H(x,z)$ is $M^2R^2=-4$, 
while the vector fields $A_M$ should get zero mass if conformal symmetry holds in the UV brane.
The dynamical breaking from $SO(5)$ to $SO(4)$ happens in the scalar sector due to a function $f(z)$ appearing in 
the nonlinear parameterization of $H$
\be
H(x,z)=\xi\Sigma\xi^{\dagger},\quad \Sigma(x,z)=\begin{pmatrix}
          0_{4\times4} & 0\\
          0 & f(z)\\
         \end{pmatrix}+iT^a\sigma^a(x,z),
         \quad \xi(x,z)=\exp\left(\frac{i\pi^i(x,z)\widehat{T}^i}{\sqrt 2 f(z)}\right).
\ee
We enumerate the $SO(5)$ generators denoting the ones of the unbroken $SO(4)$ sector as $T^a, \ a=1,...,6$ and
the rest which are broken $\widehat{T}^i,\ i=1,...,4$. For the vector fields we have $A_M=A_M^AT^A=A_M^aT^a+A_M^i\widehat{T}^i$.

The matrix field $D$ is introduced in Eq.~(\ref{5Daction}) to provide an explicit soft breaking that is used in order 
to fine-tune to zero the masses of the would-be Goldstone bosons $\pi^i$, as the boundary conditions make them naturally 
massive. It is parameterized by a function $b(z)$ as  
\be
D=\begin{pmatrix}
		0_{4\times4} & 0\\
		0 &  b (z)\\
		\end{pmatrix}.
\ee
		
The summary of ans\"atze functions is: $\Phi(z)=\kappa^2z^2$, $f(z)=f\cdot\kappa z$, $b(z)/f(z)=\mu_1+\mu_2\cdot\kappa z,$ 
where we determine $\mu_1=\mu_2=-1$ from the massless condition for the Goldstone bosons, while $f$ and $\kappa$ are 
the parameters of the model. We choose the $A_z=0$ gauge which is standard for SW models in QCD. 

From the quadratic part of the $5D$ Lagrangian we can get the masses of the composite resonances in $4D$ and the two-point
correlators of the composite operators. For the properties of the $4D$ resonances we look for the normalizable solutions
of equations of motion subject to the Dirichlet boundary condition at $z=\varepsilon$. For the correlators we can use the 
AdS/CFT prescription on the variation of the on-shell $5D$ action that is holographically connected to the $4D$ partition
function $\ln \mathcal Z_{4D}$ defined via
\be
\mathcal Z_{4D}[\phi_{\mathcal O}]=\int [{\cal D} s] \exp\{ i\int d^4x [\mathcal L_{str. int.}(x)+ 
g_V\phi_{\mathcal O \mu}^A(x)\mathcal O_{vec}^{A\ \mu}(x)
+g_S\phi_{\mathcal O}^{\alpha\beta}(x)\mathcal O_{sc}^{\beta\alpha}(x)]\}.\label{Zqft}
\ee
The constants $g_V$ (ditto for $g_S$) describe the relative strength of the coupling of the sources representing
strongly interacting states with respect to the (perturbative) gauge fields (not displayed above, see Eq.\ref{lagr1}). For
two-point functions $g_V$ and $g_S$ can be absorbed into other constants and play no relevant role.

Consider e.g. the vector sector. The $5D$ vector fields can be represented as Kaluza--Klein (KK)  infinite towers 
of $4D$ massive states with specific $z$-profiles holographically provided. The $4D$ masses explicitly depend on the model,
and in the present context we have ($V$ and $A$ correspond to unbroken and broken directions in resemblance 
to vectors and axial-vectors of QCD)
\be
M_V^2(n)=4\kappa^2(n+1),\quad M_A^2(n)=4\kappa^2\left(n+1+\frac{(g_5Rf)^2}{2k_s}\right),\quad n=0,1,2,....
\ee

The vector correlators, after subtracting the generic ambiguities of a form $C_0+C_1q^2$, take the following form
\begin{align}
 \Pi_{V}(q^2)=\sum\limits_{n}\frac{q^4F_V^2}{M^2_V(n)(-q^2+M^2_V(n))},\quad 
 \Pi_{A}(q^2)=\sum\limits_n\frac{q^4 F_A^2(n)}{M^2_A(n)(-q^2+M^2_A(n))}-F^2;\\
 F_V^2=\frac{2R\kappa^2}{g_5^2},\quad F_A^2(n)=\frac{2R\kappa^2}{g_5^2}\frac{n+1}{n+1+\frac{(g_5Rf)^2}{2k_s}},\quad 
F^2=\frac{2R\kappa^2}{g_5^2}\sum\limits_n\frac{\frac{(g_5Rf)^2}{2k_s}}{n+1+\frac{(g_5Rf)^2}{2k_s}}
\end{align}
A similar analysis applies for the part of the Lagrangian with the scalar fields. The masses of the KK radial
excitations in the unbroken scalar and broken Goldstone sectors are
\be
 M_\sigma^2(n)=4\kappa^2(n+1),\quad M_\pi^2(n)=4\kappa^2 n,\quad n=0,1,2,....
\ee
\begin{align}
&i\int d^4x e^{iqx}\langle \mathcal O_{s/p}^a(x) \mathcal O_{s/p}^b(0)\rangle=\delta^{ab}\Pi_{S/G}(q^2),\\
\Pi_S(q^2)=\sum\limits_{n}\frac{F^2_\sigma}{q^2-M^2_\sigma(n)},&\
\Pi_G(q^2)=\sum\limits_{n}\frac{F^2_\pi}{q^2-M^2_\Pi(n)};\quad F^2_\sigma=\frac{16\kappa^2R}{k_s} ,\
F^2_\pi=\frac{16\kappa^2R}{k_s}.
\end{align}

The free parameters $g_5^2$ and $k_s$ can be matched to a single parameter of the $4D$ strongly interacting sector. 
The large $Q^2$ limit of the listed correlators should be compared with the one obtained by  the usual field theory methods 
in $4D$. We find
\be
\frac{k_s}{R}=\frac{64 \pi^2}{5N_{tc}},\quad \frac{g_5^2}{R}=\frac{8 \pi^2}{5N_{tc}}.
\ee

\subsection{Two point functions and mixings}
In the effective Lagrangian~(\ref{lagr1}) a subgroup $SU(2)'\times U(1)'\subset SO(4)'$ is already gauged
because the SM fields $W_\mu^{a}$ and $B_\mu$  couple to the particular currents of the strongly interacting sector. 
 They are among the vector currents that are holographically connected to the vector composite fields. Let us name
 the first three operators of the unbroken vector sector $\mathcal O_\mu^{a_L}(x)$ and the last three -- $\mathcal O_\mu^{a_R}(x)$.
 Then $W_\mu^a$ couples to $J^a_\mu=\frac g{\sqrt2} O_\mu^{a_L}$ and $B_\mu$ to $J^Y_\mu=\frac {g'}{\sqrt2} O_\mu^{3_R}$, 
 as we assume the hypercharge to be $Y=T_{3_R}$.
Hence, we may include to the $4D$ partition function the following terms quadratic in the external sources $W$ and $B$:
${W}^\mu\langle \widetilde J^L_\mu(q) \widetilde J^L_\nu(-q)\rangle {W}^\nu $,
${W}^\mu  \langle \widetilde J^L_\mu(q) \widetilde J^R_\nu(-q)\rangle {B}^\nu $,
${B}^\mu  \langle \widetilde J^R_\mu(q) \widetilde J^R_\nu(-q)\rangle {B}^\nu $.
 The relevant quadratic contribution of the $4D$ effective action is
\begin{align}
\notag
S_{4D}^{eff}\supset\int d^4q\bigg[&\left(\frac{q^\mu q^\nu}{q^2}-
\eta^{\mu\nu}\right)\frac{1}{4}\Pi_{diag}(q^2)(g^2{W}_\mu^{1}{W}_\nu^{1}+g^2{W}_\mu^2{W}_\nu^2+g^2{W}_\mu^3{W}_\nu^3+g'^2{B}_\mu{B}_\nu)
\\ \label{gaugedL}
&+\left(\frac{q^\mu q^\nu}{q^2}-\eta^{\mu\nu}\right)\frac{1}4 \Pi_{LR}(q^2) gg'{W}_\mu^{3}{B}_\nu\bigg],\\
 &\Pi_{diag}(q^2)=\frac{1+\cos^2\theta}2 \Pi_{V}(q^2)+\frac{\sin^2\theta}2 \Pi_{A}(q^2),\\
&\Pi_{LR}(q^2)=\sin^2\theta \left(\Pi_{V}(q^2)-\Pi_{A}(q^2)\right)
\end{align}
The diagonal self-energies result in the mass terms for the gauge fields in a small $q^2$ limit
\be
M^2_W=\frac{g^2}4\sin^2\theta F^2,\quad M^2_Z=\frac{g^2+g'^2}4\sin^2\theta F^2,\quad M^2_\gamma=0. \label{Gmasses}
\ee
The left-right two-point function defines the $S$ parameter of Peskin and Takeuchi \cite{Peskin:1990zt}.
In terms of the masses and decay constants
of the vector composite states it gets a form  (in our description $F_V(n)=F_V$ for all values of $n$)
\be \label{Spar}
S=4\pi\sin^2\theta \bigg[\sum\limits_n\frac{F^2_V(n)}{M^2_V(n)}-\sum\limits_n\frac{F^2_A(n)}{M^2_A(n)}\bigg].
\ee
All other electroweak oblique parameters are vanishing or naturally small in the considered model.

At the same time, the structure of the correlation functions provides the mixing between gauge bosons and composite resonances.
For instance, for the $W$ field we have ($\mathcal D^{\mu\nu}=\Box\eta^{\mu\nu}-\partial^\mu\partial^\nu$)
\begin{align}\nn
+\frac g{\sqrt2}W_\mu^a(x) \mathcal D^{\mu\nu}&\sum\limits_n \frac{F_V}{M_V(n)}  \left[\frac{1+\cos\theta}2 A_{L\ \nu(n)}^a(x) +
\frac{1-\cos\theta}2 A_{R\ \nu(n)}^a(x)\right]\\ 
-\frac g{\sqrt2} W_\mu^a(x)\mathcal D^{\mu\nu}&\sum\limits_n \frac{F_A(n)}{M_A(n)} \frac{\sin\theta}{\sqrt2} A_{br\ \nu(n)}^a(x) . 
\end{align}
Mixing is not very significant numerically.

\begin{figure}[t]
	\centerline{\includegraphics[scale=0.5]{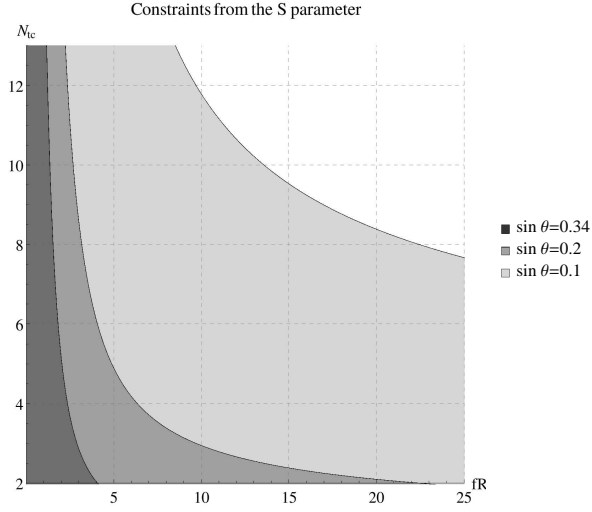}}
	\caption{\label{Spar_fig} The $(\sin\theta, fR, N_{tc})$ parameter region allowed by the $S$ parameter constraints.
From \cite{katanaeva}}.
\end{figure}

\begin{table}[b]\center
\caption{Different predictions of the minimal vector masses for $\sin\theta=0.25$ and $0.30$. From \cite{katanaeva}}.
{\begin{tabular}{@{}ccc|c|c@{}} 
$\sin\theta$ & $N_{tc}$ & $fR$  & $M_V(0)$, $\text{TeV}$ & $M_A(0)$, $\text{TeV}$ \\
\hline
$0.25$& $2$ & $9.1$ & $0.89$ & $2.20$  \\
$0.25$& $3$ & $5.2$ & $1.21$ & $1.99$  \\
$0.25$& $4$ & $3.9$ & $1.37$ & $1.92$ \\
$0.25$& $10$ & $2.0$ & $1.66$ & $1.86$  \\
\hline
$0.30$& $2$ & $5.5$ & $1.26$ & $2.14$  \\
$0.30$& $3$ & $3.7$ & $1.50$ & $2.03$ \\
$0.30$& $4$ & $2.9$ & $1.61$ & $1.99$ \\
$0.30$& $10$ & $1.6$ & $1.81$ & $1.96$ \\
\end{tabular}\label{tab-0}}
\end{table}
The masses of the composite resonances are governed by the scale of parameter $\kappa$. The latter can be constrained 
from the experimental values known for quantities of Eqs.~(\ref{Gmasses}) and (\ref{Spar}). The first provides 
a particular equation that combines the model parameters $\kappa,\ \sin\theta,\ fR,\ N_{tc}$.  The latter should 
be considered as an expression of the $S$ parameter in terms of $\sin\theta,\ fR,\ N_{tc}$. Assuming \cite{Gfitter_2014} 
$ -0.06\leq S\leq 0.16 $
one obtains the areas these parameters may span as depicted in Fig.~\ref{Spar_fig}. By saturating the $S$ bound. 
one can see from Table~\ref{tab-0} that the proposed programe 
can actually accommodate rather light values for the ground states $M_V(0)$ and $M_A(0)$ of order $1-2$~TeV, but higher masses 
are certainly not excluded.

Consider fixing any two parameters among $(\sin\theta,$ $fR,\ N_{tc})$, then 
the growth of the third parameter results in smaller $\kappa$ and a possibility of lower masses. Indeed, an unlimited 
growth in $fR$ results in unlikely small masses for $\sin\theta \lesssim 0.1$. However, higher values of other 
two parameters soon face the upper experimental limit of the $S$ parameter.

\subsection{Three point functions}
We focus now on some particular  couplings. Let us consider {\it e.g.} the $hWW$ vertex. To get the direct coupling of 
this kind we modify the $5D$ Lagrangian with a redefined covariant derivative
\be
D_\mu H(x,z)=\partial_\mu H(x,z)-i[A_\mu(x,z), H(x,z)]-i[\widetilde{X}_\mu(x), H(x,z)]
\ee
where we include gauge boson fields in the bulk through the rotated field  
$\widetilde{X}_\mu=X_\mu^{a}r^{-1}(\theta)T^{a}(0)r(\theta)$. It is important that $X_\mu$ is assumed to be $z$-independent. 
$W$ is $X_\mu^{L\ \alpha}=\frac g{\sqrt2} W_\mu^{\alpha}$.
This modification results in a particular three point vertex $WW\pi^4$ in $5D$. The $4D$ vertex is obtained integrating 
out the $z$-dimension using the Kaluza--Klein representation of $\pi^4$ and flat profiles of $W$. Recognizing $n=0$ mode 
of $\pi^4$ as the Higgs field we get
\be
\mathcal L_{4D}\supset\frac{g_{hWW}}2 h W^{1,2}_{\mu}W^{1,2\mu},\quad 
g_{hWW}=\frac{g^2\sin\theta}2 \kappa (fR) \sqrt{\frac R{k_s}} \cdot \sqrt{\frac\pi2}\cos\theta.
\ee
We shall not dwell further into this matter here, but we note that this expression differs from the naive 
one.

\subsection{QCD-like models}
In the rest of this section we turn to briefly review some recent analysis that shows how difficult it is to get 
realistic composite Higgs models that fulfill all the phenomenological constraints with 
scaled-up versions of QCD or similar models. The following summarizes a recent 
interesting discussion \cite{Belyaevetal}. The stuation is very different from the MCHM just described.

The holographic model in question is the Dynamic AdS/QCD model which is described in detail in \cite{AHLO}. 
The action is
\begin{equation} 
S_{5D} = -\int d^4x d z \sqrt{-g}{\rm{Tr}} 
\left[  {1 \over k_s} |D H|^2 +  M^2(r) |H|^2   + {1 \over 4g_5^2} (F_V^2 + F_A^2) \right], 
\end{equation}
The theory lives in a geometry described by the metric (Poincar\'e coordinates are
used to parametrize de Sitter space)
\be
ds^2 = \frac{r^2}{R^2} dx^2 + \frac{R^2}{r^2} dz^2,\quad r^2 = z^2+ |Tr H|^2R^4 
\ee
Here $z$ is the holographic coordinate dual to energy scale, and $H$ is a field containing the meson 
excitations, as above, dual to the fermion condensate. Fluctuations of $H$ around its vacuum expectation 
value describe the pseudoscalar 
and scalar excitations as in the previous model. The vector and axial vector 5D fields $V$ and $A$ will
eventually describe towers of vector and axial resonances, dual to the vector
and axial-vector currents, expressed in terms of the fermionic degrees of freedom.

$H$ is included in the metric definition in order to provide a back-reaction on the metric 
and communicates the mass gap to the mesonic spectrum as it breaks conformal invariance, which
is restored if $H=0$. $M^2(r)$ is a renormalization group scale/radially dependent
mass term whose running can be fixed from the running of the gauge coupling in the theory of
interest. The ansatz eventually includes IR fixed points for appropriate choices of $N_c$
and the number of fermion species $N_f$,computed at the two-loop order. Notice that
according to the usual holographic procedure, $M^2R^2= -3$, so departures from this value
indicate an additional breaking of conformal invariance. 

The above action is similar to the ones used to describe holographic QCD within a bottom-up approach.
For such a description including a soft-wall cut-off, see e.g. \cite{sergeya}.

The spectrum of the theory is found by looking at linearized fluctuations of the fields about the vacuum.
By substituting the corresponding wave functions back
into the action and integrating over the variable $z$, the decay constants
can also be determined, just as in the previous model. The normalization constants $g_5$ and
$k_s$  are determined, as before, by matching to the gauge theory expectations for the vector-vector, 
axial-axial and scalar-scalar correlators in the UV of the theory.

The results of the analysis, borrowed from \cite{Belyaevetal}, in summary are the following: after
imposing the necessary EW precision bounds, the technicolor
theories that emerge enter the strong coupling at a very large scale before
settling on an IR fixed point that triggers symmetry breaking at the 1 TeV  range or so. It was found that the IR theory
constructed in the way described is largely independent of the UV theory. The bound states are in the 4 TeV range. 
Such theories are therefore beyond the reach
of the current LHC searches. In a sense such theories display in an extreme way the issues that any extension
of the standard model that addresses the hierarchy problem must now encounter - to make the Higgs light there
must be tuning at one part in 100 or so and new states must be pushed to high scale.

In conclusion, it seems fair to say that no bottom-up holographic model is fully satisfactory. 
For one, soft wall models
do not have a satisfactory operator product expansion as they exhibit a would-be dimension 2 condensate 
that is forbidden in a gauge-invariant field theory. Hard wall models typically do not show Regge-like trajectories 
for the tower of resonances, a
behavior that is well established in non abelian gauge theories. In any case, there are several free parameters and, worse, 
the modelization itself is not unique. It is therefore difficult to attach realistic error bars to the predictions 
that are derived by means of holographic techniques.  

%%%%%%%%%%%%%%%%%%%%%%%%%%%%%%%%%%%%%%%%%%%%%%%%%%%%%%%%%%%%%%%%%%%%%%%%%%%%%%%%%%%%%%%%%%%%%%%%%%%%%%%%%
%%%%%%%%%%%%%%%%%%%%%%%%%%%%%%%%%%%%%%%%%%%%%%%%%%%%%%%%%%%%%%%%%%%%%%%%%%%%%%%%%%%%%%%%%%%%%%%%%%%%%%%%%

\section{The input from lattice field theory}
Technicolor theories \cite{Weinberg:1975gm,Susskind:1978ms} provide an elegant mechanism for dynamical 
electroweak symmetry breaking. A fermion condensate 
is generated in a QCD-like manner. This is coupled to EW currents and the Fermi scale $v$ can be generated much like
the pion decay constant $f_\pi$ appears in QCD. As it is well known, the EW bosons get a small part of their mass
from the familiar chiral symmetry in QCD. In this rather economic mechanism the longitudinal degrees of freedom
of the $W,Z$ eat the only three Goldstone bosons that are produced in the spontaneous breaking of $SU(2)_L$. 

The idea of EW symmetry breaking being due to the existence of a new strongly interacting sector at the TeV scale
beyond the SM was proposed many years ago, but the
simplest models obtained by a naive rescaling of QCD are inadequate since they are in contradiction with the 
experimental evidence of precision EW tests~\cite{TCexcl}. More importantly,
realistic models should contain an additional light scalar degree of freedom--the Higgs. Therefore, to be acceptable 
candidates for phenomenology, such theories need to be different from  scaled-up versions of QCD.

Walking and conformal technicolor theories have been proposed 
(see e.g.\cite{Holdometal}) as possible candidates. These theories may exhibit an IR fixed point, or at least an ample region
where the beta function is close to zero and the theory does not `run', at least significantly. Such
theories may then also provide a convenient arena for the Extended Technicolor (ETC) ideas, suggesting that
the mechanisms of generation of a mass for SM fermions and gauge bosons are analogous. Such a possibility 
is ruled out in the simplest technicolor due to obvious current algebra arguments. 

The dynamics of these theories should be sufficiently different from QCD so that they
would not violate the experimental constraints. In particular, the idea of using models with matter fields in
representations other than the fundamental near the onset of the
conformal window has been advocated~\cite{Sanninoetal}. As an example 
one can mention the so-called Minimal Walking Technicolor theory, based on the gauge group SU(2)
with two Dirac fermions in the adjoint representation.  This is 
``minimal'' in the sense that having only 2 flavors provides the estimated minimal value of the
Peskin-Takeuchi $S$ parameter among all theories of this type, with matter fields in
one representation only.

In such a model, fermion condensation triggers electroweak symmetry breaking. In addition, if 
a nearly conformal sector is present, the spectrum of states at the electroweak scale
could then contain a narrow scalar resonance, the pseudo-Goldstone boson of conformal symmetry
breaking, with Higgs-like properties. If the conformal sector is strongly coupled, this pseudo-dilaton
may be the only new state accessible at high energy colliders. 

In Sect. \ref{sec:geometry} we already discussed some properties of the resulting model and 
the prospects for distinguishing this mode from a minimal Higgs boson at the LHC and ILC lay 
mostly in the cubic self-interactions and a potential enhancement of couplings to massless SM gauge bosons \cite{Grinstein}.

In the previous section we described a QCD-like model where the running and the corresponding mass anomalous
dimension was gotten from  two-loop calculations (and we saw that the spectrum of possible vector resonances was
moved to the high UV region). However, in order to get more accurate predictions, numerical lattice work
is still required. See e.g. \cite{picagranada} and references therein.

%%%%%%%%%%%%%%%%%%%%%%%%%%%%%%%%%%%%%%%%%%%%%%%%%%%%%%%%%%%%%%%%%%%%%%%%%%%%%%%%%%%%%%%%%%%%%%%%%%%%%%%%%%%
%%%%%%%%%%%%%%%%%%%%%%%%%%%%%%%%%%%%%%%%%%%%%%%%%%%%%%%%%%%%%%%%%%%%%%%%%%%%%%%%%%%%%%%%%%%%%%%%%%%%%%%%%%

\section{Quantization. The Equivalence Theorem}
With few exceptions, in the previous sections we have been dealing with a classical gauged Lagrangian. 
Perturbative quantization of gauge theories can be accomplished 
by the Faddeev-Popov method. This requires the introduction of a set of appropriate gauge fixing functions $f^i$ with $i=1,2,3,4$ 
and adding to the classical Lagrangian the gauge fixing terms
\begin{equation}
{\cal L}_{GF} =-\frac{1}{2\xi_W} \sum_{i=1}^3 (f^i)^2 
-\frac{1}{2\xi_B}(f^4)^2 
\end{equation}
where $\xi_W$ and $\xi_B$ are the so called gauge-fixing parameters. In the case of electroweak gauge theories with 
SSB it is particularly convenient to choose the so called  
 t'Hooft, $R_\xi$ or renormalizable
 gauges, which have the form
\begin{eqnarray}
f^i & = &   \partial^\mu W^i_\mu -
\frac{g v \xi_W}{2} \  \pi^i , \;\;\; i=1,2,3. \nonumber \\
f^4 & = &   \partial^\mu B_\mu +
\frac{g' v \xi_B}{2} \ \pi^3 .
\label{gfun}
\end{eqnarray}
The addition of these gauge-fixing terms with these gauge-fixing functions have the virtue of canceling the bilinear 
terms appearing in the classical Lagrangian connecting NGB and gauge-boson fields. This cancellation makes the physical 
interpretation of the Lagrangian much more transparent.

The  Faddeev-Popov ghost term is given by
\begin{equation}
 {\cal L}_{FP} =\sum_{i,j=1}^{3} c_i^\dagger (x) \frac{\delta f^i}{\delta
 \epsilon_L^j} c_j(x)+ c_4^\dagger (x) \frac{\delta f^4}{\delta
 \epsilon_Y} c_4(x) ,
\end{equation}
where $\epsilon^¡_L$ and $\epsilon_Y$ are the gauge transformation parameters and $c_i(x)$ are
 the ghost fields (in fact $c_4(x)$ is decoupled). After adding the gauge-fixing and the Faddeev-Popov
terms, the gauge $SU(2)_L \times U(1)_Y$ invariance is lost but
it is replaced by BRST symmetry \cite{BeRoSt75}. This invariance gives rise to  the Slavnov-Taylor
identities, which encode the original gauge symmetry of the
classical action and ensures the gauge invariance of the
physical observables, in particular the S-matrix elements. After adding the gauge-fixing and Faddeev-Popov terms the 
Lagrangian is no longer $M$ covariant. However it is possible to keep the covariant formalism by defining  appropriate 
(non-linear) gauge-fixing functions (see \cite{Dobado:1997jx} for details). In any case, the S-matrix elements are independent 
of the coordinates used to parametrize the scalar space $M$ and thus we can use the above linear gauge-fixing function to obtain 
the Feynman rules needed for perturbative calculations.

One of the most important consequences of the Faddeev-Popov method used for spontaneously broken gauge theories with $R_\xi$ gauge 
fixing functions is the following. In this gauge one is doing the path integral on the gauge fields by forcing, for example, 
the constraint: $f_i=0$ or  $\partial^\mu W_\mu^i = M_W \xi_W \pi^i$. On the other hand the longitudinal polarization vectors 
of the massive gauge bosons are given by
\be
\epsilon_L^\mu(k) = \frac{k^\mu}{M}+ v^\mu(k),
\ee
where $k^\mu = ( E_k, \vec k)$, $M = M_W, M_Z$, $v^\mu= O(M/E_k)$ and $E_k = \sqrt{\vec k^2 + M^2}$. Therefore, at 
high energies $(E >>M)$, the path integral is done under the constraint $W_L^i \sim i \xi_W \pi^i$.  Thus, when computing the 
on-shell $S$ matrix elements of processes containing longitudinal components of the electroweak gauge bosons $W_L$ or $Z_L$, both in 
the initial or the final state, at high energies compared with $M_W$ and $M_Z$, one can perform the substitution of the 
these longitudinal components by the corresponding would-be NGB. This general result is known as the Equivalence 
Theorem (ET) \cite{ET}.  
In spite of its apparently simple explanation, a formal proof  is somewhat complicated when more than one 
longitudinal gauge boson is involved or we go beyond the tree level (see \cite{esma} for details). 
The ET is based on the Slavnov-Taylor identities 
resulting from the BRST  invariance of the 
Lagrangian used in the Faddeev-Popov method (the gauge invariant classical Lagrangian plus gauge fixing and Faddeev-Popov terms).  
A typical example of the applicability of the ET is the following relation for the longitudinal gauge boson 
on-shell scattering amplitude 
in the CM frame
\begin{equation} \label{EqTh}
T\left(W_L ^+, W_L^- \rightarrow Z Z \right) =  T\left(\pi^+  \pi^- \rightarrow \pi^0  \pi^0\right) + \mO\left(\frac{M}{\sqrt{s}}\right).
\end{equation}
where $\pi^{\pm}=(\pi^1  \mp i \pi^2)/\sqrt{2}$ and $\pi^0 = \pi^3$. In the following we will make use in some cases of the the ET 
for the computation of the amplitudes of different processes at CM energies larger than $M_W$ and $M_Z$ (and hence $M_h$ too). This 
is because the computation of amplitudes involving NGB is much simpler than the corresponding computation of amplitudes involving 
longitudinal components of gauge bosons. In turn, these amplitudes are supposed to be dominant at high energies  
(compared with the ones involving transverse gauge bosons) if the SBS of the SM is strongly interacting, in particular 
giving rise to resonances that could be probed at the LHC in the next years.

%%%%%%%%%%%%%%%%%%%%%%%%%%%%%%%%%%%%%%%%%%%%%%%%%%%%%%%%%%%%%%%%%%%%%%%%%%%%%%%%%%%%%%%%%%%%%%%%%%%%%%%%%%%%%%%%%%%%
%%%%%%%%%%%%%%%%%%%%%%%%%%%%%%%%%%%%%%%%%%%%%%%%%%%%%%%%%%%%%%%%%%%%%%%%%%%%%%%%%%%%%%%%%%%%%%%%%%%%%%%%%%%%%%%%%%%%

\section{Electroweak Chiral Perturbation Theory with a light Higgs for  VV, hh and Vh scattering}\label{sec:ecpt}

In this section we address the problem of the computation of the relevant scattering amplitudes for electroweak gauge bosons 
$V=W^\pm, Z$ and the Higgs $h$ for the study of the SBS of the SM at the LHC energies.  Assuming a strongly interacting 
SBS and $s>> M^2_V, M^2_H$, the longitudinal components $V_L$ of the gauge bosons dominate the amplitudes and also it is possible 
to use the Equivalence Theorem. Then we can identify the $W^+_L, W^-_L$ and $Z_L$ with the GB $w^+, w^-$ and $w^0=z$ and, assuming 
custodial symmetry, we can set the gauge couplings $g=g'=0$ by using the Landau gauge to avoid NGB-Ghost coupling.  
Finally we neglect the effect of fermions. Under these conditions the relevant Lagrangian for $ww, hh$ and $wh$ scattering up to next to leading order (NLO) is
\ber \label{bosonLagrangian}
{\cal L} & = & \frac{1}{2}\left(1 +2 a \frac{h}{v} +b\left(\frac{h}{v}\right)^2\right)
\partial_\mu \omega^a
\partial^\mu \omega^b\left(\delta_{ab}+\frac{\omega^a\omega^b}{v^2}\right)   
\nonumber +\frac{1}{2}\partial_\mu h \partial^\mu h \nonumber  \\
 & + & \frac{4 a_4}{v^4}\partial_\mu \omega^a\partial_\nu \omega^a\partial^\mu \omega^b\partial^\nu \omega^b +
\frac{4 a_5}{v^4}\partial_\mu \omega^a\partial^\mu \omega^a\partial_\nu \omega^b\partial^\nu \omega^b  +
\frac{g}{v^4} (\partial_\mu h \partial^\mu h )^2  \nonumber   \\
 & + & \frac{2 d}{v^4} \partial_\mu h\partial^\mu h\partial_\nu \omega^a  \partial^\nu\omega^a
+\frac{2 e}{v^4} \partial_\mu h\partial^\nu h\partial^\mu \omega^a \partial_\nu\omega^a,
\eer 
where  we have used the squared root $SU(2)$ coset parametrization
\be
U(x) = \sqrt{1-\frac{\omega^2}{v^2}}+ i \frac{\omega^a\tau^a}{v}.
\ee
instead of the more common exponential $U(x)=\exp( i \omega^a\tau^a/v)$. As discussed in detail \cite{Delgado:2014jda} , 
this election gives rise to different Feynman rules and diagrams. However, in agreement with the reparametrization theorem of QFT \cite{Kamefuchi:1961sb}, the on-shell  S-matrix elements are the same, being the computations with the squared root parametrization 
much simpler. Also we have added three new operators with chiral couplings $d,e$  and  $g$. Those operators are order $O(p^4)$ 
and must be introduced to absorb the divergencies appearing in the one-loop amplitudes for the $ww \rightarrow hh$ and  
$hh \rightarrow hh$ processes.

Starting from $w^a w^b \to w^c w^d$ scattering, due to the custodial symmetry, the amplitude has the general form
\be
\label{deco1}
T(w^a w^b \to w^c w^d)= A(s,t,u)\delta_{ab}\delta_{cd}+A(t,s,u)\delta_{ac}\delta_{bd}+A(u,t,s)\delta_{ad}\delta_{bc}\ .
\ee 
By defining the charge states $w^{\pm}=(w^1\mp i w^2)/\sqrt{2}$ and $z=w^0$ we have the relations
\begin{eqnarray}
A(w^+ w^- \rightarrow zz) & = & A(s,t,u)   \\ \nonumber
A(w^+w^- \rightarrow w^+w^-) & = & A(s,t,u)+A(t,s,u)   \\ \nonumber
A(zz \rightarrow zz) & = & A(s,t,u)+A(t,s,u)+A(u,t,s)   .
\end{eqnarray}
Now we can expand $A(s,t,u)$   in a similar way to ordinary ChPT.  Up to the one-loop level one has
\be 
\label{loopexpansion}
A = A^{(0)} + A^{(1)} \dots =  A^{(0)} + A^{(1)}_{\rm tree} + A^{(1)}_{\rm loop} \dots ,
\ee
where $ A^{(0)}$ is order $O(p^2)$ and  $A^{(1)}$ is order $O(p^4)$. From the Lagrangian above it is easy to compute the tree level part
\begin{equation} \label{Atree}
A^{(0)}(s,t,u) + A^{(1)}_{\rm tree}(s,t,u) = (1-a^2)\frac{s}{v^2} + \frac{4}{v^4}\left[2a_5 s^2 + a_4(t^2 + u^2)\right].
\end{equation}
The one-loop part is much more involved because of the large number of Feynman graphs. Using dimensional regularization with  
$D=4-\epsilon$ and following \cite{Delgado:2013hxa} one has
 \begin{equation} \label{Aloop}
A^{(1)}_{\rm loop}(s,t,u) = \frac{1}{36 (4\pi)^2 v^4}[F_1(s,t,u)s^2 +(a^2-1)^2( g(s,t,u) t^2 + F_2(s,u,t) u^2)]
\end{equation}
with auxiliary functions
\begin{eqnarray}
F_1(s,t,u) &=& 
[20 - 40 a^2 + 56 a^4 - 72 a^2 b + 36 b^2] \nonumber\\
& + &% 
[12 - 24 a^2 + 30 a^4- 36 a^2 b + 18 b^2] N_\varepsilon\nonumber\\  \label{ffunction}
& + & %
[-18 + 36 a^2 - 36 a^4 + 36 a^2 b - 18 b^2] \log\left(\frac{-s}{\mu^2}\right) \nonumber\\
& + & %
3 (a^2-1)^2 \left[\log\left(\frac{-t}{\mu^2}\right) +
\log\left(\frac{-u}{\mu^2}\right)\right] \\ \label{gfunction}
F_2(s,t,u) &=& 
26 + 12 N_\varepsilon 
-9 \log\left[-\frac{t}{\mu^2}\right]
-3 \log\left[-\frac{u}{\mu^2}\right]
\end{eqnarray}
where
\be
N_\epsilon =\frac{2}{\epsilon} + \log 4\pi -\gamma\ .
\ee
This result agrees with the one   in~\cite{Espriu:2013fia} when the limit $M_H$ going to zero is taken and can be obtained by using
algebraic manipulation software such as FeynRules, FeynArts and Formcalc in succession, or equivalent codes.

Next we can consider the process $w^a w^b \to hh$. In this case custodial symmetry dictates the amplitude's general form
\be
M(s,t,u)_{ab}=M(s,t,u) \delta_{ab}
\ee
A similar computation to the one described above for the elastic case gives
\begin{equation} \label{Mtree}
M^{(0)}(s,t,u) + M^{(1)}_{\rm tree}(s,t,u)
= (a^2-b)\frac{s}{v^2}+ 
\frac{2 d}{v^4} s^2+ \frac{e}{v^4}(t^2+u^2)
\end{equation}
that takes a one-loop correction
\begin{equation} \label{Mloop}
M^{(1)}_{\rm loop}(s,t,u) = \frac{a^2-b}{576\pi^2 v^2}\left[F_1'(s,t,u)\frac{s^2}{v^2}  
+\frac{a^2 - b}{v^2}[ F_2'(s,t,u)t^2 + F_2'(s,u,t)u^2]\right]
\end{equation}
where
\begin{eqnarray}
F_1'(s,t,u) &=& 
-8 [-9 + 11 a^2 - 2 b]
-6  N_\varepsilon [-6 + 7 a^2 - b]  \\ \nonumber
&  + &
36 (a^2 - 1)\log\left[-\frac{s}{\mu^2}\right] +
3 (a^2 - b) \left(\log\left[-\frac{t}{\mu^2}\right] + 
\log\left[-\frac{u}{\mu^2}\right]\right) 
\end{eqnarray}
and  $F_2'(s,t,u)=F_2(s,t,u)$. Due to the time reversal symmetry of the interaction considered 
here the same amplitudes describes also the process $hh \to w^a w^b $.

Finally, the tree level $hh\to hh$ elastic amplitude is 
\be \label{Ttree}
 T^{(0)}(s,t,u) + T^{(1)}_{\rm tree}(s,t,u)=
\frac{2g}{v^4}(s^2+t^2+u^2)\ ,
\ee
and the one-loop one is:
\be \label{Tloop}
T^{(1)}_{\rm loop}(s,t,u) = \frac{3(a^2-b)^2}{2(4\pi)^2 v^4}\left[T(s)s^2 + T(t)t^2 + T(u)u^2\right] \ .
\ee
where 
\begin{equation}
T(s) = 2 + N_\varepsilon - \log\left(-\frac{s}{\mu^2}\right).
\end{equation}
All those amplitudes have to be renormalized. This can be done in a relatively easy way due to the 
fact that all the particles involved in the approximation considered here are massless. That implies that 
no wave function,  vacuum expectation value $v$,  or masses need renormalization. In fact, it is not difficult to 
realize that all the divergences can be absorbed by renormalizing the bare chiral couplings: $a_4, a_5, g, d$ 
and $e$. For example, by using the $\overline{MS}$ scheme, the renormalized couplings are defined as:
\begin{eqnarray}
a_4^r & = & a_4+ \frac{N_\epsilon}{192\pi^2}(1-a^2)^2   \nonumber \\
a_5^r & = & a_5+\frac{N_\epsilon}{768 \pi^2} (2+5  a^4-4 a^2-6 a^2 b+3 b^2)\nonumber \\
g^r & = & g+\frac{3N_\epsilon}{64 \pi^2 }(a^2-b)^2   \nonumber  \\
d^r & = & d -\frac{N_\epsilon}{192 \pi^2  }(a^2-b)(7 a^2- b-6)  \nonumber  \\
e^r  & = & e+ \frac{N_\epsilon}{48 \pi^2 }(a^2-b)^2.
\end{eqnarray}
Notice that the couplings $a$ and $b$ do not require renormalization. A trivial consistency check 
of these equations is the minimal SM where $a=b=1$ which, being renormalizable, does not require any additional 
chiral coupling. Also the renormalization of $a_4$ and $a_5$ agrees with \cite{Espriu:2013fia}.

In terms of the renormalized couplings, the amplitudes read \\

$ww \to ww$
\begin{eqnarray} \label{Arenorm}
A(s, t, u) & = & \frac{s}{v^2} (1 - a^2)+\frac{ 4}{v^4} [2 a^r_5(\mu) s^2 + a^r_4(\mu) (t^2 + u^2)] \\  \nonumber    
& + &\frac{1}{16 \pi^2 v^4}\left(\frac{1}{9} (14 a^4 - 10 a^2 - 18 a^2 b  + 9 b^2 + 
        5 ) s^2
     + \frac{13}{18} (a^2 - 1)^2 (t^2 + u^2) \right.  \\ \nonumber  
       &  - & \frac{1}{2}  (2 a^4 - 2 a^2 - 
        2 a^2 b  + b^2 + 
        1)   s^2 \log\frac{-s}{\mu^2} \\  \nonumber   
       &  +& \frac{1}{12} (1-a^2 )^2 (s^2 - 3 t^2 - 
        u^2) \log\frac{-t}{\mu^2}        \\ \nonumber  
       &  + & \left.
  \frac{1}{12}   (1-a^2 )^2 (s^2 - t^2 - 3 u^2) \log\frac{-u}{\mu^2}
    \right)
\end{eqnarray}
$ww  \rightarrow h h$ and $hh \to ww$
\begin{eqnarray} \label{Mrenorm}
M(s,t,u)  & = & \frac{a^2-b}{v^2}s  + \frac{2 d^r(\mu)}{v^4}s^2+ \frac{e^r(\mu)}{v^4}(t^2+u^2) \nonumber  \\
& + & 
\frac{(a^2-b)}{576\pi^2v^4}
\left\{\left[72 -  88 a^2+ 16 b  + 36 (a^2-1)\log\frac{-s}{\mu^2}\right.\right. \nonumber \\
& + &
\left.\left. 3  (a^2-b)\left(\log\frac{-t}{\mu^2}+\log\frac{-u}{\mu^2}\right)\right]s^2 \right. \nonumber  \\
& + &  (a^2-b)\left(26-9\log\frac{-t}{\mu^2}-3\log\frac{-u}{\mu^2}\right)t^2  \nonumber  \\
& + & \left. (a^2-b)\left(26-9\log\frac{-u}{\mu^2}-3\log\frac{-t}{\mu^2}\right)u^2 \right\}
\end{eqnarray}
and finally $h h \rightarrow  h h$
\begin{eqnarray} \label{Trenorm}
T(s,t,u) & = & \frac{2g^r(\mu)}{v^4}(s^2+t^2+u^2) \\ \nonumber
 &+& 
\frac{3(a^2-b)^2 }{32\pi^2v^4}
\left[ 2(s^2+t^2+u^2)-s^2\log\frac{-s}{\mu^2}-t^2\log\frac{-t}{\mu^2}
-u^2\log\frac{-u}{\mu^2}\right]\ .
\end{eqnarray}
where we have explicitly exhibited the renormalization scale $\mu$ dependence of the renormalized couplings. On the other hand,
as we do not have wave-function renormalization, the amplitudes above are just S-matrix elements and they must be 
renormalization scale ($\mu$) invariant. This means that the explicit  $\mu$ dependence has to be compensated by 
the $\mu$ dependence of the chiral couplings. From this simple fact it not difficult to find the renormalization 
group  equations:
\begin{eqnarray} \label{RGE}
a_4^r (\mu)& = & a_4^r(\mu_0)- \frac{1}{192 \pi^2}(1-a^2)^2    \log\frac{\mu^2}{\mu_0^2}   \nonumber \\
a_5^r(\mu) & = & a_5^r(\mu_0)- \frac{1}{768 \pi^2}\left[3(a^2-b)^2+2(1-a^2)^2\right] \log\frac{\mu^2}{\mu_0^2}\nonumber \\
g^r(\mu) & = & g^r(\mu_0)-\frac{3}{64\pi^2}(a^2-b)^2  \log\frac{\mu^2}{\mu_0^2}  \nonumber  \\
d^r(\mu) & = & d^r(\mu_0) +\frac{1}{192 \pi^2}(a^2-b)\left[(a^2-b)-6(1-a^2)\right]  \log\frac{\mu^2}{\mu_0^2} \nonumber  \\
e^r(\mu)  & = & e(\mu_0)- \frac{1}{48 \pi^2}(a^2-b)^2 \log\frac{\mu^2}{\mu_0^2}\ .
\end{eqnarray}
From the above result for  $ww \to hh$, using crossing symmetry, it is possible to obtain the renormalized amplitude 
for the $wh \to wh$ process too. In this case we have
\begin{equation}
N(\omega^{I_3} h \rightarrow \omega^{I'_3} h)=N(s,t,u)\delta_{I_3I'_3}
\end{equation}
where $I=1$ and $I_3$ are the custodial isospin and the corresponding third component and
\begin{eqnarray} \label{invAmp}
N(s,t,u) & =  &\frac{a^2-b}{v^2}t+ \frac{2d^r(\mu)}{v^4}t^2+\frac{e^r(\mu)}{v^4}(s^2+u^2)\\   \nonumber
& + & \frac{a^2-b}{576\pi^2v^4}[(72-88a^2+16b+36(a^2-1)\log \frac{-t}{\mu^2}   \\  \nonumber
&+ & 3 (a^2-b       )    (\log \frac{-s}{\mu^2}+\log \frac{-u}{\mu^2}      ))t^2   \\   \nonumber
& + & (a^2-b       )    (26 - 9\log \frac{-s}{\mu^2}-3\log \frac{-u}{\mu^2}      ))s^2   \\   \nonumber
& +&  (a^2-b       )    (26 - 9\log \frac{-u}{\mu^2}-3\log \frac{-s}{\mu^2}      ))u^2 ] . \\   \nonumber
\end{eqnarray}

%%%%%%%%%%%%%%%%%%%%%%%%%%%%%%%%%%%%%%%%%%%%%%%%%%%%%%%%%%%%%%%%%%%%%%%%%%%%%%%%%%%%%%%%%%%%%%%%%%%%%%%%%%%%%%%%%%%%%%%%%%%%
%%%%%%%%%%%%%%%%%%%%%%%%%%%%%%%%%%%%%%%%%%%%%%%%%%%%%%%%%%%%%%%%%%%%%%%%%%%%%%%%%%%%%%%%%%%%%%%%%%%%%%%%%%%%%%%%%%%%%%%%%%%%

\section{NLO partial waves}\label{sec:nlo}
For further study of the unitarity and analytic properties of the amplitudes it is useful to project them on the different $I$ 
(custodial isospin) and $J$ (angular momentum ) partial waves. For $ww \to ww$ elastic scattering there are 
three custodial-isospin $A_I$ amplitudes ($I=0,1,2$) which are defined as
\begin{eqnarray}\label{deco2}
T_0(s, t, u)  & = &  3 A(s, t, u) + A(t, s, u) + A(u, t, s)  \\ \nonumber
T_1(s, t, u)  & = & A(t, s, u) - A(u, t, s)     \\ \nonumber
T_2(s, t, u)  & = & A(t, s, u) + A(u, t, s)\ .
\end{eqnarray}
The partial waves are given by
\begin{equation} \label{Jprojection}
t_{IJ}(s)=\frac{1}{64\,\pi}\int_{-1}^1\,d(\cos\theta)\,P_J(\cos\theta)\, T_I(s,t,u)\ .
\end{equation}
The  corresponding chiral expansion is
\be
t_{IJ}(s)=t^{(0)}_{IJ}(s)+t^{(1)}_{IJ}(s)+\dots ,
\ee
where the first two terms (NLO) have the general form
\begin{eqnarray}\label{expandpartialwave}
\nonumber   t^{(0)}_{IJ}(s)     & = & K_{IJ} s   \\
    t^{(1)}_{IJ}(s) & = & \left( B_{IJ}(\mu)+D_{IJ}\log\frac{s}{\mu^2}+E_{IJ}\log\frac{-s}{\mu^2}\right) s^2\ .
\end{eqnarray}
where $B_{IJ}(\mu)$ depends on the $a_4(\mu)$ and $a_5(\mu)$ renormalized chiral constants (in the following we will 
omit the $r$ superindex in the renormalized chiral couplings). The $t_{IJ}(s)$ independence on the renormalization scale
implies
\begin{equation} \label{Bruns}
B_{IJ}(\mu)=
 B_{IJ}(\mu_0)+(D_{IJ}+E_{IJ})\log\frac{\mu^2}{\mu_0^2}\  .
 \end{equation}
 
The values of the different parameters for all non-vanishing different $IJ$ channels up to NLO can be obtained easily 
from the $ww \to ww$ amplitudes \cite{Delgado:2013hxa} \\
 
Scalar-isoscalar $IJ=00$:
\begin{eqnarray}{}\label{partial00}
\nonumber   K_{00} & = & \frac{1}{16 \pi v^2} (1-a^2) \\
 \nonumber   B_{00}(\mu) & = &\frac{ 1}{9216 \pi^3 v^4}
  \left[101(1-a^2)^2 + 68(a^2-b)^2 + 
   768 \{7 a_4(\mu) + 11 a_5(\mu)\}\pi^2\right]\\
 \nonumber   D_{00} & = & -\frac{1}{4608\pi^3v^4}
  \left[7(1-a^2)^2 + 3(a^2-b)^2\right]\\
 E_{00} & = & -\frac{1}{1024\pi^3v^4} \left[4(1-a^2)^2  + 3(a^2-b)^2\right]\ .
\end{eqnarray}
Vector-isovector $IJ=11$:
\begin{eqnarray}{}\label{partial11}
\nonumber   K_{11} & = & \frac{1}{96 \pi v^2} (1-a^2) \\
 \nonumber   B_{11}(\mu) & = & \frac{1}{110592\pi^3 v^4}
  \left[8(1-a^2)^2 - 75(a^2-b)^2 + 
   4608 \{a_4(\mu) - 2 a_5(\mu)\} \pi^2 \right] \\
 \nonumber   D_{11} & = & \frac{1}{9216\pi^3v^4}
  \left[(1-a^2)^2 + 3(a^2-b)^2\right] \\
  E_{11} & = & -\frac{1}{9216\pi^3v^4}  (1-a^2)^2\ .
\end{eqnarray}
Scalar-isotensor $IJ=20$:
\begin{eqnarray}{}\label{partial20}
\nonumber   K_{20} & = & -\frac{1}{32 \pi v^2} (1-a^2) \\
 \nonumber   B_{20}(\mu) & = & \frac{1}{18432 \pi^3 v^4}
 \left[91(1-a^2)^2 + 28(a^2-b)^2 +
  3072 \{2 a_4(\mu) + a_5(\mu)\} \pi^2 \right] \\
 \nonumber   D_{20} & = & -\frac{1}{9216\pi^3v^4}
  \left[11(1-a^2)^2 + 6(a^2-b)^2\right]\\
  E_{20} & = & -\frac{1}{1024\pi^3v^4}(1-a^2)^2  \ 
\end{eqnarray}
Tensor-isoscalar $IJ=02$:
\begin{eqnarray}{}\label{partial02}
\nonumber   K_{02} & = & 0 \\
 \nonumber   B_{02}(\mu) & = & \frac{1}{921600 \pi ^3 v^4}
 \left[320(1-a^2)^2 + 77(a^2-b)^2 +
  15360\{2 a_4(\mu) + a_5(\mu)\} \pi^2 \right] \\
 \nonumber   D_{02} & = & -\frac{1}{46080\pi^3v^4}
  \left[10(1-a^2)^2 + 3(a^2-b)^2\right] \\
  E_{02} & = & 0\ .
\end{eqnarray}
And tensor-isotensor $I=J=2$: 
\begin{eqnarray}{}\label{partial22}
\nonumber   K_{22} & = & 0 \\
 \nonumber   B_{22}(\mu) & = & \frac{1}{921600 \pi ^3 v^4}
 \left[71(1-a^2)^2 + 77(a^2-b)^2 +
  7680\{a_4(\mu) + 2 a_5(\mu)\} \pi^2 \right] \\
 \nonumber   D_{22} & = & -\frac{1}{46080\pi^3v^4}
  \left[4(1-a^2)^2 + 3(a^2-b)^2\right] \\
  E_{22} & = & 0\ .
\end{eqnarray}
Partial waves with angular momentum $J=3$ and higher are at least  $O(s^3)$ and vanish at NLO. Other
combinations are impossible when custodial and Bose  symmetry are taken into account, which forbids various couplings that would be
allowed otherwise.

The partial waves for the  $ww \rightarrow hh$ and $hh \rightarrow hh$ processes are computed in a similar way but now  
only in the isospin zero channel $I=0$ since the Higgs $h$ is an isosinglet. 

The proper normalization  of the $ww$  isoscalar state $I=0$ introduces a  $1/\sqrt{3}$ factor and the sum over 
the three contributing charge combinations $(+-,-+,00)$ a factor 3, so that for the isoscalar amplitude is  
$M_0(w w  \rightarrow h h)=\sqrt{3} M(s,t,u)$. Then the partial waves have again the general  form
\be
   m_{J}(s)      =  K'_J s+  \left( B'_J(\mu)+D'_J\log\frac{s}{\mu^2}+E'_J\log\frac{-s}{\mu^2}\right)s^2+ \dots 
\ee
and the non-vanishing parameters up to NLO are \\

Scalar-isoscalar $IJ=00$:
\begin{eqnarray}{} \label{Mscalar}
\nonumber   K'_{0} & = & \frac{\sqrt{3}}{32 \pi v^2} (a^2-b) \\
 \nonumber   B'_{0}(\mu) & = & \frac{\sqrt{3}}{16\pi v^4} \left[d(\mu)+\frac{e(\mu)}{3}\right]
    +\frac{\sqrt{3}}{18432\pi^3 v^4}(a^2-b)\left[72(1-a^2) + (a^2-b)\right]    \\
 \nonumber   D'_{0} & = & - \frac{\sqrt{3}(a^2-b)^2}{9216\pi^3 v^4}   \\  
  E'_{0} & = &  -\frac{\sqrt{3}(a^2-b)(1-a^2)}{512\pi^3 v^4}                                   
\end{eqnarray}

Tensor-isoscalar    $IJ=02$:

\begin{eqnarray}{} \label{Mtensor}
\nonumber   K'_{2} & = & 0 \\
 \nonumber   B'_{2}(\mu)& = & \frac{e(\mu)}{160\sqrt{3}\pi v^4}  +\frac{83(a^2-b)^2}{307200\sqrt{3}\pi^3 v^4}    \\
 \nonumber   D'_{2} & = & - \frac{(a^2-b)^2}{7680\sqrt{3}\pi^3 v^4}.\\ 
  E'_{2} & = &  0\  .                                
\end{eqnarray}
For the $hh \to hh$ channel $T_0(h h \rightarrow h h)= T(s,t,u)$ and the partial waves $a_J$ have again the same form
\be
 a_{J}(s)      =  K''_J s  +  \left( B''_J(\mu)+D''_J\log\frac{s}{\mu^2}+E''_J\log\frac{-s}{\mu^2}\right) s^2+\dots
\ee
with the scalar-isocalar channel $IJ=00$:
\begin{eqnarray}{}\label{Tscalar}
\nonumber   K''_{0} & = & 0 \\
 \nonumber   B''_{0}(\mu) & = & \frac{10g(\mu) }{96\pi v^4}   +  \frac{(a^2-b)^2}{96\pi^3 v^4}   \\
 \nonumber   D''_{0} & = & - \frac{(a^2-b)^2}{512\pi^3 v^4}  \\ 
  E''_{0} & = &  -   \frac{3(a^2-b)^2}{1024\pi^3 v^4}   
\end{eqnarray}{}  
and the tensor-isoscalar $IJ=02$:
\begin{eqnarray} \label{Ttensor}
\nonumber   K''_{2} & = & 0 \\
 \nonumber   B''_{2}(\mu) & = & \frac{g(\mu) }{240\pi v^4}   +  \frac{77(a^2-b)^2}{307200\pi^3 v^4}   \\
 \nonumber   D''_{2} & = &  -\frac{(a^2-b)^2}{5120\pi^3 v^4}  \\ 
  E''_{2} & = &    0 \ .
\end{eqnarray}         

Finally, for the $wh \to wh$ reaction we only have the isovector-(axial)vector channel $I=J=1$ up to NLO. 
The partial-wave is defined as
 \begin{equation}
n_{11}(s)= \frac{1}{32\pi}\int_{-1}^1 x \, M(s,t,u) \, dx.
\end{equation}
Notice that we have now a divisor 32 instead of 64 since the particles in the initial and final states are not identical.
In any case the partial wave have the same form
\be
   n_{11}(s)      =  K'''_{11}s+  \left( B'''_{11}(\mu)+D'''_{11}\log\frac{s}{\mu^2}+E'''_{11}\log\frac{-s}{\mu^2}\right)s^2+ \dots 
\ee
with \cite{Dobado:2017lwg}:
\begin{eqnarray} \label{constsAmp}
K'''_{11} & = & \frac{a^2-b}{96 \pi v^2}    \\   \nonumber
B'''_{11}(\mu) & = & \frac{e^r(\mu)-2d(\mu)}{96 \pi v^4}  -\frac{a^2-b}{110592 \pi^3 v^4}\left(150(1-a^2)-83(a^2-b)\right)       \\   \nonumber
D'''_{11} & = & \frac{a^2-b}{4608 \pi^3 v^4}\left(3(1-a^2)-(a^2-b)\right)       \\   \nonumber
E'''_{11}& = & -\frac{(a^2-b)^2}{9216 \pi^3 v^4}      \    .
\end{eqnarray}
As it can be seen in all the cases the functions $B'(\mu)$, $B''(\mu)$ and $B'''(\mu)$ are in all analogous to
$B(\mu)$ depending on different combinations of the renormalized chiral couplings and with similar evolution equations 
given in Eq.\ref{Bruns}. By using these evolution equations it is possible to check that all the  partial waves above 
are $\mu$ independent.

%%%%%%%%%%%%%%%%%%%%%%%%%%%%%%%%%%%%%%%%%%%%%%%%%%%%%%%%%%%%%%%%%%%%%%%%%%%%%%%%%%%%%%%%%%%%%%%%%%%%%%%%%%%%%%%%%%%%%%%%%%%%%
%%%%%%%%%%%%%%%%%%%%%%%%%%%%%%%%%%%%%%%%%%%%%%%%%%%%%%%%%%%%%%%%%%%%%%%%%%%%%%%%%%%%%%%%%%%%%%%%%%%%%%%%%%%%%%%%%%%%%%%%%%%%%

\section{ Analyticity and unitarity}
The partial waves $t_{IJ}(s), m_J(s)$, $a_J(s)$ and $n_{11}(s)$  have to fulfill a number of properties coming from 
well established  first principles of QFT. In particular (micro)causality and probability conservation 
lead to analiticity and unitarity, which set important conditions on the mathematical form  on the partial 
waves\cite{Delgado:2015kxa}. First  of all, the partial waves are analytical functions of the complex Maldelstam-$s$  and  
must have an unitarity or right cut (RC)  along the positive real axis starting at the threshold at $s=0$. The physical 
values are then obtained just on the upper side of this cut $s=E_{CM}^2+i \epsilon$. Also the $\cos\theta$ integration appearing in 
the partial-wave definition (\ref{Jprojection} gives rise to a left cut (LC) on the negative part of the real axis. 
For physical $s$ values 
unitarity requires a set of non-trivial relations between the different partial waves. Because of the angular momentum and weak isospin  
symmeties, the reaction matrix is block-diagonal 
\be
T(s)=\begin{pmatrix}
T_{00} & 0&0&0&0 \\
0& T_{02} &0&0&0 \\
0&0&T_{11} & 0&0 \\
0&0& 0&T_{20} & 0 \\
0 & 0&0&0&... \\
\end{pmatrix},
\ee
where $T_{IJ}(s)$ are the partial-waves matrices. For example, for $I=0$ and $J=0,2$ we have
\be
T_{0J}(s)=\begin{pmatrix}
t_{0J}(s) & m_J(s) \\
m_J(s)& a_J(s) \\
\end{pmatrix}.
\ee
For $I \ne 0$ there is no mixing with the $hh$ channel so that
\be
 T_{IJ}(s)=t_{IJ}(s)
\ee
i.e. the reaction matrices have just one single element.
Now unitarity requires
\be
\Imag T(s) = T(s)T^\dagger(s).
\ee
in  the physical region, i. e. on the RC.

From this equation it is possible to obtain a set of relations involving different partial waves. 
For $I=0$ and $J=0$ or $J=2$ we have
\begin{eqnarray}{}
 \Imag t_{0J} &  = & \lvert t_{0J} \rvert ^2 +  \lvert m_J \rvert ^2 \\
 \nonumber    \Imag m_J &  = & t_{0J} m_{J}^*+  t_J a_J^* \\
 \nonumber    \Imag a_J &  = & \lvert m_J \rvert ^2+  \lvert a_J \rvert ^2\   .
\end{eqnarray}         
However these relations are not respected by perturbation theory. For example, at the one-loop level one has
\begin{eqnarray}{}
\nonumber \Imag t^{(1)}_{0J} &  = &\lvert t^{(0)}_{0J}\rvert^2+  \lvert m^{(0)}_{J}\rvert ^2 \\
 \nonumber   \Imag m^{(1)}_{J} &  = & a^{(0)}_{0J} m^{(0)}_{J}+ 
 M^{(0)}_{J}T^{(0)}_{J} \\
 \nonumber    \Imag t^{(1)}_{J} &  = & \lvert m^{(0)}_J\rvert ^2 +  \lvert a^{(0)}_{J}\rvert^2   
\end{eqnarray}    
thus indicating a breakdown of unitarity at high energies. This fact sets very strong limitations to the applicability 
of pertubative results in the interesting physical region around or higher one TeV of center of mass energy $E_{CM}$.
     
For the elastic $ww \rightarrow ww$  channels with $I=J=1$ and $I=2$, $J=0$ the unitarity condition is just
\be \label{unitarity1channel}
 \Imag t_{I J}   =  \lvert t_{IJ} \rvert ^2  \ \ \ I\neq 0
\ee
and at the one-loop or NLO  level
\be
 \Imag t^{(1)}_{I J}   =  \lvert t^{(0)}_{IJ} \rvert ^2  \ \ \ I\neq 0\ .
\ee

Due to the manifest loss of unitarity in the TeV region for many processes, if one solely relies on perturbation
theory, the experimental bounds that can be derived from experiment concerning higher dimensional operators that typically 
give rise to contributions that violate unitarity pretty quickly are potentially overestimated. Sometimes severely
overestimated.  To avoid this pitfall one must make comparison with experiment using amplitudes that do exhibit 
a unitary behavior. Because after all we know that the fundamental theory that gives rise to those low-energy coefficients
{\em is} unitary. Several methods to restore unitarity will be presented in the coming sections.

%%%%%%%%%%%%%%%%%%%%%%%%%%%%%%%%%%%%%%%%%%%%%%%%%%%%%%%%%%%%%%%%%%%%%%%%%%%%%%%%%%%%%%%%%%%%%%%%%%%%%%%%%%%%%%
%%%%%%%%%%%%%%%%%%%%%%%%%%%%%%%%%%%%%%%%%%%%%%%%%%%%%%%%%%%%%%%%%%%%%%%%%%%%%%%%%%%%%%%%%%%%%%%%%%%%%%%%%%%%%%

\section{The Inverse Amplitude Method}
Let us concentrate on the following in these elastic channels. The fact that the partial waves are analytic functions 
of the variable $s$ in the whole complex plane makes it possible to write the dispersion relation (DR)
\be
t(s)=K s+\frac{s^2}{\pi}\int_0^\infty  \frac{ds' \Imag t(s')}{s'^2(s'-s-i\epsilon)} +\frac{s^2}{\pi}\int_{-\infty}^{0}  
\frac{ds' \Imag t(s')}{s'^2(s'-s-i\epsilon)}+
\frac{s^2}{2 \pi i}\int_{C_\infty}  \frac{ds'  t(s')}{s'^2(s'-s)}
\ee
where $C_\infty$ is the circunference at the infinite oriented anticlockwise and we have omitted the $I$ and $J$ 
indices for simplicity \cite{Dobado:1997jx,Dispers2}.

On the other hand, the  DR for the partial-wave amplitude expanded up to NLO, that is, truncated 
up to order $s^2$, $t^{\rm NLO}(s)=t^{(0)}(s) +t^{(1)}(s)$, can be written as
\be \label{disprelNLO}
t^{\rm NLO}(s)= Ks + \frac{s^2}{\pi}\int_0^{\Lambda^2}  \frac{ds' \Imag t^{(1)}(s')}{s'^2(s'-s-i\epsilon)} 
+\frac{s^2}{\pi}\int_{-\Lambda^2}^0  \frac{ds' \Imag t^{(1)}(s')}{s'^2(s'-s-i\epsilon)}+
\frac{s^2}{2 \pi i}\int_{C_\Lambda}  \frac{ds'  t^{(1)}(s')}{s'^2(s'-s)}.
\ee
where $C_{\Lambda}$ is a circumference of radius $\Lambda^2$  and $\Lambda$ is an UV regulator. Then the integrals 
can be easlly computed
\begin{eqnarray}
 \frac{1}{\pi}\int_0^{\Lambda^2}  \frac{ds' \Imag t^{(1)}(s')}{s'^2(s'-s-i\epsilon)}  & = &  E \log\frac{-s}{\Lambda^2} \nonumber \\
 \frac{1}{\pi}\int_{-\Lambda^2}^0  \frac{ds' \Imag t^{(1)}(s')}{s'^2(s'-s-i\epsilon)}  & = &  D \log\frac{s}{\Lambda^2} \nonumber \\
\frac{1}{2 \pi i}\int_{C_\Lambda}  \frac{ds'  t^{(1)}(s')}{s'^2(s'-s)} & = &  B(\mu)+D\log\frac{\Lambda^2}{\mu^2}+E\log\frac{\Lambda^2}{\mu^2}
\end{eqnarray}     
so that the dispersion relation for $t^{\rm NLO}(s)$ in Eq.~(\ref{disprelNLO})
reproduces Eq.~(\ref{expandpartialwave}). Notice the consistency check of the dispersion relation and the interplay 
with renormalized chiral couplings, since the integral over the circle trades the UV-cutoff scale $\Lambda$ by the 
renormalization scale $\mu$.

The Inverse Amplitude Method (IAM)~\cite{Truong:1988zp} provides a way to improve perturbation theory in order to 
fulfill the unitarity and analytical constraints exactly.  It was developed for ordinary ChPT for mesons~\cite{Dobado:1992ha} 
and later applied to the unitarization of the one-loop would-be NGB scattering amplitudes (see \cite{DHD} and third 
reference in~\cite{Truong:1988zp}) long before the Higgs was discovered.

In order to see how it works one considers the auxiliary function
\be \label{gdef}
  W(s)\equiv \frac{(t^{(0)}(s))^2}{t(s)}\ .
\ee 
By construction it has the  same analytic structure as $t(s)$ up to possible poles coming from $t(s)$ zeroes. 
Also one has $W(s)= K s + O(s^2)$ and
\be
\Imag W(s)=-[t^{(0)}(s)]^2
\ee
on the RC. Neglecting the possible pole contribution an excellent approximation is:
\be
W(s)=K s+\frac{s^2}{\pi}\int_0^{\Lambda^2}  \frac{ds' \Imag W(s')}{s'^2(s'-s-i\epsilon)} 
+\frac{s^2}{\pi}\int_{-\Lambda^2}^{0}  \frac{ds' \Imag W(s')}{s'^2(s'-s-i\epsilon)}  
+ \frac{s^2}{2 \pi i}\int_{C_\Lambda}  \frac{ds' W (s')}{s'^2(s'-s)}\ .
\ee
Now the essential point is that the definition of $W(s)$ in Eq.~(\ref{gdef}) makes it  possible to compute the 
unitarity (RC) integral exactly since $\Imag W(s)=-K^2s^2= E \pi s^2$ there. The LC integral cannot be computed 
exactly but  it can be computed perturbatively using
\be
 \Imag W(s) \simeq - \Imag t^{(1)}(s) .
\ee
to find
\be
 W(s) \simeq K s-D s^2\log\frac{s}{\Lambda^2}-E s^2\log\frac{-s}{\Lambda^2} 
+ \frac{s^2}{2 \pi i}\int_{C_\Lambda}  \frac{ds'  W(s')}{s'^2(s'-s)}  .      
\ee
This equation is trivially solved by $W(s)=t^{(0)}(s)-t^{(1)}(s)$ which is quite  remarkable since $W(s)$ in 
Eq.\ref{gdef} is defined from the exact partial  wave $t(s)$.  Therefore one finally has the IAM amplitude written  as
\be \label{IAM1channel}
 t(s)\simeq t^{\rm IAM}(s) = \frac{(t^{(0)}(s))^2}{t^{(0)}(s)-t^{(1)}(s)} \ .
\ee

This amplitude obtained with the IAM has a number of interesting properties which are important to remember:
\begin{itemize}

\item 
First, it has the proper analytical structure with the right cuts. In particular unitarity on the RC produces a 
second Riemann sheet. Poles in this so-called unphysical sheet have a natural interpretation as dynamically generated 
resonances. However poles in the first or physical sheet (ghosts) are not allowed and they must be understood 
as artifacts of the model if they appear.

\item 
The IAM partial waves are obtained entirely from the one-loop or NLO approximation. Thus the results depend only on 
the renormalized chiral couplings and are UV and  IR finite and renormalization scale independent.

\item 
They  satisfy exact elastic unitarity on the RC,
\be
 \Imag  t^{\rm IAM} = t^{\rm IAM}(t^{\rm IAM})^*\ .
 \ee

\item 
When expanded at low energy, they match the NLO amplitude
\be
t^{\rm IAM}(s) =t^{\rm NLO}(s) + O(s^3). 
\ee

\item 
Finally, the IAM method can be extended to the coupled channel case too 
\cite{JRA} particularly if  the particles appearing in the different channels are all of them massless to 
avoid overlapping left and right cuts. This is the case here since we are considering the NGB and the $h$ particle as  
massless. From the perturbative expansion
\be
T_{IJ}=T_{IJ}^{(0)}+T_{IJ}^{(1)}+\dots \\ 
\ee
a natural generalization of the IAM method gives
\be \label{IAMforF}
 T_{IJ}^{\rm IAM}=T_{IJ}^{(0)}(T_{IJ}^{(0)}-T_{IJ}^{(1)})^{-1}T_{IJ}^{(0)}
\ee
which satisfies exact multichannel elastic unitarity on the RC
\be \label{multichannelUnit}
\Imag  T_{IJ}^{\rm IAM} = T_{IJ}^{\rm IAM}(T_{IJ}^{\rm IAM})^\dagger .
\ee
In addition the matrix elements of $T_{IJ}^{\rm IAM}$ enjoy all the already mentioned desirable properties of the elastic IAM. 
This coupled-channel IAM is particularly useful in the isoscalar channels where the $\omega\omega$ and $hh$ pairs can 
be strongly coupled.
\end{itemize}

The IAM has been extensively used to describe the low-energy meson-meson scattering where it has proven to be extremely 
successful. With a very small set of parameters, it is able to  describe many different channels including their first 
resonances \cite{JRA}.

%%%%%%%%%%%%%%%%%%%%%%%%%%%%%%%%%%%%%%%%%%%%%%%%%%%%%%%%%%%%%%%%%%%%%%%%%%%%%%%%%%%%%%%%%%%%%%%%%%%%%%%%%%%%%%%%%%%%%%%%%%%
%%%%%%%%%%%%%%%%%%%%%%%%%%%%%%%%%%%%%%%%%%%%%%%%%%%%%%%%%%%%%%%%%%%%%%%%%%%%%%%%%%%%%%%%%%%%%%%%%%%%%%%%%%%%%%%%%%%%%%%%%%%

\section{The N/D method}
There are other possibilities for implementing full 
unitarity and analyticity in the one-loop results. In particular a well-known alternative that we also consider here is the 
N/D method \cite{N/D}. There are many ways for applying it depending on the problem at hand. Here we will follow an approach 
which is particularly suited to perturbation theory. To start with we will consider first the elastic $w w$ scattering. 
The main assumption of the N/D method is that the corresponding partial waves can be written as
\be
t(s)=\frac{N(s)}{D(s)}\
\ee
where the numerator function $N(s)$ has only a LC and the denominator function $D(s)$ only a RC in such a way that  
$t(s)$ has the expected analytical structure. Thus $\Imag N(s)=0$ on the RC and $\Imag D(s)=0$ on the LC. 

Now elastic  unitarity requires $\Imag D(s)=-N(s)$ on the RC and we also have $\Imag N(s)=D(s)\Imag A(s)$ on the LC. 
It is then possible in principle to write two coupled dispersion relations for $N(s)$ and $D(s)$. In the simplest case, 
using the normalization $D(0)=1$ one has
\begin{eqnarray}\label{NsobreD}
D(s) & = & 1- \frac{s}{\pi}\int_0^\infty  \frac{ds' N(s')}{s'(s'-s-i\epsilon)}\\ \label{NsobreD2}
N(s) & = & \frac{s}{\pi}\int_{-\infty}^{0}  \frac{ds' D(s') \Imag A(s')}{s'(s'-s-i\epsilon)}   \ .
\end{eqnarray}
More generally, one may need  DR with a number of subtractions.

One possibility for exploiting  the above coupled equations for $N(s)$ and $D(s)$ is using some recursive method. 
For example, starting from some approximate $N_0(s)$ function featuring a LC (for example a tree level result) 
we can obtain $D_0(s)$ by integration on the RC. 
Then a first approximation for the partial wave would be $t_0(s)=N_0(s)/D_0(s)$. To continue the procedure one can 
now insert $D_0(s)$ in the second coupled equation to get the new a $N_1(s)$ yielding $t_1(s)=N_1(s)/D_1(s)$  and so on. 
Under some conditions depending on the $N_0(s)$ choice and  not to be discussed here, this procedure will converge to a 
solution of the DR.  Even more in many cases the simplest approximation $t(s) \simeq N_0(s)/D_0(s)$ could be considered appropriate enough. 

However, introducing the NLO results of HEFT in the N/D method is far from trivial for various reasons. 
If one wants to use the N/D method starting from perturbation theory the first problem to solve is how to choose $N_0(s)$ 
in a renormalization scale invariant way. To solve this problem we split $t^{(1)}(s)$ in two pieces, one having only 
a RC and the other only a LC and being both $\mu$-independent. Thus we define
\begin{eqnarray}\label{splitLR}
\nonumber             t_L(s)      & \equiv & \left(\frac{B(\mu)}{D+E}+\log\frac{s}{\mu^2}\right) D s^2 \\
                       t_R(s)      & \equiv & \left(\frac{B(\mu)}{D+E}+\log\frac{-s}{\mu^2}\right) E s^2   .
\end{eqnarray}
Obviously  $t^{(1)}(s)= t_L(s)+t_R(s)$  and it is not difficult to check that  both, $t_L$ and $t_R$, are renormalization 
scale ($\mu$) independent.  In addition, on the RC (the physical region), perturbative unitarity reads $\Imag  t^{(1)} =\Imag t_R= (t^{(0)})^2$.
Another useful definition is
\be \label{defgwithB}
g(s)=\frac{1}{\pi}\left(\frac{B(\mu)}{D+E}+\log\frac{-s}{\mu^2}\right)    \ .
\ee
This function is $\mu$-independent and analytical on the whole complex plane but for a RC. On this RC (i.e. for $s=E_{CM}^2+i\epsilon$) 
we have $\Imag g(s)= -1$.  Then it is not difficult to find the perturbative expansion
\be
t(s)=t^{(0)}(s)+t_L(s)-(t^{(0)}(s))^2g(s)+O(s^3) \ .
\ee
Now we can apply perturbation theory to the $N/D$ method. Our starting point is
\be \label{numeratordef}
N_0(s)\equiv t^{(0)}(s)+t_L(s) \ . 
\ee
This function has a LC but no RC, it has  information about the chiral couplings  and, finally,  it is $\mu$ independent. 
Now to obtain a UV-finite integral three subtractions are required. Thus we can write
\be
D_0(s) = 1+h_1s+h_2s^2- \frac{s^3}{\pi}\int_0^\infty \frac{ds' [t^{(0)}(s)+t_L(s)]}{s'^3(s'-s-i\epsilon)}.
\ee
The N/D partial wave in lowest order approximation can be written as
\be\label{N/D_A}
t(s) \simeq t^{\rm N/D}(s)  = \frac{N_0(s)}{D_0(s)}. 
\ee
Then it is possible to show that an appropriate election of the constants $h_1$ and $h_2$ allows us to get \cite{Delgado:2015kxa}:
\be\label{N/D_D0}
D_0(s)=1-\frac{t_R(s)}{t^{(0)}(s)}+\frac{\pi}{2} (g(s)^2 D s^2=1-\frac{t_R(s)}{t^{(0)}(s)}-\frac{t_L(-s)t_R(s)}{2(t^{(0)}(s))^2}.
\ee

This amplitude is UV and IR finite,  it is $\mu$ independent,  it has the right analytical structure, 
it satisfies elastic unitarity exactly
\be
\Imag t^{\rm N/D}(s) = \lvert t^{\rm N/D}(s) \rvert^2
\ee
on the RC and finally it is compatible with the NLO computation since
\be
t^{\rm N/D}(s) =t^{(0)}(s)+t^{(1)}(s)+O(s^3).
\ee
Thus these properties are the same that we found for  the IAM. In fact it is possible to  show that this amplitude 
converges to the IAM amplitude whenever $t_L \ll t^{(0)}$.

As for the IAM it is possible to generalize the N/D method to the multichannel case needed for the $I=0$ ($J=0, 2$) 
processes where the $w w $ state couples to the $hh$ channel. According to \cite{Bjorken} we write the matrix relation
\be
T_{IJ}(s)= [D_{IJ}(s)]^{-1}N_{IJ}(s)\ .
\ee 
To generalize our previous result for the single channel case, we start again from the perturbative expansion at NLO. 
By suppressing the $I$ and $J$ indices and by using an obvious matrix notation we define 
\be
N_0(s)=T^{(0)}(s)+T_L(s)
\ee
and then it is possible to find the DR solution
\be
D_0(s)=
1- T_R(s)[T^{(0)}(s)]^{-1}- \frac{1}{2}T_R(s) [T^{(0)}(s)]^{-2}T_L(-s) 
\ee
It is not difficult to check that the partial waves in this equation fulfill exact elastic coupled-channel unitarity on the RC,
\be
\Imag T^{\rm N/D}=T^{\rm N/D}\left(T^{\rm N/D}\right)^\dagger 
\ee
and reproduce the low-energy expansion up to the one-loop level
\be
T^{\rm N/D}(s)=T^{(0)}(s)+T^{(1)}(s)+...
\ee

The $T^{\rm N/D}(s)$ partial-wave amplitudes have all the required properties including unitarity and analyticity. 
They have a LC and RC and also they  can be extended to the second Riemann sheet.  For some regions of the coupling 
space some they  have poles there that can  be understood as dynamical resonances. 

%%%%%%%%%%%%%%%%%%%%%%%%%%%%%%%%%%%%%%%%%%%%%%%%%%%%%%%%%%%%%%%%%%%%%%%%%%%%%%%%%%%%%%%%%%%%%%%%%%%%%%%%%%%%%%%%%%%%%%%%%%%
%%%%%%%%%%%%%%%%%%%%%%%%%%%%%%%%%%%%%%%%%%%%%%%%%%%%%%%%%%%%%%%%%%%%%%%%%%%%%%%%%%%%%%%%%%%%%%%%%%%%%%%%%%%%%%%%%%%%%%%%%%%

\section{Comments on the K-matrix method}
Finally we will comment on another popular unitarization method known as the K-matrix method \cite{Kmatrix} (see also
\cite{Kilian:2014zja} for a recent review in the  context of this work). 
The K-matrix is defined in terms of the $S$ matrix as
\be\label{SofK}
S=\frac{1- i K/2}{1+i K/2}.
\ee
Obviously the unitarity of the  $S$ operator is equivalent to  $K$ Hermiticity. In practice the $S$ matrix is   
obtained in the form of some expansion:
$S= 1 + S^{(1)} +S^{(2)}+\dots$. The truncation of this expansion typically  produces an approximate $S$ matrix 
which is not unitary. However one can instead expand $K$ as $K=K^{(1)}+K^{(2)}+\dots\ $ and introduce this (truncated Hermitian) 
series into Eq.~(\ref{SofK}) to find a new series for $S$,
\be
S=1+ \tilde S^{(1)}+\tilde S^{(2)}+\dots\ ,
\ee
which is exactly unitary at any order. Normaly the K-matrix method is used in terms of partial waves. Given  some 
unspecified elastic process $t(s)$ one starts from some approximate estimation $t_0(s)$  real in the physical region and 
therefore not unitary (typically the tree level result). Then the K-matrix unitarized partial wave is
\be \label{oldKamplitude}
t_0^K(s)=\frac{t_0(s)}{1-i t_0(s)}.
\ee
In this way unitarity is satisfied again in the physical region,
\be
\Imag t_0^K= \left\lvert t_0^{K} \right\rvert^2 \ .
\ee

 However it is very important to stress that this K-matrix partial wave has not the appropriate analytical structure and 
consequently it is lacking the RC and the second Riemann (unphysical) sheet. Therefore it cannot produce poles that could be 
understood as resonances as we will see below.
 Previous experience in hadron physics clearly shows  one can describe hadronic resonances by using for example 
the IAM method \cite{Dobado:1992ha,JRA}. The original K-matrix method cannot reproduce these hadronic resonances and should 
be considered as less appropriate than other methods that are, not only unitary,  but also analytical, as is the 
case of the IAM or N/D methods. 

The K-matrix method can be improved as follows by using  the analytical function $g(s)$ defined above in the context 
of the N/D method. The proper RC is introduced by performing the formal substitution: $-i \rightarrow g(s)$ in the 
K-matrix method to get the so called improved K-matrix (IK) amplitude \cite{Delgado:2015kxa}
\be
t^{\rm IK}(s)  = \frac{t_0(s)}{1+g(s)t_0(s)}.
\ee
This new amplitude is, not only unitary, but also analytical on the whole complex plane but for a RC, which  allows 
for analytical continuation to the second Riemann sheet, making possible the existence of poles 
as in the IAM or N/D methods \cite{Barducci:2015oza}. However, this improved IK method can be considered as a particular instance of the N/D and therefore 
it will not be discussed anymore in this work. 

\begin{figure}
\includegraphics[width=0.32\textwidth]{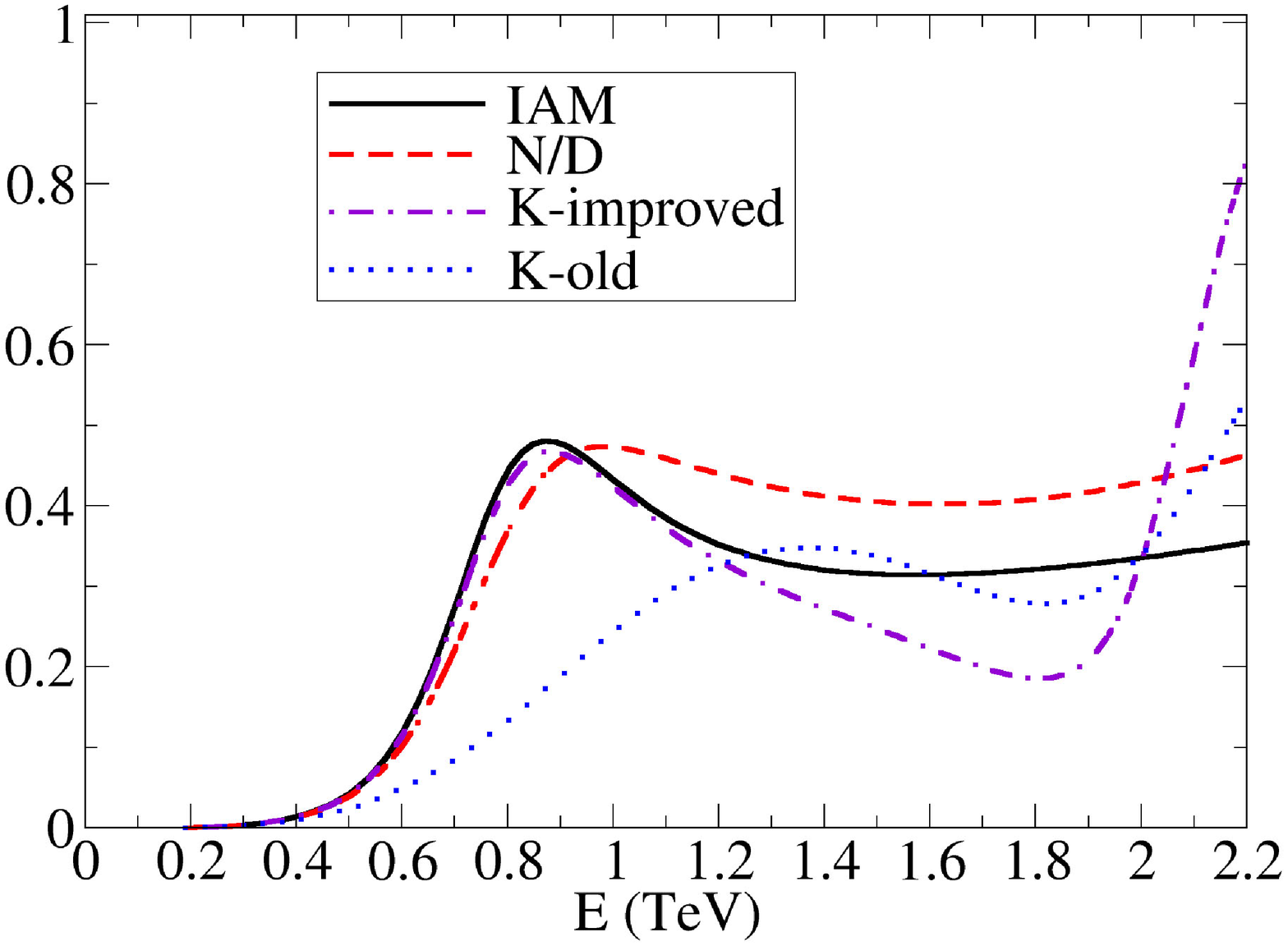}
\includegraphics[width=0.32\textwidth]{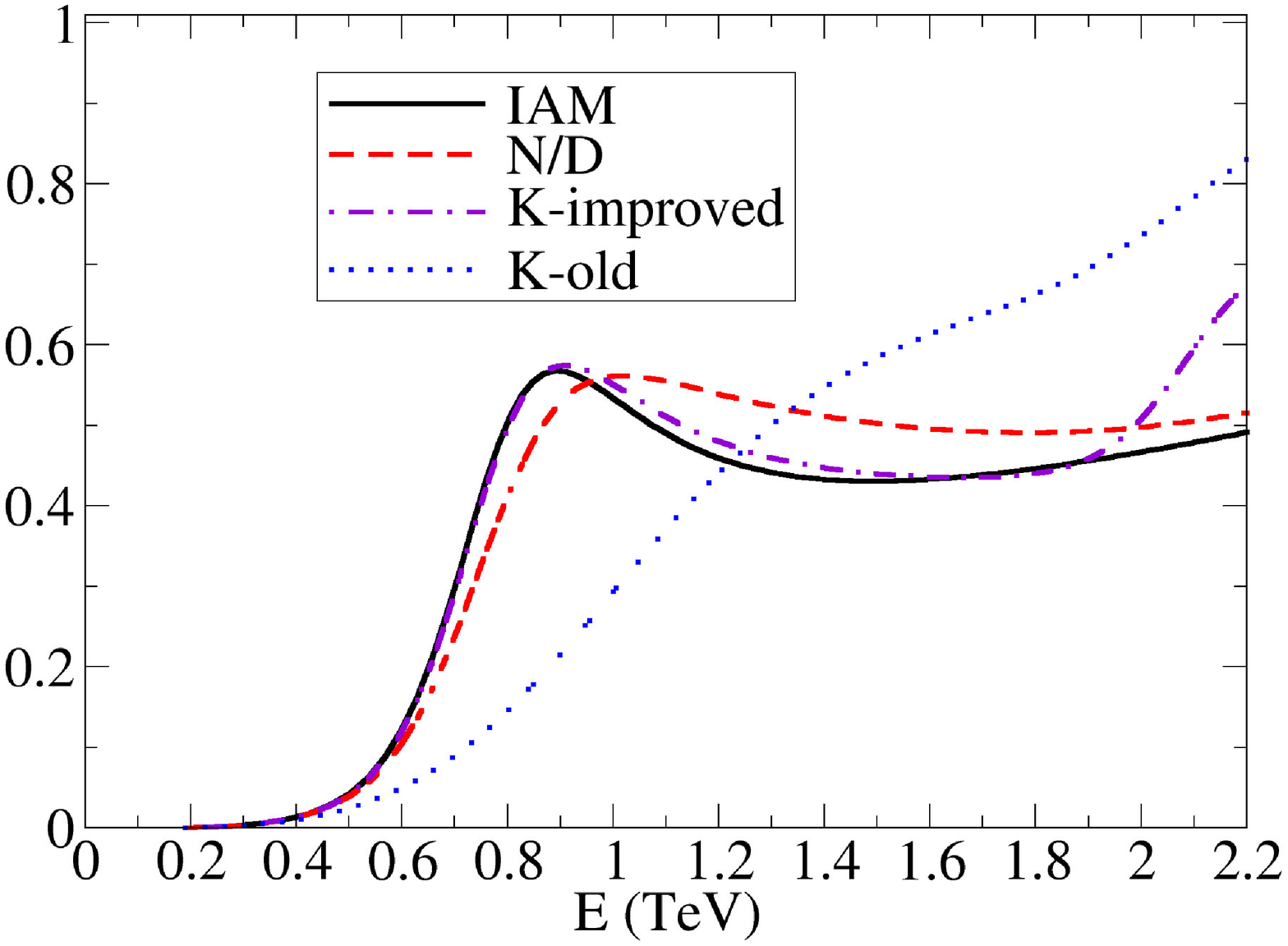}
\includegraphics[width=0.32\textwidth]{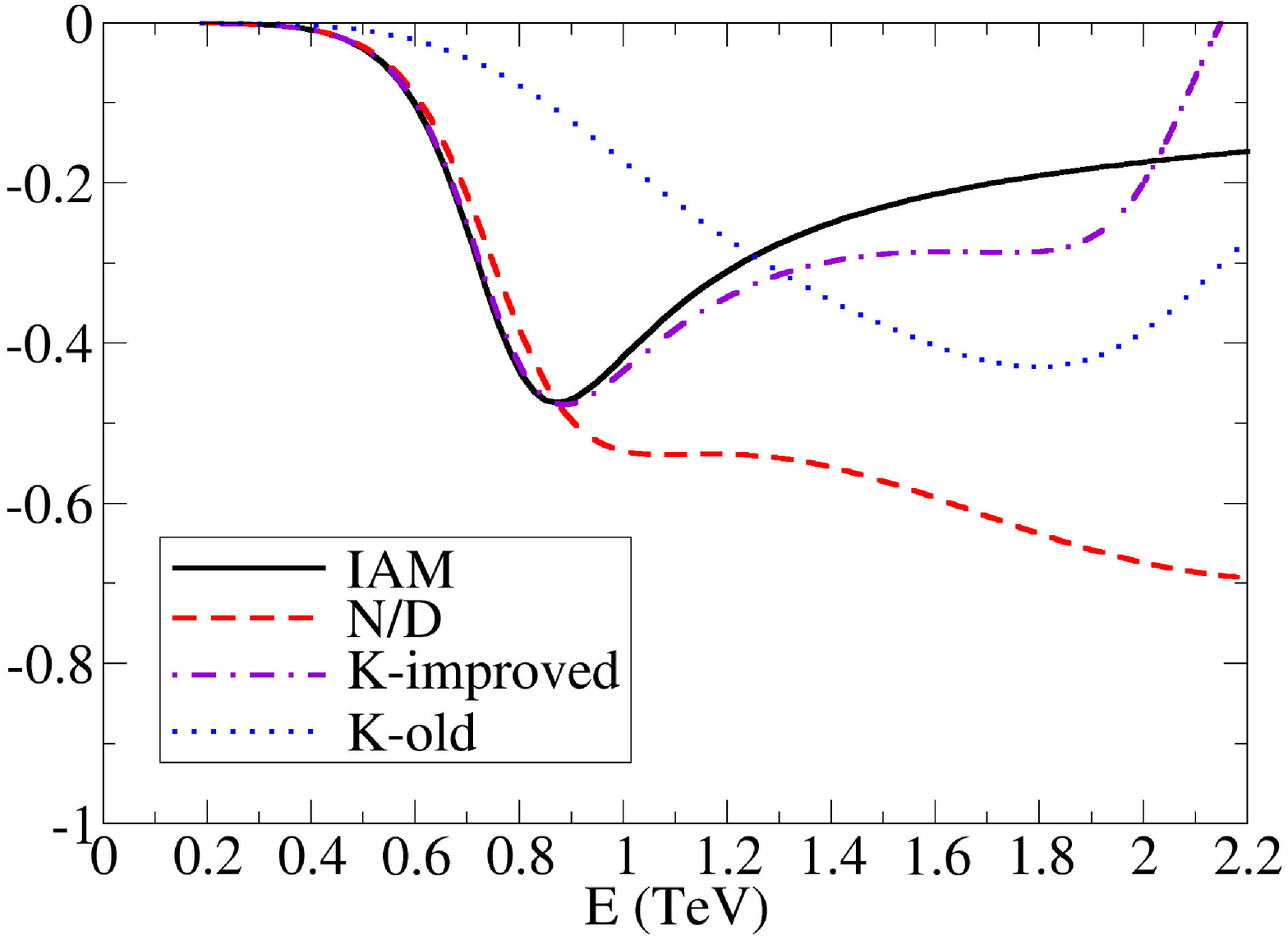}
\caption{\label{fig:compJ0}%
Resonant Scalar-isoscalar amplitudes (from left to right, elastic $ww$, elastic $hh$, and cross-channel $ww\to hh$), for $a=0.88$, $b=3$, and all NLO parameters set to 0 at a scale $\mu=3$ TeV. Note that, as explained in the text, the old K-matrix method gives different results because its complex-s plane analytic structure is not the correct one. The IAM, N/D and IK methods agree quite well for the three channels, at least up to the resonance energy. Figure taken from \cite{Delgado:2015kxa}. }
\end{figure}

In order to show an example of the agreement of the different unitarizations methods we show in   Fig.\ref{fig:compJ0} the $I=J=0$ amplitudes for parameter values giving rise to a scalar resonance at 0.9 TeV. The resonance appears for all the channels: elastic  $ww$, elastic $hh$ and cross-channel $ww\to hh$ for the IAM, N/D and IK methods. However the old K-matrix produces a completely different results because the lack of proper analytical behavior.

%%%%%%%%%%%%%%%%%%%%%%%%%%%%%%%%%%%%%%%%%%%%%%%%%%%%%%%%%%%%%%%%%%%%%%%%%%%%%%%%%%%%%%%%%%%%%%%%%%%%%%%%%%%%%%%%%%%%%%%%%
%%%%%%%%%%%%%%%%%%%%%%%%%%%%%%%%%%%%%%%%%%%%%%%%%%%%%%%%%%%%%%%%%%%%%%%%%%%%%%%%%%%%%%%%%%%%%%%%%%%%%%%%%%%%%%%%%%%%%%%%%

\section{Dynamically generated resonances}
One of the more interesting properties of the IAM and N/D  partial waves is the possibility of finding poles in the 
second or unphysical Riemann sheet just under the real axis for some regions of the parameter (chiral coupling) space. This 
is because those poles have a natural interpretation as dynamically generated resonances which are expected in any strongly 
interacting system such as the SBS of the SM model considered in this work. On the other hand it is clear that the non-trivial 
analytical behavior appearing around the physical region in our amplitudes (cuts, branches and sheets) is coming exclusively 
from the $\log(-s/\mu)$ factors. With the usual definition of the first Riemann sheet $log(z)=\log (\lvert z \rvert) + i\arg(z)$ 
(with the $\arg(z)$ cut lying along the negative real axis), the second Riemann sheet below the RC definition becomes
\be
\log^{II}(-z)=\log (\lvert z \rvert) + i[\arg(z)-\pi] .
\ee

This formula allows us to continue the IAM and the N/D partial waves to the second Riemann sheet and then to look for 
poles by locating zeroes in the corresponding denominators (see for example ~\cite{Delgado:2015kxa}). Another possibility 
is based in the realization  that, given some analytical elastic amplitude defined in principle on the first (physical) 
Riemann sheet $t(s)=t^{I}(s)$, the second Riemann sheet in the region below the physical region can be obtained as \cite{Gribov}
\be
\label{anapro}
t^{II}(s)=\frac{t(s)}{1- 2 i t(s)}.
\ee 
Therefore the position of the resonances under the real physical $s$ axis (RC) are located at points $s_R$ solving the  equation
\be
\label{resoneq}
t(s_R)+\frac{i}{2}=0.
\ee
Now we can define the mass $M$ and the width $\Gamma$ of the corresponding resonance as $s_R=M^2- i \Gamma M$ which 
implies $s_R= \lvert s_R \rvert e^{-i \theta}$ with $\theta > 0$ and $\tan\theta = \gamma = \Gamma/M$. Clearly the 
particular form of the resonance Eq.\ref{resoneq} depends on the 
unitarization method. Thus one has  for the IAM method
\be \label{IAMRES}
t^{(0)}(s_R)-t^{(1)}(s_R)- 2 i (t^{(0)}(s_R))^2=0
\ee
and for the N/D method:
\be
t^{(0)}(s_R)-t_R(s_R)+ \frac{1}{2}g(s_R)t^{(0)}(s_R)t_L(-s_R)- 2 it^{(0)}(s_R) (t^{(0)}(s_R)+t_L(s_R))  =0.
\ee
Both resonance equations are renormalization scale ($\mu$) independent. Also, as expected, they are different, but they 
converge in the  limit $t_L(s_R) \ll 1$, since $t^{(1)}(s_R)=t_R(s_R)+t_L(s_R)$. 

If we find a solution $s_R$ for some given channel $IJ$ and some given unitarization method  (IAM or  N/D) in the appropriate 
region $M, \Gamma > 0$, this solution will be a $\mu$ invariant function of the $a, b$ and the renormalized chiral parameters, i.e.
\be
M  =  M(a,b, a_4(\mu),  a_4(\mu), d(\mu), e(\mu), g(\mu) ; \mu) 
\ee
and
\be
\Gamma  =  \Gamma (a,b, a_4(\mu),  a_5(\mu), d(\mu), e(\mu), g(\mu) ; \mu).
\ee 
Then these functions must fulfill the renormalization group equations
\be
\frac{d M}{d  \mu} = \frac{\partial  M}{\partial  \mu}+\frac{\partial  M}{\partial  a_4}\frac{d a_4}{d  \mu}
+\frac{\partial  M}{\partial  a_5}\frac{d a_5}{d  \mu}+... =0  
\nonumber
\ee
and
\be
\frac{d \Gamma}{d  \mu} = \frac{\partial  \Gamma}{\partial  \mu}+\frac{\partial  \Gamma}{\partial  a_4}\frac{d a_4}{d  \mu}
+\frac{\partial  \Gamma}{\partial  a_5}\frac{d a_5}{d  \mu}+...=0.
\ee

In the case of coupled channels, the different amplitude matrix elements (partial waves) $(T_{IJ})_{ij}(s)$ correspond 
to different reactions having the same quantum numbers $IJ$. Clearly, if there is a resonance in one of the channels it should appear 
also in all the others since physically these resonances can be produced in any of the $j \rightarrow i$ reactions. 
It is very easy to check that this is precisely the case in the IAM and N/D unitarization methods.  The reason is that both 
of them require the inversion of some matrix. Thus they always contain a common denominator factor which is a determinant. 
The roots of these determinants in the second Riemann sheet will define the resonance position common for all 
the different processes having the same $I$ and $J$.

In addition to the poles in the second Riemann sheet (resonances) it is also possible that the  IAM or the N/D could produce 
poles in the physical (first) Riemann sheet for some regions of the chiral coupling space. In this case, due to the 
Schwarz's reflection principle, these spurious poles always come in pairs, one above and one below, the real axis. 
However causality requires the partial waves to be analytic in the upper half-plane (absence of tachyonic ghosts). 
Thus these poles are not acceptable and therefore the corresponding regions of chiral parameter space are not allowed. 
May be there is not a well-defined UV completion of the corresponding effective theory \cite{Adams:2006sv} or perhaps it is just a 
failure of the 
different approximations considered to get the unitarized partial waves.

When dealing with resonances a particularly interesting situation happens in elastic channels when the resonance is narrow. 
In this case the pole is close to real axis  and the parameter $\gamma = \Gamma/M$ is small $\gamma << 1$. Then, by neglecting 
higher powers of $\gamma$, it is possible to solve the resonance equations above to compute $M$ and $\Gamma$. This turns 
to be equivalent to finding the zero of the real part of the partial wave denominator to determine $M$
and the usual formula for $\Gamma$ in terms of the phase shift derivative obtained by assuming a Breit-Wigner behavior 
of the partial wave close to the resonance position. More in detail, any elastic partial wave $t(s)$ fulfilling 
$ \Imag t= \left\lvert t \right\rvert^2$ in the physical region can be written as
\be
t(s) = e^{i \delta(s)} \sin \delta(s) 
\ee
where $\delta(s)$ is the phase shift. Close to the the region $s=M^2$ the Breit-Wigner form of the partial wave on the physical region is
\be
t(s) \simeq \frac{-M \Gamma}{s-M^2 + i \Gamma M}
\ee
which implies:
\be
 \Gamma^{-1}= M \frac{d \delta(s)}{d s}\biggr\rvert_{s=M^2}.
\ee
Then, in the elastic case and for $\gamma <<1$ we can find closed formulae for the resonance parameters $M$ and $\Gamma$. 
Using the IAM method we have
\be
M^2= \frac{K}{B(M)}
\ee
and
\be
\gamma=\frac{K^2}{B(M)+D+E}.
\ee
For example, in the case of the vector channel $V$  ($I=J=1$) for $ww$ elastic scattering it is possible to find

\begin{align}
M_V^2 &=\dfrac{1152 \pi^2v^2 (1-a^2)}{8(1-a^2)^2-75(a^2-b)^2+4608\pi^2(a_4(M_V)-2a_5(M_V))}\,,\\
%\Gamma_0&=\dfrac{(1-a^2)}{96\pi v^2}M_V^3\,;~~
\Gamma_V &=\dfrac{(1-a^2)}{96\pi v^2}M_V^3\,\left[1+\dfrac{(a^2-b)^2}{32\pi^2 v^2 (1-a^2)}M_V^2\right]^{-1},
%\dfrac{1}{1+\dfrac{(a^2-b)^2}{32\pi^2 v^2 (1-a^2)}M_V^2}
\label{MVGVET}
\end{align}
with $a_4(M_V)$ and $a_5(M_V)$ being the scale dependent parameters evaluated at the resonance scale $\mu=M_V$. Thus, at least for 
the narrow resonance case, we can compute the mass and the width directly from the chiral couplings (self-consistently evaluated 
at the resonance mass). Similar formulae can be obtained for the $N/D$ method.

As we have seen so far there are at least two acceptable unitarization methods (the IAM and the N/D). Therefore one may 
ask which is the most appropriate
to be used in phenomenological applications. The answer depends on the considered channel for two reasons. First of all the 
the IAM method cannot be applied for those channels where  $K_{IJ}=0$. In particular this is the case of the $J = 2$ channels. 
Second, the N/D cannot be applied in the case $D_{IJ}+E_{IJ}=0$ because in this case it is not possible to define the 
splitting of $t^{(1)}$ into $t_L$ and $t_R$. This happens for example in the vector channel of the  elastic $ww$ scattering in 
the Higgless QCD-like model since then $a=b=0$ and
\be
D_{11}+E_{11}= \frac{3}{(96)^2\pi^3 v^4}(a^2-b)^2.
\ee
Thus the N/D method cannot be used in the vector channel whenever $a^2=b$. However it can be used in the coupled isoscalar 
channel $I=J=0$ and in the elastic $I=2$, $J=0$, and then both methods can be applied giving similar results. However one could 
argue the superiority of the IAM, due to its simplicity, not requiring 
the $L$ and $R$ splitting of the NLO matrix element. Then the  elastic $ww$ and the inelastic $w h\to w h$ ($I=J=1$) cases 
(vector and axial channels) should be better unitarized using the elastic IAM method to avoid the instability in the region 
of parameter space close to $D_{11}+E_{11}=0$ (as for example in a Higgsless QCD-like theory and more importantly 
whenever $a^2  \simeq b$ as in the minimal SM or dilaton models). The coupled $IJ=02$ has $K_{02}=0$ and then cannot be 
unitarized with the IAM method but can be unitarized with the coupled-channel N/D, which is stable in the limit $K_{02}=0$. 
Finally, for the elastic case $I=2$  and $J=2$, also with $K_{22}=0$, one can use the simpler elastic $N/D$ method too. 
In this way, it is possible to unitarize all non-vanishing channels appearing in the HEFT up to the one-loop level. 
The situation can be summarized as follows:
\begin{itemize}
\item
Scalar-isoscalar, $IJ=00$: IAM or N/D (coupled channels).
\item
Vector-isovector, $IJ=11$: IAM.
\item
Axial Vector-isovector, $IJ=11$: IAM.
\item
Scalar-isotensor, $IJ=20$: IAM or N/D.
\item
Tensor-isoscalar, $IJ=02$: N/D (coupled channels).
\item
Tensor-isostensor, $IJ=22$: N/D.

\end{itemize}

%%%%%%%%%%%%%%%%%%%%%%%%%%%%%%%%%%%%%%%%%%%%%%%%%%%%%%%%%%%%%%%%%%%%%%%%%%%%%%%%%%%%%%%%%%%%%%%%%%%%%%%%%%%%%%%%%%%%%%%%%%%%%%%%%
%%%%%%%%%%%%%%%%%%%%%%%%%%%%%%%%%%%%%%%%%%%%%%%%%%%%%%%%%%%%%%%%%%%%%%%%%%%%%%%%%%%%%%%%%%%%%%%%%%%%%%%%%%%%%%%%%%%%%%%%%%%%%%%%%

\section{Resonances as explicit degrees of freedom}
In the  previous section we have described how dynamically generated resonances may appear 
in the HEFT through the unitarization of the 
scattering amplitudes by using DR. Of course effective Lagrangians explicitly containing higher massive states as
explicit resonances have a long tradition. In this section we will review how to 
introduce the resonances directly in the Lagrangian of the HEFT as explicit degrees of freedom \cite{EGPR}. 

By doing so and then integrating them out one actually gets estimates for the coefficients of the higher dimensional operators
in the HEFT  (the $a_i$). In fact one can then reverse the process and, starting from these values for the $a_i$, use
any of the available unitarization methods described above to `predict' the masses and couplings of the resonances
integrated out previously. In general the masses are well reproduced \cite{gback} and this may serve to some 
extent as a consistency check of the unitarization prescriptions. 

What is then the advantage of using a unitarized HEFT in front of dealing with a Lagrangian containing explicit scalar
or vector resonances? The answer is obvious: not only the masses of the latter are predicted as dynamical resonances, but
also the widths and the couplings to the light degrees of freedom (the NGB). In addition the amplitudes so obtained
are unitary.  Yet, it is very convenient to consider explicit resonances to get a feeling of the corresponding values
for the $a_i$ coefficients. And viceversa, if these are measured in  an experiment, to guess the spectrum of 
resonances.   

In order to illustrate how it is possible to introduce resonances in this way we particularize to the coset 
$M=G/H$ where $G=SU(2)_L \times SU(2)_L$ and $H=SU(2)_{L+R}$  \cite{Pich:2012jv}. The NGB are parametrized 
through $U=u^2= \exp (i \vec \tau \, \vec \pi/ v) $ where $u(\pi)$ is an element of the coset $M$ which under 
a $g=(g_L,g_R) \in G$ transformation behaves as
\be
u(\pi) \rightarrow g_L u(\pi) h^\dagger (\pi,g) = h(\pi,g)u(\pi) g^\dagger_R.
\ee
Here $h(\pi,g) \in H$ is the compensating transformation that preserves the chosen coset representative  \cite{CoWeZu69}.
The electroweak gauge fields are introduced as usual through the covariant derivatives: 
\be
 D_{\mu} U =  \partial_{\mu}U +
\frac{1}{2} i g W_{\mu}^{i} \tau^{i} U - \frac{1}{2} i g' B_{\mu}^{i} U \tau^{3}= \partial_{\mu}U- i \hat W_{\mu} U + i U \hat B_{\mu}
\ee
and, by defining $u^\mu = i u D^\mu U^\dagger u =-i u^\dagger D^\mu U u^\dagger = u^{\mu \dagger}$, the HEFT $SU(2)_L \times U(1)_Y$ 
gauge invariant Lagrangian can be written as
\be
\label{GaugedHEFT2}
\mathcal{L}  =  - \frac{1}{2g^2}{\rm Tr} \,    \hat W_{\mu\nu}\hat W^{\mu\nu}  - \frac{1}{2g'^2}{\rm Tr}\,     \hat B_{\mu\nu}\hat B^{\mu\nu} 
+ \frac{v^{2}}{4} F(h) {\rm Tr}\, u_{\mu}u^{\mu}+ \frac{1}{2} \partial_{\mu} h \partial^{\mu} h  -V(h)
\ee
where 
\be
\hat W_{\mu\nu}= \partial_\mu  \hat W_\nu -\partial _\nu \hat W_\mu - i [\hat W_\mu, \hat W_\nu] 
\ee
and
\be
\hat B_{\mu\nu}= \partial_\mu  \hat B_\nu -\partial _\nu \hat B_\mu .
\ee
Notice the breaking of the $G$ symmetry by the gauge field $\hat B_\mu$. At the next-to-leading order (NLO) one has to 
consider the one-loop corrections to the previous Lagrangian and the $O(p^4)$ new terms. Assuming parity conservation and custodial 
symmetry those terms can be written as
\be
\label{p4Lagrangian}
\mathcal{L}_4= \sum_{i=1}^9F_i(h) \,O_i
\ee
where $F_i(h)$ are arbitrary analytical functions of $h/v$ and the $O_i$ operators can be defined for example as  \cite{Pich:2015kwa}:
\begin{eqnarray}
\label{p4operators}
O_1  & = & \frac{1}{4}\, {\rm Tr} \,(f ^{\mu\nu}_+f_{\mu\nu}^+ - f ^{\mu\nu}_- f_{\mu\nu}^-)  \nonumber \\
O_2 & =  & \frac{1}{2}\, {\rm Tr} \,(f ^{\mu\nu}_+f_{\mu\nu}^+ + f ^{\mu\nu}_- f_{\mu\nu}^-)     \nonumber \\
O_3 & =  & \frac{i}{2}\, {\rm Tr} \,(f ^{\mu\nu}_+ [u_\mu ,u_\nu])     \nonumber \\
O_4 & =  & {\rm Tr} \,(u_\mu u_\nu)\, {\rm Tr} \,(u^\mu u^\nu)     \nonumber \\
O_5 & =  & {\rm Tr} \,(u_\mu u^\mu)\,  {\rm Tr} \,(u_\nu u^\nu)     \nonumber \\
O_6 & =  & \frac{1}{v^2} \, \partial _\mu h \, \partial ^\mu h \, {\rm Tr} \,(u_\nu u^\nu)     \nonumber \\
O_7 & =  & \frac{1}{v^2} \, \partial _\mu h \partial _\nu h \, {\rm Tr} \,(u^\mu u^\nu)     \nonumber \\
O_8 & =  & \frac{1}{v^4} \, \partial _\mu h \, \partial ^\mu h \, \partial _\nu h \, \partial ^\nu h \nonumber \\
O_9 & =  & \frac{1}{v} \, \partial _\mu h \, {\rm Tr}\, (f^{\mu\nu}_- u_\nu) 
\end{eqnarray}
where for convenience the gauge  strength tensors has been written as 
$f^{\mu\nu}_{\pm}= u^\dagger \hat W^{\mu\nu} u \pm u \hat B^{\mu\nu} u^\dagger$. The $O(p^4)$ operators above 
correspond with those considered previously in this work by making the coupling constant identification
\begin{eqnarray}
  a_1 & = & F_1(0)      \nonumber \\
    a_2-a_3 & = & F_3(0)      \nonumber \\
   a_4 & = & F_4(0)      \nonumber \\
    a_5 & = & F_5(0)      \nonumber \\
        d & = & F_6(0)      \nonumber \\
         e & = & F_7(0)      \nonumber \\
     g & = & F_8(0).   
\end{eqnarray}
Now we will assume that the strongly interacting dynamics of the SBS of the SM generates heavy (of the order one or several TeV) 
resonances.  For definiteness we will consider only the case of vector $V$ and axial-vector $A$ transforming as $H$ triplets, 
i.e. $R \rightarrow h R h^\dagger $ for $h \in H$ and $R= V, A$. In principle there are several ways for introducing massive 
vector resonances in our Lagrangian, namely as Proca fields, by using the hidden symmetry formalism, as gauge fields through 
the Higgs mechanism and finally as antisymmetric tensors \cite{EGPR}. Here we will consider only this last method since 
it is probably the simplest one (comparison with other methods can be found in  \cite{Pich:2016lew}). Thus we introduce an 
antisymmetric tensor $W^{\mu\nu}= -W^{\mu\nu}$ with a free Lagrangian
\be
\label{LW}
\mathcal{L}_W=-\frac{1}{2}\, \partial^\mu W_{\mu\nu}\, \partial_\rho W^{\rho\nu}+ \frac{1}{4}\, M^2 W_{\mu\nu}W^{\mu\nu}.
\ee
Next we define the vector field:
\be
W_{\mu}= \frac{1}{M}\, \partial^\nu W_{\nu\mu}.
\ee
From this definition and the antisymmetry of $W^{\mu\nu}$ the Lorentz condition follows as an identity
\be
\partial _\mu W^\mu=0.
\ee
In addition from Eq.\ref{LW} one obtains the the Proca equation
\be
 \partial_\rho ( \partial^\rho W^\mu-\partial^\mu W^\rho) + M^2 W^\mu=0
\ee
which means that $W^\mu$ really describes a spin-1 particle of mass $M$. In the following we will apply this formalism to introduce explicitely in the HEFT Lagrangian the vector and axial triplets $V_\mu$ and $A_\mu$ with masses $M_V$ and $M_A$ respectively. Thus, at leading order, the most general parity conserving and $G$ invariant interaction with one vector or axial-vector 
can be written in terms of the Lagrangian  \cite{Pich:2012dv}:
\begin{eqnarray}
\label{Resonancecoupling}
\mathcal{L}_{R}   & =  & \frac{v^{2}}{4} F(h) {\rm Tr}\, u_{\mu}u^{\mu}   \nonumber \\ 
& + & \frac{f_V}{2\sqrt{2}} {\rm Tr}\, (V_{\mu\nu}f^{\mu\nu}_+)+
\frac{i g_V}{2\sqrt{2}} {\rm Tr}\, (V_{\mu\nu}[u^\mu u^\nu])  \nonumber \\
 & + &   \frac{f_A}{2\sqrt{2}} {\rm Tr}\, (A_{\mu\nu}f^{\mu\nu}_-) +
        \sqrt{2} \lambda_1^{hA} \partial_\mu \frac{h}{v} \, {\rm Tr}\, (A^{\mu\nu}u_\nu),
\end{eqnarray}
where we have introduced the antisymmetric gauged strength tensor triplet $V_{\mu\nu}$ as
\be
 V_{\mu\nu} = \nabla_\mu  V_\nu -\nabla_\nu  V_\mu\, , 
\ee
with
\be
 V_\mu =\frac{ \tau^a {V}_\mu^a}{\sqrt{2} } \,=\,
\left(\begin{array}{cc}
\frac{V^0_\mu }{\sqrt{2}} & V^+_\mu \\ V^-_\mu  &-\frac{V_\mu^0}{\sqrt{2}}
\end{array}
\right)\, 
\ee
and the covariant derivative is defined as:
\be
\nabla_\mu \, =\, \partial_\mu  \, +\, [\Gamma_\mu , \,\,\,\,] 
\ee
with
\be
 \Gamma_\mu =
\frac{1}{2} \Big(\Gamma_\mu^{L} +\Gamma_\mu^{R}\Big)
\ee
and
\be
\Gamma_\mu^{L} = u^\dagger \left(\partial_\mu + i\,\frac{g}{2} \vec{\tau}\vec{W}_\mu
\right) u^{\phantom{\dagger}} 
\, , \quad\;
\Gamma_\mu^{R} = u^{\phantom{\dagger}}  \left(\partial_\mu + i\, \frac{g'}{2} \tau^3 B_\mu\right)u^\dagger.
\ee
Similar definitions hold for $A_{\mu\nu}$. These $V_{\mu\nu}$ and $A_{\mu\nu}$ should not be confused with any instance of the antisymmetric $W_{\mu\nu} $ tensor mentioned above. We have also introduced the constants $f_V$, $g_V$ $f_A$ and $\lambda_1^{hA}$ measuring 
the intensity of the resonance coupling to the $\pi$ and $h$ particles. Now one can consider the NGB and $h$ interactions 
mediated by the $V$ and $A$ resonances. Those interactions are typically non local. However they can be expanded as a series 
of local terms corresponding to different power of $p^2/M_V^2$ or $p^2/M_A^2$. Thus, by integrating out these resonances 
(at the tree level), it is possible to find the contribution of the resonances to the  Lagrangian in Eq.\ref{p4Lagrangian}  
\cite{Pich:2015kwa}. The result is given by
\begin{eqnarray}
\label{DeltaF}
\Delta F_1(0) & = & \frac{f_A^2}{4M_A^2}-\frac{f_V^2}{4M_V^2}  \nonumber.  \\
\Delta F_2(0) & = & -\frac{f_A^2}{8M_A^2}-\frac{f_V^2}{8M_V^2} \nonumber   \\
\Delta F_3(0) & = & -\frac{f_V g_V}{2M_V^2} \nonumber  \\
\Delta F_4(0) & =  & \frac{g_V^2}{4M_V^2} \nonumber  \\
\Delta F_5(0)  & =  & -\frac{g_V^2}{4M_V^2} \nonumber   \\
\Delta F_6(0) & = & -\frac{(\lambda_1^{hA})^2v^2}{M_A^2} \nonumber   \\
\Delta F_7(0) & = & \frac{(\lambda_1^{hA})^2v^2}{M_A^2} \nonumber   \\
\Delta F_8(0) & = & 0 \nonumber  \\
\Delta F_9(0) & = & -\frac{f_A\lambda_1^{hA}v}{M_A^2}. 
\end{eqnarray}
The same kind of computations can be done for  resonances of other type such as scalars and pseudo-scalars (singlet or triplets), 
tensor-isoscalar, scalar-isotensor and so on. Thus the coupling constants of the HEFT may have in principle contributions 
from several of the first resonances present in the spectrum of the SBS.

%%%%%%%%%%%%%%%%%%%%%%%%%%%%%%%%%%%%%%%%%%%%%%%%%%%%%%%%%%%%%%%%%%%%%%%%%%%%%%%%%%%%%%%%%%%%%%%%%%%%%%%%%%%%%%%%%%%%%
%%%%%%%%%%%%%%%%%%%%%%%%%%%%%%%%%%%%%%%%%%%%%%%%%%%%%%%%%%%%%%%%%%%%%%%%%%%%%%%%%%%%%%%%%%%%%%%%%%%%%%%%%%%%%%%%%%%%%
 
\section{Form factors}
When dealing with resonance production at the LHC it is interesting to introduce the concept of form factors. This is because 
there exsits the possibility for resonant production of $ww$ pairs, not only through the elastic process $ww \rightarrow R \rightarrow ww$ 
(with $w= w^+, w^- $ or $ z$), but also more directly from the proton constituent quarks through the process 
$q \bar q' \rightarrow V^* \rightarrow R \rightarrow w w$ for $V= W, Z$. The same reasoning applies to  resonant  
$hw $ and $hh$ production. In this last case the amplitude can be computed from the perturbative SM result dressed 
with the corresponding form factor $F_R(s)$. For example \cite{Dobado:2015hha,Dobado:2017lwg} 
\be
T(q \bar q' \rightarrow V^* \rightarrow R \rightarrow w w) = T_{SM}(q \bar q' \rightarrow V^* \rightarrow w w)F_R(s).
\ee 
Clearly here the form factor $F_R(s)$ carry all the non-perturbative SBS physics giving rise to the resonant $ww$ rescattering. 
In the following we will concentrate on the case of vector and axial-vector resonances. This will allow us to determine the 
effective couplings of these resonances to, say, $VV$ fusion processes. For example the vector form factor $F_V(s)$ is defined as
 \be
 <W^i_L(k_1)W^j_L(k_2\mid J_V^{k\mu} \mid 0> = (k_1-k_2)^\mu F_V(s)\epsilon^{ijk}
 \ee
where $J_V^{k\mu}$ is the interpolating vector current with isospin $k$ that creates the vector resonance and $s= (k_1+k_2)^2$. 
This form factor can be computed in HEFT in a similar way as the scattering amplitudes for the NGB and the $h$ as a series 
of terms with increasing powers of $s/v^2$
\be
F_V(s) = F_V^{(0)}(s)+F_V^{(1)}(s)+\dots
\ee
where:
\begin{eqnarray}
F_V^{(0)}(s) & = & 1 \nonumber \\
F_V^{(1)}(s) & = &   s \left(  G_V(\mu) +H_V     \log \frac{-s}{\mu^2}      \right)     
\end{eqnarray}
and
\begin{eqnarray}
G_V(\mu)   & = &   \frac{1}{v^2}\, \left(-2F_3(\mu)+ \frac{1-a^2}{36 \pi^2}   \right)              \nonumber \\
H_V  & = & - \frac{1-a^2}{96 \pi^2 v^2} . 
\end{eqnarray}
Here $F_3$  is defined as $F_3(0)$, i.e. the value of the function $F_3(h)$ for $h=0$. In the formula above  this $F_3$ has 
been renormalized to absorb the usual one-loop divergence by using dimensional regularization and the $\overline{MS}$ 
renormalization prescription. Thus $F_3$ becomes the renormalized scale dependent $F_3(\mu)$. However $F_V(s)$ must be renomalization 
scale independent (since it is observable). Then it is easy to find the $\mu$ dependence of $F_3(\mu)$ which is given by
\be
F_3(\mu)= F_3(\mu_0)+\frac{1-a^2}{192 \pi^2}\,   \log \frac{\mu^2}{\mu^2_0} .  
\ee 
As required by unitarity the form factor above has a cut on the positive real axis starting at the threshold for $ww$ 
production which in this case is just $s=0$. As for scattering partial waves this cut defines a second
Riemann sheet where possibly a pole could appear representing a dynamically generated resonance if it is located close enough 
and below the real axis. Just on the cut (physical region $s= E^2+ i \epsilon$) elastic unitarity requires also
\be
\Imag F_V = F_V t_{11}^*
\ee
together with the elastic unitarity condition for $t_{11}$
\be
 \Imag t_{11}   =  \lvert t_{11} \rvert ^2.  
\ee
The NLO computation for $F_V$ shown above satisfies this condition only perturbatively as expected. Thus it is easy to check
\be
\Imag F_V^{(1)} = F_V^{(0)} t_{11}^{(0)}
\ee
which amounts to the relation: $K_{11}=-\pi H_V$.

In a similar way we can define the axial vector form factor $F_A(s)$ as
 \be
 <W^i_L(k_1)h(k_2)\mid J_A^{i\mu} \mid 0> = a(k_1-k_2)^\mu F_A(s)
 \ee
where $J_A^{i\mu}$ is the interpolating vector current with isospin $i$ that creates the resonant $A$. 
Again we can expand $F_A(s)$ by using HEFT as
\be
F_A(s) = F_A^{(0)}(s)+F_A^{(1)}(s)+\dots
\ee
where:
\begin{eqnarray}
F_A^{(0)}(s) & = & 1 \nonumber \\
F_A^{(1)}(s) & = &   s \left(  G_A(\mu) +H_A     \log \frac{-s}{\mu^2}      \right)     
\end{eqnarray}
and
\begin{eqnarray}
G_A(\mu)   & = & \frac{1}{v^2} \left( -\frac{F_9(\mu)}{a}+ \frac{a^2-b}{36 \pi^2 } \right)                \nonumber \\
H_A  & = & - \frac{a^2-b}{96 \pi^2 v^2} . 
\end{eqnarray}
where we have used the same conventions with $F_8$ than we did with $F_3$ in the vector form factor. Now the renormalization 
group evolution equation is
\be
F_9(\mu)= F_9(\mu_0)+\frac{a(a^2-b)}{96 \pi^2}\,   \log \frac{\mu^2}{\mu^2_0} .  
\ee 
The analytical and unitary properties of the the axial form factor are analogous to that of the vector form factor. 
More specifically the elastic unitarity conditions now read
\be
\Imag F_A = F_V n_{11}^*
\ee
with the elastic unitarity relation for $n_{11}$
\be
 \Imag n_{11}   =  \lvert n_{11} \rvert ^2 . 
\ee
The NLO computation for $F_A$ fulfills:
\be
\Imag F_A^{(1)} = F_A^{(0)} n_{11}^{(0)}
\ee
or $K'''_{11}=-\pi H_A$.

In the case of vector or axial-vector dominance the form factors can also be obtained from a tree level computation in HEFT  
models equipped with explicit resonances as discussed in the previous section. In this case one finds
\begin{eqnarray}
   F_V(s) & = & 1 + \frac{f_V g_V}{v^2} \, \frac{s}{M_V^2-s}         \nonumber \\
      F_A(s) & = & 1 + \frac{f_A \lambda_1^{hA}}{av^2} \, \frac{s}{M_A^2-s} .  
  \end{eqnarray}
Assuming $ww$ and $wh$ to be the dominant decay channels for $V$ and $A$ respectively, it is also possible to find the 
corresponding widths
\begin{eqnarray}
   \Gamma_V & = &  \frac{ g_V^2 M_V^5}{48 \pi v^4}        \nonumber \\
      \Gamma_A & = &  \frac{  (\lambda_1^{hA})^2 M_A^5}{48 \pi v^4}   .
  \end{eqnarray}
Then one can use this result to improve the form factors above to obtain  Breit-Wigner-like formulae
\begin{eqnarray}
   F_V(s) & = & 1 + \frac{f_V g_V}{v^2} \, \frac{s}{M_V^2-i M_V \Gamma_V-s}         \nonumber \\
      F_A(s) & = & 1 + \frac{f_A \lambda_1^{hA}}{av^2} \, \frac{s}{M_A^2-i M_A \Gamma_A-s} .  
  \end{eqnarray} 
If one expands these form factors in powers of $s/v^2$  and compares the first term with the HEFT NLO result it is very easy to 
check the values for $\Delta F_3(0)$ and $\Delta F_8(0)$ given in Eq.\ref{DeltaF}. Notice also that if we require 
the form factor to vanish for very large energies $s >> M_V^2, M_A^2$, we get the conditions
\begin{eqnarray}
\label{conditions}
          f_Vg_V & = & v^2               \nonumber   \\
          f_A \lambda^{hA}  & = & a v^2.
\end{eqnarray}

Further constraints on the parameters of the model can be obtained by considering the self-energies of the $V$ and $A$  two-point 
currents correlators $\Pi_{VV}$ and $\Pi_{AA}$. For underlying  asymptotically free theories one has the so-called Weinberg Sum Rules 
(WSR)  \cite{WSR}
\begin{eqnarray}
     \frac{1}{\pi} \, \int_0^\infty ds \left( \Imag \Pi_{VV}(s)- \Imag \Pi_{AA}(s)   \right)  & = & v^2  \nonumber \\
         \frac{1}{\pi} \, \int_0^\infty ds \, s \left( \Imag \Pi_{VV}(s)-  \Imag \Pi_{AA}(s)   \right)  & = & 0      
\end{eqnarray}
The first WSR is valid also for theories with a non-trivial UV fixed point  
  \cite{Peskin:1990zt}. However the second one could fail in some conformal or walking technicolor \cite{Holdometal}
models of the SBS of the SM. In any case in the resonance model defined in Eq.\ref{Resonancecoupling} 
the above WSR read \cite{Pich:2012dv}
\begin{eqnarray}
   f_V^2-f_A^2    & = & v^2   \nonumber \\
    f_V^2M_V^2-f_A^2M_A^2   & = & 0      
\end{eqnarray}
which implies $M_V < M_A$. In addition, at the one-loop level,   Eq.\ref{conditions} and the second WSR 
imply $a=M_V^2/M_A^2 < 1$   \cite{Pich:2013fea}. By using these high energy conditions it is possible to compute 
the $\Delta F_i(0)$ in terms of $M_V$, $M_A$ and $v$ only.

Beyond the models including resonances explicitly one may wonder which are the mathematical properties that the form 
factor $F(s)= F_V(s), F_A(s)$ must fulfill starting from first principles. Ideally a fully realistic form factor $F(s)$ would 
have the following properties \cite{Dobado:2017lwg}: a) $F(s)$ must be analytical in the whole complex plane featuring 
just a right cut along the positive real axis starting at $s=0$. b) Coincidence of any possible pole in the second Riemann 
sheet with those of the corresponding elastic partial wave $t_{11}(s)$ or $n_{11}(s)$, since they represent the same physical 
resonance. c) Elastic unitarity conditions on the physical region. d) Low-energy behavior matching the HEFT expansion
\begin{eqnarray}
F(s)= F^{(0)}(s)+F^{(1)}(s)+O(\frac{s^2}{v^4}).
\end{eqnarray}

A very simple pair of functions fulfilling some of the conditions above is:
\begin{eqnarray}
\tilde F_V=  \frac{ F_V^{(0)}}{1-  F_V^{(1)}/  F_V^{(0)}}   \nonumber \\    
\tilde F_A=  \frac{ F_A^{(0)}}{1-  F_A^{(1)}/  F_A^{(0)}} .
\label{Formfactor1}
\end{eqnarray}
Using perturbative unitarity it is easy to show that these form factors are exactly unitary and have the proper analytical structure. However they do not have the property of having the poles in the second Riemann sheet in the same place as the corresponding $ww$ and $wh$ elastic channels since they do not depend on the same couplings. 

From the elastic amplitude alone it is possible to build up a different form factor model. These new form factors are given by:
\begin{eqnarray}
\tilde F_V=  \frac{ 1}{1-  t_{11}^{(1)}/  t_{11}^{(0)}}      \nonumber \\
\tilde F_A=  \frac{ 1}{1-  n_{11}^{(1)}/  n_{11}^{(0)}} .
\label{Formfactor2}
\end{eqnarray}
These form factors are unitary and have the same poles as the ones obtained by the IAM method. Nevertheless they do not match the one-loop result at low-energies.

However a very simple function fulfilling all the conditions above can be found as follows: We start from some partial waves 
$\tilde t_{11}(s)$ and $\tilde n_{11}(s)$ satisfying the appropriate analytical and unitary properties. In particular
\begin{eqnarray}
\Imag \tilde t_{11}= \mid  \tilde t \mid ^2  \\ 
\Imag \tilde n_{11}= \mid  \tilde n \mid ^2
\end{eqnarray}
on the physical region. Then we can define the unitarized form factors
\begin{eqnarray}
\tilde F_V=   F_V^{(0)}+  F_V^{(1)} \, \frac{\tilde t_{11}}{t_{11}^{(0)}}.   \\
\tilde F_A=  F_A^{(0)}+  F_A^{(1)} \, \frac{\tilde n_{11}}{n_{11}^{(0)}}.
\end{eqnarray}
In particular, if we unitarize the partial waves by using the IAM method one has
\begin{eqnarray}
\tilde F_V= 1+  F_V^{(1)} \, \frac{ t_{11}^{(0)}}{t_{11}^{(0)}- t_{11}^{(1)}} \nonumber \\
\tilde F_A= 1+  F_A^{(1)} \, \frac{ n_{11}^{(0)}}{n_{11}^{(0)}- n_{11}^{(1)}}.  
\label{Formfactor3}
\end{eqnarray}
With this definitions it is possible to show that these unitarized form factors fulfill the four conditions above and, in particular,  
the unitary conditions
\begin{eqnarray}
\Imag \tilde F_V=  \tilde  F_V \tilde t_{11}^*  \\ 
\Imag \tilde F_A=  \tilde  F_A \tilde n_{11}^* 
\end{eqnarray}
on the physical region. In order to prove these relations it is needed to use the perturbative unitarity relations for the partial 
waves and the form factors. Then it is possible to find form factors with the correct mathematical properties out of 
the NLO computations (although they feature a LC which should not be present in form factors). In any case these unitarized form factors are very useful for phenomenology since the can be used to build up physically  
realistic amplitudes for production of $ww$ or $wh$ final states (resonant or not) parametrized in terms of the HEFT NLO 
parameters $a$, $b$ and $a_i$ or $F_i(0)$.

A further interesting point is that the three unitarizations methods in Eq.\ref{Formfactor1},  
Eq.\ref{Formfactor2} and  Eq.\ref{Formfactor3} become the same under the mathematical conditions:
\begin{eqnarray}
 t_{11}^{(1)} & = & F_V^{(1)} t_{11}^{(0)}  \nonumber  \\ 
 n_{11}^{(1)} & = &  F_A^{(1)} n_{11}^{(0)}  
\end{eqnarray}
which are equivalent to the set of relations for the vector form factor:
\begin{eqnarray}
 D_{11} & =  & 0  \nonumber  \\ 
E_{11} &=  & K_{11} H_V    \nonumber  \\
B_{11}(\mu) & = &  K_{11} G_V (\mu)
\end{eqnarray}
and for the axial form factor:
\begin{eqnarray}
 D_{11}''' & =  & 0  \nonumber  \\ 
E_{11}''' &=  & K_{11}''' H_A'''    \nonumber  \\
B_{11}'''(\mu) & = &  K_{11}''' G_A''' (\mu).
\end{eqnarray}
In both cases the first equation boils down to neglecting the LC contribution to the elastic scattering amplitudes. The second is always obeyed since it is a consequence of perturbative unitarity. The last one imposes a particular relation between $F_3$ (or $F_9$) and the rest of coplings involved in the two different channels at some given scale $\mu$.

All the previous discussion concerning vector or axial form factors can be extended to other channels like scalar-isoscalar,  
scalar-isostensor, tensor-isotensor etc. However the problem is more involved in those cases since they require including mixing 
with $hh$ states and/or using the N/D unitarization methods for the partial waves scattering amplitudes.

%%%%%%%%%%%%%%%%%%%%%%%%%%%%%%%%%%%%%%%%%%%%%%%%%%%%%%%%%%%%%%%%%%%%%%%%%%%%%%%%%%%%%%%%%%%%%%%%%%%%%%%%%%%%%%%%%%%%%%%%%%%%%%
%%%%%%%%%%%%%%%%%%%%%%%%%%%%%%%%%%%%%%%%%%%%%%%%%%%%%%%%%%%%%%%%%%%%%%%%%%%%%%%%%%%%%%%%%%%%%%%%%%%%%%%%%%%%%%%%%%%%%%%%%%%%%%

\section{Coupling NGB to the $\gamma\gamma$ system}
We shall now apply the methods discussed in the previous sections to various processes of interest at the LHC. We will study the
fusion of two longitudinal gauge vector bosons to produce a diphoton final state, $\gamma\gamma \to hh$, the production
of top pairs from longitudinal gauge vector fusion, $hh$ production from $t\bar t$, and the phenomenologically 
very relevant case of vector resonance production in $VV$ fusion.
We will start with diphoton processes.

In order to couple $W_L^+ W_L^- $ and $Z_LZ_L$ pairs to $\gamma\gamma$ states at high energies ($s >> M_W^2, M_Z^2, M_h^2$) it is 
possible to use the ET again. Then one is only interested (at least at the NLO) in NGB-$\gamma$ interactions. These interactions 
can then be obtained from the Lagrangian in Eq.\ref{GaugedHEFT}. By using again the squared root $U(x)$ parametrization
\be
U(x) = \sqrt{1-\frac{\omega^2}{v^2}}+ i \frac{\omega^a\tau^a}{v},
\ee
and keeping only the terms relevant for the NLO computation of the $\gamma\gamma\to zz$ and $\gamma\gamma\to w^+w^-$ 
processes, the Lagrangian becomes ~\cite{Delgado:2014jda}:
\begin{equation}\label{chLagr:Lagr:WBGB:gamma:spherical:param}
\begin{split}
\mL_2 ={} & %
       \frac{1}{2}\partial_\mu h\partial^\mu h
      +\frac{1}{2} F(h)(2\partial_\mu\omega^+\partial^\mu\omega^-
      +\partial_\mu\omega^0\partial^\mu\omega^0) \\
     &+\frac{1}{2v^2}F(h)(\partial_\mu\omega^+\omega^- 
      +\omega^+\partial_\mu\omega^- 
      +\omega^0\partial_\mu\omega^0)^2 \\
     &+ie F(h)A^\mu (\partial_\mu\omega^+\omega^- - \omega^+\partial_\mu\omega^-)
      +e^2 F(h)A_\mu A^\mu\omega^+\omega^-,
\end{split}
\end{equation}
where the photon field is given by $A_{\mu}= \sin \theta_W W_{\mu}^3+ \cos \theta_W B_{\mu}$ and as usual the absolute 
value of the electron charge is $e = g g' / \sqrt{g^2+ g'^2}$. By using this Lagrangian at NLO for the processes mentioned 
above one can expect to find divergences. In principle these divergences can be absorbed by renormalizing the couplings  
$a_1$, $a_2$, $a_3$ and $c_\gamma$ appearing in the dimension four counterterms (see ~\cite{Delgado:2014jda}),
\begin{multline}
\mL_4 = %
  a_1 {\rm Tr}(U \hat{B}_{\mu\nu} U^\dagger \hat{W}^{\mu\nu})
  + i a_2 {\rm Tr} (U \hat{B}_{\mu\nu} U^\dagger [V^\mu, V^\nu ]) 
  - i a_3  {\rm Tr} (\hat{W}_{\mu\nu}[V^\mu, V^\nu]) \\
 -\frac{c_{\gamma}}{2}\frac{h}{v}e^2 A_{\mu\nu} F^{\mu\nu} + \dots,
\label{eq.L4}
\end{multline}
where $ V_\mu = (D_\mu U) U^\dagger$ and
\begin{eqnarray}
\hat{W}_{\mu\nu} &=& \partial_\mu \hat{W}_\nu - \partial_\nu \hat{W}_\mu
 + i  [\hat{W}_\mu,\hat{W}_\nu ],\;\\
 \hat{B}_{\mu\nu} & = & \partial_\mu \hat{B}_\nu -\partial_\nu \hat{B}_\mu ,\label{fieldstrength}\\
F_{\mu\nu} &=& \partial_\mu A_\nu - \partial_\nu A_\mu, \
\end{eqnarray}
with $\hat{W}_\mu = g \vec{W}_\mu \vec{\tau}/2 ,\;\hat{B}_\mu = g'\, B_\mu \tau^3/2 $. In terms of the NGB fields Eq.~(\ref{eq.L4})  
can be written as
\begin{multline}
\mL_4   =  %
\frac{e^2a_1}{2v^2}F_{\mu\nu}F^{\mu\nu}\left(v^2 - 4\omega^+\omega^-\right) %
 + \frac{2e(a_2-a_3)}{v^2}F_{\mu\nu}\left[%
         i\left(\partial^\nu\omega^+\partial^\mu\omega^- - \partial^\mu\omega^+\partial^\nu\omega^- \right) %
   \right. \\ \left. %
         +  eA^\mu\left( \omega^+\partial^\nu\omega^- + \omega^-\partial^\nu\omega^+ \right)
        -eA^\nu\left( \omega^+\partial^\mu\omega^- + \omega^-\partial^\mu\omega^+ \right)
        \right] -\frac{c_{\gamma}}{2}\frac{h}{v}e^2 F_{\mu\nu} F^{\mu\nu} .
\end{multline}
From the Lagrangian $\mL = \mL_2 + \mL_4$ it is possible to compute the amplitudes for  $\gamma\gamma\to zz$ and 
$\gamma\gamma\to\omega^+\omega^-$  or the inverse
$z z \to \gamma\gamma$ and $\omega^+\omega^-   \to \gamma\gamma$ (which are the same respectively due to time reversal 
invariance of the interactions involved) up to NLO. The obtained amplitudes have the general form (see ~\cite{Delgado:2014jda})
\begin{equation}\label{A:scatter:gammagamma:lorentz}
  T(\gamma\gamma \to \omega\omega) = e^2\left(\epsilon_1^\mu\epsilon_2^\nu T_{\mu\nu}^{(1)}\right) A %
       +e^2\left(\epsilon_1^\mu\epsilon_2^\nu T_{\mu\nu}^{(2)}\right) B,
\end{equation}
where the tensors are given by:
\begin{subequations}
\begin{align}
  \left(\epsilon_1^\mu\epsilon_2^\nu T_{\mu\nu}^{(1)}\right) &= %
     \frac{s}{2}(\epsilon_1\epsilon_2) - (\epsilon_1 k_2)(\epsilon_2 k_1) %
\label{A:scatter:gammagamma:LorentzStructure:1}\\
  \left(\epsilon_1^\mu\epsilon_2^\nu T_{\mu\nu}^{(2)}\right) &= %
     2s(\epsilon_1\Delta)(\epsilon_2\Delta)-(t-u)^2(\epsilon_1\epsilon_2) %
   -2(t-u)[(\epsilon_1\Delta)(\epsilon_2 k_1)-(\epsilon_1 k_2)(\epsilon_2\Delta)] %
\label{A:scatter:gammagamma:LorentzStructure:2}
\end{align}
\end{subequations}
with  $\epsilon_i(k_i, \lambda_i)$  being the polarization vectors, $\lambda_i$ the helicity and the $k_i$ the 4-momentum 
of each photon with $i=1,\,2$; $p_i$ are the 4-momenta of the NGB ($i=1,\,2$)  and $\Delta^\mu = p_1^\mu - p_2^\mu$.

For the neutral $\gamma\gamma\to zz$ channel the LO vanishes because $z$ is a neutral particle,
\begin{equation}\label{A:scatter:gammagamma:to:zz:Atree}
  \mM (\gamma\gamma\to zz)_{\rm LO} = 0,
\end{equation}
and  the NLO contribution is
\begin{subequations}\label{A:scatter:gammagamma:to:zz:Aloop}
\begin{align}
  A(\gamma\gamma\to zz)_{\rm NLO} &= %
      \frac{2ac_\gamma^r}{v^2} + \frac{a^2-1}{4\pi^2 v^2} \equiv A_N 
\label{A:scatter:gammagamma:to:zz:Aloop:A}\\
  B(\gamma\gamma\to zz)_{\rm NLO} &= 0 .
\label{A:scatter:gammagamma:to:zz:Aloop:B}
\end{align}
\end{subequations}

For the charged $\gamma\gamma \to w^+ w^-$ channel we have at LO
\begin{equation}\label{A:scatter:gammagamma:to:ww:Atree}
  A(\gamma\gamma\to w^+  w^-)_{\rm LO} = %
    2s B(\gamma\gamma\to w^+ w^-)_{\rm LO} = %
    -\frac{1}{t} - \frac{1}{u},
\end{equation}
and at  NLO:
\begin{subequations}\label{A:scatter:gammagamma:to:ww:Aloop}
\begin{align}    
  A(\gamma\gamma\to w^+w^-)_{\rm NLO} &= %
    \frac{8(a_1^r-a_2^r+a_3^r)}{v^2} %
    +\frac{2ac_\gamma^r}{v^2} +\frac{a^2-1}{8\pi^2 v^2} \equiv A_C%
\label{A:scatter:gammagamma:to:ww:Aloop:A}  \\
  B(\gamma\gamma\to w^+  w^-)_{\rm NLO} &= 0.
\label{A:scatter:gammagamma:to:ww:Aloop:B}
\end{align}
\end{subequations}
One of the most interesting properties of these results is that, even being done at the one-loop level, they are UV finite 
in the dimensional regularization scheme. This, in particular implies that the combinations of couplings appearing 
in the amplitudes are renormalization group invariant, or in other words
\begin{subequations}
\begin{align}
  c_\gamma^r &= c_\gamma \\
  a_1^r-a_2^r+ a_3^r  &= a_1-a_2+a_3.\end{align}
\end{subequations}
This $\mu$ independence on the renormalization scale $\mu$  can be independently checked ~\cite{Delgado:2014jda}.

In the CM frame it is possible to chose the coordinate axis so that
\begin{subequations}\label{A:scatter:gammagamma:def:momenta}
\begin{gather}
  k_1     = (E,0,0,E)    \,\quad %
  k_2     = (E,0,0,-E)   \\
  p_1     = (E,\vec{p})  \,\quad %
  p_2     = (E,-\vec{p}) \,\quad %
  \Delta  = p_1 - p_2    \\
  \vec{p} = (p_x,p_y,p_z) = E (\sin\theta\cos\varphi,\sin\theta\sin\varphi,\cos\theta) .
\end{gather}
\end{subequations}
So that we have:  $  \epsilon_i(\pm)\cdot k_j = 0$ which simplifies
Eqs.~(\ref{A:scatter:gammagamma:LorentzStructure:1}) and (\ref{A:scatter:gammagamma:LorentzStructure:2}) to
\begin{equation}\label{A:scatter:gammagamma:LorentzStructure:1:tmp1}
  \epsilon_1^\mu\cdot\epsilon_2^\nu T_{\mu\nu}^{(1)} = %
       \frac{s}{2}\epsilon_1\cdot\epsilon_2\ ,
\end{equation}
and since the would-be NGB are taken to be massless we also have
\begin{equation}\label{A:scatter:gammagamma:LorentzStructure:2:tmp1}
   \epsilon_1^\mu\cdot\epsilon_2^\nu T_{\mu\nu}^{(2)} = %
       2s(\epsilon_1\cdot\Delta)(\epsilon_2\cdot\Delta) - s^2(\cos\theta)^2 (\epsilon_1\cdot\epsilon_2) \ .
\end{equation}
By choosing  $ \sqrt{2}\, \epsilon_1(\pm) =  (0, \mp 1, -i, 0)$ and  $\sqrt{2}\, \epsilon_2(\pm) = (0, \mp 1, i, 0)$ 
we have $\epsilon_1(+)\cdot\epsilon_2(-)=\epsilon_1(-)\cdot\epsilon_2(+)=0$ and 
$\epsilon_1(+)\cdot\epsilon_2(+)=\epsilon_1(-)\cdot\epsilon_2(-)=-1$.

Now we can introduce the custodial $SU(2)_{L+R}$ isospin basis by using the appropriate Clebsch-Gordan coefficients to 
obtain the matrix elements that follow. First we define the $ww$ states in the isospin basis $\mid I,I_3>$ 
as usual as $\mid 1,1>= -w^+$,  $\mid 1,0>= w^0=z$ and $\mid 1,-1>= w^{-}$.
Thus our isospin basis relevant for the processes considered here is
\be
\mid 0,0>= - \frac{1}{\sqrt{3}} \left( \mid w^+w^->  + \mid w^-w^+>+ \mid zz> \right)
\ee
and
\be
\mid 2,0>=  \frac{1}{\sqrt{6}} \left(2 \mid zz>- \mid w^+w^->  - \mid w^-w^+> \right).
\ee
In the following we will refer to $zz$ as the neutral state (N) and  $w^+w^-$ as the charged state (C) and we define 
 $T_I^{\lambda_1\lambda_2}\equiv < I,0|T|\lambda_1\lambda_2>$. Then we have
\begin{subequations}\label{T0}
\begin{align}
   T_0^{\lambda_1\lambda_2} &= -\frac{1}{\sqrt{3}}\left(2 T_C^{\lambda_1\lambda_2}+T_N^{\lambda_1\lambda_2}\right) \\ \label{T2}
   T_2^{\lambda_1\lambda_2} &= \frac{2}{\sqrt{6}}\left(T_N^{\lambda_1\lambda_2} - T_C^{\lambda_1\lambda_2}\right) .
\end{align}
\end{subequations}
And by using  Eq.~(\ref{A:scatter:gammagamma:to:zz:Aloop:A}) through 
\ref{A:scatter:gammagamma:to:ww:Aloop:B}, we find:
\begin{subequations}\label{A:scatter:gammagamma:T_I_lambda1_lambda2}
\begin{align}
   T_0^{++} &= T_0^{--} = \frac{e^2 s}{2\sqrt{3}}\left(2A_C + A_N\right) &
   T_2^{++} &= T_2^{--} = \frac{e^2 s}{\sqrt{6}}\left(A_C - A_N \right) %
 \label{A:scatter:gammagamma:T_I_lambda1_lambda2:1}\\
   T_0^{+-} &= (T_0^{-+})^* = \frac{4e^2}{\sqrt{3}} e^{2i\varphi} &
   T_2^{+-} &= (T_2^{-+})^* = \frac{4e^2}{\sqrt{6}}e^{2i\varphi}\ . %
 \label{A:scatter:gammagamma:T_I_lambda1_lambda2:2}
\end{align}
\end{subequations}

We can consider also the process $\gamma\gamma \to hh$. At the NLO, using the polarization vectors 
defined above, one finds
\begin{equation}
    R(\gamma\gamma\to hh) = \frac{e^2}{8\pi^2 v^2} (a^2-b)\delta_{\lambda_1,\lambda_2} .
\end{equation}
Notice that the LO vanishes since the Higgs is neutral. This amplitude is proportional to  $(a^2-b)$ and thus to the 
LO amplitude $w w \to hh$ so that the $hh$-$\gamma\gamma$ coupling proceeds through $ww$ loops.
As the final  $\mid hh >$ state is an isospin singlet the non-vanishing  isospin amplitudes 
\be
  R_I^{\lambda_1\lambda_2}  = < hh| T(\gamma\gamma\to hh)|\lambda_1 \lambda_2 >
  \ee
  are:
\begin{equation}\label{A:scatter:gammagamma:R}
    R_0^{++}=R_0^{--} = \frac{e^2}{8\pi^2 v^2} (a^2-b) .
\end{equation}

%%%%%%%%%%%%%%%%%%%%%%%%%%%%%%%%%%%%%%%%%%%%%%%%%%%%%%%%%%%%%%%%%%%%%%%%%%%%%%%%%%%%%%%%%%%%%%%%%%%%%%%%%%%%%%%%%%%%%%%%%%%
%%%%%%%%%%%%%%%%%%%%%%%%%%%%%%%%%%%%%%%%%%%%%%%%%%%%%%%%%%%%%%%%%%%%%%%%%%%%%%%%%%%%%%%%%%%%%%%%%%%%%%%%%%%%%%%%%%%%%%%%%%%%
\subsection{Partial waves}
As  was the case  in $w w$ and $hh$ scattering the NLO $\gamma\gamma$ processes amplitudes are unitary only at 
the perturbative level and therefore they are not physically acceptable at higher energies. In order to study the unitarity 
properties of the  $\gamma\gamma\to w w$ scattering amplitudes, it is useful to introduce the partial waves
\begin{align}
   p_{IJ}^{\lambda_1\lambda_2} &= \frac{1}{128\pi^2}\sqrt{\frac{4\pi}{2J+1}}\int d\Omega\, 
T_I^{\lambda_1\lambda_2}(s,\Omega) Y_{J,\Lambda}(\Omega), & \Lambda&=\lambda_1-\lambda_2,
\end{align}
whose inverse is
\begin{equation}\label{chLagr:PartialWavesGeneral:invers}
   T_I^{\lambda_1\lambda_2}(s,\Omega) = 128\pi^2\sum_J\sqrt{\frac{2J+1}{4\pi}}%
       p_{IJ}^{\lambda_1\lambda_2} Y_{J,\Lambda}(\Omega).
\end{equation}
 As the photon is a spin-1 massless boson, Landau-Yang's theorem forbids the partial wave with $J=1$. Thus, to NLO 
in the effective theory, the possible angular momenta are  $J=0,\,2$.  For $J=0$, and taking into account parity conservation, 
it is clear that our amplitude only couples to the positive parity state $(\mid+->+\mid -+>)/\sqrt{2}$. Then we define
\begin{equation}
   p_{I0} \equiv \frac{1}{\sqrt{2}}\left(p_{I0}^{++} + p_{I0}^{--}\right) = \sqrt{2}p_{I0}^{++} = \sqrt{2}p_{I0}^{--}.
\end{equation}
For $J=2$, the only non-vanishing contributions come from $p_{I2}^{+-}$ ($\Lambda=+2$) and $p_{I2}^{-+}$ ($\Lambda=-2$). 
The amplitudes $\Lambda=0$ vanish (see Eqs.~(\ref{A:scatter:gammagamma:T_I_lambda1_lambda2:1}) 
and~(\ref{A:scatter:gammagamma:T_I_lambda1_lambda2:2})). Now we introduce
\begin{equation}
   p_{I2} \equiv p_{I2}^{+-} = p_{I2}^{-+} .
\end{equation}
Thus the lowest order non-vanishing   partial waves are:
\begin{subequations}\label{A:scatter:gammagamma:partialWaves:F:e2}
\begin{align}
 p_{00}^{(0)} &= \frac{\alpha s}{8\sqrt{6}}(2A_C + A_N) & %
  p_{02}^{(0)} &= \frac{\alpha}{6\sqrt{2}} %
\label{A:scatter:gammagamma:partialWaves:F:e2:1}\\
 p_{20}^{(0)} &= \frac{\alpha s}{8\sqrt{3}}(A_C - A_N) &
  p_{22}^{(0)} &= \frac{\alpha}{12} %
\label{A:scatter:gammagamma:partialWaves:F:e2:2}
\end{align}
\end{subequations}
where we have introduced the fine structure constant $\alpha=e^2/4\pi$. Notice that  the $J=0$ partial waves are 
NLO while the $J=2$ ones are LO. 

The $hh$ final state is an isospin singlet, and only couples with $J=0$ and positive parity states. The 
corresponding partial waves are
\begin{equation}
   r_I^{(0)} \equiv \frac{1}{\sqrt{2}}\left(r_{I0}^{++}+r_{I0}^{--}\right) = \sqrt{2}r_{I0}^{++}.
\end{equation}
and then:
\begin{equation}\label{A:scatter:gammagamma:amp:R}
r_0^{(0)} = \frac{\alpha}{32\sqrt{2}\pi^2 v^2}(a^2-b) \ .
\end{equation}
In all the equations above $A_C$ and $A_N$ must be taken from Eqs.~(\ref{A:scatter:gammagamma:to:zz:Aloop})  
and~ \ref{A:scatter:gammagamma:to:ww:Aloop}.

%%%%%%%%%%%%%%%%%%%%%%%%%%%%%%%%%%%%%%%%%%%%%%%%%%%%%%%%%%%%%%%%%%%%%%%%%%%%%%%%%%%%%%%%%%%%%%%%%%%%%%%%%%%%%%%%%%%%%%%%%%%%%%%
%%%%%%%%%%%%%%%%%%%%%%%%%%%%%%%%%%%%%%%%%%%%%%%%%%%%%%%%%%%%%%%%%%%%%%%%%%%%%%%%%%%%%%%%%%%%%%%%%%%%%%%%%%%%%%%%%%%%%%%%%%%%%%%%

\subsection{Unitarization of the  $\gamma\gamma$ amplitudes}\label{sec:unitarity}
In order to simplify the discussion we decouple the $hh$ channel by setting $a^2=b$ (we also set other  
couplings between $ww$ and $hh$ to  $d=e=0$). The different amplitudes can be arranged in a $3 \times 3$  matrix \cite{Delgado:2016rtd}:
\begin{equation}\label{chLagr:UnitProceduresWW:Fcoupled1}
  T_J(s) = 
   \begin{pmatrix}
      t_{0J}(s)   & 0          & p_{0J}(s) \\
      0           & t_{2J}(s)  & p_{2J}(s) \\
      p_{0J}(s)   & p_{2J}(s)  & 0
   \end{pmatrix} + \mO(\alpha^2),
\end{equation}
where  $J$ can  take the values 0 or 2,  $t_{IJ}(s)$ are the (isospin conserving) elastic partial 
waves $\omega\omega\to\omega\omega$  and $p_{IJ}(s)$ are the partial-wave projected $\gamma\gamma\to\omega\omega$ amplitudes. 
Notice  that we are considering only the leading order in the electromagnetic coupling $\alpha$. For this reason 
we have taken  $<\gamma\gamma |T|\gamma\gamma> \simeq0$. The unitarity condition reads
\begin{equation}\label{unitarity}
\Imag T_J(s) = T_J(s) T_J(s)^\dagger
\end{equation}
on the RC. Working to LO in $\alpha$ this unitary condition becomes;
\begin{subequations}\label{unitarityexpanded}
\begin{align}
   \Imag t_{IJ} &= \lvert t_{IJ}\rvert^2\\
   \Imag p_{IJ} &= p_{IJ}t_{IJ}^*\ .
\end{align}
\end{subequations}
The structure of these equations allows for a sequential solving of the unitarity equation. As we have already shown, 
for the elastic $w w \to ww$ amplitude in the $J=0$  partial-wave case can be  achieved by the elastic IAM method  
from the first two orders of the perturbative expansion $t_{I0}=t_{I0}^{(0)}+t_{I0}^{(1)}+\dots$,
\begin{equation}\label{IAMpp}
  \tilde t_{I0}(s) = \frac{t_{I0}^{(0)}(s)}{1-\frac{t_{I0}^{(1)}(s)}{t_{I0}^{(0)}(s)}} .
\end{equation}
This amplitude has the correct analytic structure in the complex $s$ plane and it may show resonances in the 
second Riemann sheet below the RC, where it satisfies elastic unitarity. In addition it matches the 
chiral expansion at low energies.

In order to unitarize $p_{I0}$ we realize that the second  Eq.~(\ref{unitarityexpanded}) is the statement of Watson's theorem, 
that sets the $p_{00}$  phase to that of $ww$ final state rescattering. Then we can try the ansatz for the unitarized 
amplitude $\tilde p_{I0}(s)=f(s)\tilde t_{I0}^{(0)}(s)/ t_{I0}(s)$ with the function $f(s)$ being real for real $s$. Matching 
the low-energy perturbative result $\tilde p_{I0}=p_{I0} ^{(0)}+...$   we find   $f(s) =p_{I0}^{(0)}(s)/t_{I0}^{(0)}(s)$ and then the 
unitarized partial wave is
\begin{equation} \label{Omnes}
   \tilde p_{I0}  = \frac{p_{I0}^{(0)}}{t_{I0}^{(0)}}\tilde t_{I0}= \frac{p_{IO}^{(0)}}{1-\frac{t_{I0}^{(1)}}{t_{I0}^{(0)}}} .
\end{equation}
 Taking the imaginary part on the RC it is easy to show that it fulfills  the second unitary relation on  
Eq.~(\ref{unitarityexpanded}),
\begin{equation}\label{chLagr:UnitProceduresWW:Fcoupled1:proof}
   \Imag\tilde{p_{I0}} = %
     \frac{p_{I0}^{(0)}}{t_{I0}^{(0)}}\Imag\tilde t  = %
     \frac{p_{I0}^{(0)}}{t_{I0}^{(0)}}\lvert\tilde t \rvert^2 = %
     \tilde p_{I0} \tilde t_{I0}^*.
\end{equation}

The unitarized amplitude above has all the good properties of the IAM unitarized amplitudes. In addition, if for some values 
of the parameters there is a resonance in the second Riemann sheet of the $\tilde t_{I0} $ partial-wave, the same resonances 
appear also in the $\tilde p_{I0}$ partial wave as it must happen with dynamically generated resonances.

In the case of  the tensor $J=2$ channel it is not possible to apply the IAM method. Notice that the $p_{I2}$ are constant and 
also we have a vanishing LO elastic $ww$ scattering amplitude $t_{I2}^{(0)}=K_{I2}s$  since $K_{I2}=0$. However in this 
case we can use the $N/D$ method to unitarize the 
partial waves. Then the unitarized elastic $ww$ amplitude is defined as
\begin{equation}\label{unitar:ND:elastic}
\tilde{t} = t^{\rm N/D} = \frac{t_L(s)}{1+\frac{1}{2}g(s)t_L(-s)},
\end{equation}
where we have omitted the $I,J$ indices for simplicity and
\begin{subequations}
\begin{align}
   g(s)   &= \frac{1}{\pi}\left(\frac{B(\mu)}{D}+\log\frac{-s}{\mu^2}\right)       \label{unitar:ND:elastic:g}\\
   t_L(s) &= \left(\frac{B(\mu)}{D} + \log\frac{s}{\mu^2}\right) D s^2 = \pi g(-s)Ds^2  . \label{unitar:ND:elastic:AL}
\end{align}
\end{subequations}

The $B$ and $D$ which appear in Eq.~(\ref{unitar:ND:elastic:AL}) are the ones appearing in the corresponding 
NLO elastic $ww$ partial waves.
Once the $J=2$ elastic $ww$ waves have been unitarized, it is easy to satisfy
the second unitarity relation in  Eq.~(\ref{unitarityexpanded}) by defining
\begin{equation}\label{chLagr:UnitProceduresWW:Fcoupled3:fin}
   \tilde{p}_{I2} = \frac{p_{I2}^{(0)}}{t_{{\rm L},I2}}t_{I2}^{\rm N/D}, 
   \quad I=0,2 .
\end{equation}

In the general $a^2 \ne b $ case one has to consider also the coupling to the $hh$ channel. The  reaction matrix  
is the $4 \times 4$ matrix
\begin{equation}\label{chLagr:UnitProceduresWW:Fcoupled2}
  T_J = %
   \begin{pmatrix}
      t_{0J}   & m_J     & 0       & p_{0J} \\
      m_J      & a_J     & 0       & r_J    \\
      0        & 0       & t_{2J}  & p_{2J} \\
      p_{0J}   & r_J     & p_{2J}  & 0
   \end{pmatrix} + \mO(\alpha^2)\ ,
\end{equation}
where again, $t_{IJ}(s)$ are the partial waves $ww\to ww$; $m_J(s)$ the $w w \to hh$ 
partial wave; $a_J(s)$  the elastic $hh\to hh$ one; $p_{IJ}(s)$  the $\gamma\gamma\to\omega\omega$ ones and $r_J(s)$  
the $\gamma\gamma\to hh$. At leading order in $\alpha$, unitarity implies on the RC
\begin{subequations}\label{chLagr:sec:Unitar:gg:YEShh:J0:new_unit_relat}
\begin{align}
   \Imag t_{0J} &=  \lvert t_{0J}\rvert^2 + \lvert m_J\rvert^2 \\
   \Imag t_{2J} &=  \lvert a_{2J}\rvert^2 \\
   \Imag m_J    &=  t_{0J}m_J^* + m_J a_J^* \\
   \Imag a_{J\ }&=  \lvert m_J\rvert^2 + \lvert a_J\rvert^2 ,
\end{align}
\end{subequations} 
plus the $\gamma\gamma$ isoscalar amplitude relations
\begin{subequations}\label{chLagr:sec:Unitar:gg:YEShh:J0:new_unit_relat2}
\begin{align}
   \Imag p_{0J} &= p_{0J}t_{0J}^* + r_J m_J^* \\
   \Imag r_J &= p_{0J}m_J^* + r_J a_J^* 
\end{align}
\end{subequations}  
and finally the isotensor block decouples from the isoscalar ones
\begin{subequations}\label{chLagr:sec:Unitar:gg:YEShh:J0:new_unit_relat3}
\begin{align}
   \Imag t_{2J} &= \lvert t_{2J}\rvert^2 \\
   \Imag p_{2J} &= p_{2J}t_{2J}^* ,
\end{align}
\end{subequations}
which is identical to Eq.~(\ref{unitarityexpanded}) and to the $a^2=b$ case. Therefore we concentrate in what follows only 
in the first two blocks of unitary equations corresponding to isospin 0. Again all the previous unitary relations are fulfilled 
only at the perturbative level by our NLO computations shown above.

In order  to unitarize the NLO partial waves we first introduce the reaction submatrix for the strongly interacting 
subchannels $ww,hh\to ww,hh$,
\begin{equation}\label{chLagr:UnitProceduresWW:Fcoupled2:tmp0}
   K_J =  %
     \begin{pmatrix}
       t_{0J} & m_0 \\
       m_0    & a_0
     \end{pmatrix} \equiv %
     \begin{pmatrix}
       t & m \\
       m & a
     \end{pmatrix}.
\end{equation}
 This definition can be extended in a obvious way to $K_J^{(0)}$ and $K_J^{(1)}$.

In the $J=0$ case we can use the matricial generalization of the IAM method which  yields a unitary $\tilde K_0$ from the first 
two terms of the chiral expansion
\begin{equation}
\tilde K_0= K_0^{(0)}(K_0^{(0)}-K_0^{(1)})^{-1} K_0^{(0)}\ 
\end{equation}
with:
\begin{equation}\label{IAMpieces}
  \tilde K_0 = %
     \begin{pmatrix}
       \tilde t & \tilde m \\
       \tilde m & \tilde a
     \end{pmatrix}\ .
\end{equation}
This IAM approximation to the exact $K_0$ in Eq.~(\ref{chLagr:UnitProceduresWW:Fcoupled2:tmp0}) is unitarity in the RC, 
{\it i.e.} $\Imag  \tilde K_0 = K_0 \tilde K_0^\dagger$, it is analytical in the complex plane and it matches the NLO at low energies. 
Now introducing the doublet  $(p,r)^T\equiv (p_{00}, r_0)$, the unitary condition for these partial waves can be written as
\begin{equation}
   \Imag\begin{pmatrix} p \\ r\end{pmatrix} = K_0^*\cdot \begin{pmatrix} p \\ r\end{pmatrix}\ .
\end{equation}
This equation can be solved by generalizing Eq.~(\ref{Omnes})
\begin{equation}\label{chLagr:UnitProceduresWW:Fcoupled2:raw}
   \begin{pmatrix}\tilde{p}\\\tilde{r}\end{pmatrix} \equiv %
      \tilde K_0 (K_0^{(0)})^{-1}\begin{pmatrix}p^{(0)}\\r^{(0)}\end{pmatrix}.
\end{equation}
as can be easly checked by using the perturbative unitary relations.

As in the $a^2=b$ particular case, we cannot use the IAM method in the $J=2$ channel. However we can use the coupled 
channel version of the N/D method.  The matricial N/D formula, analogous to  Eq.~(\ref{unitar:ND:elastic}), is
\begin{equation}\label{NDpieces}
       \tilde  K_2= %
      \left[1 +\frac{1}{2}G(s)F_L(-s)\right]^{-1}K_L(s) ,
\end{equation}
where
\begin{subequations}
\begin{align}
   G(s)   &= \frac{1}{\pi}\left[B(\mu)D^{-1}+\log\frac{-s}{\mu^2}\right]		\\
   K_L(s) &= \left[B(\mu)D^{-1}+\log\frac{s}{\mu^2} \right]Ds^2 = \pi G(-s)Ds^2	
  \end{align}
\end{subequations}
are the generalizations of Eqs.~(\ref{unitar:ND:elastic:g}) and following. Notice that  we are in a coupled channel 
case in the sense that $ww \to hh\to ww$ rescattering takes place. However $hh$ states do not couple 
to $\gamma\gamma$ for $J=2$ (see Eq.~\ref{A:scatter:gammagamma:R}).
Therefore in this case  we need the matricial N/D method of Eq.~\ref{NDpieces} for unitarizing the $ww \to ww$ 
partial waves, but the coupling with $\gamma\gamma$ states can be computed by using the (scalar) 
Eq.~(\ref{chLagr:UnitProceduresWW:Fcoupled3:fin}) which fulfills the unitary relation
 \be
 \Imag \tilde p_{I2}= \tilde p_{I2}\tilde t^*_{I2}.
 \ee

%%%%%%%%%%%%%%%%%%%%%%%%%%%%%%%%%%%%%%%%%%%%%%%%%%%%%%%%%%%%%%%%%%%%%%%%%%%%%%%%%%%%%%%%%%%%%%%%%%%%%%%%%%%%%%%%%%%%
%%%%%%%%%%%%%%%%%%%%%%%%%%%%%%%%%%%%%%%%%%%%%%%%%%%%%%%%%%%%%%%%%%%%%%%%%%%%%%%%%%%%%%%%%%%%%%%%%%%%%%%%%%%%%%%%%%%%

\section{The HEFT for $t \bar t$ production}\label{sec:EWChTFermions}
In this section we address the processes $W_L^+ W_L^- \rightarrow t \bar t$ or $Z_L Z_L \rightarrow t \bar t$  
(and also $hh \rightarrow t \bar t$) at energies that are high when compared with $M_Z$, $M_W$ and $M_h$. At these high energies 
we can use the ET  and concentrate only in the would-be NGB $w^\pm, z$, $h$ and the heavy quarks $b$ and $t$. In particular we will 
consider the regime $M_t^2/v^2\ll\sqrt{s}M_t/v^2\ll s/v^2$. In this case we can consider, at least formally, $M_t/v$ as a small 
parameter and do the computations at the lowest non-trivial order in this parameter. This will simplify enormously the computation 
of the amplitudes above   (because of the significant smaller number of Feynman diagrams to be taken into account) since at linear 
order in $M_t/v$ no fermions loops need to be considered (only NGB and the $h$ internal lines appear at the one loop level). 
This in particular means that no wave-function or mass renormalization is needed and amplitude renormalization requires  only 
renormalization of the coupling of a four-dimensional operator.
Thus we make again the approximation $M_h=M_Z=M_W=0$ but with $M_t \ne 0.$

Now we introduce heavy fermions in the HEFT. This will make possible the computation of the amplitudes of processes 
such as $W_L^+ W_L^- \rightarrow t \bar t$ or $Z_L Z_L \rightarrow t \bar t$  (and also $hh \rightarrow t \bar t$) 
in the energy regime $M_t^2/v^2\ll\sqrt{s}M_t/v^2\ll s/v^2$. The HEFT Lagrangian is then
\begin{eqnarray}  \label{LagrangianI}
\mL  & = & \frac{v^2}{4} F(h)\Tr\left[\left(D_\mu U\right)^\dagger D^\mu U\right] %
      +\frac{1}{2}\partial_\mu h \partial^\mu h   -  V(h) \nonumber   \\
      & + & i\bar{Q}\slashed \partial Q 
      - v G(h)\left[\bar{Q}_L^\prime UH_Q Q_R^\prime + \text{h.c.} \right],
\end{eqnarray}
where the $U(x)\in SU(2)$ is parametrized again as 
\begin{equation}
U=\sqrt{1-\frac{\omega^2}{v^2}}+i\frac{\bar{\omega}}{v},
\end{equation}
with $\bar{\omega}=\tau_i\omega^i$.
In the fermionic  sector (last line) of the Lagrangian in Eq.~(\ref{LagrangianI}), the quark doublets are
\begin{equation}
Q^{(\prime)} = \left( %
    \begin{array}{c}
        \mU^{(\prime)} \\ 
        \mD^{(\prime)}
    \end{array} %
\right),
\end{equation}
where the two $Q$ components correspond to the up and down quark sectors
\begin{align}
\mU &=\left( u,c,t\right),  &
\mD &=\left( d,s,b\right).
\end{align}
The Yukawa-coupling matrix in Eq.~(\ref{LagrangianI}) can be written as
\begin{equation}
H_Q=\left( 
      \begin{array}{cc}
        H_U & 0 \\ 
        0   & H_D%
      \end{array}%
\right) .
\end{equation}
This matrix can be diagonalized by unitary transforming independently the right and left-handed up and down quark flavor multiplets
\begin{align}
\mU_{L,R} = V_{L,R}^U \mU_{L,R}^\prime , &&\mD_{L,R} = V_{L,R}^D\mD_{L,R}^\prime  ,
\end{align}
where $V_{L,R}^{U,D}$ are four $3\times 3$ unitary matrices. As it happens with $F(h)$,  the $G(h)$ function is an  
arbitrary analytical functions on the Higgs field $h$, which is usually parametrized as
\begin{equation}
     G\left( h\right) = 1 + c_1\frac{h}{v} + c_2\frac{h^2}{v^2}+\dots
\end{equation}
In the rest of this  work, these functions are only needed up to the quadratic terms. In the minimal SM we 
have $c_1=1$ and $c_i=0$ for $i \ge 2$. The Yukawa part of the Lagrangian can be written as
\begin{align}
\mL_Y ={}&- G\left( h\right) \left\{%
   \sqrt{1-\frac{\omega^2}{v^2}} \left( \overline{\mU}M_U\mU + \overline{\mD}M_D\mD\right)              \right. \nonumber\\  
  &\left. +\frac{i\omega^0}{v}\left( \overline{\mU}M_U\gamma^5\mU - \overline{\mD}M_D\gamma^5\mD\right) \right. \nonumber\\
  &\left. +i\sqrt{2}\frac{\omega^+}{v}\left(\overline{\mU}_L V_{CKM} M_D\mD_R-\overline{\mU}_R M_U V_{CKM}\mD_L\right)  \right. \nonumber\\
  &\left. +i\sqrt{2}\frac{\omega^-}{v}\left(\overline{\mD}_L V_{CKM}^\dagger M_U\mU_R-\overline{\mD}_R M_DV_{CKM}^\dagger \mU_L\right)  \right\} 
\end{align}
with the charged NGB states given by  $\omega^\pm =(\omega^1 \mp i \omega^2)/\sqrt{2}$,  $\omega^0 = \omega^3$. 
$V_{CKM}= V_{L,}^U V_{L,}^{D\dagger}$ is the  Cabibbo-Kobayashi-Maskawa matrix  and the new quark fields $\mU$ and $\mD$ are  the mass 
eigenstates with $M_{U}$ and $M_{D}$ being the corresponding mass-matrices which are diagonal, real and positive.

Here we are interested in a scale of energies  at which the only quark mass different from zero is the top mass $M_t $ 
and we will use the ET. Also we can take $V_{tb} \simeq 1$. Then the relevant Yukawa Lagrangian for us  is
\begin{align}
\mL_Y ={}& - G(h) \left\{%
     \left( 1-\frac{\omega^2}{2v^2}\right) M_t t\bar{t}\right. \nonumber\\
    &\left. +\frac{i\omega^0}{v}M_t\bar{t}\gamma^5 t - i\sqrt{2}\frac{\omega^+}{v} M_t\bar{t}_R b_L%
        +i\sqrt{2}\frac{\omega^-}{v} M_t\bar{b}_L t_R\right\} ,
\end{align}
where we have kept only $\mO(\omega^2/v^2)$ terms. 

Finally, the relevant HEFT Lagrangian  to describe the $w^a w^b \to t\bar{t}$ and $hh\to t\bar{t}$ processes, in 
the regime $M_t^2/v^2\ll M_t\sqrt{s}/v^2\ll s/v^2$, is given by
\begin{align} \label{EFTLagrangianexpanded}
\mL &=\frac{1}{2}\partial_\mu h\partial^\mu h -\left(1+c_1\frac{h}{v}+c_2\frac{h^2}{v^2}\right)
\left\{ \left(1-\frac{\omega^2}{2v^2}\right) M_t t\bar{t} \right.\notag\\
    &\left.+ \frac{i\sqrt{2}\omega^0}{v} M_t\bar{t}\gamma^5 t-i\sqrt{2}\frac{\omega^+}{v}M_t\bar{t}_R b_L+
i\sqrt{2}\frac{\omega^-}{v}M_t \bar{b}_L t_R\right\}  \notag\\
&+\frac{1}{2}\left(1+2a\frac{h}{v}+b\left(\frac{h}{v}\right)^2\right)
\partial_\mu\omega^i\partial^\mu\omega_j\left(\delta_{ij}+\frac{\omega_i\omega_j}{v^2}\right) .
\end{align}
As we will see below the divergencies appearing at the one-loop level in the amplitudes of these processes 
can be absorbed by renormalization of the to couplings $g_t$ and $g'_t$ of the four dimension operators
\begin{equation}
\mL_{4t} = g_t\frac{M_t}{v^4}(\partial_\mu\omega^i\partial^\mu\omega^j) t\bar{t} 
   +g'_t \frac{M_t}{v^4}(\partial_\mu h\partial^\mu h) t\bar{t}.
\end{equation}

%%%%%%%%%%%%%%%%%%%%%%%%%%%%%%%%%%%%%%%%%%%%%%%%%%%%%%%%%%%%%%%%%%%%%%%%%%%%%%%%%%%%%%%%%%%%%%%%%%%%%%%
%%%%%%%%%%%%%%%%%%%%%%%%%%%%%%%%%%%%%%%%%%%%%%%%%%%%%%%%%%%%%%%%%%%%%%%%%%%%%%%%%%%%%%%%%%%%%%%%%%%%%%%

\subsection{Tree level and one-loop contributions  \label{sec:amplitudeswTLOl}}
At tree level, the scattering amplitude for  $w^a w^b\to t^{\lambda_1}\bar{t}^{\lambda_2}$    is given by
\be
\mQ^{\text{tree}} \left(w^a w^b\to t^{\lambda_1}\bar{t}^{\lambda_2}\right)  
 = \sqrt{3}\left(1-ac_1+\frac{g_t}{2}\frac{s}{v^2}\right) \frac{M_t}{v^2} \bar{u}^{\lambda_1}(p_1)v^{\lambda_2}(p_2)\delta_{ab}, \label{GoldGold1}
\ee
where $p_1$, $p_2$ and $\lambda_1$, $\lambda_2$ are the top, antitop momenta and helicities respectively. The $\sqrt{3}$  
factor is a color factor since the $t\bar t$ pair is produced in a color singlet state. At the one-loop level and 
using dimensional regularization with dimension $D=4-\epsilon$ and at the lowest $M_t/v$ order, the  amplitude is  \cite{Castillo:2016erh}
\begin{equation}
  \mQ\left(w^a w^b\to t^{\lambda_1}\bar{t}^{\lambda_2}\right) = %
      \sqrt{3}\left(Q^{\text{tree}}+Q^{\text{1-loop}}\right)\frac{M_t}{v^2}\bar{u}^{\lambda_1}v^{\lambda_2}\delta_{ab},
\end{equation}
and:
\begin{align}
Q^{\text{tree}}(s)   &= 1-ac_1 + \frac{g_t}{2}\frac{s}{v^2}  \label{res_Q_tree}\\
Q^{\text{1-loop}}(s) &= \frac{s}{(4\pi)^2 v^2}C_t \left(N_{\varepsilon}+2-\log\frac{-s}{\mu^2}\right) \label{res_Q_NLO} 
\end{align}
where $\mu$ is an arbitrary renormalization scale and
\begin{equation}
C_t = (1-ac_1)(1-a^2)+c_2 (b-a^2) .
\end{equation}
Now it is possible to absorb the  divergence in Eq.~(\ref{res_Q_NLO}) by renormalizing the $g_t$ coupling, 
for example by   using  the  $\overline {MS}$ renormalization scheme. Now we define the renormalized coupling $g_t^r$ as
\begin{equation}
   g_t^r=g_t+\frac{C_t}{8\pi^2}N_\epsilon 
\end{equation}
and the NLO is given by 
\begin{align} \label{Qdecomposed}
Q^{\rm NLO}(s) &= Q^{\text{tree}}\left(s\right)+Q^{\text{1-loop}}\left(s\right) \\  \nonumber 
               &= 1-ac_{1}+\frac{s}{v^2}\left[\frac{g_t^r}{2}+\frac{C_t}{(4\pi)^2}\left(2-\log\frac{-s}{\mu^2}\right)\right] .
\end{align}
As we do not have any wave-function or mass renormalization, this amplitude must be observable and  
hence $\mu$-independent. Then it is very easy to find the renormalization group evolution equation
\begin{equation}
g^r_t\left(\mu\right) = g^r_t\left(\mu_0\right)-\frac{C_t}{8\pi^2}\log\left(\frac{\mu^2}{\mu_0^2}\right) .
\end{equation}
On the other hand, the non-vanishing  spinor combinations of helicities appearing in the above 
amplitudes are to the LO in $M_t/\sqrt{s}$ expansion
\be
\bar{u}^+ (p_1) v^{+}(p_2) = -\bar{u}^- (p_1) v^{-}(p_2)= \sqrt{s-4M_t^2}\simeq\sqrt{s},
\ee
where the helicity indices $+$ and $-$ refer to $\lambda = +1/2$ and $\lambda = - 1/2$, respectively.
Therefore, the  non-vanishing  $w^a w^b\to t\bar t$ amplitudes are given by
\be
\mQ\left(w^a w^b\to t^+\bar{t}^+\right)=-\mQ\left(w^a w^b\to t^-\bar{t}^-\right) 
=\sqrt{3}Q^{\rm NLO}(s)\frac{M_t\sqrt{s}}{v^2}\delta^{ab}.\label{AmplitudeGoldstone}
\ee
In a similar way  we may consider the  $hh\to t\bar{t}$ annihilation. The result is
\be
  \mN \left( hh\to t^+ \bar{t}^+\right) = -2\sqrt{3}c_{2}\frac{M_t\sqrt{s}}{v^2}   
 + \sqrt{3}\frac{s}{v^{2}}\left[ \frac{g'^r_t(\mu)}{2}-\frac{3C'_t}{32\pi^2}
\left( 2-\log\frac{-s}{\mu^2}\right) \right]  \frac{M_t\sqrt{s}}{v^2},  \label{AmplitudeHiggsII}
\ee
where the renormalized coupling $g'^r_t$ is defined as
\begin{equation}
g'^r_t = g'_t-\frac{3C'_t}{(4\pi)^2}N_\varepsilon
\end{equation}
and
\begin{equation}
C'_t=(b-a^2)(1-ac_1).
\end{equation}
The renormalization group evolution equation is
\begin{equation}
g'^r_t\left(\mu\right) = g'^r_t(\mu_0)+\frac{3C'_\text{t}}{(4\pi)^2}\log\left(\frac{\mu^2}{\mu_0^2}\right) .
\end{equation}

As in the previous reaction here we have also
\be
 \mN\left( hh\to t^-\bar{t}^-\right) = -\mN\left(hh\to t^+\bar{t}^+\right).
\ee
with vanishing amplitudes for  the rest of  helicity combinations.

%%%%%%%%%%%%%%%%%%%%%%%%%%%%%%%%%%%%%%%%%%%%%%%%%%%%%%%%%%%%%%%%%%%%%%%%%%%%%%%%%%%%%%%%%%%%%%%%%%%%%%%%%%%%%%%%%%%%%%%%%%%%%%%%
%%%%%%%%%%%%%%%%%%%%%%%%%%%%%%%%%%%%%%%%%%%%%%%%%%%%%%%%%%%%%%%%%%%%%%%%%%%%%%%%%%%%%%%%%%%%%%%%%%%%%%%%%%%%%%%%%%%%%%%%%%%%%%%%%

\subsection{Helicity amplitudes and unitarity}
In order to study the unitarity behavior of the NLO amplitudes for $t \bar t$ production from $ww$ or $hh$ pairs 
it is also useful to perform a partial wave decomposition.  As the $t$ quark is a member of a custodial isospin doublet, 
a $t \bar t$ state can couple in principle to $I=0$ or $I=1$ $ww$ or $hh$ states (only to $I=0$ in the second case). However 
the amplitudes obtained for $t \bar t $ production are parity conserving and  the $I=1$ case is excluded. Thus the 
$t^{\lambda_1}\bar t^{\lambda_2}$ states  couple only to $\mid I=0> = \sum_i\mid \omega^a\omega^a>/\sqrt{3}$ and $hh$ 
states~\cite{Castillo:2016erh}. Even more, the initial state $\mid I=0>$  couples only to 
$\mid S=1,S_Z=0>= (\mid +,+>-\mid-,->)/\sqrt{2}$ $t\bar t$ state. The corresponding partial waves are given by
\begin{align}
 q^{J}_{\lambda_{1}\lambda_{2}}(s)=\frac{1}{64\pi^{2}}\int d\Omega \mD_{0\lambda}^{J}\left(\phi,\theta,-\phi\right)
 \mQ\left( w w \to t^{\lambda_{1}}\bar{t}^{\lambda_{2}}\right)
\end{align}   
and
\begin{align}
 n^{J}_{\lambda_{1}\lambda_{2}}(s)=\frac{1}{64\pi^{2}}\int d\Omega \mD_{0\lambda}^{J}\left(\phi,\theta,-\phi\right)
 \mN\left( h h \to t^{\lambda_{1}}\bar{t}^{\lambda_{2}}\right),
\end{align}   
where we have to consider the case $J=0$ and $\lambda=\lambda_1-\lambda_2=0$ only. Then, the partial wave corresponding 
to the $\mid S=1,S_z=0 >$ $t\bar t$ state  is given by
\begin{equation}
   q = \frac{1}{\sqrt{2}}(q^0_{++}-q^0_{--})=\sqrt{2}q^0_{++},
\end{equation} 
which can be expanded as:
\begin{equation}
   q=q^{(0)}+q^{(1)}+\dots,
\end{equation}
that have the form:
\begin{align}
q^{(0)}\left( s\right) &= K^q\sqrt{s}M_t, \label{CoeffI} \\
q^{(1)}\left( s\right) &= \left( B^q\left( \mu \right) + E^q\log\frac{-s}{\mu^2}\right) s\sqrt{s}M_t, \label{CoeffII}
\end{align}
with
\begin{align}
K^q                 &=  \frac{3}{16\pi v^2}\left( 1-ac_1\right) , \\
B^q\left(\mu\right) &=  \frac{3}{16\pi v^4}\left[\frac{g_t(\mu)}{2}+\frac{C_t}{8\pi^2}\right] , \\
E^q                 &= -\frac{3}{16\pi v^4}\frac{C_t}{%
16\pi ^{2}}.
\end{align}
In a similar way it is possible to obtain the $J=0$ partial wave for the  $hh\to t\bar{t}$ reaction
\begin{equation}
n=\sqrt{2}n^0_{++}=n^{(0)}+n^{(1)}+\dots,
\end{equation}
where the first two terms have the same form than in the $ww$ case. Using an  obvious notation one gets in this case
\begin{align}
K^n                 &= -\frac{\sqrt{3}c_2}{8\pi v^2}, \\
B^n\left(\mu\right) &=  \frac{\sqrt{3}}{16\pi v^4}\left(\frac{g'_t(\mu)}{2}-\frac{3C'_t}{16\pi^2}\right), \\
E^n                 &=  \frac{\sqrt{3}}{16\pi v^4}\frac{3C'_t}{32\pi^2}.
\end{align}
Now we can collect all the $ww$, $hh$ and  $t \bar t$ states $J=0$ partial waves in the amplitude matrix: 
\begin{equation}
\label{Tmatrix}
T=T_{00} = \left( 
\begin{array}{ccc}
   t_{00} & m_{0} & q \\ 
   m_{0} & a_{0} & n \\ 
   q & n & r \\ 
\end{array}%
\right) .
\end{equation}
Obviously $r$ is the appropriate $t \bar t \to t \bar t $ partial wave. However this partial wave is of order $M_t^2/v^2$ and,  
according to our approximation, it will be set to zero. The amplitude above is symmetric since all interactions are 
time-reversal invariant. As all the particles involved are considered massless (in accordance with the use of the ET) 
the matrix entries are analytical functions on the Mandelstam variable $s$ with a unitarity RC starting at $s=0$ and a 
LC starting also at this point.  
As usual  the physical partial waves are found on the RC along $s=E_{CM}^2+i\epsilon$, where $E_{CM}$ is the reaction's  
CM energy. For this physical $s$ values the unitarity condition for the $T$ matrix reads
\begin{equation} \label{matrixunitarity}
\Imag T=TT^\dagger 
\end{equation}
which, in terms of the matrix elements, translates to:
\begin{subequations}\label{eq:unitar:general}
\begin{align}
   \Imag t &= \lvert t\rvert^2 + \lvert m\rvert^2 + \dots \label{eq:unitar:first}\\
   \Imag m &= t m^* + m t^* + \dots\\
   \Imag a &= \lvert m\rvert^2 + \lvert a\rvert^2 + \dots \\
   \Imag q &= t q^* + mn^* + \dots \label{chLagr:UnitProceduresWW:Qcoupled2:unitarQ}\\
   \Imag n &= mq^* + an^* + \dots \label{newunitarity2}\\
   \Imag r &= 0 + \dots \label{newunitarity3}
\end{align}
\end{subequations}
where we neglect terms that are higher order in $M_t/v$. On the other hand the $T$ matrix can be expanded according to 
our previous computations as
\be
T = T^{(o)}+T^{(1)}+...
\ee
Up to NLO the relations we get are the following: 
\begin{subequations}
\begin{align}
\Imag t_{00}^{(1)} &= \left\lvert t_{00}^{(0)}\right\rvert^2 + \left\lvert m_{0}^{(0)}\right\rvert^2 \label{pertunit1}\\
\Imag m_{0}^{(1)}  &= t_{00}^{(0)} m_{0}^{(0)} \label{pertunit2}\\
% + M_{0}^{(0)} T_{0}^{(0)}  
\Imag a_{0}^{(1)}  &=  \left\lvert m_{0}^{(0)} \right\rvert^2 \label{pertunit3}\\
% + \left\lvert T_{0}^{(0)}\right\rvert^2 
\Imag q^{(1)}      &= t_{00}^{(0)} q^{(0)}  + m^{(0)}_0 n^{(0)}  \label{pertunit4}\\
\Imag n^{(1)}      &= m_{0}^{(0)} q^{(0)} \label{pertunit5}\\
\Imag r^{(0)}       &= 0 , \label{pertunit6}
\end{align}
\end{subequations}
as it can be checked case by case.  The first three equations were obtained in \cite{Delgado:2013hxa}. The last two are equivalent to
\begin{align}
 -\pi E^Q &= K K^Q + K_0' K^N, \nonumber \\
 -\pi E^N &=   K_0'  K^Q
\end{align}
which can be explicitly obtained. This is a very non-trivial check of the computations since all the equations above can be 
collected in
\be
\Imag T^{(1)}=T^{(o)}T^{(o)}
\ee
i.e. we have unitarity at the perturbative level but it is broken as we rise the energy, limiting strongly the applicability of 
these kind of computations in their simplest version.

However, at it happened in previous considered cases like $ww$, $hh$ and $\gamma\gamma$ scattering, it is possible to 
improve the situation by means of using an appropriate unitarization procedure. This unitarization method will make it 
possible, not only to extend the validity of the NLO computations to higher energies, but also to generate poles in the 
second Riemann sheet that will play the role of dynamical resonances. In order to introduce the unitarization method 
we will start by considering the simpler particular case $a^2=b$. Then  the $hh$ state decouples and the $T$ matrix simplifies to
 \begin{equation}
   T= %
     \begin{pmatrix}
        t_{00} & q_0 \\
        q_0 & r_0
     \end{pmatrix}
     \equiv %
     \begin{pmatrix}
        t & q \\
        q & r
     \end{pmatrix}.
\end{equation}
In the physical region unitarity reads:
\begin{subequations}
\begin{align}
   \Imag t &= \lvert  t \rvert^2 + \mO\left(\frac{M_t^2}{v^2}\right) \label{elasticunit}\\
   \Imag q &= t q^* + \mO\left(\frac{M_t^3}{v^3}\right) \label{chLagr:UnitProceduresWW:Qcoupled1:unitarQ}\\
   \Imag r &= 0 + \mO\left(\frac{M_t^2}{v^2}\right) .
\end{align}
\end{subequations}
In order to solve these unitarity constraints we can proceed as follows: first we solve  Eq.\ref{pertunit4} by using the 
elastic IAM method  defining the unitarized amplitude
\begin{equation}\label{onechannelIAM}
\tilde{t}=\frac{(t^{(0)})^2}{t^{(0)}-t^{(1)}}.
\end{equation}
Next we introduce the   unitarized $w w  \to  t \bar t$ partial wave as 
\begin{equation}\label{chLagr:UnitProceduresWW:Qcoupled1:unitar:raw}
   \tilde{q} = q^{(0)} + q^{(1)}\frac{\tilde{t}}{t^{(0)}} .
\end{equation}
The amplitudes $\tilde t$ and $\tilde q$ fullfil the unitarity equations  $\Imag\tilde{t} = \tilde{t}\tilde{t}^*$ and 
$\Imag \tilde q = t q^*$ in the physical region. They are analytical functions on the complex variable $s$ with a unitarity 
RC and also a LC. They match the perturbative computation at low energies
\begin{align}
 \tilde t &= t^{(0)} + t^{(1)} + \dots \nonumber\\
 \tilde q &= q^{(0)} + q^{(1)} + \dots
\end{align}
Finally, both functions can show a pole in the second Riemann sheet (resonance) for some regions of the coupling space. 
A simple inspection of the $\tilde t(s)$ and $\tilde q(s)$ partial waves shows that, if that is the case, the position 
of the pole is exactly the same for both processes as it must be (corresponding to the same $I=J=0$ physical resonance).

With some more effort the unitarization method for $t \bar t$ production can be extended to the $a^2 \ne b$ case where the 
$hh$ state is also involved. Now we come back to the $3\times 3$ $T$ matrix in Eq.~(\ref{Tmatrix})
\begin{equation}
T= \left( 
\begin{array}{ccc}
   t & m & q \\ 
   m & a & n \\ 
   q & n & r  \\ 
\end{array}%
\right) 
\end{equation}
where we have simplified the notation an obvious way. Now we introduce the submatrix
\begin{equation}
   K \equiv \begin{pmatrix}
           t & m \\
           m & a
       \end{pmatrix},
\end{equation}
with chiral expansion
\be
K=K^{(0)}+K^{(1)}+\dots
\ee
 with $\Imag K^{(1)}= K^{(0)}K^{(0)}$ on the RC (perturbative unitarity). By using the IAM method we introduce the matrix
 \begin{equation}
\tilde{K}=K^{(0)}(K^{(0)}-K^{(1)})^{-1} K^{(0)} \ 
\end{equation}
which fulfills exact unitarity on the RC:  $\Imag\tilde{K} = \tilde{K}\tilde{K}^\dagger$. The rest of the unitary conditions 
can be written in a condensed way
\begin{equation}
\label{Unitaryconditions}
 \Imag 
  \begin{pmatrix}
             q \\
             n
       \end{pmatrix} = K
  \begin{pmatrix}
             q^* \\
             n^*
       \end{pmatrix} .       
\end{equation}
At the perturbative level we have
\begin{equation}
 \Imag 
  \begin{pmatrix}
             q^{(1)} \\
             n^{(1)}
       \end{pmatrix} = K^{(0)}
  \begin{pmatrix}
             q^{(0)} \\
             n^{(0)}
       \end{pmatrix}        .
\end{equation}
By using this equation it is not difficult to show that the amplitudes
\begin{equation}
   \begin{pmatrix}
             \tilde q \\
             \tilde n
       \end{pmatrix} =
  \begin{pmatrix}
             q^{(0)} \\
             n^{(0)}
       \end{pmatrix} + \tilde K  K^{(0)-1}
  \begin{pmatrix}
             q^{(1)} \\
             n^{(1)}
       \end{pmatrix}        
\end{equation}
fulfill the unitarity conditions in Eq.~(\ref{Unitaryconditions}). 
In addition these partial waves have all the good properties mentioned above as analyticity in the whole 
complex plane, LC, RC, the possibility for developing poles in the second Riemann sheet and they match the low-energy NLO result
\begin{equation}
  \begin{pmatrix}
           \tilde  q \\
            \tilde n
       \end{pmatrix} =
  \begin{pmatrix}
             q^{(0)} \\
             n^{(0)}
       \end{pmatrix}        +
  \begin{pmatrix}
             q^{(1)} \\
             n^{(1)}
       \end{pmatrix}   +...            
\end{equation}

%%%%%%%%%%%%%%%%%%%%%%%%%%%%%%%%%%%%%%%%%%%%%%%%%%%%%%%%%%%%%%%%%%%%%%%%%%%%%%%%%%%%%%%%%%%%%%%%%%%%%%%%%%%%%%%%%%%%%%
%%%%%%%%%%%%%%%%%%%%%%%%%%%%%%%%%%%%%%%%%%%%%%%%%%%%%%%%%%%%%%%%%%%%%%%%%%%%%%%%%%%%%%%%%%%%%%%%%%%%%%%%%%%%%%%%%%%%%%

\section{Resonance production in $VV$ fusion}
In this section we will describe the basic tools to study vector boson fusion beyond the MSM.
Unlike in the previous processes, we shall try to avoid using the ET as much as
possible in the study of this process for reasons that we will discuss below.

Let us begin by reviewing the experimental bounds on various parameters of the HEFT that play a role
in this process. The parameters $a$ and $b$ control the coupling of the Higgs to the gauge sector~\cite{composite}.
Couplings containing higher powers of $h/v$ or having more than four fields do not enter $WW$ scattering 
at this order\footnote{$a_4$ and $a_5$ could for instance be functions 
of $h/v$ but only their constant part is relevant at this order} and they have not
been included in (\ref{eq:1}). We have also introduced two additional parameters
 $d_{3}$, and $d_{4}$ that parameterize the three-
and four-point interactions of the Higgs field. We bear in mind that this is not the most
general form of the Higgs potential and in fact additional counter-terms are needed beyond the
Standard Model\cite{Delgado:2013hxa}
but this does not affect $W_LW_L$ scattering (see Sect \ref{sec:ecpt}).

\begin{figure}[t]
	\centerline{\includegraphics[scale=0.5]{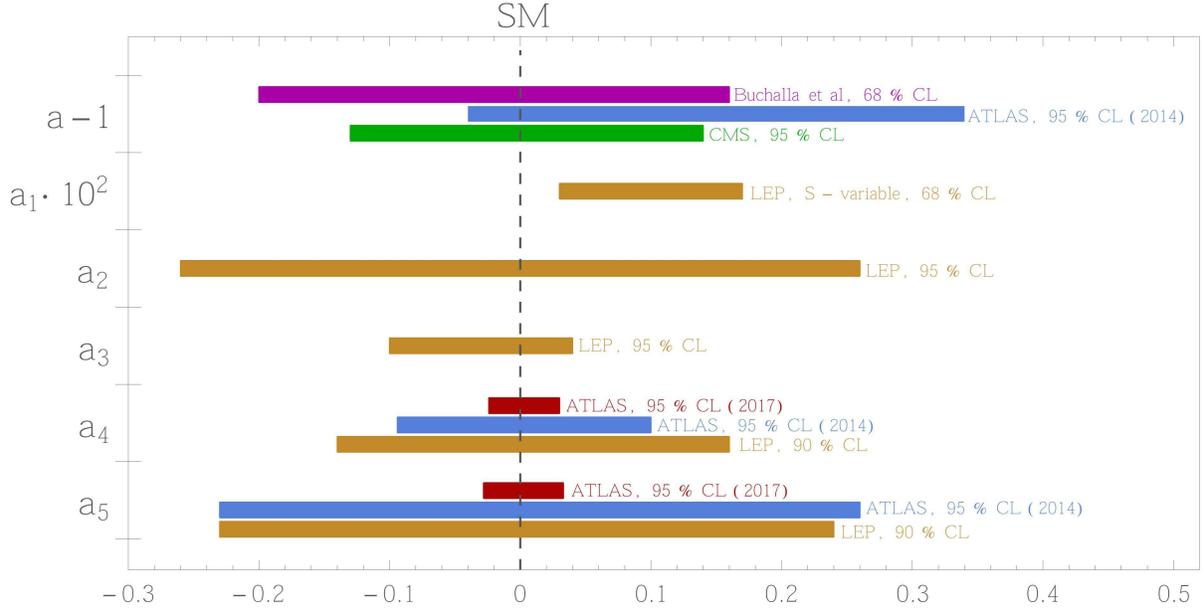}}
	\caption{\label{constraint_chiral} The experimentally allowed  parameter region of the HEFT allowed by current
and past experiment. From \cite{DDEGHMS}.}
\end{figure}

In the custodial limit the other two low energy constants that enter in the calculation
are the coefficienients $a_4$ and $a_5$ of the $O(p^4)$ HEFT, previously defined.
The MSM  case corresponds to setting $a=b=d_{3}=d_{4}=1$ and $a_4=a_5=0$.
Current experimental analysis gives the bounds for $a$, $a_{4}$ and $a_5$ shown in Fig.\ref{constraint_chiral}
(bounds on $a_1$, $a_2$ and $a_3$ are also given even if they are not very relevant for $VV$ fusion; see below)
The present data clearly favours values of $a$ close to the MSM value, while the $a_4$ and
$a_5$ are still largely unbounded. The parameter $b$ is almost totally
undetermined at present. Our   $a$ and $a_{4,5}$ coefficients stand for $a=1-\xi c_H/2$, $a_4=\xi^2 c_{11}$ and
$a_5=\xi c_{6}$ of ref.~\cite{Concha}. $c_H$ range comes from the values of Set A in table 4 and $c_{6,11}$
are from table 8 of ref.~\cite{Concha}.
 
The parameters $a_1$, $a_2$ and $a_3$ enter the oblique and triple
gauge coupling, respectively, as discussed before. Bounds on the oblique corrections, henece on $a_1$ 
are quite constraining~\cite{Pich:2013fea}, while the
triple electroweak gauge coupling has already been measured with a level of precision\cite{triple} similar to LEP.
Some results on the $\gamma\gamma W^+W^-$ coupling are also available\cite{gammagamma}.

When $a$ and $b$ depart from their MSM values $a=b=1$  the theory becomes
unrenormalizable in the conventional sense. As we already saw in previous sections,
at the one-loop level $W_LW_L$ scattering can be rendered
finite by a suitable redefinition of the coefficients $a_4$ and $a_5$  (together with $v$, $M_H$, $d_3$ and $d_4$).
This renormalization procedure could be systematically extended, in principle to all orders, at the price
of introducing more and more higher dimensional counterterms.

\begin{figure}
\centerline{\includegraphics[scale=0.5]{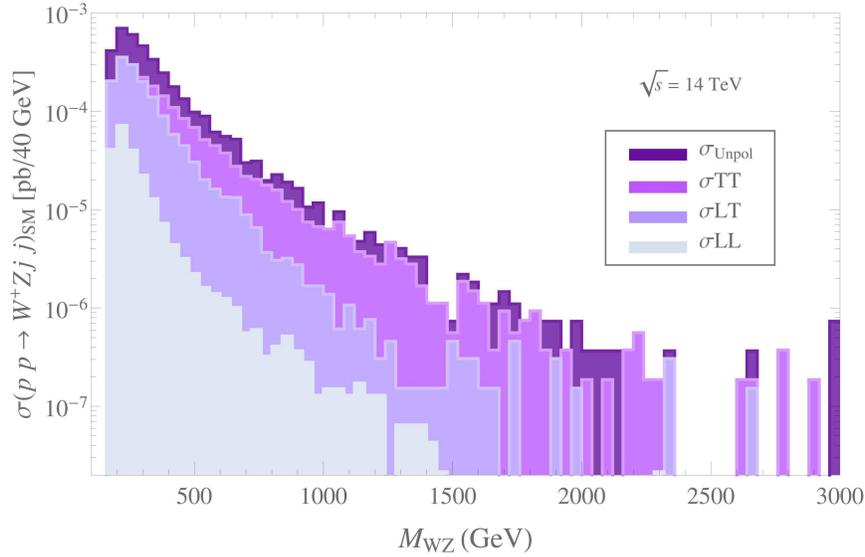}}
\caption{The plot shows the relative importance of the various polarizations in $WZ$ scattering. Distributions 
of the SM-EW background with the invariant mass of the $WZ$ pair. The imposed cuts are $|\eta_{j_1,j_2}|< 5$ , 
$ \eta_{j_1}\cdot \eta_{j_2} < 0$ and $|\eta_{W,Z}| < 2$. The predictions
for the various polarizations $\sigma_{AB}$ of the final $WZ$ pair as well as the total unpolarized cross
section are displayed separately. Starting from the upper to the lower lines they
correspond respectively to: $\sigma_{Unpol}, \sigma_{TT}, \sigma_{LT}, \sigma_{LL}$.
The calculation is done in the MSM. From \cite{DDEGHMS}.\label{fig:pp_WZjj_invmass_sm}}
\end{figure}

If a resonance is present, in that region the process will be 
dominated by the scattering of the longitudinal components $W_LW_L$ but
this does not mean at all that the scattering of transverse polarizations could be neglected. 
In Fig.\ref{fig:pp_WZjj_invmass_sm}
we show the contribution of the various polarizations to the $WZ\to WZ$ scattering process. As it can be seen
the contribution from transverse or mixed polarization initial states is very relevant, dominant in
fact. This is one of the reasons
why a blind use of the ET may result in too qualitative results for this important process.

Another reason for not trusting the ET in this particular process 
is that the lowest order partial wave $t^{(0)}(s)$ vanishes for $a=1$. This is a relevant point, as we 
know experimentally that $a$ does not depart significantly from that point, and it makes perfect sense to 
explore a range of values for $a_4$ and $a_5$ while keeping $a$ 
very close or equal to 1. Recall that if $t^{(0)}_{IJ}$ vanishes the IAM method cannot be applied. 

As emphasized in \cite{Espriu:2012ih} when dealing with longitudinally polarized amplitudes,
as opposed to using the ET approximation, caution must be exercised to account
for an ambiguity introduced by the longitudinal polarization vectors that do not transform under
Lorentz transformations as 4-vectors. Expressions involving the polarization vector $\epsilon_{L}^\mu$ 
cannot be cast in terms of the Mandlestam variables $s$, $t$, and $u$
until an explicit reference frame has been chosen, as they cannot themselves be written
solely in terms of covariant quantities. Obviously, amplitudes still satisfy crossing
symmetries when they remain expressed in terms of the external 4-momenta. A short discussion
on this point is placed in appendix~\ref{sec:appendix_crossing}. This subtlety makes
the discussion and decomposition in fixed-isospin partial waves slightly more involved than the one 
presented above in Sects. \ref{sec:ecpt} and \ref{sec:nlo}.

A generic  amplitude, $A(W^{a}(p^{a})+W^{b}(p^{b})\to W^{c}(p^{c}) + W^{d}(p^{d}))$,  can be written
using isospin and Bose symmetries as
\bea
A^{abcd}(p^{a},p^{b},p^{c},p^{d})&=&  \delta^{ab} \delta^{cd} A(p^{a},p^{b},p^{c},p^{d}) +
\delta^{ac} \delta^{bd} A(p^{a},-p^{c},-p^{b},p^{d})\label{eq:isospin_general}\\
&+&
\delta^{ad} \delta^{bc} A(p^{a},-p^{d},p^{c},-p^{b}),\no
\eea
with
\bea
\label{eq:amplitudes_general}
A^{+-00} & = & A(p^{a},p^{b},p^{c},p^{d}) \\ \no
A^{+-+-} & = & A(p^{a},p^{b},p^{c},p^{d}) + A(p^{a},-p^{c},-p^{b},p^{d}) \\
A^{++++} & = & A(p^{a},-p^{c},-p^{b},p^{d}) + A(p^{a},-p^{d},p^{c},-p^{b}).\no
\eea
The fixed-isospin amplitudes are given by
\bea
\label{eq:fixed_isospin}
T_{0}(s,t,u) & = & \langle 00\vert S\vert 00 \rangle  =   3 A^{+-00} +   A^{++++} \\ \no
T_{1}(s,t,u) & = & \langle 10\vert S\vert 10 \rangle  =   2 A^{+-+-} - 2 A^{+-00} - A^{++++} \\ \no
T_{2}(s,t,u) & = & \langle 20\vert S\vert 20 \rangle  =   A^{++++} \, . \nn
\eea
We shall also need the amplitude for the process $W^+W^-\to hh$. Taking into account that the final state
is an isospin singlet and defining
\be
A^{+-}= A(W^{+}(p^{+})+W^{-}(p^{-})\to h(p^{c}) + h(p^{d})) \, ,
\ee
the projection of this amplitude to the $I=0$ channel gives
\be
\label{eq:fixed_isospin_hh}
T_{H, 0}(s,t,u)= \sqrt{3} A^{+-}.
\ee
The partial wave amplitudes for fixed isospin $I$
and total angular momentum $J$ are given by Eq. \ref{Jprojection}. As customary, in the analysis we 
will consider tree-level and one-loop corrections; namely
\bea\label{eq:tloop}
t_{IJ}(s) = t_{IJ}^{(0)}(s) + t_{IJ}^{(1)}(s)\,.
\eea
$t_{IJ}^{(0)}(s)$  can be constructed from the above expressions
by using crossing and isospin relations for the tree level contributions of $A^{+-00}$ (Fig.~\ref{tree-diagrams}).
The analytic results of $A^{+-00}$ at tree-level are in appendix~\ref{sec:appendix_amplitudes}.  
$t_{IJ}^{(0)}(s)$ contains the anomalous coupling $a$ but $b$ does not enter
at tree-level.  $t_{IJ}^{(1)}(s)$  includes tree-level contributions from $a_i$  counter-terms
(see appendix~\ref{sec:appendix_amplitudes} for analytic result) and
the one-loop corrections to the diagrams  in Fig.\ref{tree-diagrams}.
At one-loop level, the $b$ parameter enters $t_{IJ}^{(1)}(s)$ through the one-loop expression of $A^{+-00}$ 
calculated in \cite{Espriu:2013fia}.
\begin{figure}[h!]
\begin{center}
\includegraphics[clip,width=0.20\textwidth]{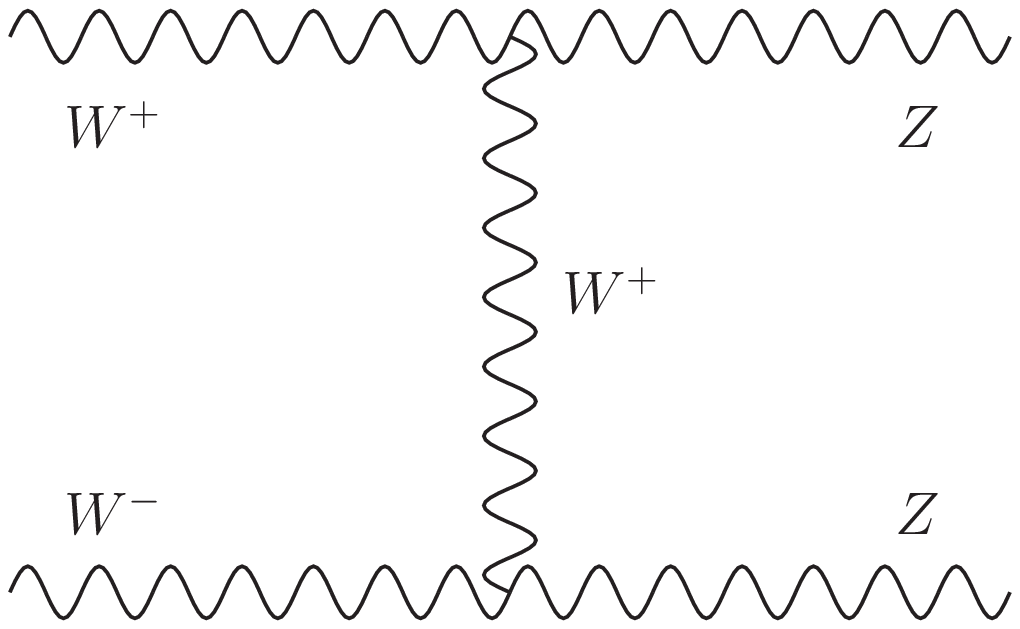} \hspace{0.70cm}
\includegraphics[clip,width=0.20\textwidth]{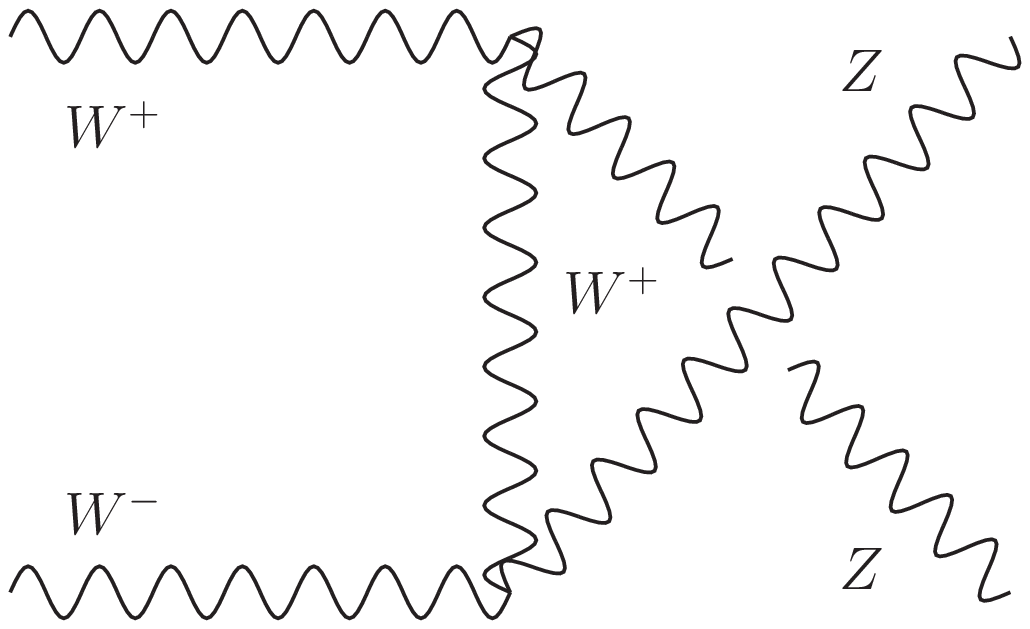} \\ \vspace{0.15cm}
\includegraphics[clip,width=0.20\textwidth]{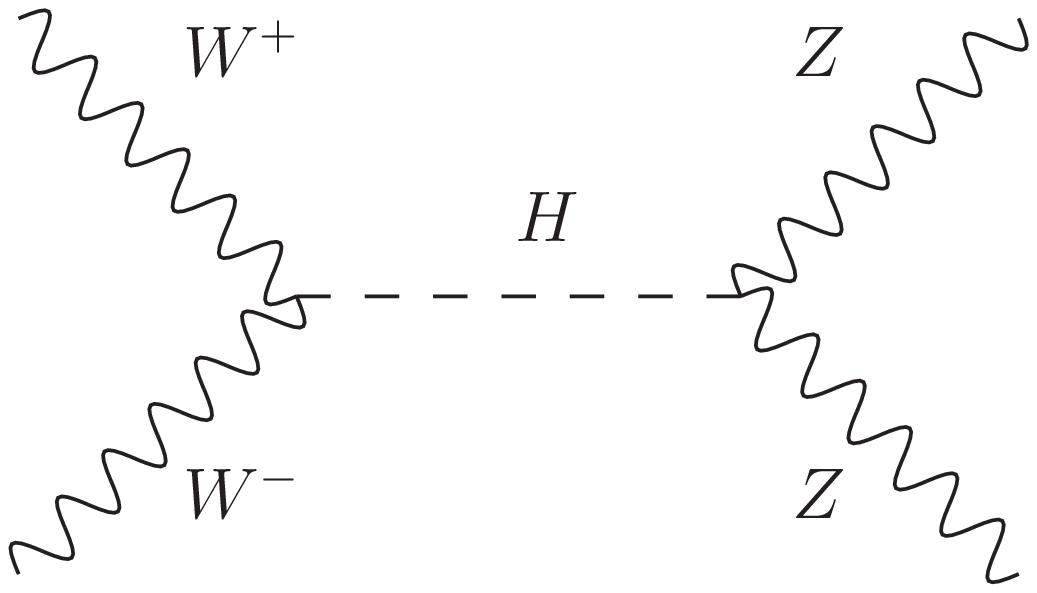} \hspace{0.70cm}
\includegraphics[clip,width=0.20\textwidth]{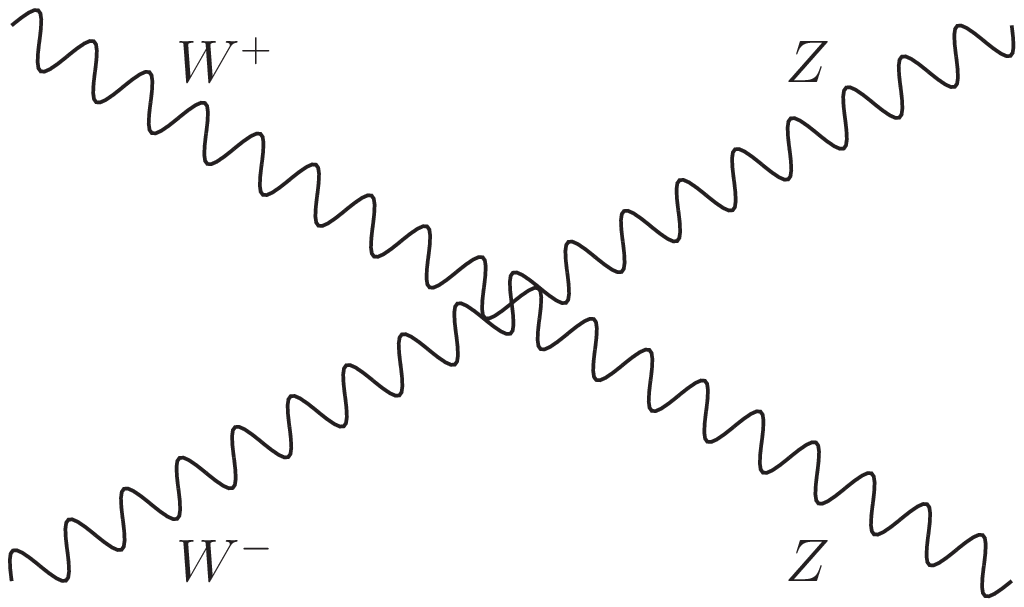}
\caption{Diagrams contributing to $A(s,t,u)$ at tree level.\label{tree-diagrams}}
\end{center}
\end{figure}

\subsection{Scrutiny of the tree-level amplitudes}
For values of $a$ different from $1$, the behavior of the tree-level
$W_LW_L$ scattering amplitudes differs from the MSM case $a=1$ and the $|t_{IJ}|<1$ unitarity bound is violated 
pretty quickly. We shall see later how to restore unitarity with the help of higher loops and 
counter-terms but in this subsection we concentrate on the peculiarities of the tree level 
amplitudes $t_{00}^{(0)}$,  $t_{20}^{(0)}$ and  $t_{11}^{(0)}$.
These partial wave amplitudes will be studied without making use of the ET approximation.
This is a key point since there are interesting kinematical features of  $t_{IJ}^{(0)}$ that are missed
in the ET approximation.

\subsubsection{Case $a=1$}
\begin{figure}[h!]
\begin{center}
\includegraphics[scale=0.80]{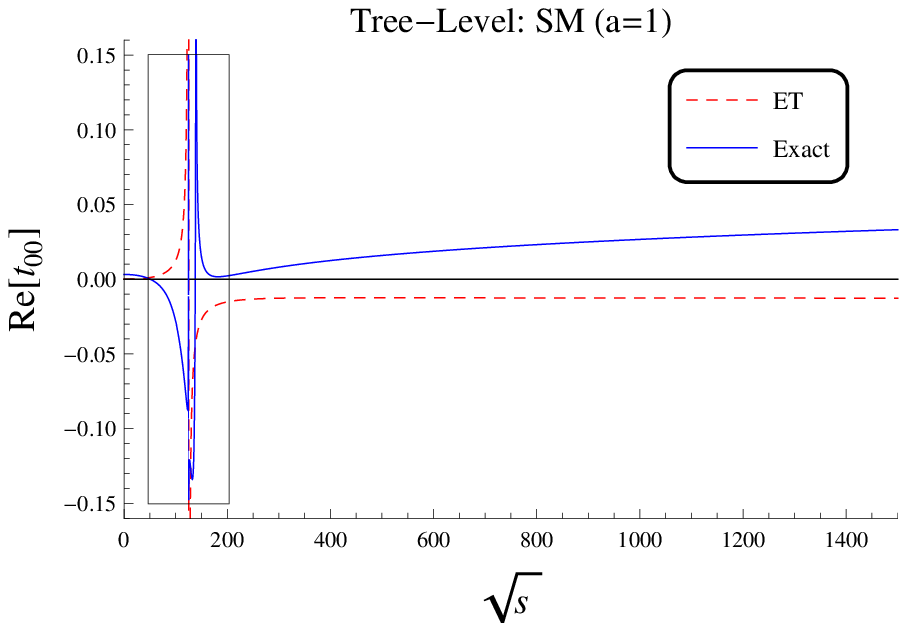}
\includegraphics[scale=0.80]{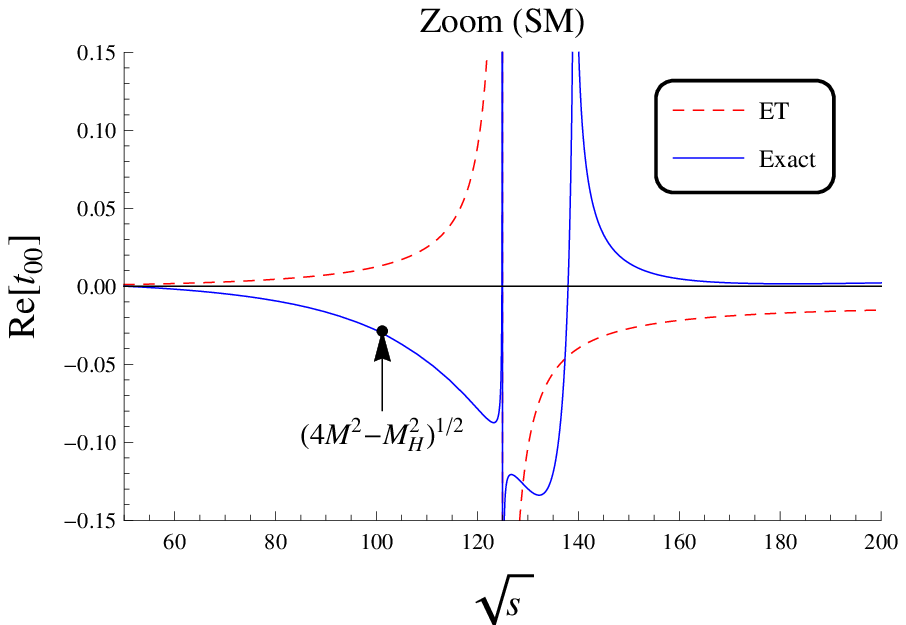}\\
\includegraphics[scale=0.80]{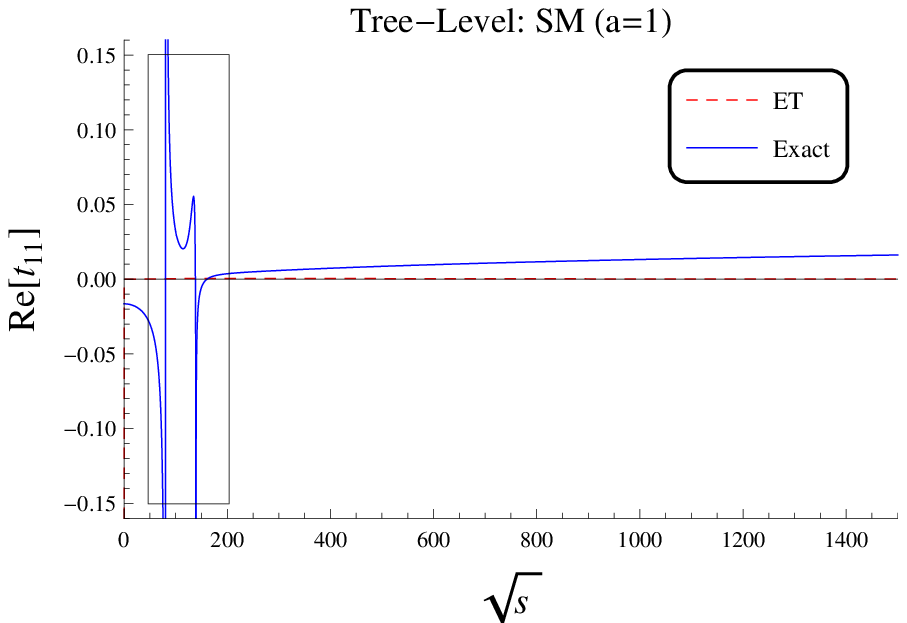}
\includegraphics[scale=0.80]{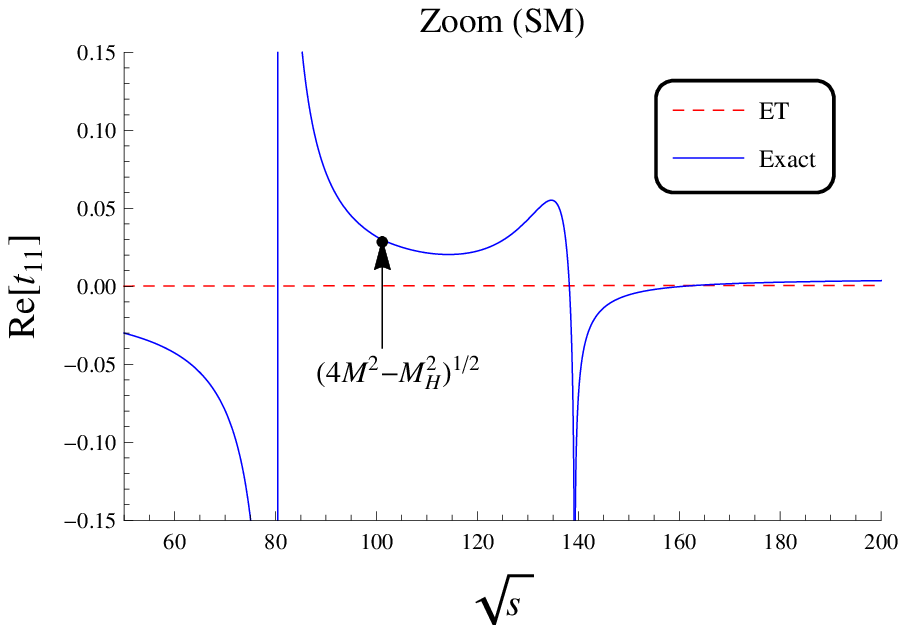}
\caption{Plot of $t_{00}^{(0)}$ (above) and $t_{11}$ (below) for $a=1$. A zoom on the lowest
values of $s$ shows the complete analytic structure. The arrow indicates the position
of one of the sub-threshold singularities that is invisible at the scale of the
plot.\label{t00-SM}}
\end{center}
\end{figure}

In Fig.~\ref{t00-SM} we plot the tree-level  isoscalar partial wave amplitude $t_{00}^{(0)}(s)$ 
for $W_LW_L\to Z_LZ_L$ as a function of $s$. The external $W$ legs are taken on-shell ($p^2=M^2=M_W^2=M_Z^2$).
As we see from Fig.~\ref{t00-SM} the partial wave amplitude has a rather rich analytic structure:
a pole at $s=M_H^2$,  a second singularity at the value $s= 3M^2$, and a third one at $s= 4M^2-M_H^2$,
invisible in the  Fig.~\ref{t00-SM} as it happens
to be multiplied by a very small number. These singularities correspond to poles
of the $t$ and $u$ channel diagrams in Fig.~\ref{tree-diagrams} that after
the angular integration to obtain the partial wave
amplitudes behave as logarithmic  divergences. The $t$ and $u$ channels are absent in the ET approximation.
Note that both singularities are below the physical threshold at $s=4M^2$. Beyond the $s=3M^2$ singularity
the amplitude for $a=1$ is always positive.

In Fig.~\ref{t00-SM} we also plot the tree-level partial wave amplitude $t_{11}^{(0)}(s)$.
Here, a pole at $s=M^2$ is visible, as expected, along with the two kinematical sub-threshold singularities
already mentioned. In Fig.~\ref{t00-SM} the $t_{00}^{(0)}$ and $t_{11}^{(0)}$ amplitudes
are also compared with the respective amplitudes obtained in ET approximation (computed assuming $M=0$).
As can be seen the ET is inadequate at low energies as it fails in
reproducing the rich analytic structure of the amplitudes.
The non-analyticity at $s= 3M^2$  and $s= 4M^2-M_H^2$ due to sub-threshold singularities
is also present in the $t_{20}^{(0)}$ partial wave amplitude (not depicted), 
corresponding like in the other two cases to a (zero width) logarithmic
pole. These sub-threshold singularities are genuine effects in the $W_LW_L\to ZZ$ amplitudes,
independent from the value of $a$, but they should be hardly visible at the LHC due 
to the off-shellness of
the $W_LW_L\to Z_LZ_L$ amplitude on $pp\to WW jj$.   The experimental process spreads 
the logarithmic poles over a range of
invariant masses. For instance, the singularity at $s=M^2$ appears actually
at $s=\sum q_i^2 -M^2$ if $W$ legs are off-shell.

\subsubsection{Case $a<1$}
\begin{figure}[h!]
\begin{center}
\includegraphics[scale=0.80]{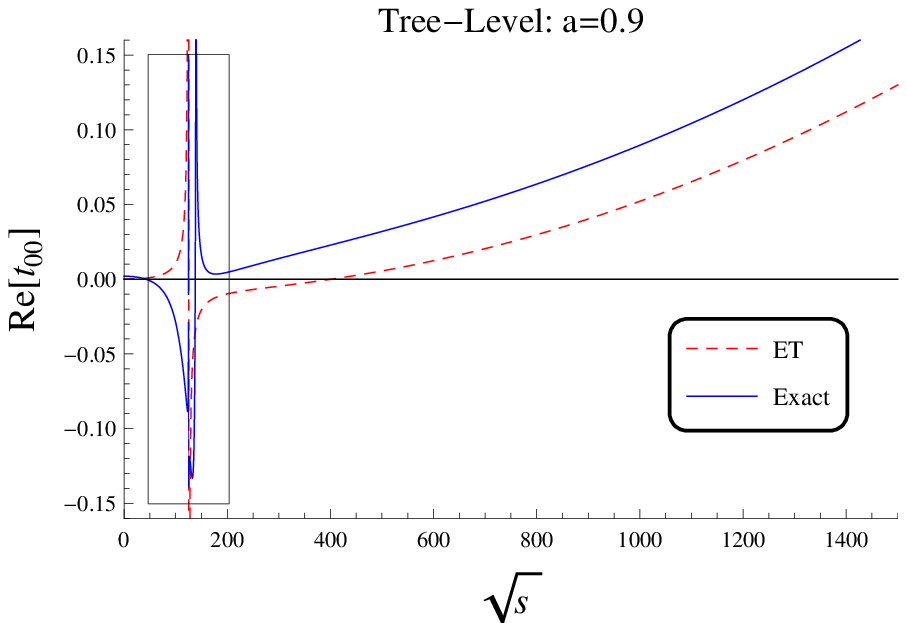}
\includegraphics[scale=0.80]{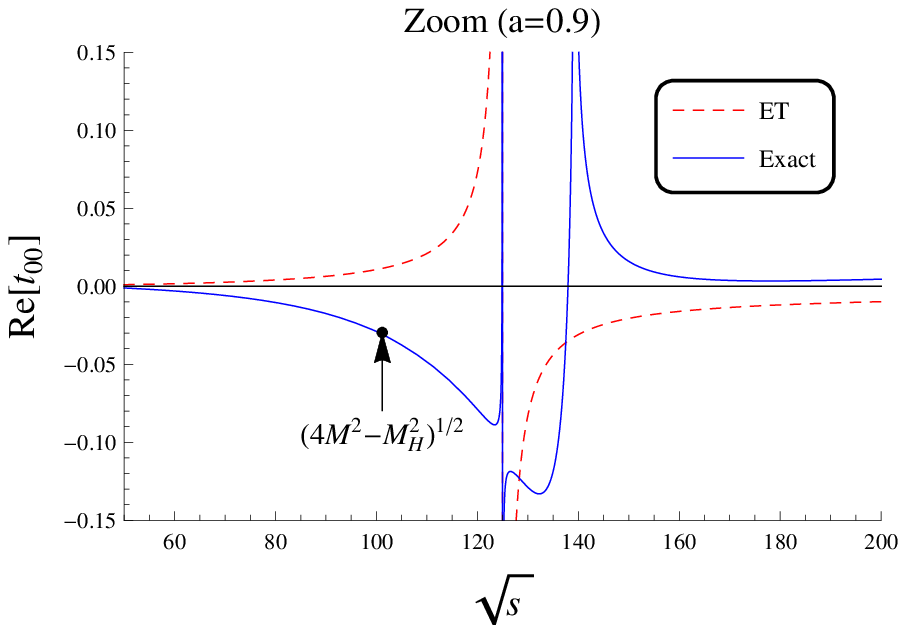}\\
\includegraphics[scale=0.80]{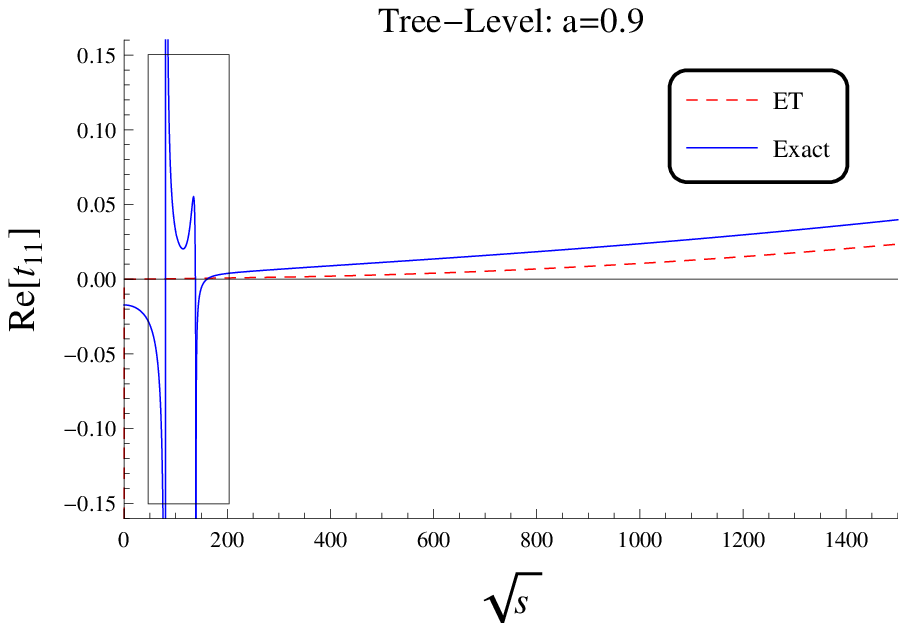}
\includegraphics[scale=0.80]{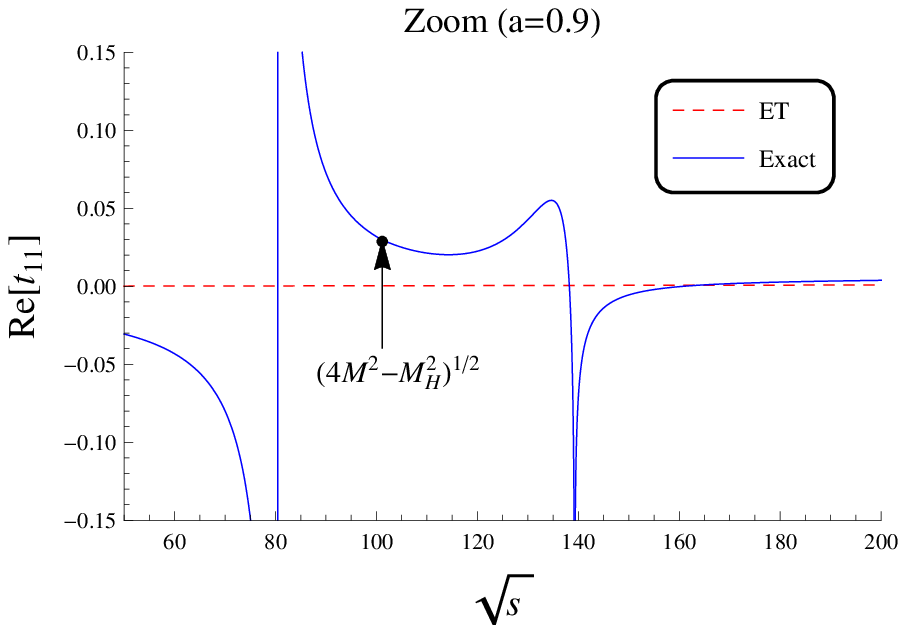}
\caption{Plots of $t_{00}^{(0)}$ and $t_{11}^{(0)}$ for $a=0.9$ and a zoom of the region
at low $s$ where the amplitudes are very small. No additional zero appears
and the amplitudes also show a non-unitary behaviour at large $s$. The nearly
invisible logarithmic singularity at $s=4M^2-M_H^2$ is indicated. The results in the ET
approximation are also indicated by a dotted line.\label{t00-a09}}
\end{center}
\end{figure}

The behaviour of the tree apmplitudes for $a<1$ shows no zeroes beyond two sub-threshold
singularities at $s= 3M^2$  and $s= 4M^2-M_H^2$. Amplitudes are positive and go to $\infty$ as $s$ 
increases. This clearly reflects the non-unitary character of  $t_{IJ}^{(0)}$ amplitudes for $a\ne1$.
In Fig.\ref{t00-a09}, we show as an example the
$t_{11}^{(0)}(s)$ and $t_{20}^{(0)}(s)$ amplitudes in the case $a=0.9$.
The equivalent amplitudes computed by making use of the ET are also shown in Fig.\ref{t00-a09}.
Both in this $a<1$ case and in the $a>1$ one we see that the ET works reasonably well
for large values of $s$, but fails at low and moderate values.

\subsubsection{Case $a>1$}
\begin{figure}[h!]
\begin{center}
\includegraphics[scale=0.80]{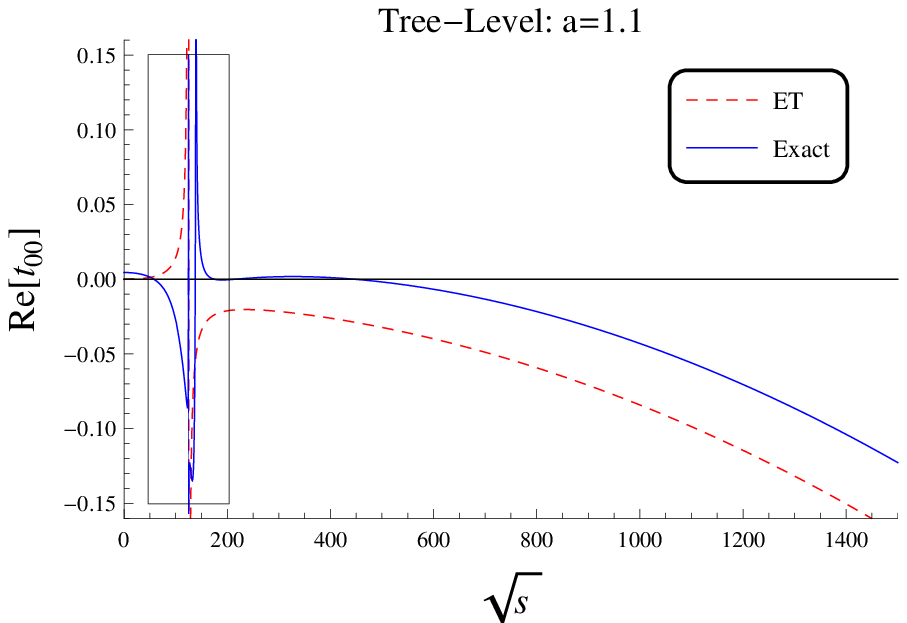}
\includegraphics[scale=0.80]{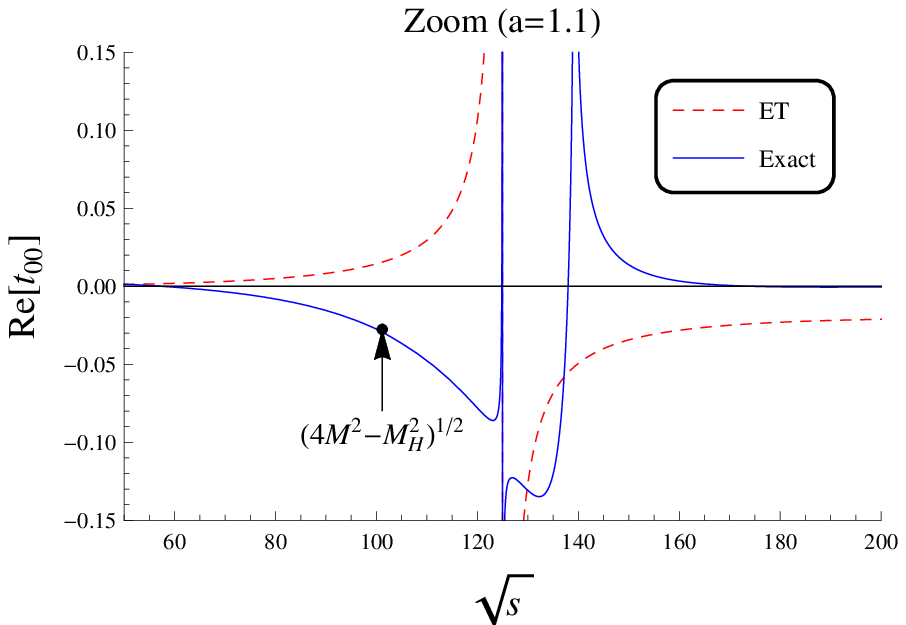}\\
\includegraphics[scale=0.80]{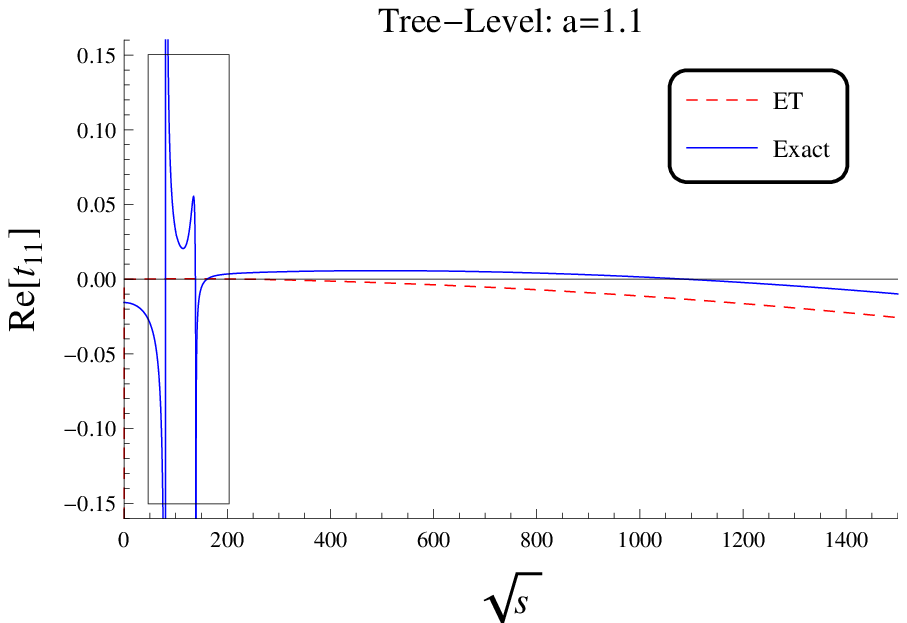}
\includegraphics[scale=0.80]{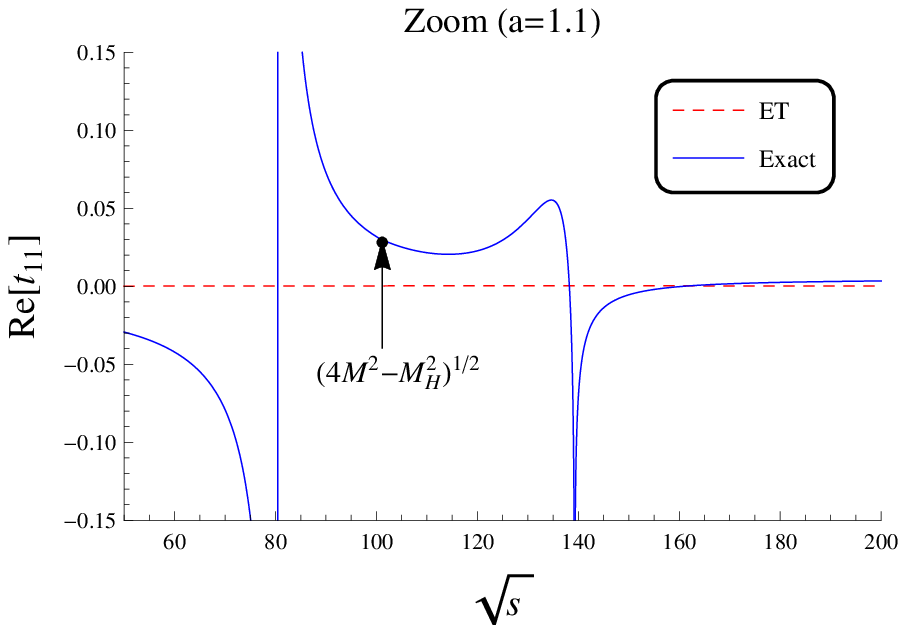}
\caption{Plot of $t_{00}^{(0)}$ and $t_{11}^{(0)}$ for $a=1.1$ and a zoom on
the low $s$ region where the amplitude is very small. Several additional zeroes appear above threshold
and is not unitary. The ET result is shown by (red) a dotted line.\label{t00-a11}}
\end{center}
\end{figure}

For $a>1$ the partial wave amplitudes show  new features. As shown
in Fig.\ref{t00-a11} for $a=1.1$ the amplitudes $t_{00}^{(0)}(s)$ and $t_{11}^{(0)}(s)$ exhibit
clearly non-unitary behaviours. In addition, for $a>1$
the tree-level partial wave amplitudes for $t_{IJ}^{(0)}(s)$  have zeroes 
for values of $s$ above threshold and well below the cut-off scale ($3$ TeV) 
of our effective Lagrangian. For example for $a=1.3$,
the zeroes of $t_{11}^{(0)}(s)$ and $t_{20}^{(0)}(s)$
are at $\sqrt{s}$ around $450$ GeV. The presence of zeroes for the tree-level amplitudes
at low values of $\sqrt{s}$ is an interesting point as it means that around these zeroes the $W_LW_L\to Z_LZ_L$ amplitudes are strongly suppressed for these values of $a$
It may be relevant to note that the $t_{00}^{(0)}$ and $t_{11}^{(0)}$  amplitudes are very small over a fairly extended
range of values of $s$ for a range of values of $a>1$ (particularly so in the isovector channel). These facts
could perhaps be used to set rather direct bounds on this particular coupling. This issue may deserve further
phenomenological study (however we will see that $a>1$ leads to some inconsistencies that might rule
out this range of values).

\subsection{Unitarization of the amplitudes}
We will now proceed to use the IAM in order to unitarize the various partial waves
we have just discussed. For this we need, in addition to the tree-level amplitudes,
the one-loop result $t_{IJ}^{(1)}(s)$.  We refer
to~\cite{Espriu:2012ih} and references therein for a more detailed discussion.
We will ue the $W_LW_L \to Z_LZ_L$ as our workhorse.

Up to now, a full fledged calculation of the one-loop $t_{IJ}^{(1)}(s)$ contribution in Eq.~(\ref{eq:tloop})
is not available for arbitrary $a$ and $b$. This would require the
evaluation of over one thousand diagrams. A numerical calculation is only available in~\cite{denner}
for the case $a=b=1$ but it is not very useful for our purposes.
For this reason, to estimate the $t_{IJ}^{(1)}(s)$ contribution in Eq.~(\ref{eq:tloop}) we proceed in the following way.
The analytic contribution from $a_{4}$, $a_5$ terms is calculated exactly with  longitudinally polarized $W$ and $Z$
(appendix~\ref{sec:appendix_amplitudes}) like the tree-level contribution $t_{IJ}^{(1)}(s)$ .
The real part of $t_{IJ}^{(1)}(s)$ will however be determined using the ET~\cite{ET,esma}; i.e.
we replace this loop amplitude by the corresponding process $w^+ w^- \to zz$.
For this part of the calculation we take $q^2=0$ for external legs and set $M=0$ but the Higgs mass
is kept. The relevant diagrams of $A(ww \to zz)$ entering  $t_{IJ}^{(1)}(s)$
were calculated in~\cite{Espriu:2013fia} where explicit expressions for the different
diagrams for arbitrary values of the couplings $a$ and $b$ can be found.
As to the imaginary part of $t_{IJ}^{(1)}(s)$ we can take advantage of the optical theorem to circumvent
the problem of using the ET approximation.
In the $I=1, J=1$ and $I=2,J=0$ cases we can use the relations
\be
\label{eq:part_unitarity}
{\rm Im \,} t_{IJ}^{(1)}(s) = \sigma(s) |t_{IJ}^{(0)}(s)|^{2} \, ,
\ee
While for the $I=0$ amplitude we also have  a contribution from a two-Higgs
intermediate state. Then
\be
\label{eq:full_unitarity_hh}
{\rm Im \,} t_{00}(s) = \sigma(s) |t_{00}(s)|^{2} + \sigma_{H}(s) |m_0(s)|^{2} \, ,
\ee
with
\be
\sigma(s) = \sqrt{1 - \frac{4 M^{2}}{s}} \hspace{1cm},\hspace{1cm}  \sigma_{H}(s) = \sqrt{1 - \frac{4 M_{H}^{2}}{s}}.
\ee
where $M=M_W, M_Z$. 
We believe that for the purpose of identifying dynamical resonances, normally occurring at $s\gg M_H^2$ the approximation
of relying on the ET for the real part of the loops is fine. Note that the dominant contribution
to the real part for large $s$, of order $s^2$, is controlled by the contribution coming from couplings $a_{4,5}$. 

There is no really unambiguous way of applying the IAM to the case where there are coupled channels
with different thresholds. This will be relevant to us only in the $t_{00}$ case as there is an intermediate state consisting
of two Higgs particles. Here we shall adhere to the simplest choice that
consists in assuming to remain valid also in this case. In addition, there
is decoupling of the two $I=0$ channels in the case $a^2=b$. We have checked our results for 
different values of $b$, in particular we see that
setting $b=a^2$ does not give results for the resonances eventually found that are
noticeably different from those obtained for other values of $b$. Finally, we have also explicitly checked  
the unitarity of the results using the optical theorem, which also validates {\em a posteriori} the previous 
approximations.

\subsection{Dynamical resonances} 
Making use of the IAM amplitudes one observes that new resonances appear in ample regions of
the parameter space of $a_{4}$, $a_5$. We will make a distinction between the cases $a=b=1$, where the difference between 
the HEFT and the MSM lies exclusively in the presence of $a_4$ and $a_5$, and the case $a\neq 1$, where the lowest 
dimensional operator already differs from the MSM (and, in addition, one may have non-zero values for $a_4$, $a_5$). 

The case $a=b=1$ was discussed extensively in \cite{Espriu:2012ih} using the method presented here. 
Let us summarize these findings.
While any strongly interacting theory is expected to exhibit an infinite number of resonances,
including contributions up to $\mathcal{O}(p^4)$, the  expression of $t_{IJ}(s)$ consists of
polynomials (modulo logs) of to order $s^2$ at most. Therefore, in each channel we expect  
to find one or two resonances at best ---the lowest lying ones. 

It should also be borne in mind that non-zero values for $a_4$ or $a_5$ do not necessarily 
signal the presence of  a non-perturbative, strongly interacting,
EWSBS.  If instead the EWSBS is of perturbative nature but very massive,
(the simplest possibility could be an extended scalar sector or two Higgs-doublet models with 
large masses for all but the physical Higgs),
integrating the massive states out would yield no-vanishing values for the coefficients $a_4$ and $a_5$~\cite{escia}.
The IAM unitarization method would in this case actually approximately reproduce the masses of the 
particles that were originally
integrated out \cite{gback}.

We perform a scan for the values $|a_4|<0.01$ and $|a_5|<0.01$ and $a$ and $b$ fixed  looking for the possible
presence of resonances. In the search of these dynamical 
resonances we use several methods. First we look for a zero of the real part
of the denominator of the partial wave and use the optical theorem to determine the imaginary part ---i.e. the width--- 
at that location. A second method consists in searching directly for a pole in the complex plane. Both methods
give very similar results because the widths are generally quite small. 
It should be stated that because of the way we compute the full amplitude, with separate derivations of the real and the
imaginary parts, the analytic continuation to the whole complex plane for $s$ could be problematic had the imaginary parts
turned out to be large. In general proper resonances tend to reveal themselves in a
rather clear way. Some difficult cases present themselves for $a>1$ when the putative
resonance is close to one of the zeroes of the tree-level amplitude that appear in
this case and we had to study these situations carefully.

Physical resonances must have a positive width and are only accepted as genuine resonances if $\Gamma < M/4$.
Theories with resonances having a  negative width violate causality and the corresponding values of
the low energy constants in the effective theory  are to be rejected as leading to unphysical theories as
no meaningful microscopic theory could possibly lead to these values for the effective couplings \cite{Adams:2006sv}.

\begin{figure}[tb]
\centering
\includegraphics[clip,width=0.45\textwidth]{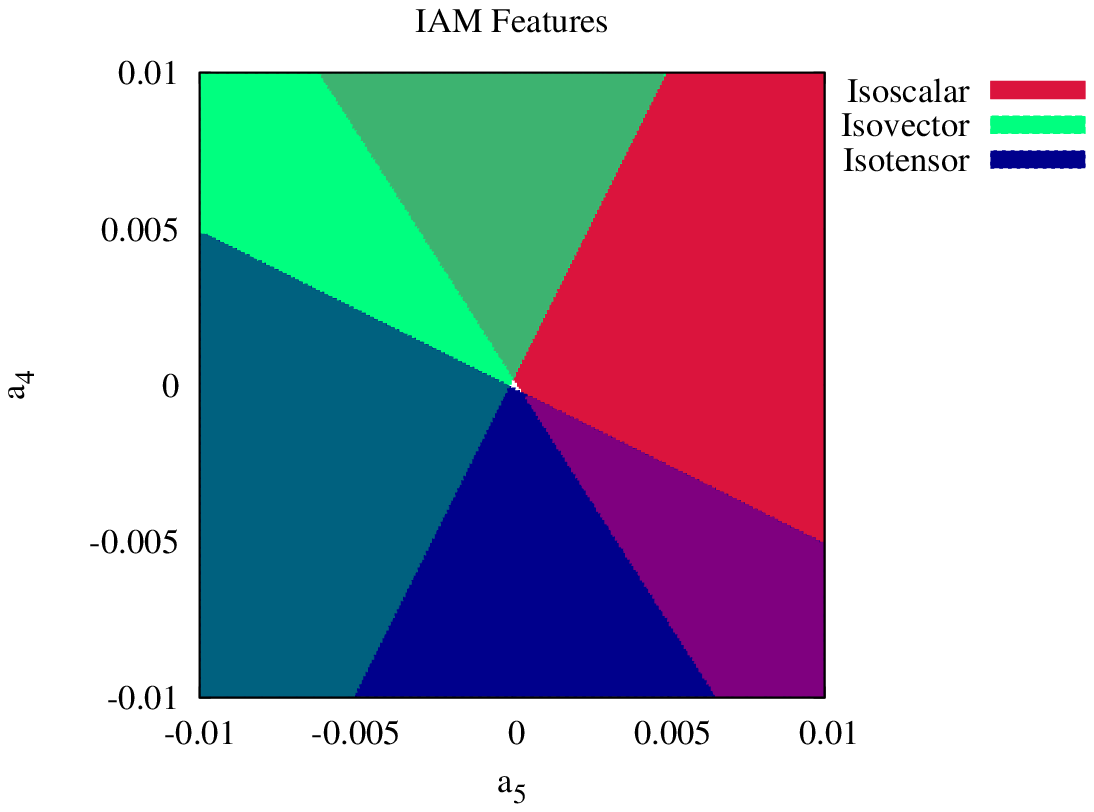}
\includegraphics[clip,width=0.45\textwidth]{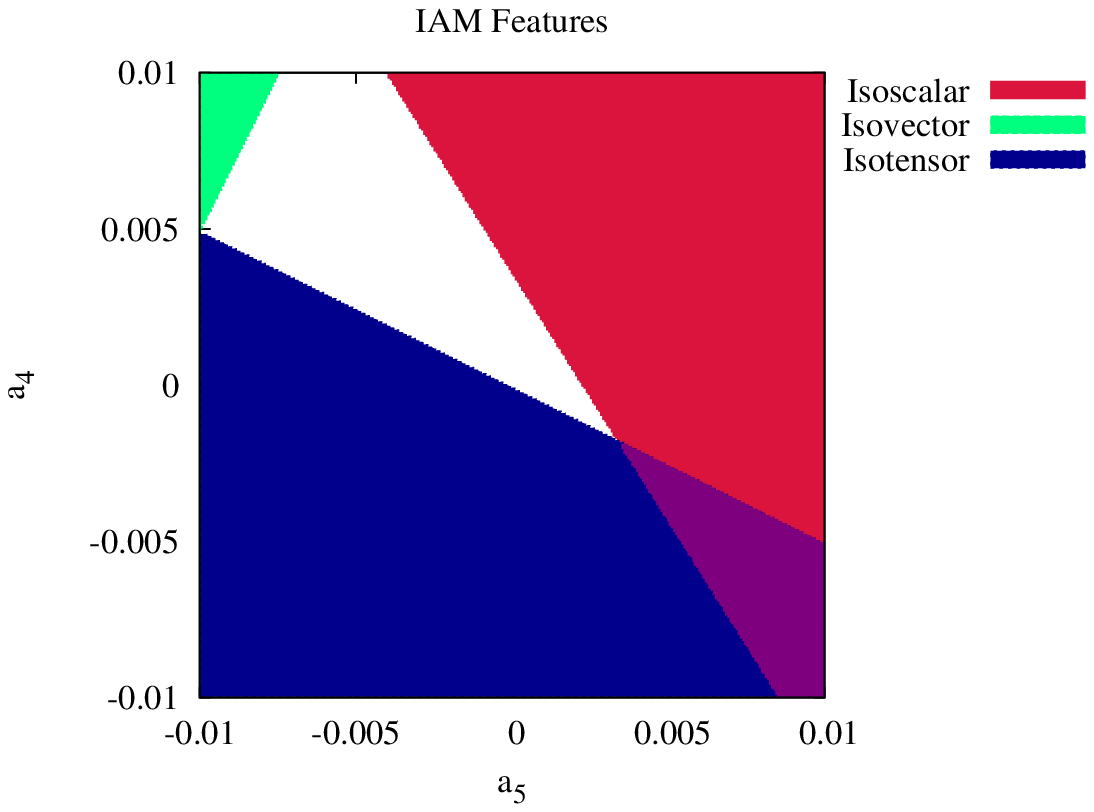} 
\caption{Left: Regions with isoscalar and isovector resonances (and the isotensor \textit{exclusion} region) up to 
a scale $4\pi v \approx 3$~TeV.  Right: same as in the left plot, but only showing regions of parameter space where 
isoscalar/isovector resonances have $M_{S,V}<600$~GeV. These coloured regions
of parameters can be considered excluded for all practical purposes as explained in the text. Note that the range
excluded even with this rather loose condition is much larger than the one resulting from direct 
bounds shown in Fig.\ref{constraint_chiral}.}
\label{fig:resonances}
\end{figure}

In Fig.~\ref{fig:resonances} (left), we present the results for our search in the case $a=b=1$; that is, a point
in parameter space where the $hWW$ and $hhWW$ couplings are identical to the SM, but we allow for non-zero
values for $a_4$ and $a_5$.  The usual cutoff $\sqrt{s} = 4\pi v\simeq 3$~TeV has been imposed.  We find that there 
is a region (shown in red) where there are only scalar resonances, a region (in green) where there are only vector resonances, 
an overlapping region where there are both, and finally a large region (in blue) in which the isotensor amplitude develops 
unphysical and therefore must be excluded.  There is also a region, centered around $a_{i}=0$, in which there 
are no resonances. This point  corresponds to 
the MSM with a light Higgs boson, a theory which suffers no problems of unitarity 
and therefore  should not be expected to feature dynamical 
resonances from this method.  The absence of any features in this region 
is a nice consistency check that the IAM is not introducing them when it should not.

Fig.~\ref{fig:resonances}~(left) has to be understood in the following way: typically an extended symmetry breaking 
scenario has more resonances than just a light ``Higgs''. There could be additional scalars, vector resonances, 
or even higher spin states. The low-energy contribution from these 
states is parametrized by the $a_{i}$. Fig.~\ref{fig:resonances}~(left), then, addresses the following question: 
what range of parameters do we exclude if we assume that {\it no} additional resonance is seen anywhere 
between the Higgs at 125 GeV and 
$4\pi v\simeq 3$~TeV? The excluded region, then, in $a_{4}$, $a_{5}$ parameter space looks very dramatic, because 
only values for $a_4$ and $a_5$ extremely close to zero are acceptable, reflecting that the new states must be quite heavy, 
perhaps beyond the consistency cut-off of our method ($\sim 3$ TeV, as mentioned). 

Conversely, if $a_4$ and/or $a_5$ are different from zero, except for the above set of very small values, 
and do not belong to the {\em exclusion} region, it
is unavoidable that unitarity forces the existence resonances, even if $a=1$. 

Let us relax the condition of no resonance being observed and examine which is the exclusion region 
for $a_{4}$ and $a_{5}$ that can be obtained by assuming that no new resonances 
exist, say, below 600 GeV (but resonances may still exist above this energy). This is shown in Fig.~\ref{fig:resonances}~(right);
the white area is the region of parameter space where existing resonances are heavier than $600$ GeV.

However note that the fact that resonances may be present implies nothing about their visibility. In 
fact the dynamical resonances need not have signals with strengths comparable to that of a MSM Higgs boson of 
the same mass or anything like that.  This will
be discussed in the next subsection.

The requirement that possible resonances originating from an EWSBS (either scalar or vector) should be heavier
than 600 GeV stems from the analysis of dilepton pair production using the full 8 TeV data. 
Even if the would-be resonances produced are very narrow  and only produced via $VV$ fusion (and, 
assumed to decay only in this channel) strict searches and analysis\cite{grecoliu}, indicate the absence of 
such resonances below this mass with high certainty. In fact, this lower limit is very conservative because
precision observables\cite{Pich:2013fea}, including the study of holographic models such as the one 
presented in Section \ref{sec:holog}, indicate that vector resonances should
be in all cases heavier than 1 TeV. This would reduce further the (white) area still
allowed for $a_4$ and $a_5$. 

\begin{figure}[tb]
\centering
\includegraphics[clip,width=0.45\textwidth]{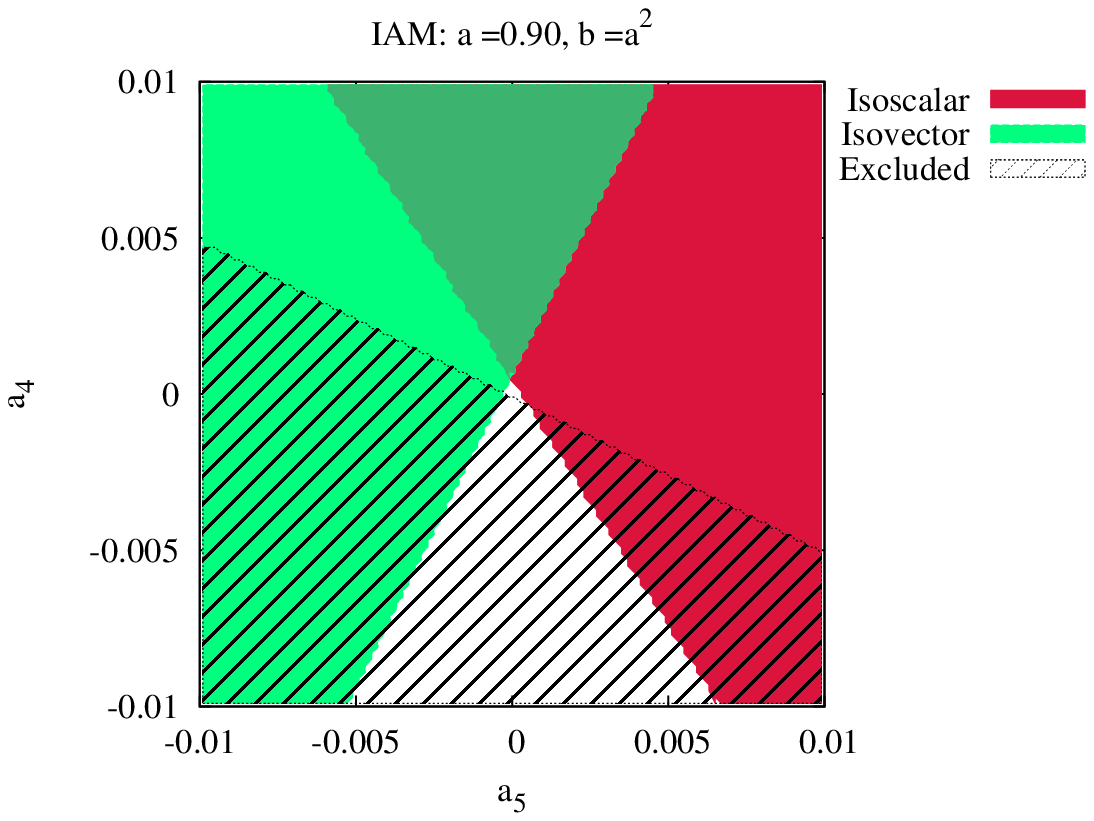}
\includegraphics[clip,width=0.45\textwidth]{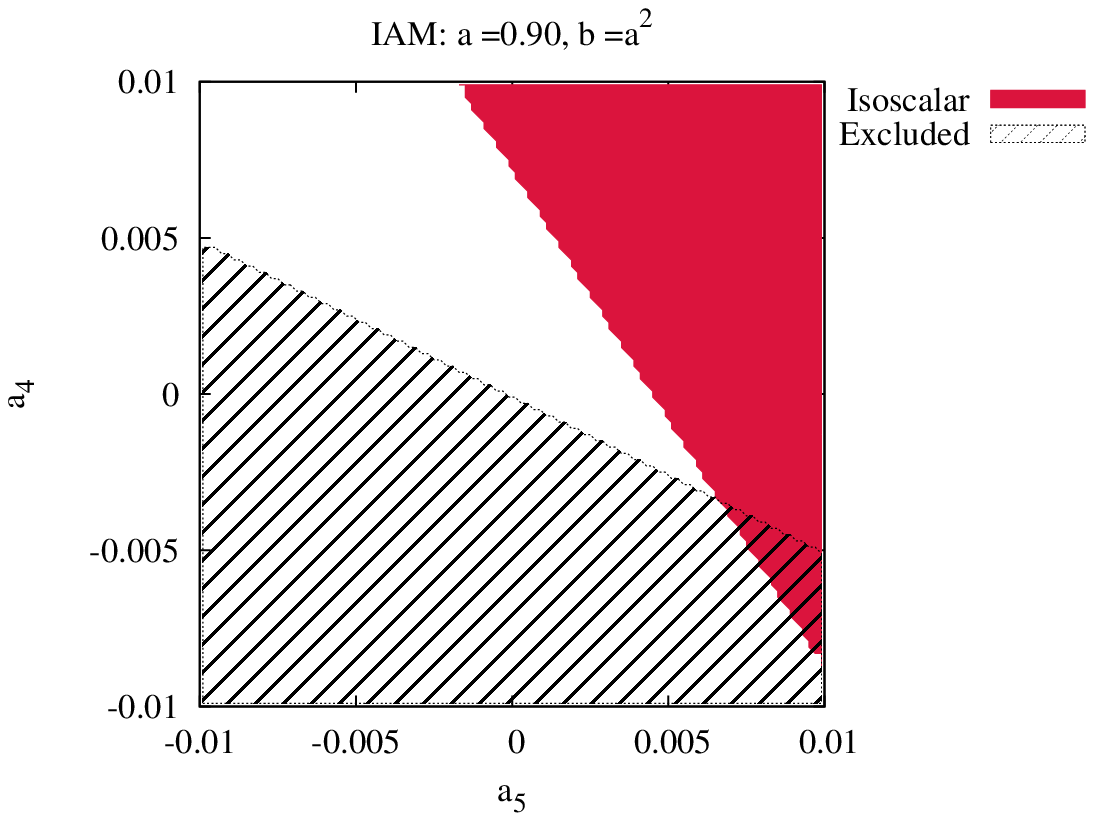}
\caption{For $a=0.9$ and $b=a^2$:  (left) Regions with isoscalar and isovector resonances
 (and the isotensor \textit{exclusion} region) up to a scale $4\pi v \approx 3$~TeV.  (right) Same 
but only showing isoscalar/isovector resonances in which $M_{S,V}<600$~GeV. \label{explot09ba2}}
\end{figure}

Let us now consider the case where $a<1$ ($b$ is largely irrelevant in the discussion and we set $b=a^2$ to exclude
coupled channels, even if this does not affect the $IJ=11$ channel). Several values of $a$ have also 
been studied but we here present results only for $a=0.9$, 
compatible with the experimental bounds on $a$ and enough to see the general trend. In fact as long as one remains 
relatively close to $a=1$ the situation is qualitatively similar. Like for $a=1$ it is easy to find resonances in various channels.
Most of them have the right causality properties that make the theory acceptable. However, in the $I=2,J=0$ 
channel we see that there is a region in the $a_4-a_5$ plane where causality is violated. This corresponds to the shaded
region in the lower part of Fig. \ref{explot09ba2} and microscopic theories giving rise to these values for the 
low-energy parameters $a_4$, $a_5$ are again not acceptable. 

In Fig. \ref{explot09ba2} we show the region of parameter space in $a_4$, $a_5$
where isoscalar and isovector resonances exist for the value $a=0.9$
along with the isotensor exclusion region. The pattern here has some analogies with the case
$a=1$ studied in \cite{Espriu:2012ih} but proper\footnote{Recall that resonances are required to
have, in addition to the correct causal properties, $\Gamma < M/4$.} resonances are somewhat harder to form, in particular in
the vector channel no resonance is found below 600 GeV for $a=0.9$ in contrast to
the $a=1$ case. If no resonances are found at the LHC all the way up to 3 TeV, the values
of $a_4$ and $a_5$ in the coloured regions could be excluded and then $a_4$ and $a_5$ should lie within
the small central region in the left plot. Small as this regions is, it is noticeably larger than
the one corresponding to $a=1$, which was virtually non-existent. This is true even
for $a=0.95$ which is very close to the MSM value  $a=1$.

In order to see what order of magnitude one should expect for the $a_4$ and $a_5$ coefficients
we show how the reverse process works: assuming that a resonance in the 1.8 to 2 GeV region
is found, what is the allowed range of coefficients so as to reproduce the properties 
of such resonance via the IAM. This exercise was made on occasion of tentative evidence of a small 
enhancement in $VV$ scattering at the LHC, later not confirmed.

\begin{figure}[ht!]
\centering
\includegraphics[clip,width=0.4\textwidth]{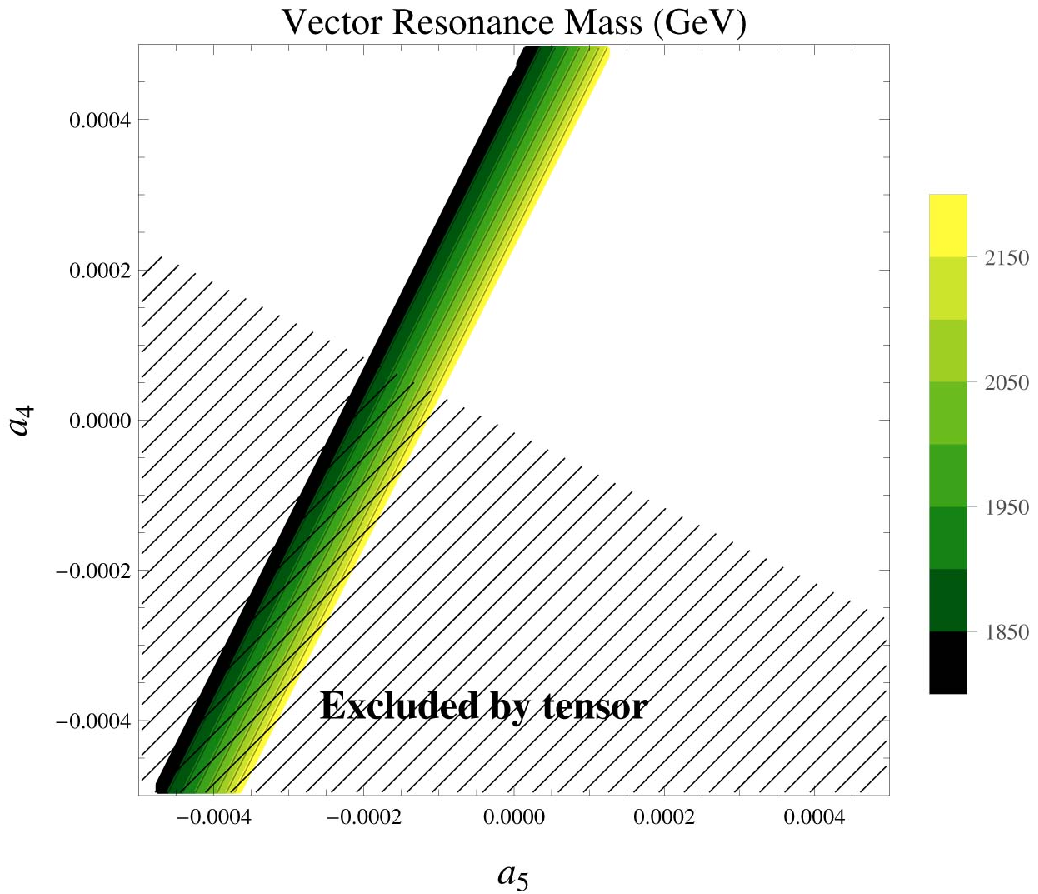} 
\includegraphics[clip,width=0.4\textwidth]{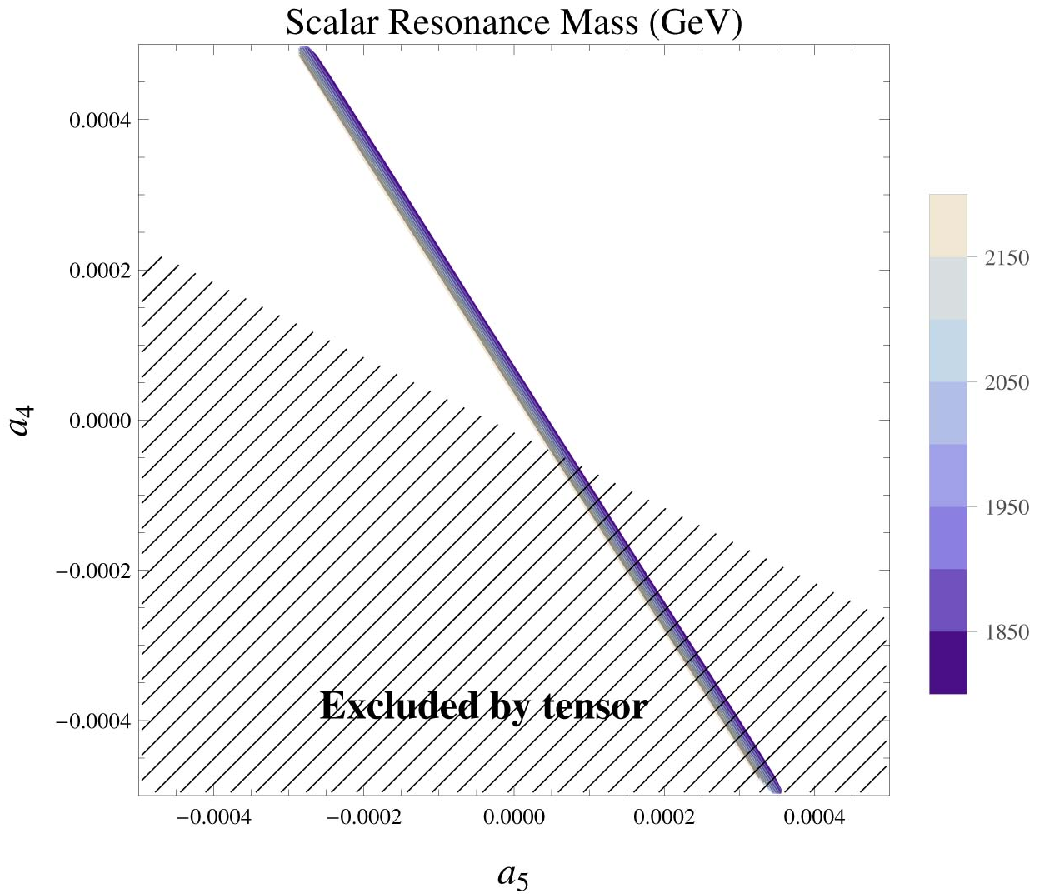} 
\caption{For $a= 1$ and $b=1$:  (a) allowed values for $a_4$, $a_5$ corresponding to a 
vector resonance
with a mass between 1.8 TeV and 2.2 TeV. Note the extremely limited range of variation that
is allowed in the figure for the low-energy constants. (b) Same for a scalar resonance. 
The corresponding widths as predicted by unitarity are very small; a characteristic value in the vector 
channel is 20 GeV --- quite narrow for such
a large mass. In the scalar channel are in the 70 to 100 GeV range. 
The dashed area is excluded on causality grounds stemming from the $I=2$ channel.\label{fig:1820}}
\end{figure}

After requiring a resonance in the vector channel with a mass in the quoted range one gets in 
a $a_4-a_5$ plane the region shown on the left in Fig.\ref{fig:1820} for $a=1$.
An analogous procedure but assuming that the resonance 
is the $I=0,J=0$ channel results in the allowed region in the $a_4-a_5$ plane depicted on the right
in the same figure

We would like to emphasize the very limited range allowed for the parameters that is 
shown in the above figures. The constants $a_4$ and $a_5$ lay in the small 
region $|a_4|, |a_5| < 5\times 10^{-4}$ (this region includes of course the MSM value
$a_4=a_5=0$ where obviously there are no resonances).
Note that the much broader range $|a_4|,|a_5| < 0.02$ is considered in most studies
as still being phenomenologically acceptable. 
Indeed, setting even a relatively loose bound for the mass of the resonance restricts the range of 
variation of the relevant low-energy constants enormously.

\begin{figure}[tb]
\centering
\includegraphics[clip,width=0.45\textwidth]{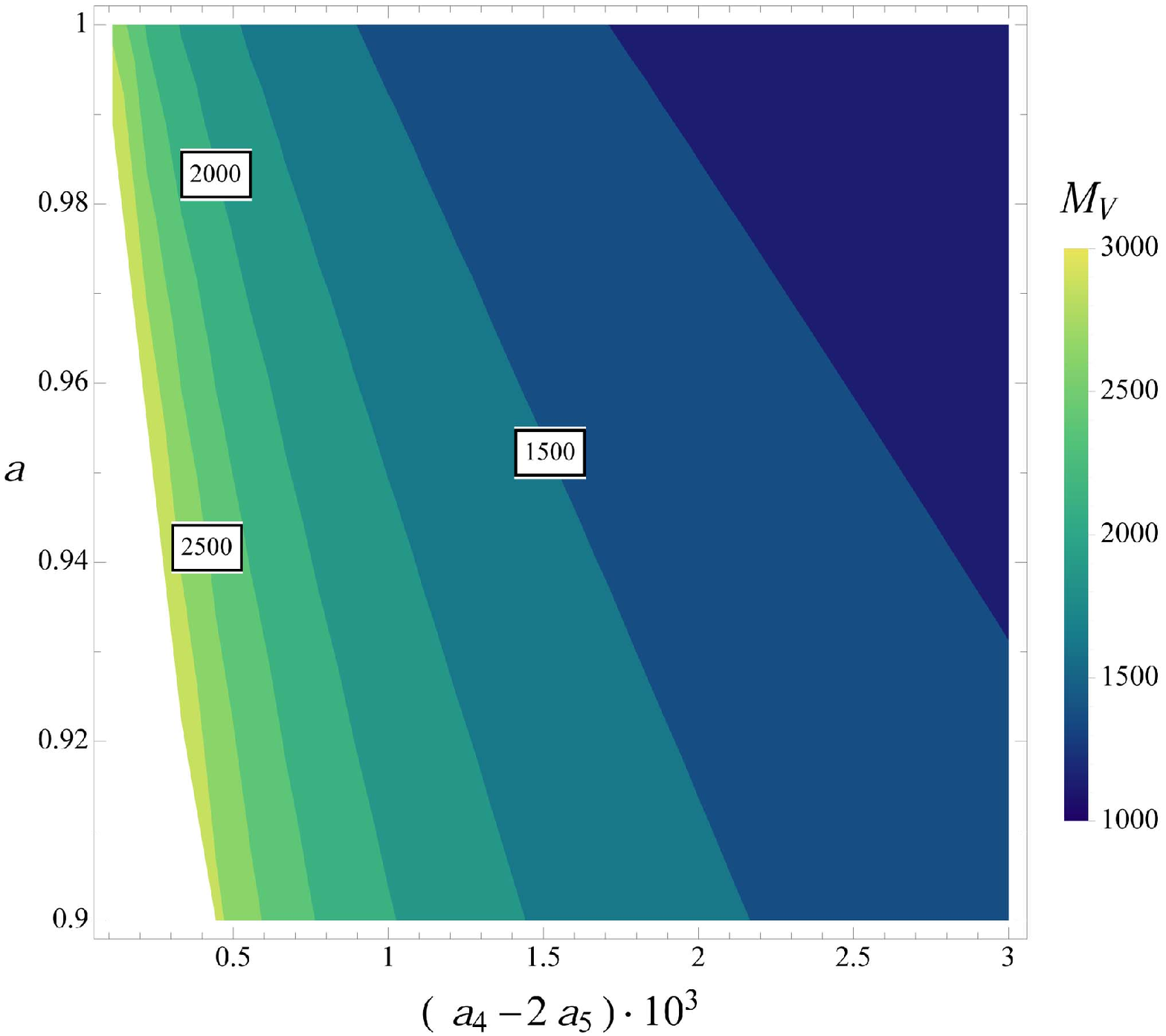} \hspace{0.5cm}
\includegraphics[clip,width=0.45\textwidth]{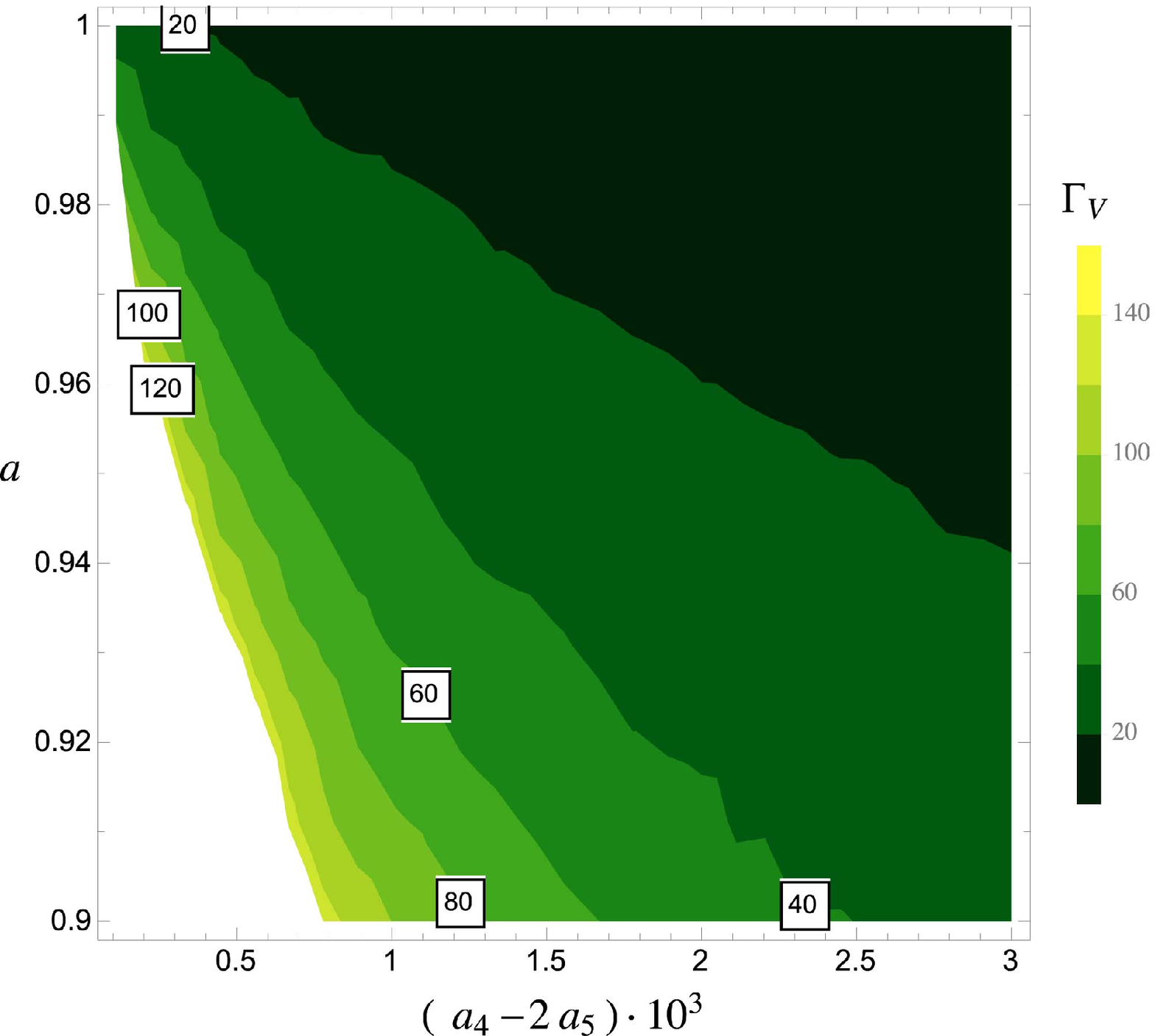} 
\caption{Masses (left) and widths (right) in GeV for vector resonances predicted from the unitarized
partial wave amplitudes of $WW \to WW$ scattering. They are plotted as a  
function of $a$ and the combination $a_4-2a_5$ that is the relevant one  in this channel. 
All studied cases with vector resonances are such that no corresponding scalar or tensor
resonances appear. The white area denotes the region with resonances heavier than 3000 GeV,
beyond the validity of the effective theory.
From \cite{DDEGHMS}.\label{fig:masses}}
\end{figure}

In Fig.\ref{fig:masses}, we give contours for the predicted masses and widths of the isovector 
resonances over the $a_{4}-a_{5}$ parameter space and various values of the $hWW$ coupling $a$.  
To estimate the widths, we continue our amplitudes into their second Riemann sheet 
and solve for the complex pole such that 
\be
t_{IJ}^{-1}(s_{R}) = 0 \, ,
\ee
where $s_{R}$ is interpreted as 
\be
s_{R} = \left( M_{R}^{2} - i \, M_{R} \Gamma_{R} \right) \, .
\ee

Fig.\ref{fig:masses} shows the masses and widths of the scalar and vector resonances
obtained for $a=0.9$. As we see, in general they
tend to be slightly heavier and broader when $a$ departs from the $a=1$ case.
We emphasize that the resonance in the scalar channel is additional to the Higgs at 125 GeV.

We note that for large estimated widths, these pole masses may separate slightly from those predicted by the 
location where the phase shifts $\delta_{IJ}$ pass  through $(\pi / 2)$.  While vector resonance masses 
range from $\sim 550$~GeV to $\sim 3$~TeV. 
The widths are particularly interesting: except for the largest masses, they are $\mathcal{O}(10$~GeV) 
to $\mathcal{O}(100$~GeV);  
that is, {\em very} narrow. Obviously the inclusion of a light Higgs-like state substantially alters the characteristics of the 
resonances produced by the IAM as compared to the old higssless case.

The case $a>1$ appears to be pathological for nearly all values of $a_4, a_5$ investigated.
One indeed finds zeroes of the denominator of the IAM amplitude that would correspond to
resonances provided that the numerator does not vanish. This comment is relevant because many of
the resonances present, particularly in the vector channel, appear in  region near the last
(as $s$ increases) zero of the amplitude and this requires particular care. In fact for a set of values
of $a_4$ and $a_5$ the determination as to whether a pole exists or not becomes ambiguous.

When we continue our amplitudes into their second Riemann sheet to estimate
the width and solve for the complex pole we find that in various channels the imaginary part is
such that it corresponds to a negative width. When two poles in a given channel are found, one is acceptable
but then the other one leads to acausal behavior (this can be proven analytically). For other
values of the coupling the resonances are however perfectly acceptable.

The result is that a very sizable part of the space of parameters is ruled out.
In particular we have been unable to find a bona fide
$I=2$ resonance for $a=1.1$ and $a=1.3$ and this seems to be the generic situation for $a>1$. This result is
at odds with some dispersion relation analysis \cite{sumrule} claiming that theories having $a>1$
must show a dominance of the $I=2$ channel and even a model with a $I=2$ resonance is suggested. An explanation
for the discrepancy is given in appendix~\ref{sec:appendix_sumrule}.

\subsection{Are these resonances visible?}
One thing is having a resonance and a very different one is being able to detect it. Searching for new particles in the LHC
environment is extremely challenging and the statistics for $VV$ fusion limited. As we will see below
the actual signal strength of the new resonances predicted is such that most cases are not currently being probed in
LHC Higgs searches ---a situation that will change when substantially more luminosity is accummulated.
The previous considerations emphasize the importance of indirect measures of the couplings $a_4$ and $a_5$
by searching for deviations in the cross-section that could be attributed to these anomalous coefficients.
Measuring these anomalous couplings will be one of the relevant tasks of future LHC runs.
However, this will not be easy either as the sensitivity to them with current statistics is low.  In addition, a
proper determination may require unitarization in order to obtain realistic bounds. To make
things worse, as we just saw, heavy resonances
imply small values for $a_4$ and $a_5$. If this is the range 
of values chosen by nature, their direct determination at the LHC will be unfeasible. 

The IAM method is able not only of predicting masses and widths but
also their couplings to the $W_LW_L$ and $Z_LZ_L$ channels. In \cite{Espriu:2012ih} the experimental signal of 
the different resonances was compared to that of a MSM Higgs 
with an identical mass. Because the decay modes are similar (in the vector boson channels that is) and limits on different
Higgs masses are well studied this is a very intuitive way of presenting the results.

\begin{figure}
\centering
\includegraphics[clip,width=0.65\textwidth]{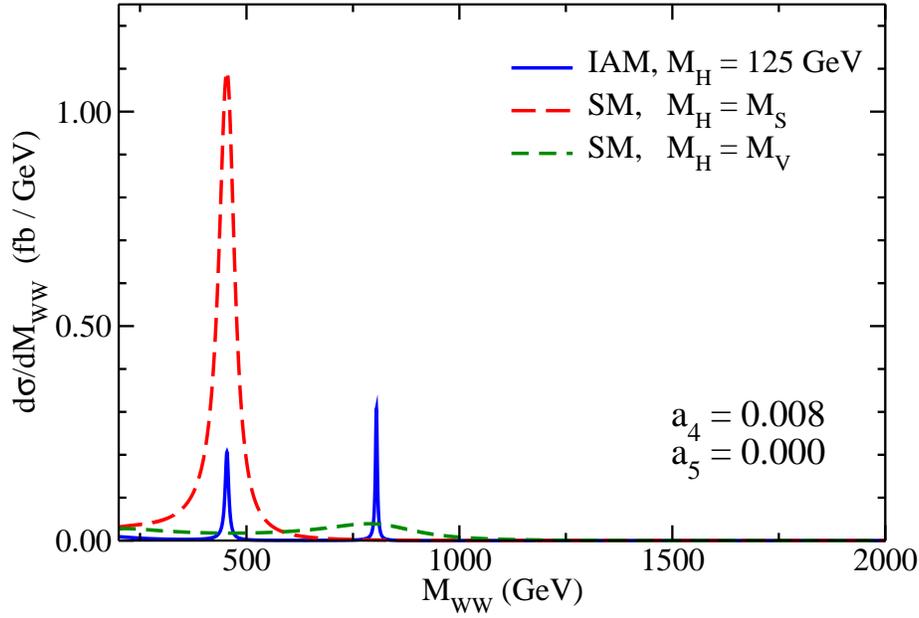}
\caption{The figure shows the relative contribution to the $W_LW_L\to W_LW_L$ cross section of a vector and a scalar resonance
obtained via the IAM method from a HEFT with values $a=1, a_4=0.008, a_5=0$. These values generate 
after unitarization both a scalar resonance at
$M_S\simeq 500$ GeV and $M_V\simeq 800$ GeV. The corresponding cross-sections are compared with that of a SM Higgs with masses
$M_H=M_S$ and $M_H=M_V$. \label{fig:resplots}}
\end{figure}

Let us, first of all, show a plot that will set the reader in the right frame of mind. In Fig.\ref{fig:resplots}
we show a comparison of a vector and a scalar resonances generated via the IAM and the unitarization process. The figure
corresponds to the values $a=1$, $a_4=0.008$ and $a_5=0$. With those values both a scalar and a vector resonance 
are present. The plot depicts the cross-section for the process $W_LW_L\to W_LW_L$ showing the
corresponding peaks at the respective resonances. As we see, they are really narrow (for comparison we 
also show the case where $M_H=M_V$, the mass of the resonance, where the Higgs is very broad, melting as a resonance). It is most
interesting to compare the cross-section due to the scalar resonance present in this model at a mass of approximately
500 GeV with the one due to the exchange of a (theoretical) SM Higgs of the same mass. The experimental signal of the
scalar resonance is much smaller. It is clear that this type of resonances produced in $WW$ scattering will be hard to see.

It should be borne in mind that the comparison is done only at the level of the $t_{00}$ and $t_{11}$ partial waves and
that the contribution from other partial waves to the process is neglected. Near the respective resonances the former are 
dominant anyway.

One can be slightly more precise and use the effective W approximation (EWA \cite{ewa}) to determine the following quantities
that are convenient for our purposes; namely we define the cross section coming from the resonance region for a 
given resonance of mass $M_{R}$ and width $\Gamma_{R}$ as
\be
\label{eq:peak_r}
\sigma^{peak}_{R} \equiv \int_{M_{R} - 2 \Gamma_{R}}^{M_{R} + 2 \Gamma_{R}} \left[d M_{WW} \times \frac{d\sigma_{R}}{d M_{WW}} \right] \, ,
\ee
where $\sigma_{R}$ is the cross section resulting from the amplitudes unitarized by the IAM.  
For a SM Higgs with mass $M_{H}$ set to $M_{R}$ and corresponding SM decay width $\Gamma_{H}$, we also calculate
\be
\label{eq:peak_sm}
\sigma^{peak}_{SM} \equiv \int_{M_{H} - 2 \Gamma_{H}}^{M_{H} + 2 \Gamma_{H}} \left[ d M_{WW} \times \frac{d\sigma_{SM}}{d M_{WW}} \right] \, ,
\ee
where here $\sigma_{SM}$ is calculated at tree level with the appropriate Higgs mass, whose width 
is included via the replacement $M_{H}^{2} \to \left(M_{H}^{2} - i M_{H} \Gamma_{H} \right)$.  Using this information, we then define the ratio
\be
R^{peak} \equiv \fracp{\sigma^{peak}_{R}}{\sigma^{peak}_{SM}},
\ee
which is a function of the coefficients $a_{i}$.

The corresponding LHC cross-sections are given by 
\be
\frac{d \sigma}{d M^{2}_{WW}} = \sum_{i,j} \int_{M^{2}_{WW}}^{1} \int_{M^{2}_{WW}/(x_{1} s)}^{1} \frac{dx_{1} dx_{2}}{x_{1}x_{2}s}
f_{i}(x_{1},\mu_{F}) f_{j}(x_{2},\mu_{F}) \frac{dL_{WW}}{d\tau} \int_{-1}^{1} \frac{d \sigma_{WW}}{d \cos\theta} d\cos\theta \, ,
\ee
where $\tau = \hat{s}/s = M^{2}_{WW}/(x_{1}x_{2} s)$ and where we assumed for this comparison $\sqrt{s}=8$~TeV.  
We set the factorization scale, $\mu_{F}$, to the $W$-boson mass and use the CTEQ6L1 parton distribution functions.
The effective luminosity for longitudinal $W$ and $Z$ bosons is given as
\be
\frac{d L_{WW}}{d\tau} = \fracp{g}{4\pi}^{4} \fracp{1}{\tau}\left[ 
(1+\tau) \ln\fracp{1}{\tau} - 2 (1-\tau) 
\right] \, .
\ee
A factor of $(1/2)$ should also be included in the final expression for the $ZZ \to ZZ$ amplitude to account 
for the identical particle in the final state.

Finally, the $W W$ scattering amplitude is defined in the $W W$ rest frame as
\be
\frac{d \sigma_{WW}}{d \cos\theta} = \frac{|A|^{2}}{32\pi M^{2}_{WW}} \, ,
\ee
where, for instance,
\be
A(W^{+}W^{-} \to W^{+}W^{-}) = \frac{1}{3} T_{0} + \frac{1}{2} T_{1} + \frac{1}{6} T_{2} 
\ee
and assuming dominance in the resonance region of a single partial wave,\footnote{We emphasize that this approximation 
is only valid near a resonance, in the continuum a large number of partial waves contribute -the partial wave 
expansion converges very slowly.}
\bea
T_{0} & \approx & 32\pi t_{00} \\ \nn
T_{1} & \approx & 32\pi (3 t_{11} \cos\theta) \\ \nn
T_{2} & \approx & 32\pi t_{20} \, . 
\eea

\begin{figure}[ht!]
\centering
\includegraphics[clip,width=0.45\textwidth]{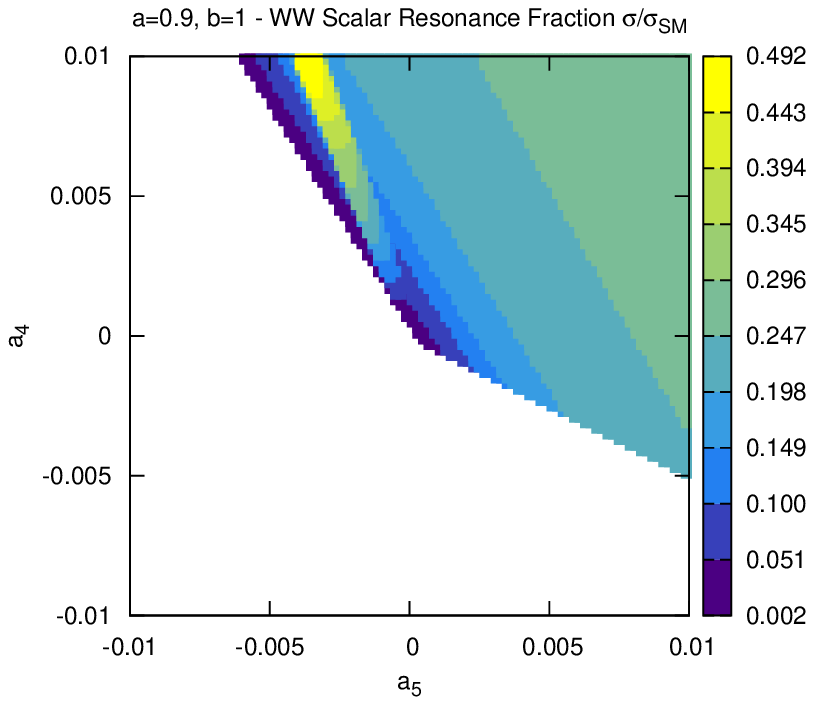} \hspace{0.5cm}
\includegraphics[clip,width=0.45\textwidth]{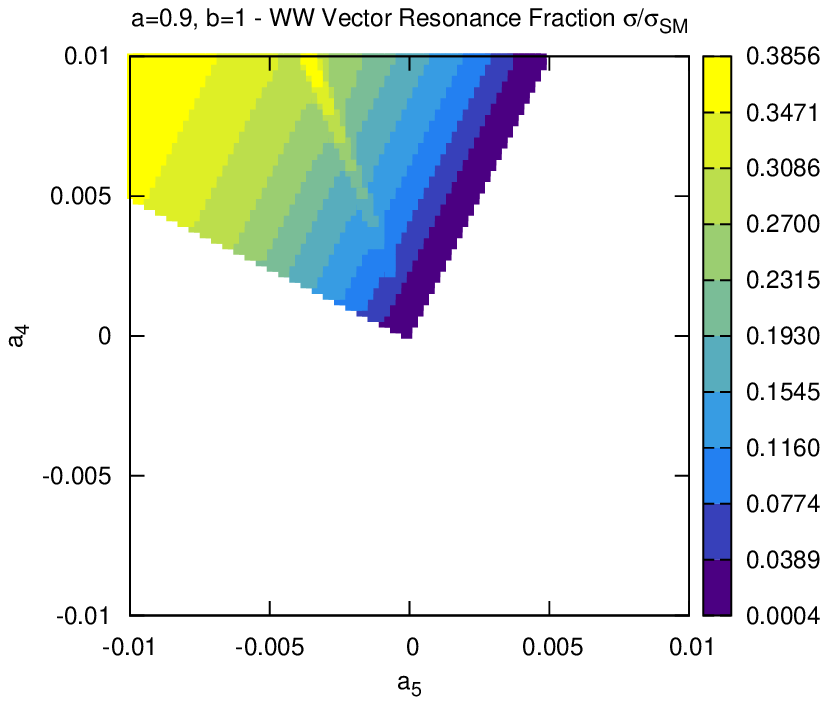}
\caption{Ratio of WW scattering cross section due to dynamical resonances with that of SM with Higgs 
boson of the same mass for (a) scalar and (b) vector resonances, taken in the ``peak'' region as defined in the text. 
\label{fig:signal_ww}}
\end{figure}

The results for $R^{peak}$ over the $a_{4} - a_{5}$ range considered are presented once again at the same benchmark point as 
the previous figure (but with $a=0.9$) and are shown in Fig. \ref{fig:signal_ww}. The figure gives the comparison 
with the SM calculation using Higgs boson masses set to those of the scalar and vector resonances.  
At lighter masses (in this case, that of the scalar), a corresponding Higgs signal would still be 
much more visible than that of these new dynamical resonances.  It should be noted, however, that at 
higher masses, such as that of the vector resonance in this figure, the Higgs width becomes very broad, 
making the direct comparison less obvious as its signal is more diluted.
It is of course difficult to draw firm conclusions from the previous analysis, but it is clear that it is
going to be difficult to detect this type of resonances coupling only to $VV$ pairs at the LHC. Although the
comparison with the Higgs is possibly misleading for the reasons that have been mentioned (basically
that resonances are expected in a region where a Higgs with an analogous mass has a huge width, while
the resonances under discussion are generally very narrow), we are tempted to provisionally conclude  
that detection of resonances in $VV$ scattering will be unfeasible 
at the LHC until a sufficiently large number of events are collected. This important issue will be discussed 
in considerably more detail in the next section.

%%%%%%%%%%%%%%%%%%%%%%%%%%%%%%%%%%%%%%%%%%%%%%%%%%%%%%%%%%%%%%%%%%%%%%%%%%%%%%%%%%%%%%%%%%%%%%%%%%%%%%%%%%%%%%%%%%%%
%%%%%%%%%%%%%%%%%%%%%%%%%%%%%%%%%%%%%%%%%%%%%%%%%%%%%%%%%%%%%%%%%%%%%%%%%%%%%%%%%%%%%%%%%%%%%%%%%%%%%%%%%%%%%%%%%%%%

\section{LHC phenomenology}

\begin{figure}[th]
\begin{center}
\includegraphics[width=0.7\textwidth]{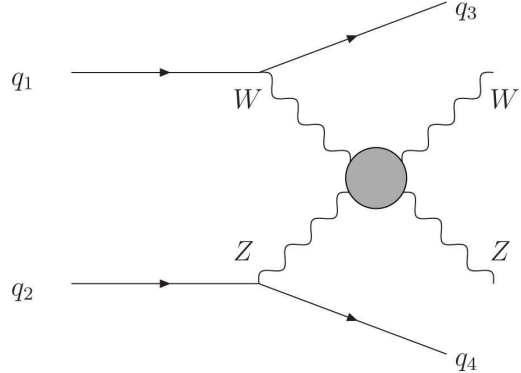}
\caption{Graphical representation of the $pp \to WZjj$ process at the LHC, at the parton level,  by means of $WZ \to WZ$ scattering. Figure taken from  \cite{DDEGHMS}.}
\label{fig:diagppWZjj}
\end{center}
\end{figure}

In order to see one example of how the methods shown in this work can be applied efficiently to the phenomenological description of processes where the dynamics of a possible strongly interacting SBS of the SM is involved, we will consider here the resonant elastic WZ scattering at the LHC (see Fig.\ref{fig:diagppWZjj}). 

As discussed previously there are two different ways for introducing resonances in the HEFT. First; one can use an analytical unitarization method like the IAM or N/D and look for the region of the coupling parameter space where poles are developed in the second Riemann sheet close enough to the real axis to be understood as dynamical resonances. The second way is introducing the resonance fields explicitly in the HEFT Lagrangian as independent degrees of freedom. In this case one is forced to introduce also new couplings and parameters describing the main properties of these resonances.

A possible mixed approach is the one considered in \cite{DDEGHMS} for the study of WZ elastic scattering through vector ($I=J=1$) resonance production. In that reference the authors use the IAM method in the region of the $a, b, a_4$ and $a_5$ parameter space  that gives rise to a dynamical vector resonance. Then the position of the pole determines the mass and the width of the resonance and the residue the coupling to the WZ initial and final states. In Table \ref{tablaBMP} it is possible to see selected benchmark points (BP) corresponding to different values of the relevant parameters and the masses and widths of the produced $I=J=1$ resonance (obtained with some approximations to be explained bellow). Given these parameters  one can build up a HEFT model with explicit resonances. Notice that this is the inverse procedure to that of obtaining the couplings by integrating out the resonances explicitly introduced in the HEFT Lagrangian. In any case this mixed approach is very convenient to be incorporated in Monte Carlo algorithms (MC) to be used to make contact with LHC  phenomenology. This is because usually MCs work with Feynman diagrams and not with analytical and unitary partial waves on which the IAM or the N/D methods are based. In the following the mixed method will be refered to as IAM-MC. 

\begin{figure}[t!]
\begin{center}
\includegraphics[width=.6\textwidth]{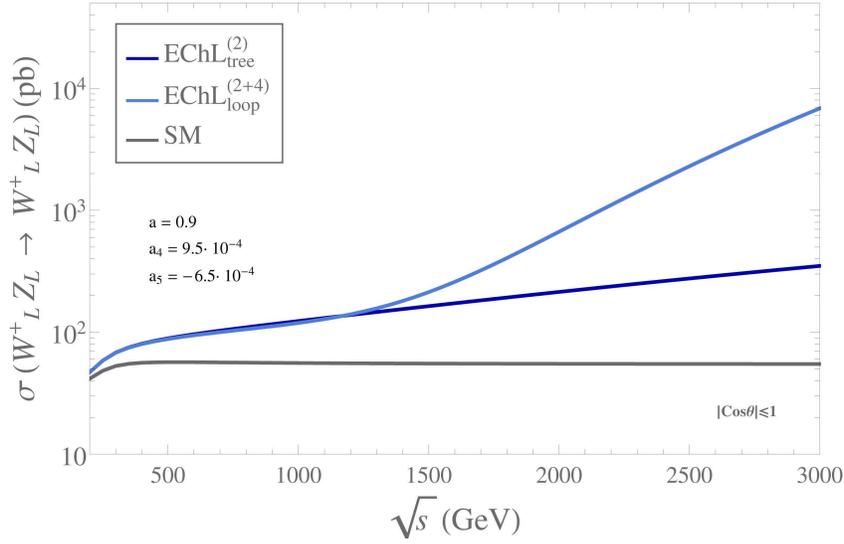}
\caption{Cross section $\sigma(W_LZ_L \to W_LZ_L)$  from the EChL (HEFT). The predictions at leading order, ${\rm EChL}^{(2)}_{\rm tree}$,
and next to leading order, ${\rm EChL}^{(2+4)}_{\rm loop}$,
are displayed separately. The HEFT coefficients
are set here to $a=0.9$, $b=a^2$, $a_4=9.5 \times 10^{-4}$ and $a_5=-6.5 \times 10^{-4}$.
The prediction of the SM cross section is also included, for comparison. Taken from \cite{DDEGHMS}.}
\label{Crosssection}
\end{center}
\end{figure}

Starting from the gauged HEFT Lagrangian it is possible in principle to compute de $W_LZ_L \to W_LZ_L$ amplitude directly without using the ET, i.e. by considering Feynman diagrams with external electroweak bosons lines equipped with the corresponding polarization vectors and taking the physical values for $M_W$; $M_Z$ and $M_H$ different from zero. This amplitude is then obtained as a formal expansion in powers of the external momenta:
\be
A(W_LZ_L \to W_LZ_L)= A^{(0)}(W_LZ_L \to W_LZ_L)+A^{(1)}(W_LZ_L \to W_LZ_L)+...
\ee
Here $A^{(0)}=A^{(2)}_{tree}$ is the tree level LO ($\mathcal{O}(p^2))$. $A^{(1)}$ is the tree level NLO ($\mathcal{O}(p^4))$ plus the one-loop corrections coming from the $\mathcal{O}(p^2))$ HEFT Lagrangian:
\be
A^{(1)}(W_LZ_L \to W_LZ_L)= A^{(4)}_{tree}(W_LZ_L \to W_LZ_L)+A^{(2)}_{1-loop}(W_LZ_L \to W_LZ_L).
\ee
The tree contribution $A^{(2)}_{tree}=$EChL$^{(2)}_{tree}$ can be computed in terms of $a, a_4$ and $a_5$ (notice that $b$ does not contribute to the tree level result) by using for example  FeynArts  \cite{Hahn:2000kx}  and FormCalc \cite{Hahn:1998yk}. As far as  we know the complete one-loop $A^{(1)}_{1-loop}$  computation is not yet available. However, following  \cite{Espriu:2012ih, Espriu:2013fia}, it is possible to estimate this contribution as follows. The real part is computed by using the  ET and the imaginary part is computed from the tree level result by making use of the optical theorem.   In the following we will refer to this estimation of the NLO result as 
EChL$^{(2+4)}_{loop}$.  In Fig.\ref{Crosssection} we show the $W_LZ_L \rightarrow W_LZ_L$ cross section corresponding to EChL$^{(2)}_{tree}$, EChL$^{(2+4)}_{loop}$ and the MSM for some choice of the $a, b, a_4$ and $a_5$ parameters. As can be seen, the MSM, being a weakly interacting and unitary theory, predicts an almost flat cross section. However, departure from the MSM by using HEFT (EChL) gives rise to very strongly interacting models with increasing cross sections with energy \cite{Delgado:2013loa}.
Both the LO and the NLO eventually break unitarity bounds (about 2 TeV in the example chosen in the figure). This shows the need for complementing the HEFT predictions with unitarization methods based in dispersion relations, as described in this work,   
 in order to provide realistic predictions for the LHC.

From the these HEFT amplitudes  $A^{(0)}(W_LZ_L \to W_LZ_L)$ and $A^{(1)}(W_LZ_L \to W_LZ_L)$    it is also possible to apply the IAM method (the preferred unitarization method for the vector channel) to obtain the resonance parameters corresponding to the different benchmark points shown in Table.\ref{tablaBMP}. There it is possible to see that for different sets of  experimentally allowed couplings one can obtain relatively narrow vector resonances with masses $M_V$ in the range 1.5 TeV to 2.5 TeV. For these benchmark points no other resonances (isoscalar or isotensor) are present. However notice that vector resonances can be present even in the case $a=b=1$ provided $a_4$ and $a_5$ are different from zero, thus indicating departure from the MSM and at the same time strongly interacting SBS.

\begin{table}[th]
\begin{center}
\vspace{.2cm}
\begin{tabular}{ |c|c|c|c|c|c|c| }
\hline
%\rowcolor{gray! 50}
\rule{0pt}{1ex}
{\footnotesize {\bf BP}} & {\footnotesize {\bf $M_V ({\rm GeV})$}}  & {\footnotesize  {\bf $\Gamma_V ({\rm GeV)}$}}  & {\footnotesize {\bf $g_V(M_V^2)$}}  & {\footnotesize {\bf $a$}}  & {\footnotesize {\bf $a_4 \cdot 10^{4}$}}  & {\footnotesize {\bf $a_5\cdot 10^{4}$}}
\\[5pt] \hline
\rule{0pt}{1ex}
BP1  & $\quad 1476 \quad $ & $\quad 14 \quad $ & $ \quad 0.033  \quad $ & $ \quad 1 \quad $ & $ \quad 3.5 \quad $ & $ \quad -3 \quad $
\\[5pt] \hline
\rule{0pt}{1ex}
BP2  & $\quad 2039 \quad $ & $\quad 21 \quad $ & $ \quad 0.018  \quad $ & $ \quad 1 \quad $ & $ \quad 1 \quad $ & $ \quad -1 \quad $
\\[5pt] \hline
\rule{0pt}{1ex}
BP3  & $\quad 2472 \quad $ & $\quad 27 \quad $ & $ \quad 0.013  \quad $ & $ \quad 1 \quad $ & $ \quad 0.5 \quad $ & $ \quad -0.5 \quad $
\\[5pt] \hline
\rule{0pt}{1ex}
BP1' & $\quad 1479 \quad $ & $\quad 42 \quad $ & $ \quad 0.058  \quad $ & $ \quad 0.9 \quad $ & $ \quad 9.5 \quad $ & $ \quad -6.5 \quad $
\\[5pt] \hline
\rule{0pt}{1ex}
BP2'  & $\quad 1980 \quad $ & $\quad 97 \quad $ & $ \quad 0.042  \quad $ & $ \quad 0.9 \quad $ & $ \quad 5.5 \quad $ & $ \quad -2.5\quad $
\\[5pt] \hline
\rule{0pt}{1ex}
BP3'  & $\quad 2480 \quad $ & $\quad 183 \quad $ & $ \quad 0.033  \quad $ & $ \quad 0.9 \quad $ & $ \quad 4\quad $ & $ \quad -1 \quad $
\\[5pt] \hline
\end{tabular}
\caption{\small Choice of  benchmark points (BP) of dynamically generated vector resonances.
The mass, $M_V$, width, $\Gamma_V$, coupling to gauge bosons, $g_V(M_V)$,
and relevant corresponding HEFT couplings $a$, $a_4$ and $a_5$.
$b$ is fixed to $b=a^2$. This table is generated using the FORTRAN code that implements
the EChL+IAM framework from \cite{Espriu:2012ih,Espriu:2013fia}. }
\label{tablaBMP}
\end{center}
\end{table}

 \begin{figure}[th]
\begin{center}
   \includegraphics[width=0.40\textwidth]{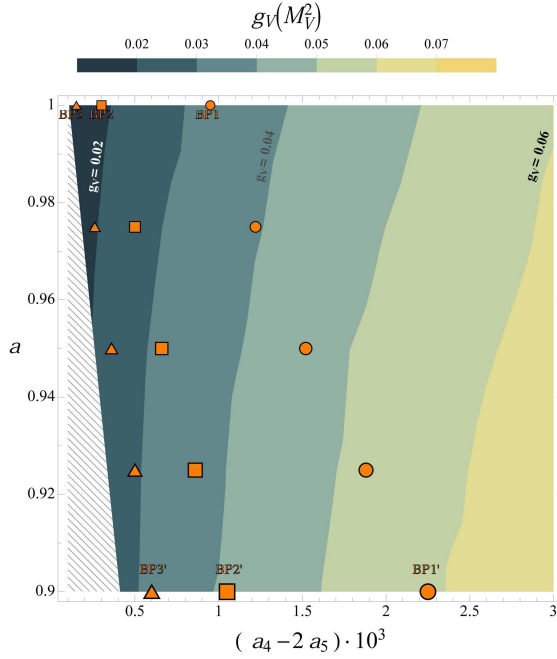}
    \caption{The effective coupling of a vector resonance of mass $M_V$ in the $WZ$ channel as a function
of the parameters $a$ and the combination $a_4 - 2a_5$, relevant in the vector channel. Our fifteen selected scenarios lay approximately over the contour lines of fixed $M_V$, 1500 GeV (circles), 2000 GeV (squares), and 2500 GeV (triangles), and have values for $a$ fixed, respectively,  to 0.9 (biggest symbols, corresponding to BP1', BP2' and BP3'), 0.925, 0.95, 0.975 and 1 (smallest symbols, corresponding to BP1, BP2, and BP3). All studied cases with vector resonances are such that no corresponding scalar or tensor resonances appear. The stripped area denotes the region with resonances heavier than 3000 GeV.}
\label{contourgv}
 \end{center}
\end{figure}

As commented above, the use of  Monte Carlo generators like MadGraph requires introducing a Lagrangian language. Therefore, instead of 
working with scattering amplitudes reconstructed from the unitarized partial waves one can work, for example, with the Lagrangian 
in Eq.\ref{Resonancecoupling}
which includes explicitly a vector resonance. In the case of a strongly interacting SBS modification of the MSM we expect the 
longitudinal modes of the electroweak bosons to be dominant over the transverse modes. Then we can set $f_V=0$ and also neglect 
the axial vector couplings. In this case the only parameters needed are $M_V$, $\Gamma_V$ and $g_V$. These three 
parameters can be obtained from the IAM method as explained above. 
The coupling $g_V$ can be obtained by comparison of the $I=J=1$ partial wave amplitude computed with the IAM  
method and with the Lagrangian in  Eq.\ref{Resonancecoupling} on the resonance. 
For example, by using the ET one would find $g_V^2= 2(a_4-2 a_5)$. Alternatively one 
can use the relations found by integrating out the vector resonance field, namely $a_4=-a_5= g_V^2/4$. Both methods disagree by 
about  50\%. In any case, using a constant value for $g_V$, even though  it may give a satisfactory answer above and on 
top of the resonance, would produce a bad behavior of the amplitude at higher energies $s> M_V$.  In fact the $I=J=1$ 
partial wave breaks unitarity in the TeV region, or even below, depending on the precise value of the couplings. 
Of course this is a consequence of using the HEFT plus vector resonance Lagrangian at the tree level. It is possible 
to fix this problem partially in the IAM-MC model by allowing $g_V$ to be a function on $s$, i.e. by introducing 
the running effective coupling 
$g_V(s)$. In order to get the effective coupling $g_V$ of the effective Lagrangian Eq.\ref{Resonancecoupling}  we impose 
the matching at the level of the partial waves by identifying the tree level predictions
from the HEFT theory equipped with resonances with the unitarized amplitude 
\begin{equation}
\Big|t_{11}^{{\rm HEFT}_{\rm tree}^{(2)}}(s=M_V^2)\Big|=\Big|t_{11}^{\rm unit.}(s=M_V^2)\Big|\,.
\label{a11MV}
\end{equation}
Solving (numerically) this  Eq.\ref{a11MV} for the given values of $(a,a_4,a_5)$
and the corresponding values of $(M_V,\Gamma_V)$ leads to the wanted solution for $g_V$.
In order to have a feeling of the sort of values one gets, we show below in Table.\ref{tablaBMP} and Fig.\ref{contourgv} some values obtained
for the coupling $g_V$ (see Eq.\ref{Resonancecoupling} from unitarization in $WZ$ scattering. In addition, in  \cite{DDEGHMS} it is shown that a good description of the IAM results in the context of the IAM-MC framework can be obtained by introducing the $g_V$ coupling function in terms of the $t$ and $u$ variables:
\begin{align}
g_V^2(z)&=g_V^2(M_V^2) \frac{M_V^2}{z} \,\,\, {\rm for} \,\, s< M_V^2 \nn\,, \\
g_V^2(z)&=g_V^2(M_V^2) \frac{M_V^4}{z^2} \,\,\, {\rm for} \,\, s> M_V^2\,,
\label{gvenergytu}
\end{align}
with $z=t,u$ corresponding to the $t,u$ channels, respectively, in which the resonance is propagating.With this choice it is possible to simulate the IAM results with the Lagrangian formalism at the tree level which is more adapted for Monte Carlo algorithms for LHC event generators. This modelization matches well with chiral perturbation results and fulfills the Froissart bound too.

As an example it is possible to generate two runs for each benchmark point. One with $W^+Zjj$ as final state and the other one including 
the final leptonic decays  $W^+ \rightarrow l^+ \nu, Z \rightarrow l^+ l^-$. We set the cuts: $2<\lvert\eta_{j_1,j_2}\rvert<5$, $\eta_{j_1}\cdot\eta_{j_2}<0$, $p_T^{j_1,j_2}>20\,{\rm GeV}$ and $M_{jj}>500\,{\rm GeV}$ and an additional cut $\lvert\eta_{W,Z}\rvert < 2$ is used  for the $W^+Zjj$ final state run. We also set the  cuts $M_Z-10\,{\rm GeV} < M_{\ell^+_Z \ell^-_Z} < M_Z+10\,{\rm GeV}$, $M^T_{WZ}\equiv M^T_{\ell\ell\ell\nu}>500\,{\rm GeV}$,  $\slashed{p}_T>75\,{\rm GeV}$ and $p_T^\ell>100\,{\rm GeV}$ for the leptonis decays run. One can also compute two SM backgrounds: pure SM-EW background  $q_1q_2\to q_3q_4W^+Z$ scattering at order $\mO(\alpha^2)$ and mixed SM-QCDEW, at order $\mO(\alpha\alpha_S)$.

\begin{figure}[th]
\begin{center}
\includegraphics[width=.49\textwidth]{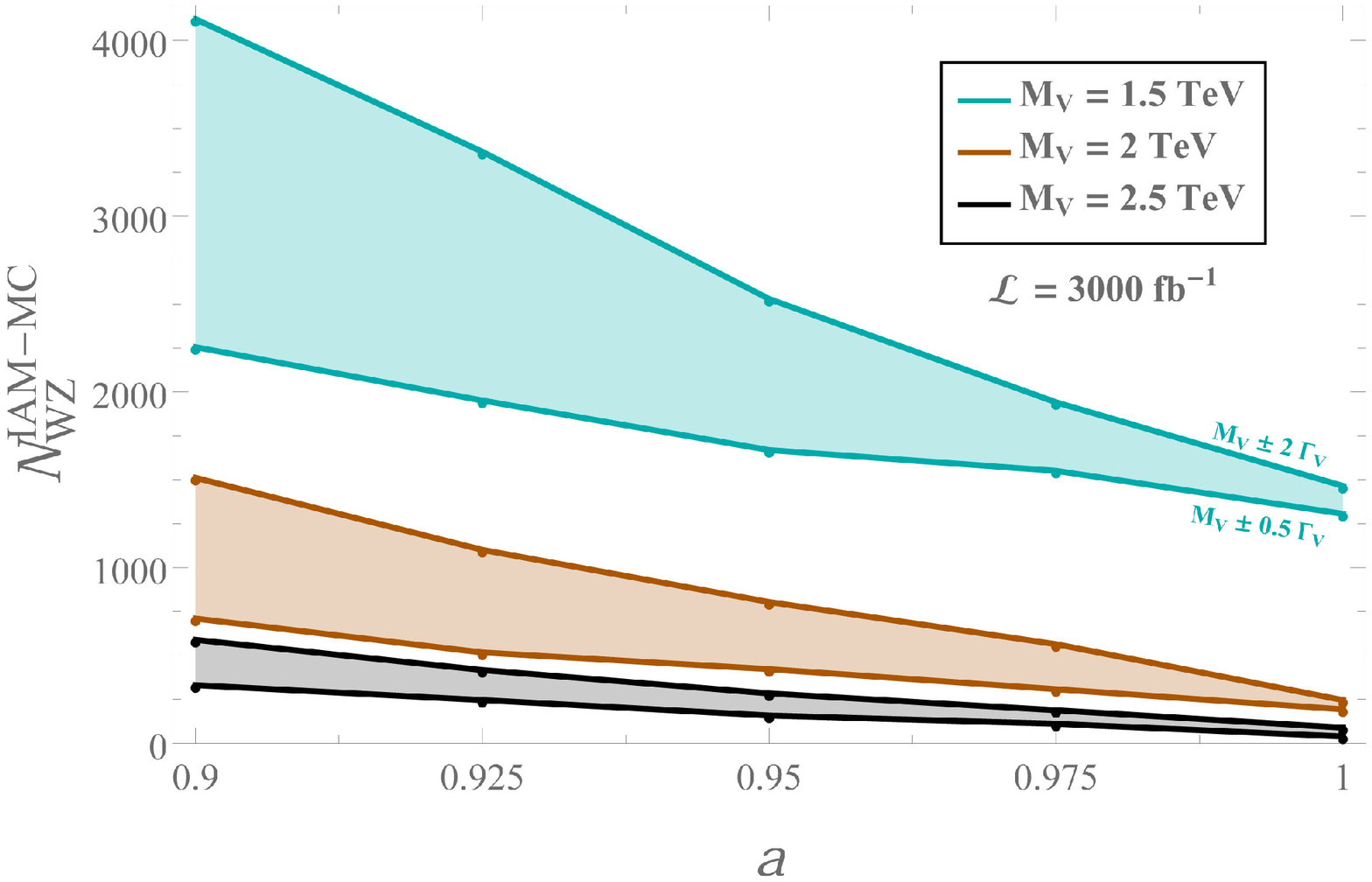}
\includegraphics[width=.48\textwidth]{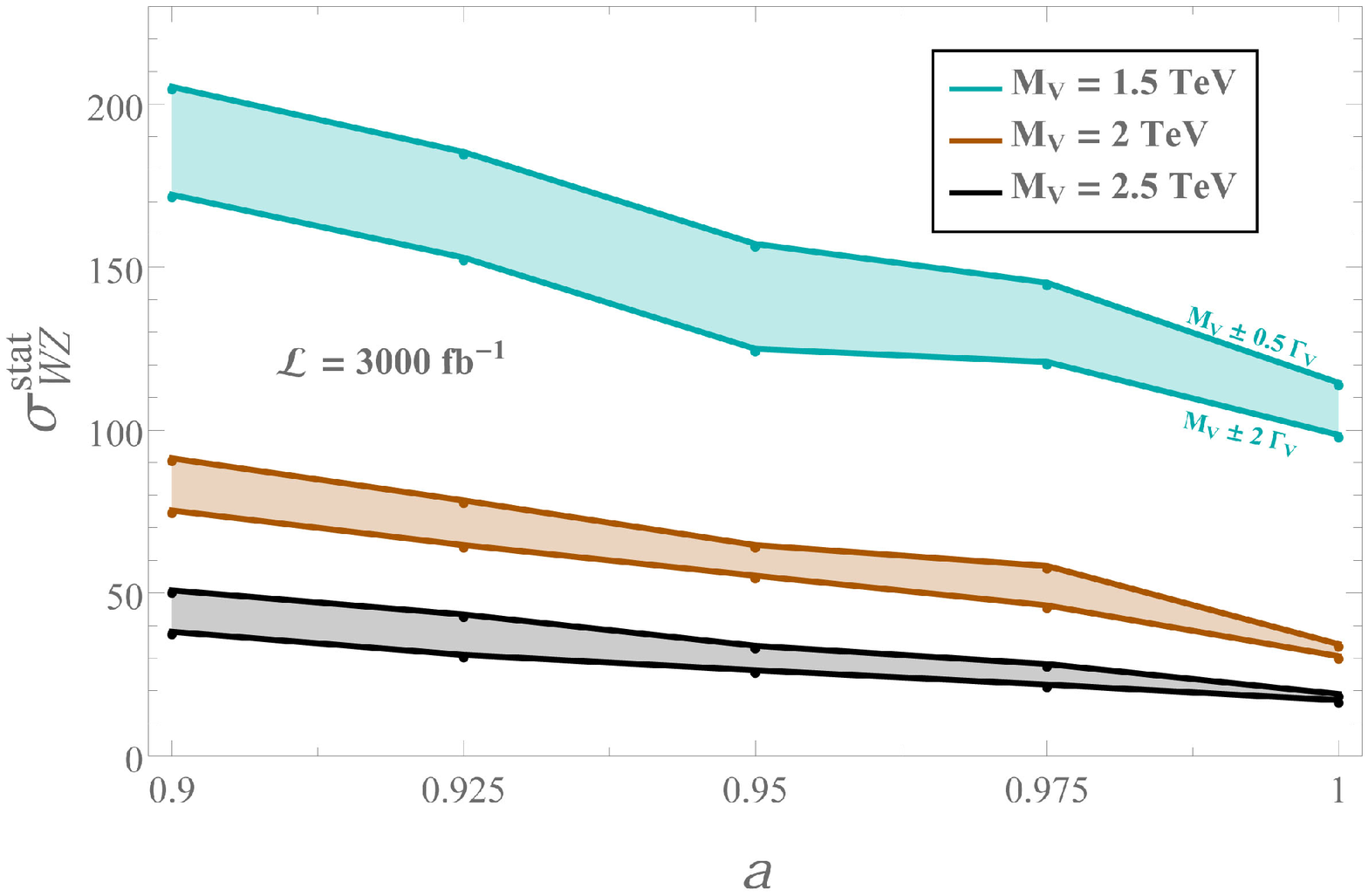}
\caption{Predictions for the number of events, ${\rm N}^{\rm IAM-MC}_{WZ}$ (left panel), and the statistical significance, $\significance_{WZ}$ (right panel), as a function of the parameter $a$ for $\mL=3000~{\rm fb}^{-1}$. The marked points correspond to our selected BPs in Fig.~\ref{contourgv}. The two lines for each mass are computed by summing events within $\pm 0.5\,\Gamma_V $ and $\pm 2\,\Gamma_V$, respectively.}\label{figevents}
\end{center}
\end{figure}

\begin{figure}[th]
\null%
\hfill\includegraphics[width=.49\textwidth]{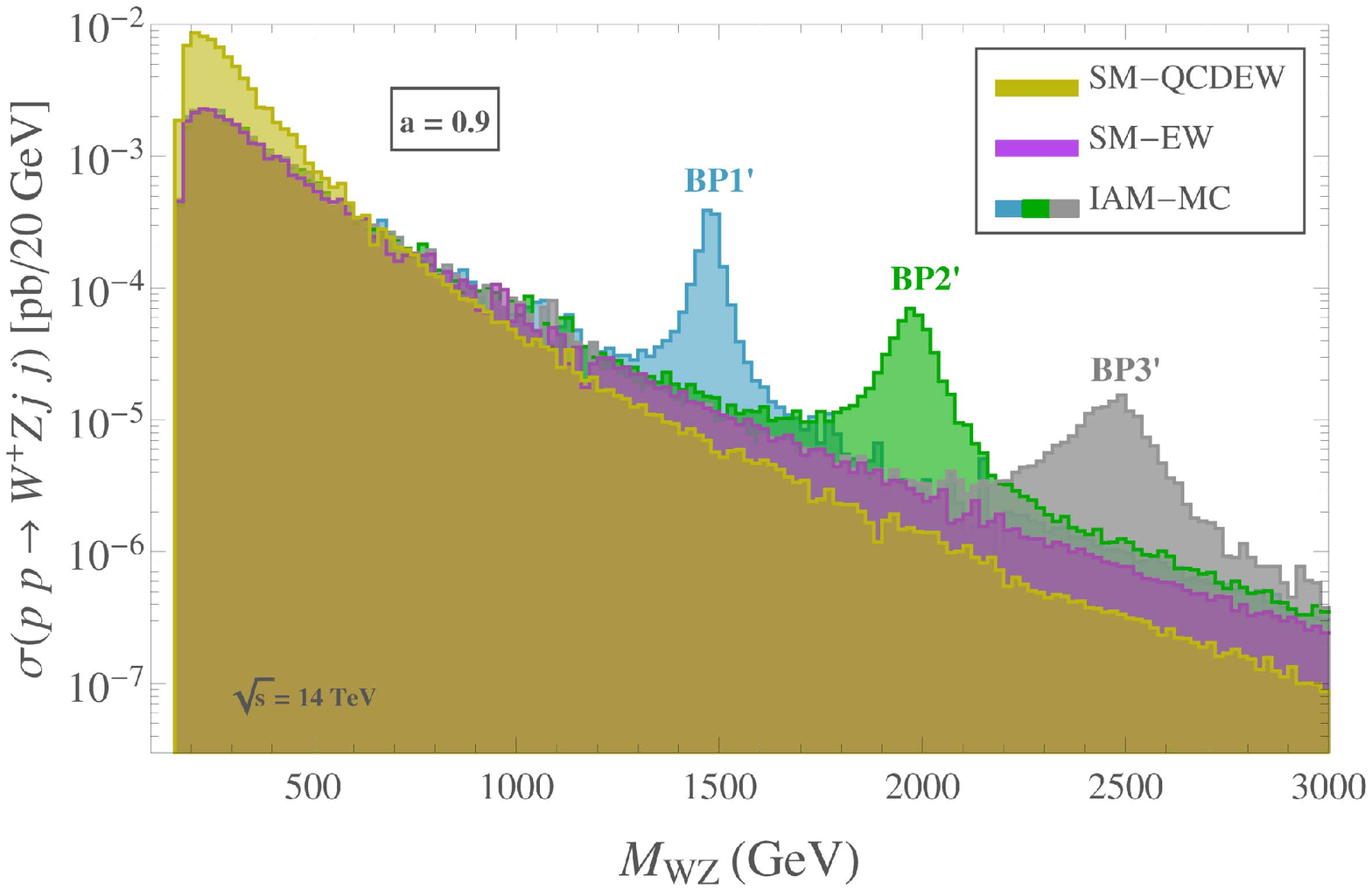}%
\hfill\includegraphics[width=.49\textwidth]{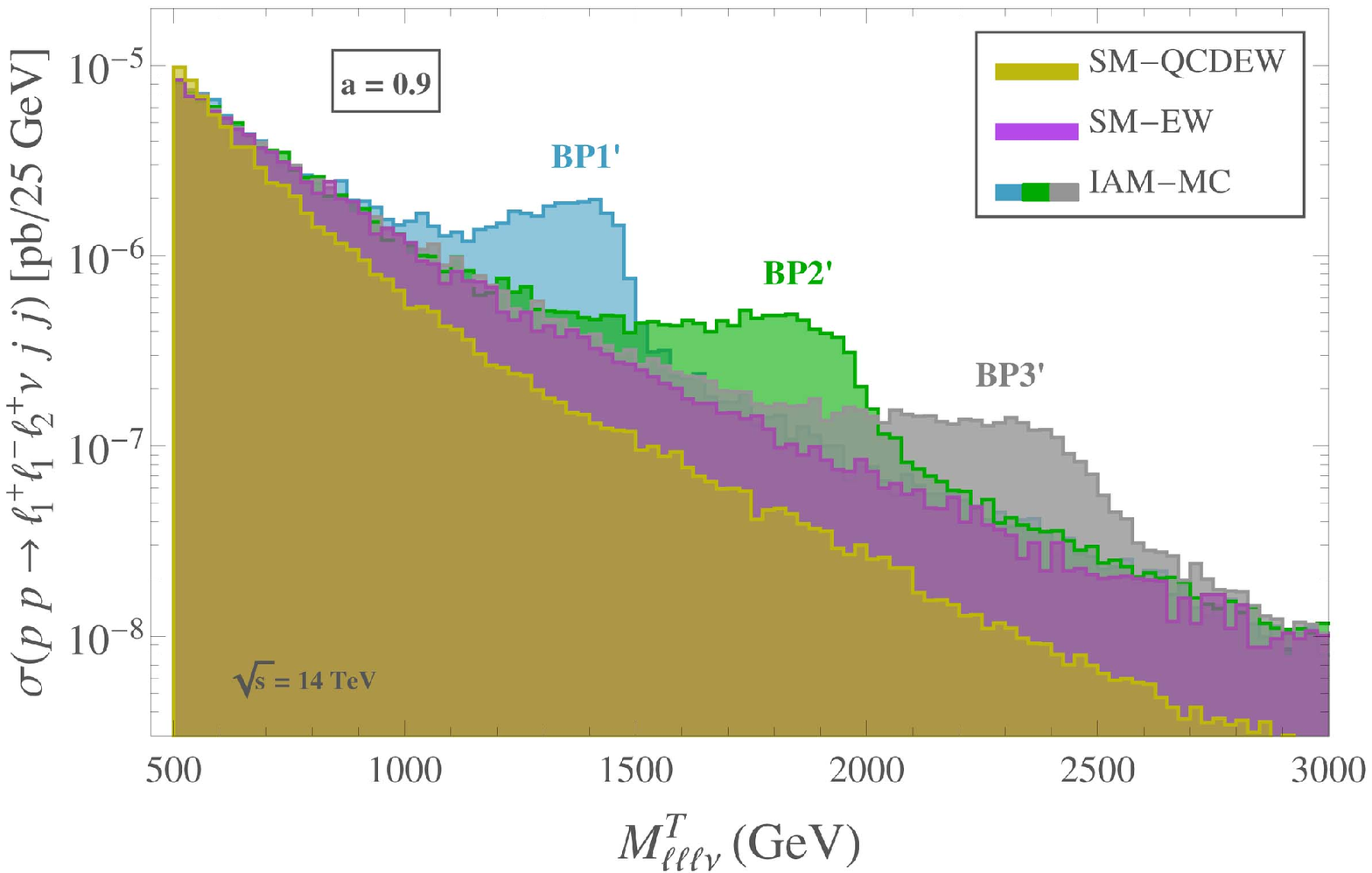}%
\hfill\null
\caption{\small BP1' (see table~\ref{tablaBMP}). $W^+Zjj$ in final state (left) vs. leptonic decay (right).}
\label{BPa9}
\end{figure}

Now it is possible to make a prediction on the number of events expected   at 14 TeV for different LHC luminosities \cite{Delgado:2018nnh} (see Fig. \ref{figevents}) and the cross-sections can be found in Fig.\ref{BPa9} for the value $a=0.9$. 

 In order to estimate the statistical significance it is possible to use the standard expression $\significance_\ell=S_\ell/\sqrt{B_\ell}$, where the signal is $S_\ell=N^{\rm IAM-MC} - N^{\rm SM}$ and  the background is $B_\ell=N^{\rm SM}$. Considering the optimal ranges:
\begin{align}
&{\rm BP1:}~1325-1450~{\rm GeV}\,, && {\rm BP2:}~1875-2025~{\rm GeV}\,, && {\rm BP3:}~2300-2425~{\rm GeV}\,,\nn \\
&{\rm BP1':}~1250-1475~{\rm GeV}\,, && {\rm BP2':}~1675-2000~{\rm GeV}\,, &&{\rm BP3':}~2050-2475~{\rm GeV}\,.
\label{MTintervals}
\end{align}
for $M^T_{lll\nu}$ one finds that the cases with $a=1$ have smaller significances, and only the lightest resonances $M_V=1.5\,{\rm TeV}$ (BP1) could be seen at $\sim 3\sigma$ with the highest luminosity ($3000\,{\rm fb}^{-1}$). For heavier $M_V\sim 2.5\,{\rm TeV}$ it seems difficult to observe resonances  due to the poor statistics in the leptonic channels. Only for BP3' it is possible to get a  significance $>2\sigma$ (for $3000\,{\rm fb}^{-1}$). Therefore semileptonic and fully hadronic channels seem  necessary to improve these statistical significances. The largest significances are obtained for $a=0.9$ and the lightest resonances, which corresponds to  BP1'. In this case significances of $\sim 2.8\sigma$, $5.1\sigma$ and $8.9\sigma$ are predicted for LHC luminosities $\mL=300\,{\rm fb}^{-1}$, $1000\,{\rm fb}^{-1}$ and $3000\,{\rm fb}^{-1}$, respectively.

Notice the enormous dependence on $a$ of the results. The equivalent plot of Fig. 21 for $a=1$ would show a 
lot less prominent signal. 
The interested reader can see  \cite{DDEGHMS} for more details.  Recent work  on the $W^+W^-\to W^+W^-$ channel is also 
available \cite{MJHDG}. In this case, the pure QCD background is overwhelming but work still in progress 
reveals that specific kinematical 
cuts may help in isolating the relevant signal.

\section{Outlook}
So far all experimental evidence points at the Minimal Standard Model, with an elementary scalar field responsible
for the mass generation mechanism both for electroweak bosons and matter fields, as the ultimate description of 
fundamental interactions. We know that this cannot be true: massive neutrinos and dark matter do not fit in this 
picture. Let alone more theoretical issues such as the naturalness problem or the unification with gravity. Were
the MSM to be the ultimate answer, fundamental issues such as the final source of $CP$ violation or the flavour puzzle would 
remain unexplained too. There is then a widespread belief that there is physics beyond the MSM. Hopefully this
new physics may be not too far away.

Exploring a strongly interacting EWSBS is an obvious possibility. Some of the theoretical difficulties 
in going BSM are alleviated
if new strong interactions at the multi-TeV region exist, providing for some scale of compositeness, both for the
longitudinal degrees of freedom of the electroweak bosons and for the Higgs particle itself, and providing us,
may be, with new dark matter candidates and a plethora of new states.  We know that this new scale cannot be
too high, lest new fine tuning and hierarchy problems appear.

In the previous sections we have tried to provide tools to analyze, then verify or falsify, the existence of this
putative new strong sector. This is challenging as only the lightest part of its spectrum may be known ---the Nambu-Goldstone
bosons associated to the spontaneous breaking of the new ``flavor'' sector. Yet, a combination of old and new techniques
allows us to corner quite precisely the possible new physics. The main conclusion, at the phenomenological level, is that
if an EWSBS exists at all, at least in its simplest form ---not connected to the matter sector---, it will be hard
to find evidence of its presence at the LHC until substantially more statistics in $VV$ fusion or $hh$ production
accumulates. We should always bear in mind that non-existence of proof is not the same as proof of non-existence. 
A strongly interacting EWSBS cannot be ruled out as of today by any means.

The high energy physics community is debating today several options for a next machine, let it be an $e^+ e^-$ 
linear collider capable of exploring the trilinear Higgs coupling in the potential or the top couplings, a 
100 km circular electron/hadron collider, or the high energy option for the LHC that would basically double the 
energy reach. It is not clear which option is more competitive in terms of exploring BSM physics. Except 
possibly for the linear collider in its lowest energy version, the methods presented in this report will be 
needed in all cases.

\section*{Acknowledgements} We would like to thank our collaborators P. Arnan, A. Castillo, G. D'Ambrosio, R. L. Delgado, 
C. Garc\'\i a-Garc\'\i a, M. J. Herrero, A. Katanaeva, F. Llanes-Estrada, X. Marcano, F. Mescia, J. R. Pel\'aez, C. Quezada, 
J. J. Sanz-Cillero, T. N. Truong and B. Yencho who participated in various parts of the research presented in this report and from whom we 
have learned a lot. We would also like to thank J. Bernabeu for various early discussions on the subject
This research is partly supported by the Ministerio de Ciencia, Inovaci\'on y Universidades under research grants: 
FPA2016-75654-C2-1-P and FPA2016-76005-C2-1-P, by grant MDM-2014-0369 of ICCUB (Unidad de Excelencia ‘Maria de Maeztu’),  
and grant 2017SGR0929 (Generalitat de Catalunya).
We are thankful to the CERN TH Division where parts of this report were prepared for hospitality and support.

\vfill

%%%%%%%%%%%%%%%%%%%%%%%%%%%%%%%%%%%%%%%%%%%%%%%%%%%%%%%%%%%%%%%%%%%%%%%%%%%%%%%%%%%%%%%%%%%%%%%%%%%%%%%%%%%%%%%%%%%%%%
\appendix

%%%%%%%%%%%%%%%%%%%%%%%%%%%%%%%%
\section{Tree-level  scattering amplitudes}
\label{sec:appendix_amplitudes}
%%%%%%%%%%%%%%%%%%%%%%%%%%%%%%%%%
\noindent
In the isospin limit, $M=M_{Z} = M_{W}$, and with massive $W$, the tree-level and $a_{4,5}$-dependent
amplitude for $\wplus_{L} \wminus_{L} \to Z_{L} Z_{L}$ scattering is given by
\bea\label{fullamp}
A_{W^{+}W^{-} \to ZZ}^{ {\rm tree} \, + \, a_{i}} \left( p_{1},p_{2},p_{3},p_{4} \right)
& = &
\;
-2 g^{2}(1  - g^{2}a_{5} )
\eeee{1}{2}{3}{4} \\ \nn
&& +  g^{2}(1 + g^{2}a_{4}) \Big[
\eeee{1}{4}{2}{3} + \eeee{1}{3}{2}{4}
\Big]
\\ \nn
& &
+g^{2} \Bigg\{ \left(\frac{1}{(p_{1}-p_{3})^{2}-M^{2}}\right)
\Bigg[ \\ \nn
& &
-4 \Big(
\eepepe{1}{2}{1}{3}{2}{4} + \eepepe{1}{4}{1}{3}{4}{2} + \\ \nn
& & \hspace{0.75cm}
\eepepe{2}{3}{3}{1}{2}{4} + \eepepe{3}{4}{3}{1}{4}{2}
\Big) \\ \nn
& &
+2\Big(
\edote{2}{4} \Big(
\pdote{1}{3}(p_{2}+p_{4})\cdot\epsilon_{1} +
\pdote{3}{1}(p_{2}+p_{4})\cdot\epsilon_{3}
\Big) + \\ \nn
& & \hspace{0.75cm}
\edote{1}{3} \Big(
\pdote{2}{4}(p_{1}+p_{3})\cdot\epsilon_{2} +
\pdote{4}{2}(p_{1}+p_{3})\cdot\epsilon_{4}
\Big) \Big)  \\ \nn
& & \hspace{0.45cm}
- \eeee{1}{3}{2}{4} \Big(
(p_{1}+p_{3})\cdot p_{2} + (p_{2}+p_{4})\cdot p_{1}
\Big)
\Bigg] \\ \nonumber
& &
+ \hspace{0.5cm} (p_{3} \leftrightarrow p_{4})
\Bigg\}
-g^{2} M^{2}\left(\frac{\edote{1}{2} \edote{3}{4}}{(p_{1}+p_{2})^{2}-M_{H}^{2}} \right) \, , \nn
\eea
where $\epsilon_{i} = \epsilon_{L}(p_{i})$. The analogous expression in the ET approximation is much simpler
\be\label{etamp}
A_{w^{+} w^{-} \to z z}^{ {\rm tree } \, + \, a_{i}} (s)
 = - \fracp{s}{v^{2}} \left(\frac{(a^{2}-1)s + M_{H}^{2}}{s-M_{H}^{2}}
 -2 \fracp{s}{v^{2}}\left(a_4 (1+\cos^2\theta) + 4 a_5\right) \right)
 \ee
For completeness, we also give the amplitude for the $\wplus_{L} \wminus_{L} \to h h$ scattering
\bea
\!\!A_{W^{+}W^{-} \to hh}^{{\rm tree }}(p_1,p_1,q_3,q_4) &=&
g^2 \left(\frac{b}{2} \edote{1}{2} - \frac{3 a M_H^2}{2(M_H^2-(p_1+p_2)^2)}\edote{1}{2}\right. \\
\no
&+&\!\!\!\!\left.a^2  \frac{g^2 v^2}{4 M^2}\left(\frac{\edotqmp{1}{3}{1}\edotqmp{2}{3}{1}
 -M^2 \edote{1}{2}}{M^2-(q_3-p_1)^2} + (q_3\leftrightarrow q_4)\right)\right),
\eea
In the CM reference frame the expression for
$A_{W^+W^-\to ZZ}^{\rm{tree}\, + \, a_{i}}(s,t,u)$ becomes
\bea\label{eq:CMS-A}
A_{W^+W^-\to ZZ}^{\rm{tree}\, + \, a_{i}}(s,t,u) &=&
\frac{a^2 \left(s-2 M^2\right)^2}{v^2 \left(M_H^2-s\right)}\\
&+&\frac{768
   M^{10}-128 M^8 (5 s+4 t)+32 M^6 \left(7 s^2+8 s t+4 t^2\right)}
{v^2 \left(s-4 M^2\right)^2 \left(M^2-t\right) \left(-3M^2+s+t\right)}\nn\\
&-& \frac{8 M^4 s
   \left(5 s^2+11 s t+4 t^2\right)+
   M^2 s^2 \left(3 s^2+18 s t+14 t^2\right)-s^3 t
   (s+t)}
   {v^2 \left(s-4 M^2\right)^2 \left(M^2-t\right) \left(-3 M^2+s+t\right)}\nn  \\
      &+&\frac{8 a_5 (s-2 M^2)^2 + 2 a_4 (16 M^4 - 8 M^2 s + (1 + \cos^2\theta) s^2))}{v^4}\nn
   \eea
Recall that $A_{W^+W^-\to ZZ}^{\rm{tree} \, + \, a_{i}}(s,t,u)$
for the scattering of longitudinally polarized $W$ is not Lorentz invariant. The expression
above is valid in CM frame only.

%%%%%%%%%%%%%%%%%%%%%%%%%%%%%%%%%%%%%%%%%%%%%
\section{Isospin decomposition and crossing}
\label{seec:appendix_matrices}
%%%%%%%%%%%%%%%%%%%%%%%%%%%%%%%%%%%%%%%%%%%%%

Let us write the amplitude for the process $a b \to c d$, where $w^a$, $a=1,2,3$ are Goldstone fields in the I=1 representation
of the isospin group  in the following way
\be
T^{abcd}=A\delta_{ab}\delta_{cd}+B\delta_{ac}\delta_{bd}+C \delta_{ad}\delta_{bc} ,
\ee
where
$A= A(s,t,u)$, $B=A(t,s,u)$ and $C=A(u,t,s)$. This decomposition is the same as used in 
Eqs. (\ref{deco1}), (\ref{deco2}) and (\ref{eq:isospin_general}), 
with slight changes in the notation.

The fixed isospin amplitudes can be expressed in terms of these functions as
\bea
T_{0} & = & 3A + B + C \nn \\ 
T_{1} & = & B - C \\
T_{2} & = & B+ C \nn
\eea
or conversely,
\bea
A & = &\frac13(T_0-T_2) \nn \\ 
B & = & \frac12 (T1+T2) \\
C & = &  \frac12 (T2-T1) . \nn
\eea
Then
\bea
T(w^+ w^- \to  zz ) &=& A = \frac13(T_0-T_2) \nn \\ 
T(w^+w^- \to w^+ w^-) &=& A+B = \frac13 T_0 + \frac12 T_1 + \frac16 T_2 \nn\\ 
T(zz \to zz) &=& A+B+C = \frac13 T_0 +\frac23 T_2\\ 
T(w^\pm z \to w^\pm z) &=& B = \frac12(T_1+T_2) \nn \\ 
T(w^\pm w^\pm \to w^\pm w^\pm) &=& B+ C = T_2 \nn \\
\eea
While it is obvious what crossing implies at the level of physical amplitudes involving the $w^\pm, z$ fields (see 
Appendix \ref{sec:appendix_crossing}), at the level of fixed isospin amplitudes the relations are somewhat more 
cumbersome and better expressed in terms of the so-called crossing matrices. If we introduce the vector
\be
\bar T = \begin{pmatrix} T_0 \\
                                T_1 \\
                                T_2 \end{pmatrix}
\ee
then we have
\be
\bar T(s, u) = C_{tu} \bar T(s, t), \qquad
\bar T(t, s)= C_{st} \bar T(s,t), \qquad
\bar T(u, t)= C_{su} \bar T(s,t),
\ee
where
\be
C_{tu}= \begin{pmatrix}  1 & 0 & 0 \\
                                    0 & -1 & 0 \\
                                   0 & 0 & 1 \end{pmatrix}, \quad
C_{st}= \begin{pmatrix}  \frac13 & 1 & \frac53 \\
                                    \frac13 & \frac12 & -\frac56 \\
                                   \frac13 & -\frac12 & \frac16 \end{pmatrix}, \quad
C_{su}= \begin{pmatrix}  \frac13 & -1 & \frac53 \\
                                    -\frac13 & \frac12  & \frac56 \\
                                   \frac13 & \frac12 & \frac16 \end{pmatrix}.  \ee

%%%%%%%%%%%%%%%%%%%%%%%%%%%%%%%%%%%%%%%%%%%%%
\section{The issue of crossing symmetry for $W_L$}
\label{sec:appendix_crossing}
%%%%%%%%%%%%%%%%%%%%%%%%%%%%%%%%%%%%%%%%%%%%%

Let us clarify the issue of crossing symmetry of amplitudes
with external $W_L$'s. To this end let us consider
just the tree-level contribution in the MSM to the processes $W_L^+ W_L^-\to W_L^+ W_L^-$ and
$W_L^+ W_L^+\to W_L^+ W_L^+$, respectively.

To keep the formulae simple while making the point let us consider the
limit $s\to \infty$, $-t\to \infty$ in the first process, which is consistent except for $\cos\theta \simeq 1$,
and expand in $M^2/s$ and $M^2/t$. We borrow the results from \cite{esma}.
The resulting amplitude is
\be
-g^2\left(\frac{M_H^2}{4M^2}\left[\frac{t}{t-M_H^2}+\frac{s}{s-M_H^2}\right]
+\frac{s^2+t^2+st}{2st}
-\frac{M_H^2}{s}\frac{2M_H^2t-s(s+t)}{(M_H^2-s)(M_H^2-t)}\right)+ \ldots.
\ee
In the second process we expand in powers of $M^2/u$ and $M^2/t$. One then gets
\be
-g^2\left(\frac{M_H^2}{4M^2}\left[\frac{t}{t-M_H^2}+\frac{u}{u-M_H^2}\right]
+\frac{u^2+t^2+ut}{2ut}
+\frac{M_H^2}{t+u}\frac{(t-u)^2}{(M_H^2-u)(M_H^2-t)}\right)+ \ldots.
\ee
The two processes are related by crossing and one would naively think that the two amplitudes can be
related by simply exchanging $s$ and $u$. While this is correct for the first two terms in both equations,
it fails for the third.
If the reader is worried about the approximations made, more lengthy complete results are given in \cite{esma} and
they show the same features.

The reason is that while crossing certainly holds when exchanging the external four vectors, the reference frame
in which the two above amplitudes are expressed are different. In both cases they correspond to center-of-mass
amplitudes, but after the exchange of momenta the two systems are boosted one with respect
to the other. Writing the amplitudes in terms of $s,t,u$ gives
the false impresion that these expressions hold in any reference system but this is not correct
because the polarization vectors are no true four-vectors.

On the contrary, the amplitudes computed via the ET, containing
only Golstone bosons,  are manifestly crossing symmetric when expressed
in terms of Mandelstam variables because they are obtained, according to the ET,
by replacing $\epsilon_L^\mu \to k^\mu$, which is obviously a covariant 4-vector.

%%%%%%%%%%%%%%%%%%%%%%%%%%%%%%%%
\section{The origin of the logarithmic poles}
\label{sec:appendix_logarithmic}
%%%%%%%%%%%%%%%%%%%%%%%%%%%%%%%%%

Here we discuss the origins of the $3$ singularities at $s_0=M^2_H$, $s_1=4 M^2 - M_H^2$ and  $s_2=3 M^2$)
entering the $t_{IJ}(s)$ amplitudes. These singularities  can be tracked back from the terms
$1/(s-M^2_H)$, $1/(t-M^2)$ and $1/(u-M^2)$ in the $W^+_LW^-_L\to Z_L Z_L$ amplitude in Eq.~\ref{eq:CMS-A}.
The origin of the pole at $s_0$  is fairly obvious and needs no justification.

As for the other two singularities, the term $1/(t-M^2)=1/((-1 + \cos\theta) (-4 M^2 + s)/2 -M^2)$ has
a pole at $s_3$ for $\cos\theta=-1$ which under integration in $\cos\theta$ to derive the partial
wave amplitude $t_{IJ}(s)$ becomes a logarithmic pole as well as for $1/(u-M^2)=1/((1 + \cos\theta) (-4 M^2 + s)/2 -M^2)$
at $\cos\theta=-1$.  This explains the presence of $s_2$ pole for $t_{IJ}(s)$ amplitudes.

The origin of the pole at $s_1$ for  $t_{IJ}(s)$ amplitudes is more complicated to see. First of all, let us notice that
the fixed-isospin amplitudes  $T_I$ in Eq.~\ref{eq:fixed_isospin} are combinations of
the $A^{++00}=A(W^+_LW^-_L\to Z_L Z_L)$ in Eq.~\ref{eq:CMS-A} and its crossed amplitude  $A^{++++}=A(W^+_LW^+_L\to W^+_LW^+_L)$.
At this point, the term $1/(s-M^2_H)$  in $A^{++00}$, Eq.~\ref{eq:CMS-A},
trasforms for the crossed amplitude $A^{++++}$ into  $1/(t-M^2_H)= (1/(-1 + \cos\theta) (-4 M^2 + s)/2 -M_H^2)$.
Then, for $\cos\theta=-1$ we have a pole at $s_1$ and under integration on $\cos\theta$ the amplitude $t_{IJ}(s)$
gets a logarithmic pole at $s_1$.

Note that these singularities are all below threshold. Note too that except
for $s_0$ they are absent in the ET treatment. For the LHC they appear at values of $s$ corresponding to
the replacement $4 M^2 \to \sum_i q_i^2$ as the external $W$ are typically off-shell.

%%%%%%%%%%%%%%%%%%%%%%%%%%%%%%%%
\section{Sum rule}
\label{sec:appendix_sumrule}
%%%%%%%%%%%%%%%%%%%%%%%%%%%%%%%%%
In \cite{sumrule} the following sum rule was derived
\be\label{sum}
\frac{1-a^2}{v^2} = \frac{1}{6\pi}\int_0^\infty \frac{ds}{s}
\left(2\sigma_{I=0}(s)^{tot} + 3 \sigma_{I=1}(s)^{tot} -5
\sigma_{I=2}(s)^{tot}\right),
\ee
where $\sigma_I^{tot}$ is the total cross section in the isospin channel $I$. This interesting result
was derived making full use of the Equivalence Theorem and setting $M=0$.
As we have seen, at low $s$ there are some relevant deviations with respect to the ET predictions when using
the proper longitudinal vector boson amplitudes and they affect the analytic properties of the amplitude.
We would like to see how this sum rule could be affected by these deviations.

The technique used in \cite{sumrule} to derive the previous result was to define
$F(s,t,u) \equiv A_{W^+W^-\to ZZ}^{\rm{tree}}(s,t,u)/s^2$, consider the
case $t=0$, corrresponding to the forward amplitude, and compute the integral
\be
\oint ds F(s,t,u)
\ee
using two different circuits: one around the origin and another one along the cuts in the real axis and
closing at infinity (this last contribution actually drops if the amplitudes are assumed to grow slower than
$s$).

Applying the strict ET, each order in perturbation theory contributes to a given order in an expansion
in powers of $s,t,u$. Therefore if the integral is done
in a small circle around the origin only the tree-level amplitude Eq.~(\ref{etamp}) contributes and
taking both contributions into account results in the
result on the left hand side of Eq.~(\ref{sum}). On the other hand, the integral along the left cut
can be related using crossing symmetry to the one on the right cut and eventually leads to the right
hand side of Eq.~(\ref{sum}).

Formulae (\ref{fullamp}) and (\ref{etamp}) show clearly that the analytic structure of the full result and the ET
one are quite different at low values of $s$. In the exact case and for the tree-level amplitude
we have four poles for $F(s,t,u)$. We assume that $s$ and $t$ are independent variables
and to make this visible we replace $t\to \bar t$
\bea
\!\!\!\!\!\!\!\!\!\!\!\!\!\!\!\!\!\!s_0 = 0 &\to& \text{Res}(F(s_0,\bar t,u))=
\frac{4 a^2 M^2 (M^2-M_H^2)}{M_H^4
   v^2}+\frac{2 \bar t \left(8 M^4-7 M^2\bar  t+\bar t^2\right)}{v^2
   \left(M^2-\bar t\right) \left(\bar t-3 M^2\right)^2}\nn\\
s_1 = M_H^2 &\to& \text{Res}(F(s_1,\bar t,u))=
-\frac{a^2 \left(M_H^2-2 M^2\right)^2}{M_H^4 v^2}\nn\\
s_2 = 3 M^2 - \bar t
&\to& \text{Res}(F(s_2,\bar t,u))=
-\frac{-27 M^8+52 M^6 \bar t+M^4 \bar t^2+2 M^2\bar  t^3}{v^2 \left(\bar t-3
   M^2\right)^2 \left(M^2+\bar t\right)^2}\nn\\
   s_3 = 4 M^2 &\to& \text{Res}(F(s_3,\bar t,u))=
   \bar t\frac{10 M^4 - 3 M^2 \bar t - 3 \bar t^2}{(M^2 - \bar t) (M^2 + \bar t)^2 v^2}\\
&& \sum_{i=0,3} \text{Res}(F(s_i,\bar t,u)) =
 \frac{(3 - a^2) M^2 - (1 - a^2)\bar  t}{(M^2 - \bar t) v^2} \label{B2}
 \eea
Note however that for $s=s_3=4 M^2$, the $t$ variable is always zero, being
$t= -(1-\cos\theta)(s - 4 M^2)/2$, and $u= -(1+\cos\theta)(s - 4 M^2)/2$.
This shows that $s$ and $t$ are dependent for some exceptional kinematical points,
for example when the inicial states are at rest ($s=s_3=4 M^2$).
Therefore when $s\to s_3$,  $t\to 0$. If we set $\bar t=0$ at the outset the sum of residues leads to
\be\label{residuesfinal}
\sum_{i=0,3} \text{Res}(F(s_i,\bar t=0,u)) =
 \frac{(3 - a^2)}{v^2}\,.
\ee
which differs from the result quoted in \cite{sumrule}. The reason is clear when looking at 
Eq.~(\ref{B2}): if we take the limit $M\to 0$ at the outset as is done in the strict ET approximation, we get one
result, while if $ \bar t$ is set to zero with $M\neq 0$, we get a different one.

In addition, in a complete calculation (as opposed to the simpler ET treatment) it is not true that a
given order in the chiral expansion corresponds
to a definite power of $s$. Therefore,  when $M$ is not neglected the order $s$ contribution will have corrections
from all orders in perturbation theory. The contribution to the left hand side of the integral, obtained after
circumnavigating all the poles will then be of the form
\be
 \frac{3 - a^2 + \mathcal{O}(g^2)}{v^2}\,.
\ee
Actually, the right cut changes too when $M$ is taken to be non-zero; it starts at $s=4M^2$ (which is not a pole
as we have just discussed because it has a vanishing residue). The left cut is not changed as for $t=0$ the
$u$ channel has a cut for $s<0$ corresponding to $u> 4M^2$.

Finally, as we have seen, crossing symmetry is not manifest
(see appendix~\ref{sec:appendix_crossing}) for the full amplitudes
and it is not possible to relate exactly the contribution along the left cut to
the analogous integral along the right one.

%%%%%%%%%%%%%%%%%%%%%%%%%%%%%%%%%%%%%%%%%%%%%%%%%%%%%%%%%%%%%%%%%%%%%%%%%%%%%%%%%%%%%%%%%%%%%%%%%%%%%%%%%%%


\newpage

\begin{thebibliography}{99}

\bibitem{atlas} G. Aad et al. [The ATLAS collaboration], \Journal{\PLB}{716}{1}{2012}.

\bibitem{cms} S. Chatrchyan et al. [The CMS collaboration], \Journal{\PLB}{716}{30}{2012}.

\bibitem{scaleofcompositeness} M. Redi and A. Tesi, \Journal{\JHEP}{1210}{166}{2012}; G. Panico, M. Redi, A. Tesi and
A. Wulzer, \Journal{\JHEP}{1303}{051}{2013}.

\bibitem{degrassietal} D. Buttazzo, G. Degrassi, P. P.  Giardino, G. F. Giudice, F. Sala, A. Salvio
and  A. Strumia, \Journal{\JHEP}{1312}{089}{2013}.

\bibitem{hierarchyprob} See e.g. the discussion in N. Arkani Hamed, S. Dimopoulos and G. Dvali, \Journal{\PLB}{429}{263}{1998}.

\bibitem{Hi64} F. Englert and R. Brout,  \Journal{\PRL}{13}{321}{1964}; P. W. Higgs, \Journal{\PRL}{13}{508}{1964}; 
 P. W. Higgs, \Journal{\PREV}{145}{1156}{1966}.

\bibitem{superfluidity} P.W. Anderson, \Journal{\PREV}{130}{439}{1963}.

\bibitem{Weinberg:1975gm} S. Weinberg, \Journal{\PRD}{13}{974}{1976}.

\bibitem{Susskind:1978ms} L. Susskind, \Journal{\PRD}{20}{2619}{1979}.

\bibitem{TCexcl} See e.g. P. G. Langacker, {\em The Standard Model and Beyond}, CRC Press (2017). 
 
\bibitem{Pich:2018ltt} A.~Pich,
  %``Effective Field Theory with Nambu-Goldstone Modes,''
  arXiv:1804.05664 [hep-ph].
 
\bibitem{Manohar:2018aog} 
  A.~V.~Manohar, 
  %``Introduction to Effective Field Theories,''
  arXiv:1804.05863 [hep-ph]. 
 
\bibitem{EGPR} G. Ecker, J. Gasser, A. Pich and E. de Rafael, \Journal{\NPB}{321}{311}{1989}.

\bibitem{VMD} J.J. Sakurai, \Journal{\ANNP}{11}{1}{1960} 1; H. B. O'Connell, B. C. Pearce, A. W. Thomas and A. G. Williams,
\Journal{\PPNP}{39}{201}{1997}.

\bibitem{hidden} M. Bando, T. Kugo and K. Yamawaki, \Journal{\PREP}{164}{217}{1988}.

\bibitem{EspMatPL} D. Espriu and J. Matias, \Journal{\NPB}{418}{494}{1994}.

\bibitem{Kamefuchi:1961sb} 
  S.~Kamefuchi, L.~O'Raifeartaigh and A.~Salam,
  %``Change of variables and equivalence theorems in quantum field theories,''
  \Journal{\NPHYS}{28}{529}{1961}. 

\bibitem{LeutLect} H. Leutwyler, \Journal{\PRD}{49}{3033}{1994}.

\bibitem{custodial_evidence} R. A. Diaz and R. Martinez, \Journal{\RMF}{47}{489}{2001}.


\bibitem{Pokorski} S.~Pokorski, {\em Gauge Field Theory}, Cambridge University Press, (2000).

\bibitem{CoWeZu69} S. Coleman, J. Wess and B. Zumino, \Journal{\PREV}{177}{2239}{1969};
C. Callan, S. Coleman, J. Wess and B. Zumino, \Journal{\PREV}{177}{2247}{1969};
L. Alvarez-Gaum\'e and P. Ginsparg, \Journal{\NPB}{262}{439}{1985}.

\bibitem{We79}  S. Weinberg, \Journal{\PHYS}{96A}{327}{1979}.

\bibitem{Cha70} J. Charap, \Journal{\PRD}{2}{1115}{1970};
 I. S. Gerstein, R. Jackiw, B.W. Lee and S. Weinberg, \Journal{\PRD}{3}{2486}{1971};
J. Honerkamp, \Journal{\NPB}{36}{130}{1972}.

\bibitem{Ta75} L. Tararu, \Journal{\PRD}{12}{3351}{1975};
T. Appelquist and C. Bernard, 
\Journal{\PRD}{23}{425}{1981}; J. Zinn-Justin, {\em  Quantum Field Theory and Critical Phenomena},
Oxford University Press, New York, (1989);
D. Espriu and J. Matias, \Journal{\NPB}{418}{494}{1994}.

\bibitem{Alonso:2016oah} R. Alonso, E. E. Jenkins and A. V. Manohar,
  %``Geometry of the Scalar Sector,''
\Journal{\JHEP}{1608}{101}{2016}
 
\bibitem{SO(5)} 
K. Agashe, R. Contino and A. Pomarol, \Journal{\NPB}{719}{165}{2005}; 
R. Contino, L. Da Rold and A. Pomarol, \Journal{\PRD}{75}{055014}{2007};
D. Barducci {\it et al.}  \Journal{\JHEP}{1309}{047}{2013}.

\bibitem{Appelquist}
T. Appelquist and C. Bernard, \Journal{\PRD}{22}{200}{1980};
A. Longhitano, \Journal{\PRD}{22}{1166}{1980}, \Journal{\NPB}{188}{118}{1981};
A. Dobado, D. Espriu and  M.J. Herrero, \Journal{\PLB}{255}{405}{1991};
B. Holdom and J. Terning, \Journal{\PLB}{247}{88}{1990};
M. Golden and L. Randall, \Journal{\NPB}{361}{3}{1991};
R. Alonso, M. B. Gavela, L. Merlo, S. Rigolin and J. Yepes,
  %``The Effective Chiral Lagrangian for a Light Dynamical "Higgs Particle",''
\Journal{PLB}{722}{330}{2013} [Erratum-ibid. 726 (2013)926].

\bibitem{Grinstein} E. Halyo, \Journal{\MPLA}{8}{275}{1993}; 
W. D. Goldberger, B. Grinstein and W. Skiba, \Journal{\PRL}{100}{111802}{2008}.

\bibitem{Alonso:2016btr} R. Alonso, E. E. Jenkins and A. V. Manohar,
  %``Sigma Models with Negative Curvature,''
\Journal{\PLB}{756}{358}{2016}.

\bibitem{ECHL} A. Dobado, D. Espriu and M.J. Herrero, \Journal{PLB}{255}{405}{1991};
D. Espriu and M.J. Herrero, \Journal{\NPB}{373}{117}{1992};
M.J. Herrero and E. Ruiz-Morales, \Journal{\NPB}{418}{431}{1994}, \Journal{\NPB}{437}{319}{1995};
D. Espriu and J. Matias,  \Journal{\PLB}{341}{332}{1995};
A. Dobado, M.J. Herrero, J.R. Pel\'aez and E. Ruiz-Morales, \Journal{\PRD}{62}{05501}{2000};
R. Foadi, M. Jarvinen and F. Sannino,
  %``Unitarity in Technicolor,''
\Journal{\PRD}{79}{035010}{2009}.

\bibitem{Espriu:2012ih} D. Espriu and B. Yencho, \Journal{\PRD}{87}{055017}{2013}.

\bibitem{debate}
J.~Elias-Miro, J.~R.~Espinosa, E.~Mass\'o and A.~Pomarol,
  %``Higgs windows to new physics through d=6 operators: constraints and one-loop anomalous dimensions,''
\Journal{\JHEP}{1311}{066}{2013},
  %``Renormalization of dimension-six operators relevant for the Higgs decays $h\rightarrow \gamma\gamma,\gamma Z$,''
\Journal{\JHEP}{1308}{033}{2013};
C.~Grojean, E.~E.~Jenkins, A.~V.~Manohar and M.~Trott,
  %``Renormalization Group Scaling of Higgs Operators and \Gamma(h -> \gamma \gamma),''
\Journal{\JHEP}{1304}{016}{2013};
E.~E.~Jenkins, A.~V.~Manohar and M.~Trott,
  %``Renormalization Group Evolution of the Standard Model Dimension Six Operators I: Formalism and lambda Dependence,''
\Journal{\JHEP}{1310}{087}{2013},
  %``Renormalization Group Evolution of the Standard Model Dimension Six Operators I: Formalism and lambda Dependence,''
\Journal{\JHEP}{1310}{087}{2013};  
G.~Buchalla, O.~Cat\`a and C.~Krause,
  %``Complete Electroweak Chiral Lagrangian with a Light Higgs at NLO,''
\Journal{\NPB}{880}{552}{2014}.

\bibitem{gellmannlevy} M. Gell-Mann and M. L\'evy, \Journal{\NCIM}{16}{705}{1960}.

\bibitem{weinbergNL} S. Weinberg, \Journal{\PREV}{166}{1568}{1968}.

\bibitem{EHRM} D. Espriu and M.J. Herrero, \Journal{\NPB}{373}{117}{1992};
M.J. Herrero and E. Ruiz-Morales, \Journal{\NPB}{418}{431}{1994}, \Journal{\NPB}{437}{319}{1995};

\bibitem{DEH} A. Dobado, D. Espriu and  M.J. Herrero, \Journal{\PLB}{255}{405}{1991}.

\bibitem{Peskin:1990zt} 
  M.~E.~Peskin and T.~Takeuchi,
  %``A New constraint on a strongly interacting Higgs sector,''
\Journal{\PRL}{65}{964}{1990}, \Journal{\PRD}{46}{381}{1992}. 

\bibitem{DeRuj} A. De Rujula, M.B. Gavela, P. Hernandez and E. Mass\'o, \Journal{NPB}{384}{3}{1992}.

\bibitem{yellowrep} D. de Florian et al. (ed.) {\em Handbook of LHC Higgs Cross Sections: 4. Deciphering the Nature 
of the Higgs Sector},
e-Print: arXiv:1610.07922.

\bibitem{Cacciapaglia:2014} G. Cacciapaglia and F. Saninno, \Journal{\JHEP}{1404}{111}{2014}.

\bibitem{Maldacenaetal} J. Maldacena, {\em Int. J. Theor.
Phys.} 38 (1999) 1113;
S. Gubser, I. Klebanov and A. Polyakov, \Journal{\PLB}{428}{105}{1998};
E. Witten, {\em Adv. Theor. Math. Phys.} 2 (1998) 253.

\bibitem{HW_2005} J. Erlich, E. Katz, D. T. Son and M. A. Stephanov, \Journal{\PRL}{95}{261602}{2005}.

\bibitem{SW_2006} A Karch,  E. Katz, D. T. Son and M. A. Stephanov, \Journal{\PRD}{74}{015005}{2006}.

\bibitem{ACP_2005} K. Agashe, R. Contino and A. Pomarol, \Journal{\NPB}{719}{165}{2005}; K. Agashe and
R. Contino, \Journal{\NPB}{742}{59}{2006}. 

\bibitem{Bellazzini2014} B. Bellazzini, C. Cs\'aki and J. Serra, \Journal{\EPJC}{74}{2766}{2014}.

\bibitem{Panico2016} G. Panico and A. Wulzer, {\em The Composite Nambu-Goldstone Higgs}, Springer (2016).

\bibitem{katanaeva} D. Espriu and A. Katanaeva, e-Print: arXiv:1706.026, e-Print: arXiv:1812.01523. 

\bibitem{gasiorowicz} S. Gasiorowicz and D. A. Geffen, \Journal{\RMP}{41}{531}{1969}; A. H. Fariborz, R. Jora,
and J. Schechter, \Journal{\PRD}{72}{034001}{2005}; D. Parganlija, F. Giacosa, and D. H.
Rischke, \Journal{\PRD}{82}{054024}{2010}.

\bibitem{Falkowski2008} A. Falkowski and M. Perez-Victoria, \Journal{\JHEP}{0812}{107}{2008}.

\bibitem{Strumia} M. Farina, D. Pappadopulo and A. Strumia, \Journal{\JHEP}{1308}{022}{2013}. 

\bibitem{Hirn2005} J. Hirn and V. Sanz, \Journal{\JHEP}{0512}{030}{2005}; J. Hirn,
N. Rius, and V. Sanz, \Journal{\PRD}{73}{085005}{2006}; L. Cappiello, G. D’Ambrosio, and D. Greynat, 
\Journal{\EPJC}{75}{465}{2015}.

\bibitem{Gfitter_2014} THE GFITTER GROUP collaboration (M. Baak et al.), \Journal{\EPJC}{74}{3046}{2014}.

\bibitem{Belyaevetal} A. Belyaev, A. Coupe, N. Evans, D. Locke and M. Scott, \Journal{\PRD}{99}{095006}{2019}. 

\bibitem{AHLO} T. Alho, N. Evans and K. Tuominen, \Journal{\PRD}{88}{105016}{2013}; N. Evans and K. Tuominen,
\Journal{\PRD}{87}{086003}{2013}.

\bibitem{sergeya} S. Afonin, \Journal{\PRC}{83}{048202}{2011}. 

\bibitem{Holdometal} B. Holdom, \Journal{\PLB}{150}{301}{1985}; M. Yamawaki, M. Bando and K. Matumoto, 
\Journal{\PRL}{56}{1335}{1986}; T. W. Appelquist, D. Karabali and L.~C.~R. Wijewardhana,
\Journal{\PRL}{57}{957}{1986}; T. Appelquist and F. Sannino,
  %``The Physical spectrum of conformal SU(N) gauge theories,''
\Journal{\PRD}{59}{067702}{1999}; M. A. Luty and T. Okui, \Journal{\JHEP}{0609}{070}{2006});
C. T. Hill and E. H. Simmons, \Journal{\PREP}{381}{235}{2003});
A. Orgogozo and S. Rychkov,
  %``Exploring T and S parameters in Vector Meson Dominance Models of Strong Electroweak Symmetry Breaking,''
\Journal{\JHEP}{1203}{046}{2012}.

\bibitem{Sanninoetal} F. Sannino and K. Tuominen, 
\Journal{\PRD}{71}{051901}{2005}; D.~D. Dietrich  and F. Sannino,
\Journal{\PRD}{75}{085018}{2007}.

\bibitem{picagranada} M. Hansen, T. Janowski, C. Pica and  A. Toniato, {\em EPJ Web Conf.} 175 (2018) 08010,  
e-Print: arXiv:1710.10831.

\bibitem{BeRoSt75} C. Becchi, A. Rouet and R. Stora, \Journal{\CMP}{42}{127}{1975}.
127

\bibitem{Dobado:1997jx} A.~Dobado, A.~Gomez-Nicola, A.~L.~Maroto and J.~R.~Pel\'aez, {\em Effective lagrangians 
for the standard model}, Springer-Verlag (Texts and Monographs in Physics), (1997)

\bibitem{ET} J.M. Cornwall, D.N. Levin and G. Tiktopoulos, \Journal{\PRD}{10}{1145}{1974};
C.E. Vayonakis, {\em Lett. Nuovo Cim. } 17 (1976) 383;
B.W. Lee, C. Quigg and H. Thacker, \Journal{\PRD}{16}{1519}{1977};
M.S. Chanowitz and M.K. Gaillard, \Journal{\NPHYS}{261}{379}{1985};
 G.J. Gounaris, R. Kogerler and H. Neufeld, \Journal{\PRD}{34}{3257}{1986};
M. S. Chanowitz, M. Golden and H. Georgi, \Journal{\PRD}{36}{1490}{1987};
A. Dobado and J. R. Pel\'aez, \Journal{\NPB}{425}{110}{1994}, \Journal{\PLB}{329}{469}{1994} 
[Addendum, ibid, 335 (1994) 554]; H.J.He,Y.P.Kuang and X.Li, \Journal{\PLB}{329}{278}{1994};
 C. Grosse-Knetter and I.Kuss, \Journal{\ZPC}{66}{95}{1995}.

\bibitem{esma} D. Espriu and J. Matias, \Journal{\PRD}{52}{6530}{1995}.

\bibitem{Delgado:2014jda} R. L. Delgado, A. Dobado, M. J. Herrero and J. J. Sanz-Cillero,
  %``One-loop $\gamma\gamma \to$ W$_{L}^{+}$ W$_{L}^{-}$ and $\gamma\gamma \to$ Z$_{L}$ Z$_{L}$ from the Electroweak Chiral Lagrangian with a light Higgs-like scalar,''
\Journal{\JHEP}{1407}{149}{2014}.
  
\bibitem{Delgado:2013hxa}  R. L. Delgado, A. Dobado and F. J. Llanes-Estrada,
  %``One-loop $W_LW_L$ and $Z_LZ_L$ scattering from the electroweak Chiral Lagrangian with a light Higgs-like scalar,''
\Journal{\JHEP}{1402}{121}{2014}

\bibitem{Espriu:2013fia} 
  D. Espriu, F. Mescia and B. Yencho,
  %``Radiative corrections to WL WL scattering in composite Higgs models,''
\Journal{\PRD}{88}{055002}{2013}; D. Espriu and B. Yencho,
  %``Longitudinal WW scattering in light of the 'Higgs' discovery,''
\Journal{\PRD}{87}{055017}{2013};
  D. Espriu and F. Mescia,
  %``Unitarity and causality constraints in composite Higgs models,''
\Journal{\PRD}{90}{015035}{2014}.
 
\bibitem{Dobado:2017lwg} A. Dobado, F. J. Llanes-Estrada and J.J. Sanz-Cillero,
  %``Resonant production of Wh and Zh at the LHC,''
\Journal{\JHEP}{1803}{159}{2018}.
  
\bibitem{Delgado:2015kxa} 
  R. L. Delgado, A. Dobado and F. J. Llanes-Estrada,
  %``Unitarity, analyticity, dispersion relations, and resonances in strongly interacting $W_LW_L$, $Z_LZ_L$, and hh scattering,''
\Journal{\PRD}{91}{075017}{2015}
 
\bibitem{Dispers2} R.J. Eden, P.V. Landshoff, D.1. Olive and J.C. Polkingorne, {\em The Analytic S-matrix}, 
Cambridge University Press (1966); A.O. Barut, {\em The Theory of the Scattering Matrix}, Macmillan. New York (1967);
K. Nishijima, {\em Fields and Particles: Field Theory and Dispersion Relations}, W.A. Benjamin Inc. New York (1969).

\bibitem{Truong:1988zp} T. N. Truong,
  %``Chiral Perturbation Theory and Final State Theorem,''
\Journal{\PRL}{61}{2526}{1988}; A. Dobado, M. J. Herrero and T. N. Truong,
  %``Unitarized Chiral Perturbation Theory for Elastic Pion-Pion Scattering,''
\Journal{\PLB}{235}{134}{1990}; A. Dobado, M. J. Herrero and T. N. Truong,
  %``Study of the Strongly Interacting Higgs Sector,''
\Journal{\PLB}{235}{129}{1990}.
  
\bibitem{Dobado:1992ha} A. Dobado and J. R. Pelaez,
  %``A Global fit of pi pi and pi K elastic scattering in ChPT with dispersion relations,''
\Journal{\PRD}{47}{4883}{1993}; A. Dobado and J. R. Pelaez,
  %``The Inverse amplitude method in chiral perturbation theory,''
\Journal{\PRD}{56}{3057}{1997}
   
\bibitem{DHD} A. Dobado and M.J. Herrero, \Journal{\PLB}{228}{495}{1989}, {\em ibid} 233 (1989) 505;
J. Donoghue and C. Ramirez, \Journal{\PLB}{234}{361}{1990}.
  
\bibitem{JRA}
J. R. Pelaez, J. A. Oller and E. Oset, \Journal{\PRL}{80}{3452}{1998};
A. G\'omez  Nicola and J.R. Pel\'aez, \Journal{\PRD}{65}{54009}{2002}.
  
\bibitem{N/D} G. F. Chew and S. Mandelstam, \Journal{\PREV}{119}{467}{1960};
B. W. Lee, C. Quigg and H. B. Thacker, \Journal{\PRD}{16}{1519}{1977};
K. Hikasa and K. Igi,  \Journal{\PLB}{261}{285}{1991};
D. A. Dicus and V. L. Teplitz, \Journal{\PRD}{49}{5735}{1994}.

\bibitem{Bjorken} J. D. Bjorken,
  %``Construction of Coupled Scattering and Production Amplitudes Satisfying Analyticity and Unitarity,''
\Journal{\PRL}{4}{473}{1960}.
  
\bibitem{Kmatrix} W. Heitler, Math. {\em Proc. Camb. Phil. Soc.} 37 (1941) 291;
J. S. Schwinger, \Journal{\PREV}{74}{1439}{1948}; S. N. Gupta, {\em  Quantum Electrodynamics}, Gordon and Breach (1981);
 A. Dicus and W. W. Repko, \Journal{\PRD}{42}{3660}{1990}.

\bibitem{Kilian:2014zja} W. Kilian, T. Ohl, J. Reuter and M. Sekulla,
  %``High-Energy Vector Boson Scattering after the Higgs Discovery,''
\Journal{\PRD}{91}{096007}{2015}


\bibitem{Barducci:2015oza} 
  D.~Barducci, H.~Cai, S.~De Curtis, F.~J.~Llanes-Estrada and S.~Moretti,
  %``Unitarity in composite Higgs boson approaches with vector resonances,''
  Phys.\ Rev.\ D {\bf 91}, no. 9, 095013 (2015)
  
\bibitem{Gribov}  Y.V. Novozhilov,
  {\em Introduction to Elementary Particle Theory}, Pergamon Press (1975);
V.~Gribov (Prepared by Y.~ L.~Dokshitzer and  J.~Nyiri), {\em Strong Interactions of Hadrons at high Energies},
  Cambridge University Press (2009).

\bibitem{Adams:2006sv}  A. Adams, N. Arkani-Hamed, S. Dubovsky, A. Nicolis and R. Rattazzi,
  %``Causality, analyticity and an IR obstruction to UV completion,''
\Journal{\JHEP}{0610}{014}{2006}.

\bibitem{gback} T. Corbett, O.J.P. \'Eboli and  M.C. Gonzalez-Garcia \Journal{\PRD}{93}{01500}{2016}.

\bibitem{Pich:2012jv} A. Pich, I. Rosell and J. J. Sanz-Cillero,
  %``One-Loop Calculation of the Oblique S Parameter in Higgsless Electroweak Models,''
\Journal{\JHEP}{1208}{106}{2012}.
 
\bibitem{Pich:2015kwa} A. Pich, I. Rosell, J. Santos and J. J. Sanz-Cillero,
  %``Low-energy signals of strongly-coupled electroweak symmetry-breaking scenarios,''
\Journal{\PRD}{93}{055041}{2016}.
 
\bibitem{Pich:2016lew}  A. Pich, I. Rosell, J. Santos and J. J. Sanz-Cillero,
  %``Fingerprints of heavy scales in electroweak effective Lagrangians,''
\Journal{\JHEP}{1704}{012}{2017}.
  
\bibitem{Pich:2012dv} A.~Pich, I.~Rosell and J.~J.~Sanz-Cillero,
  %``Viability of strongly-coupled scenarios with a light Higgs-like boson,''
\Journal{\PRL}{110}{181801}{2013},

\bibitem{Dobado:2015hha} A.~Dobado, F.~K.~Guo and F.~J.~Llanes-Estrada,
  %``Production cross section estimates for strongly-interacting Electroweak Symmetry Breaking Sector resonances at particle colliders,''
{\em Commun. Theor. Phys.} 64 (2015) 701.
 
\bibitem{WSR}  S. Weinberg, \Journal{\PRL}{18}{507}{1967}.
  
\bibitem{Pich:2013fea} A. Pich, I. Rosell and J. J. Sanz-Cillero,
\Journal{\JHEP}{1401}{157}{2014}.
  
\bibitem{Delgado:2016rtd} R. L. Delgado, A. Dobado and F. J. Llanes-Estrada,
  %``Coupling WW, ZZ unitarized amplitudes to $\gamma\gamma$ in the TeV region,''
\Journal{\EPJC}{77}{205}{2017}.

\bibitem{Castillo:2016erh} A. Castillo, R. L. Delgado, A. Dobado and F. J. Llanes-Estrada,
  %``Top–antitop production from $W^+_L W^-_L$ and $Z_L Z_L$ scattering under a strongly interacting symmetry-breaking sector,''
\Journal{\EPJC}{77}{436}{2017}

\bibitem{composite} G. F. Giudice, C. Grojean, A. Pomarol and R. Rattazzi,
  %``The Strongly-Interacting Light Higgs,''
\Journal{\JHEP}{0706}{045}{2007}; A. Falkowski, F. Riva and A.Urbano,
\Journal{\JHEP}{1311}{111}{2013};
 R. Contino, M. Ghezzi, C. Grojean, M. Muhlleitner and M. Spira,
  %``Effective Lagrangian for a light Higgs-like scalar,''
\Journal{\JHEP}{1307}{035}{2013};  B. Dumont, S. Fichet and G. von Gersdorff,
  %``A Bayesian view of the Higgs sector with higher dimensional operators,''
\Journal{\JHEP}{1307}{065}{2013}; R. Alonso, M. B. Gavela, L.Merlo, S. Rigolin and J. Yepes,
  %``The Effective Chiral Lagrangian for a Light Dynamical 'Higgs',''
\Journal{\PLB}{722}{330}{2013};  M. Montull, F. Riva, E. Salvioni and R. Torre,
  %``Higgs Couplings in Composite Models,''
\Journal{\PRD}{88}{095006}{2013}; A. Pomarol and F. Riva,
  %``Towards the Ultimate SM Fit to Close in on Higgs Physics,''
\Journal{\JHEP}{1401}{151}{2014};  T. Alanne, S. Di Chiara and K. Tuominen,
  %``LHC Data and Aspects of New Physics,''
\Journal{\JHEP}{1401}{041}{2014}; P. P. Giardino, K. Kannike, I. Masina, M. Raidal and A. Strumia, 
\Journal{\JHEP}{1405}{046}{2014}. 
  
\bibitem{DDEGHMS} R.L. Delgado, A. Dobado, D. Espriu, C. Garcia-Garcia, M.J. Herrero, X. Marcano and J.J. Sanz-Cillero, 
\Journal{\JHEP}{1711}{098}{2017}.


\bibitem{Concha}
  I. Brivio, T. Corbett, O. J. P. \'Eboli, M. B. Gavela, J. Gonzalez-Fraile, M. C. Gonzalez-Garcia, L. Merlo and S. Rigolin,
  %``Disentangling a dynamical Higgs,''
\Journal{\JHEP}{1403}{024}{2014}.

\bibitem{triple} S. Chatrchyan et al. [The CMS collaboration], CMS-PAS-SMP-13-005;
G. Aad et al. [The ATLAS collaboration], ATLAS-CONF-2013-020.

\bibitem{gammagamma}  S. Chatrchyan et al. [The CMS collaboration], \Journal{\JHEP}{1307}{116}{2013}.

\bibitem{denner} A. Denner, S. Dittmaier and  T. Hahn, \Journal{\PRD}{56}{117}{1997};
A. Denner and T. Hahn, \Journal{\NPB}{525}{27}{1998}.

\bibitem{escia} D. Espriu and P. Ciafaloni, \Journal{\PRD}{56}{1752}{1997}.

\bibitem{grecoliu} D. Greco and D. Liu, \Journal{\JHEP}{1412}{126}{2014}.

\bibitem{ewa} G. L. Kane, W. W. Repko and W. B. Rolnick, \Journal{\PLB}{148}{367}{1984}; S. Dawson,
{em Phys. }B 249 (1985) 42; M. S. Chanowitz and M. K. Gaillard, \Journal{\NPB}{261}{379}{1985}.

\bibitem{Hahn:2000kx} 
  T.~Hahn,
  %``Generating Feynman diagrams and amplitudes with FeynArts 3,''
  Comput.\ Phys.\ Commun.\  140 (2001) 418.

\bibitem{Hahn:1998yk} 
  T.~Hahn and M.~Perez-Victoria,
  %``Automatized one loop calculations in four-dimensions and D-dimensions,''
  Comput.\ Phys.\ Commun.\  118 (1999) 153.
  
  
\bibitem{Delgado:2013loa} 
  R.~L.~Delgado, A.~Dobado and F.~J.~Llanes-Estrada,
  %``Light ‘Higgs’, yet strong interactions,''
  J.\ Phys.\ G {\bf 41}, 025002 (2014)
  
  
  
  
  

\bibitem{Delgado:2018nnh} 
  R.~L.~Delgado, A.~Dobado, D.~Espriu, C.~Garcia-Garcia, M.~J.~Herrero, X.~Marcano and J.~J.~Sanz-Cillero,
  %``Collider phenomenology of vector resonances in WZ scattering processes,''
  arXiv:1811.08720 [hep-ph].

\bibitem{MJHDG} 
 M.~J.~Herrero, R.~L.~Delgado and  C.~Garcia-Garcia, arXiv:1907.11957 [hep-ph].


\bibitem{sumrule} A. Falkowski, S. Rychkov and A. Urbano, JHEP 1204 (2012) 073; A.~Urbano,
  %``Remarks on analyticity and unitarity in the presence of a Strongly Interacting Light Higgs,''
  arXiv:1310.5733 [hep-ph].
  %%CITATION = ARXIV:1310.5733;%%
  %2 citations counted in INSPIRE as of 27 Mar 2014


\end{thebibliography}
\end{document}